\documentclass[12pt]{article}
\usepackage{a4wide}
\usepackage{fullpage}
\usepackage{lmodern}
\usepackage[utf8]{inputenc}
\usepackage[T1]{fontenc}
\usepackage[sc]{mathpazo}
\usepackage{color,palatino,amsmath,amsthm,natbib,amssymb,graphicx,setspace,bm}
\usepackage[multiple]{footmisc}
\usepackage{multirow,bigdelim,bigstrut}
\usepackage{chngcntr}
\usepackage{lscape}
\usepackage{tikz}
\usepackage{url}
\usepackage[toc,page]{appendix}
\usepackage{xr}
\usepackage{xr-hyper}
\usepackage{hyperref}
\usepackage{afterpage}
\usepackage{clipboard}

\definecolor{darkblue}{rgb}{0, 0, 0.4}
\hypersetup{
  colorlinks=true,
  citecolor=darkblue,
  linkcolor=darkblue,
  urlcolor=darkblue,
  filecolor=darkblue
}
\usepackage{enumerate}
\usepackage{caption}
\usepackage{booktabs,caption,rotating}
\usepackage[capposition=top]{floatrow}
\usepackage{bbm}
\usepackage{epstopdf}
\usepackage{longtable, tabularx, array}
\usepackage{colortbl}

\usepackage[british]{babel}
\usepackage[useregional]{datetime2}
\DTMlangsetup[en-GB]{showdayofmonth=false}

\usepackage{subcaption}
\usepackage{eurosym}

\usepackage[
  colorinlistoftodos,
  backgroundcolor=white,
  prependcaption,
  textsize=tiny,
]{todonotes}

\theoremstyle{plain}
\newtheorem{result}{Result}
\newtheorem{assumption}{Assumption}

\newcommand{\bubblenotemc}{One bubble represents one of the 120 detailed occupations in the SIAB SUF. The four groups show an aggregation of these detailed occupations as described in \citet[][Online Appendix, Table~A.1]{BGS}. Bubble size corresponds to the number of employed workers in an occupation averaged across years 1985 until 2010. Regression lines across all occupations (black) and within the four broad groups (colored) are weighted by the number of employed workers. }

\newcommand{\backgroundnotemc}{Shaded lines in the background represent the 120 detailed occupations in the SIAB SUF. The four groups show an aggregation of these detailed occupations as described in \citet[][Online Appendix, Table~A.1]{BGS}. The thickness of a shaded background line corresponds to the number of employed workers in an occupation averaged across years 1985 until 2010. }

\newcommand{\employmenthorizontalnotemany}{The horizontal axes in all panels show the change of the log number of employed workers within an occupation between 1985 and 2010. }

\newcommand{\broadgroupsmc}{The four groups are based on an aggregation of detailed occupations in the SIAB SUF as described in \citet[][Online Appendix, Table~A.1]{BGS}. }

\newcommand{\confidenceintervalls}{The shaded areas around the four lines are 95\% confidence intervals. }
\newcommand{\siabdescriptives}{In the switching graphs of Panels (a) and (b), the occupations are ordered horizontally by their average wage during 1985--2010. The share of incumbents or stayers corresponds to the vertical distance between the two dashed lines. The share entering or leaving an occupation to an occupation with a higher average wage is depicted by the distance above the upper dashed line and vice versa for occupations with a lower average wage. In Panel (c), the wage growth histogram was calculated using 100 equally sized bins between -0.5 and 0.5. }
\newcommand{\mcdescriptives}{\siabdescriptives The results are averages across the 100 Monte Carlo replications. }
\newcommand{\mcresults}{Crosses ``x'' represent true values used to simulate workers' wages. Solid lines are averages across the 100 Monte Carlo replications. Shimmering lines in the background represent individual Monte Carlo replications. }
\newcommand{\crossgammas}{The table shows the estimated $\Gamma_{a, k, k}$, which represents skill accumulation for age $a$.}

\newcommand{\olsresultsmc}{OLS estimates as described by Equation~\eqref{mc:eq:estimation-main}. }
\newcommand{\ivresults}{IV estimates as described in Section~\ref{mc:sub:iv-estimation}. }
\newcommand{\amenityresults}{Estimates from the extended model, which allows to control for changing amenity values of occupations as described by Equation~\eqref{mc:eq:estimation-utility}. Amenities are allowed to vary by age group and identified relative to Prod-Op-Crafts. }
\newcommand{\norandomness}{There is actually no variation in the background to be seen here, since there is no randomness included. }
\newcommand{\stintfixedeffects}{Estimates are identified from year specific occupation fixed effects while including a separate worker-occupation fixed effects for each time the worker revisits an occupation (after a possible break or after return from another occupation). Additionally, we include controls for age and occupation dependent skill accumulation following Equation~\eqref{mc:eq:wageFEspecific-stint}. }
\newcommand{\panelwidth}{0.495\linewidth}

\hypersetup{
	pdfinfo={
		Title={The Performance of Recent Methods for Estimating Skill Prices in Panel Data},
		Author={Michael Böhm, Hans-Martin von Gaudecker}
	}
}

\begin{document}

\title{The Performance of Recent Methods\\[1ex] for Estimating Skill Prices in Panel Data}
\author{%
	Michael J.\ Böhm and Hans-Martin von Gaudecker\footnote{%
		Böhm: University of Bonn and IZA, \href{mailto:michael.j.boehm@uni-bonn.de}{michael.j.boehm@uni-bonn.de}. 
		Gaudecker: University of Bonn and IZA, \href{mailto:hmgaudecker@uni-bonn.de}{hmgaudecker@uni-bonn.de}.
	}
}

\date{\today}

\maketitle

\vspace{-0.8cm}

\begin{abstract}
    \noindent
    This paper explores different methods to estimate prices paid per efficiency unit of labor in panel data. We study the sensitivity of skill price estimates to different assumptions regarding workers' choice problem, identification strategies, the number of occupations considered, skill accumulation processes, and estimation strategies. In order to do so, we conduct careful Monte Carlo experiments designed to generate similar features as in German panel data. We find that once skill accumulation is appropriately modelled, skill price estimates are generally robust to modelling choices when the number of occupations is small, i.e., switches between occupations are rare. When switching is important, subtle issues emerge and the performance of different methods varies more strongly. 
\end{abstract}

\vspace{2ex}

\\[1ex]
~

\clearpage

\setstretch{1.45}

\section{Introduction}
\label{mc:sec:introduction}
There is a growing interest in estimating sector-specific prices per efficiency unit of labor (``skill prices''). For example, researchers studying the impact of technology on the labor market often want to estimate changes of skill prices for certain occupations or tasks which are supposed to be affected by technology \citep[e.g.,][]{Firpo2013,C2016,B2018,CE2020}. In the literature on the  transformation of the employment structure away from manufacturing and toward services, changes of skill prices for different industry sectors constitute key information to distinguish between explanations \citep{Young2014,BaranySiegel2018}. This also links back to classic papers on prices versus composition differences across sectors and over the business cycle \citep{Heckman1985,McLaughlin2001}. Finally, among others, \citet{Yamaguchi2018} has studied the effect of changing prices for different work tasks on the evolution of the gender wage gap. 

It is hard to estimate such skill prices when workers \emph{endogenously} self-select into sectors according to potential wages (i.e., the outcome of interest). Fortunately, the rising interest in skill prices has coincided with an increasing availability of worker panel data with high-quality information on sector (occupations, industries, or tasks) choices and wages as well as a substantial time dimension over multiple cohorts. Recent research has therefore been able to control for at least the time-invariant part of skill differences across workers making different decisions. It has also sought to develop estimation frameworks that flexibly account for varying life-cycle wage profiles over time and to incorporate the endogenous choice in the estimation.

This paper uses extensive, and as we argue realistic, Monte Carlo experiments to study the performance of the estimator based on panel data proposed in \citet{BGS}. We explore instrumental variables as an alternative strategy to saturated OLS considered in that paper. Finally, we discuss extensions of the economic model underlying the data generating process, extending it to include switching costs, non-pecuniary aspects of jobs, forward-looking behavior, and employer learning. Finally, we compare it to alternative estimators due to \citet{C2016} and \citet{CE2020}.

\section{Baseline Model}
\label{mc:sec:theory}

\citet{BGS} propose an empirical model for estimating sector-specific wage rates (``skill prices'') in which workers choose occupations according to their individual-specific potential wages. This departs from the classic \citet{R1951} model in that it allows for changing skills, including systematic ``average accumulation'' components and idiosyncratic deviations, over the career. \citet{BGS} also develop an estimation strategy that exploits longitudinal data on workers' occupation choices and associated wage changes over time. 

This section describes their baseline model and the econometric approach. Sections~\ref{mc:sub:baseline-model} and \ref{mc:sub:ols_estimation_strategy} are directly taken from \citet{BGS} and included to keep this paper self-contained; Section~\ref{mc:sub:iv-estimation} is specific to this paper.

\subsection{Theory}
\label{mc:sub:baseline-model}

There are $K$ distinct occupations. At any time $t$ a worker $i$ would earn a potential wage $W_{i,t,k}$ in occupation $k$. This potential wage is the product of the worker's occupation-specific skill $S_{i,t,k}$ and the occupation-specific price paid for a unit of skilled labor $\Pi_{t,k}$ that prevails in the economy.  
Most of the analysis will be in relative terms and we use lowercase letters to denote the logarithm of a variable. As in \citet{R1951}, workers maximize their incomes by choosing the occupation in which they earn the highest wage:
\begin{equation}
	w_{i,t}
	= \max\{w_{i,t,1},\ldots,w_{i,t,K}\}
	= w_{i,t,k(i,t)}
	= \pi_{t,k(i,t)} + s_{i,t,k(i,t)}.
	\label{mc:eq:observed_wage_is_max_of_potential_wages}
\end{equation}
The occupation subscript's argument $(i,t)$ indicates that $k$ is $i$'s choice at time $t$.

The Roy model is very hard to estimate in its general form \eqref{mc:eq:observed_wage_is_max_of_potential_wages} and requires strong restrictions. \citet{BGS}'s goal is to estimate the evolution of skill prices $\pi_{t, k}$ only. Based on \eqref{mc:eq:observed_wage_is_max_of_potential_wages}, they employ an approximation that allows them to disentangle prices from skills based on observed data:

\begin{assumption}
	\label{mc:ass:approx}
	For a worker who switches occupations, i.e., $k(i, t - 1) \neq k(i, t)$, it must be that $w_{i, t - 1, k(i, t - 1)} - w_{i, t - 1, k(i, t)} \geq 0$ and $w_{i, t, k(i, t - 1)} - w_{i, t, k(i, t)} < 0$. One can approximate the wage at indifference, where $w_{i, \tau, k(i, t)} = w_{i, \tau, k(i, t - 1)}$ with  $\tau \in [t - 1, t)$, to be in the middle of this interval. That is, assume the switch point to be halfway between the previous and the current relative wage: 
	\begin{equation}
		\label{mc:eq:approx-in-assumption-1}
		\frac{1}{2} \Big( (w_{i, t - 1, k(i, t - 1)} - w_{i, t - 1, k(i, t)}) + (w_{i, t, k(i, t - 1)} - w_{i, t, k(i, t)}) \Big)=0
	\end{equation}
\end{assumption}

Assumption~\ref{mc:ass:approx} implies that one can write a worker's realized (log) wage change between two periods as a function of his changes in potential wages across all occupations and his choices $k(i, \tau)$ for all $k$ and $\tau$. Intuitively, knowing changes in potential wages and the share of a period that the worker stays in each occupation, overall wage growth can be computed by simply adding those up. Result~\ref{mc:result:wagegrowth} expresses realized wage changes as a function of potential wage changes in the occupations worker $i$ chooses in $t - 1$ and $t$:

\begin{result}\label{mc:result:wagegrowth}
	Under Assumption~\ref{mc:ass:approx}, worker $i$'s observed wage growth between $t - 1$ and $t$ can be written in terms of the potential wage growth in the occupation(s) he chooses in those periods: 
	\begin{equation}
		\Delta w_{i, t} = \frac{1}{2} \big(\Delta{w}_{i, t, k(i, t)} + \Delta{w}_{i, t, k(i, t - 1)}\big),
	\end{equation}
	where $\Delta {w}_{i, t, k(i, t)} \equiv {w}_{i, t, k(i, t)} - {w}_{i, t - 1, k(i, t)}$ and $\Delta {w}_{i, t, k(i, t - 1)} \equiv {w}_{i, t, k(i, t - 1)} - {w}_{i, t - 1, k(i, t - 1)}$ .
\end{result}

See \citet{BGS} for the derivation. Result~\ref{mc:result:wagegrowth} is  economically attractive as it includes workers' \textit{endogenous} switches between occupations \textit{due to} changing potential wages. Naturally, if a worker stays in an occupation during adjacent periods so $k(i, t - 1) = k(i, t)$,  his realized wage change is equal to the change in his potential wage in the chosen occupation (i.e.,  $\Delta w_{i, t} = \Delta w_{i, t, k(i, t)}$). If the worker decides to switch occupations and $k(i, t - 1) \neq k(i, t)$, half of his wage gain stems from the wage change he would have experienced had he stayed in his previous occupation ($\Delta w_{i, t, k(i, t - 1)}$). The other half is the wage change had he been in the destination occupation all along ($\Delta w_{i, t, k(i, t)}$). Section~\ref{mc:sub:no-shocks} reports on a large set of Monte Carlo experiments on the accuracy of Approximation~\ref{mc:ass:approx}. 

Using Equation~\eqref{mc:eq:observed_wage_is_max_of_potential_wages},  Result~\ref{mc:result:wagegrowth} can be decomposed into changes of prices and skills:
\begin{equation}\label{mc:eq:wagegrowth}
	\Delta w_{i, t} = \frac{1}{2}(\Delta{\pi}_{t, k(i, t)}+\Delta{s}_{i, t, k(i, t)}) + \frac{1}{2}(\Delta{\pi}_{t, k(i, t - 1)}+\Delta{s}_{i, t, k(i, t - 1)}), 
\end{equation}
\citet{BGS} put more structure on Equation~\eqref{mc:eq:wagegrowth} in order to separate changes in skill prices from changes in skills, 
modeling the skill accumulation process as learning-by-doing on the job. Its speed is occupation-specific and depends on age; working in one occupation $k$ impacts subsequent skills in all other occupations. They assume:
\begin{assumption}
	\label{mc:ass:accumulation}
	Occupation-specific skill changes are time invariant in expectation. For all $k \in \{1, \ldots, K\}$ and $i \in \{1, \ldots, N\}$
	\begin{equation}
		\label{mc:eq:accumulation}
		E\big[ \Delta s_{i, t, k}\; |\; a(i, t - 1), k(i, t - 1), k \big] = \Gamma_{a(i, t - 1), k(i, t - 1), k}
	\end{equation}
	with $a(i, t - 1)$ denoting the age group of individual $i$ in period $t - 1$.
\end{assumption}
Stacking the equations for age groups and origin occupations, the matrix $\Gamma$ maps the previous occupation choice $k(i, t - 1)$ interacted with age groups into skill changes in all potential occupations $k$ in the current period.\footnote{The subscript $a(i, t - 1)$ need not be restricted to age but can include any observable variable.} \citet{BGS} argue that this is a common assumption in the literature  and they provide some supportive evidence of constant relative wage growth across age groups that is implied by it (see their Figure~4).

Assumption~\ref{mc:ass:accumulation} identifies changes in skill prices from the difference between wage changes and expected skill changes up to a constant. One further normalization is thus needed, on skill price growth during part of the sample:
\begin{assumption}
	\label{mc:ass:const-base-period}
	Skill prices are constant during a base period, that is,
	\[
		\Delta\pi_{t, k} = 0\; \forall \; k \in \{1, \ldots K\}, t \in \{ 1, \ldots, T_{\text{base}} \}.
	\]
\end{assumption}
Under Assumption~\ref{mc:ass:const-base-period}, $\Gamma$ is identified from the base period due to its time-invariant nature. Accordingly, the estimates of $\Delta\pi_{t,k}$ can be interpreted as actual changes of skill prices for $t > T_{\text{base}}$. In case Assumption~\ref{mc:ass:const-base-period} does not hold, one can still identify \emph{accelerations or decelerations of skill price changes} relative to the base period. Proceeding with Assumption~\ref{mc:ass:const-base-period} as a normalization and focusing on occupation stayers for notational simplicity, identifies $\Gamma_{a(i,t - 1), k, k} + \overline{\Delta\pi_{k, base}}$. Accordingly, the skill price estimates $\Delta{\pi}_{t,k}-\overline{\Delta\pi_{k, base}}$ for $t > T_\text{base}$. If $\overline{\Delta\pi_{k, base}} \neq 0$, the estimated skill price changes in subsequent years are accelerations or decelerations relative to their (unknown) trends during the base period.

\subsection{OLS Estimation Strategy}
\label{mc:sub:ols_estimation_strategy}

Combining the equations for wage growth \eqref{mc:eq:wagegrowth} and skill accumulation \eqref{mc:eq:accumulation} obtains the baseline estimation equation:
\begin{equation}
	\begin{aligned}
		\label{mc:eq:estimation-main}
		\Delta w_{i,t} = & \;  \Delta{\pi}_{t, k(i, t - 1)} \cdot \frac{I_{k(i,t-1)}}{2} + \Delta{\pi}_{t, k(i,  t)} \cdot \frac{I_{k(i,t)}}{2} \\[1.5ex]
		                 & \; + \Gamma_{a(i, t - 1),  k(i, t - 1),  k(i, t - 1)} \cdot \frac{I_{a(i,t-1)} \cdot I_{k(i,t-1)}}{2}                \\[1.5ex]
		                 & \; + \Gamma_{a(i, t - 1),  k(i, t - 1), k(i, t)} \cdot \frac{I_{a(i,t-1)} \cdot I_{k(i,t-1)} \cdot I_{k(i,t)}}{2}    \\[1.5ex]
		                 & \; + \varepsilon_{i, t}
	\end{aligned}
\end{equation}
with $I_{a(i,t)}$ and $I_{k(i,t)}$ denoting indicator variables for $i$'s age group and choices of professions, respectively. In line with Assumption~\ref{mc:ass:const-base-period}, we will set $\Delta\pi_{k, t} = 0 \; \forall \; k \in \{1, \ldots K \}, t \in \{1, \ldots, T_{\text{base}}\}$ and estimate \eqref{mc:eq:estimation-main} for the whole period $t \in \{1, \ldots, T\}$ by OLS.

Regression~\eqref{mc:eq:estimation-main} is perfectly saturated in age groups, previous and current occupations. The regression error $\varepsilon_{i, t}$ reflects individuals' skill shocks or idiosyncratic deviations that are different from the average worker in the respective cell spanned by these dummies. That is, one can write $\varepsilon_{i, t}= \frac{1}{2}(u_{i, t, k(i,t-1)} + u_{i, t, k(i,t)})$ where $u_{i, t, k} \equiv \Delta s_{i, t, k} - \Gamma_{a(i, t - 1), k(i, t - 1), k}$. By Assumption~\ref{mc:ass:accumulation}, we have $E\big[u_{i, t, k}\, |\, a(i, t - 1), k(i, t - 1), k \big] =  0$ and the error term in equation~\eqref{mc:eq:estimation-main} is uncorrelated with the regressors\label{pg:endogeneity-to-prices}.
\begin{result}\label{mc:result:identification}
	Under Assumptions \ref{mc:ass:approx}--\ref{mc:ass:const-base-period}, OLS estimation of equation~\eqref{mc:eq:estimation-main} consistently identifies price changes $\{\Delta{\pi}_{t, k} | t > T_{\text{base}}\}$ and average skill changes $\{{\Gamma}_{a(i, t - 1), k(i, t - 1), k}\}$. Without Assumption~\ref{mc:ass:const-base-period}, OLS identifies these parameters up to the normalization of the base period.
\end{result}

We study this Result~\ref{mc:result:identification} under different assumptions for the true skill change process $\{{\Gamma}_{a(i, t - 1), k(i, t - 1), k}\}$ below. We already note that average  skill changes $\Gamma$  do not in general correspond to structural skill accumulation parameters. Suppose there exists a learning-by-doing production function, such that \emph{ex ante} a worker accumulates skills according to $\Gamma^\ast_{a(i, t - 1),  k(i, t - 1), k} = E\big[\Delta s_{i, t, k}\, |\, a(i, t - 1), k(i, t - 1) \big]$ with idiosyncratic innovations $u^\ast_{i, t, k}$. What regression~\eqref{mc:eq:estimation-main} identifies is \emph{ex post} accumulation in the sense that \[\Gamma_{a(i, t - 1),  k(i, t - 1), k(i, t)} = \Gamma^\ast_{a(i, t - 1),  k(i, t - 1), k(i, t)} + E\big[u_{i, t, k(i, t)}\, |\, a(i, t - 1), k(i, t - 1), k(i, t) \big]\] is conditional on the current choice $k(i, t)$ and $u_{i, t, k} = u^\ast_{i, t, k} - E\big[u^\ast_{i, t, k(i, t)}\, |\, a(i, t - 1), k(i, t - 1), k(i, t) \big]$  is the deviation of the structural shock from its ex post mean. By the nature of the data,  $\Gamma_{a(i, t - 1),  k(i, t - 1), k}$ with $k(i, t - 1) \neq k$ will be identified from switchers. For these workers, often $E\big[u^\ast_{i, t, k}\, |\, a(i, t - 1), k(i, t - 1), k \big] > E\big[u^\ast_{i, t, k}\, |\, a(i, t - 1), k(i, t - 1) \big]$. That is, we will find that the off-diagonal accumulation parameters are upward-biased because of a classic self-selection problem. More subtly, stayers' $\Gamma_{a(i, t - 1),  k(i, t - 1), k(i, t)}$ with $k(i, t - 1) = k(i, t)$ will tend to be overestimated relative to $\Gamma^\ast$, too. This is because, by self-selection, stayers are likely to have experienced more positive shocks compared to leavers. 

Nevertheless, regression~\eqref{mc:eq:estimation-main} succeeds in estimating the skill price changes as long as Assumption~\ref{mc:ass:accumulation} holds. Indeed, Section~\ref{mc:sub:idiosyncratic-shocks} brings this baseline model to its limits in our Monte Carlo simulations. The price changes are well identified using our method for all plausible parameter ranges. It will also become clear, however, that the estimated $\Gamma$ matrix needs to be interpreted in the above \emph{ex post} sense.

\subsection{IV to Estimate Structural Accumulation Parameters}
\label{mc:sub:iv-estimation}

When trying to estimate the structural skill accumulation $\Gamma^\ast_{a(i, t - 1),  k(i, t - 1), k}$, an approach to removing self-selection bias is by instrumental variables (IV). In particular, we instrument the regressors $\left\{ I_{k(i,t)}\right\}_{k=1}^{K}$ with their predetermined components $\left\{  I_{k(i,t-1)}\right\}_{k'=1}^{K}$, which are not a function of $u^\ast_{i, t, k(i, t)}$. As in dynamic panel data models \citep{AndersonHisao1982,ArellanoBond1991}, we could in principle use long occupational histories as instruments. It is well-known, however, that this leads to issues with many weak instruments \citep[e.g.,][]{NeweyWindmeijer2009}. We thus instrument $I_{k(i,t)}$ to get $\Delta{\pi}_{t, k(i,  t)}$ with $I_{k(i,t-1)}$, i.e., individual $i$'s occupation choice in the year before in order to have an instrument for skill price changes between years $t-1$ and $t$. For skill changes, we instrument $I_{k(i,t-1)} \cdot I_{k(i,t)}$ with the occupational history in the two years preceding , $I_{k(i,t-2)} \cdot I_{k(i,t-1)}$ and $I_{k(i,t-3)} \cdot I_{k(i,t-2)}$. This leads to an estimation very similar to the above-mentioned dynamic panel data models.

Our instruments are exogenous to workers' idiosyncratic skill shocks if the latter are period $t$ innovations as assumed above. In the Monte Carlo experiments, we also explore the robustness to serially correlated $u^\ast_{i, t, k(i, t)}$. Our IV strategy amounts to $(T-T_{base})\cdot K + 2\cdot K^2\cdot L$, where $L$ is the number of elements in the characteristics (age groups $a$ here). This will not be feasible for large $K$ but we employed the IV as a major alternative specification for the four broad occupation groups used below.

\section{Extensions of the Model}
\label{mc:sec:model-extensions}

In this section, we consider important extensions of the baseline theory and develop associated changes to the estimation method.  

\subsection{Costs of Switching Occupations}
\label{mc:sub:costs_of_switching_occupations}

Switching occupations, which often requires moving to a different
employer and / or city, may be associated with pecuniary and non-pecuniary
non-wage costs, e.g., financial expenses and psychological stress
of moving house to be close to the new job. These costs are potentially
as important as wage costs \citep[]{Dix-Carneiro2014,Artuc2015,Cortes2017}.

For simplicity, assume that there is a fixed cost of switching occupations
$c>0$ which is incurred if and only if a worker is moving to a different
occupation. Define the modified variable entering the decisions of
the worker as $w_{i,t,k}^{\ast}$ as:
\begin{equation}
  \label{mc:eq:wage_variable_under_switching_cost}
  w_{i,t,k}^{\ast}=\begin{cases}
    w_{i,t,k}   & \text{if}\quad k(i, t) = k(i,t-1) \\
    w_{i,t,k}-c & \text{if}\quad k(i, t) \neq k(i,t-1)
  \end{cases}
\end{equation}
That is, a worker switches occupations only if the resulting wage
is at least $c$ higher than in his origin occupation. Equipped with
the decision-relevant wage $w_{i,t,k}^{\ast}$, we can
make the derivations corresponding to Section~\ref{mc:sub:baseline-model}.
In particular, workers are indifferent in terms of $w_{i,t,k}^{\ast}$
at the switch point but realized wages will now discretely jump at
that point by a function of $c$.

After the approximation, we end up with $\Delta w^{\ast}_{i, t} = \frac{1}{2} \big(\Delta{w}^{\ast}_{i, t, k(i, t)} + \Delta{w}^{\ast}_{i, t, k(i, t - 1)}\big)$.
This can be rearranged in terms of realized wages similar to Section
\ref{mc:sub:nonpecuniary} below:
\begin{equation}
	\label{mc:eq:wagegrowth-augm-1}
	\Delta w_{i,t} = \frac{1}{2}\big(\Delta{w}_{i,t,k(i,t)}+\Delta{w}_{i,t,k(i,t-1)}\big)+\frac{1}{2}\big(\bar{c}_{i, t, k(i,t)}-\bar{c}_{i, t, k(i,t-1)}\big),\nonumber
\end{equation}
where $\bar{c}_{i, t, k}$ is the notation for $i$'s average costs incurred in the two
periods $t$ and $t-1$ because of switching into or out of occupation $k$. The observed
wage difference $\Delta w_{i,t}$ for switchers (i.e., when $\Delta I_{i, t, k}\neq0$) will
be larger than the difference for the decision variable $\Delta w_{i,t}^{\ast}$.

We deliberately employed the notation that is typically used in measurement error models
in econometrics: We would love to observe $w_{i,t}^{\ast}$, all we see in the data is
$w_{i,t}$. While there may be some hope for estimating a basic version with strong
restrictions on $c$ (like it being a constant across all occupations or a fraction of
the previous wage), we view any restrictions that may have empirical bite as too strong.
They will likely vastly differ across regions, urban/rural areas, the ``distance''
between occupations, whether the same employer offers different occupations, and
individual characteristics like family composition. There is no chance to observe a
meaningful subset of such factors in our data. We hence treat $c$ as unobservable and it
will enter the error term of the estimation Equation~\eqref{mc:eq:estimation-main} much in the
same way that $u_{i, t, k}$ does. Of course, it will not have mean zero anymore. In
particular, it will exacerbate the correlation between the unobservables and the
period-$t$-choice, leading to upward-biased skill accumulation coefficients of workers
switching occupations but otherwise not affect the estimates. We extensively examine
switching costs in Section~\ref{mc:sub:switching-costs}, showing that moderate switching cost help rationalize key empirical results.

\subsection{Non-Pecuniary Benefits}
\label{mc:sub:nonpecuniary}

In the baseline theory, individuals are myopic and maximize their current wages. In this section, we show how the model can be extended to accommodate non-pecuniary valuations of different occupations. Suppose that the utility of worker $i$ in occupation $k$ at time $t$ is:
\begin{align}
  U_{i, t, k} & = w_{i, t, k} + V_{i, t, k}\text{ with } \label{mc:eq:potutility}     \\
  V_{i, t, k} & = \Psi_{X, t, k}+v_{i, t, k}, \label{mc:eq:amenity}
\end{align}
where $V_{i, t, k}$ is occupation $k$'s amenity value; it may vary across workers. The discussion here partly follows \citet{B2018}. See, for example, \citet{Lee2006} for a full-fledged structural model that incorporates both, forward-looking behavior and non-pecuniary amenities.

Similar to the skill accumulation parameters $\Gamma$, the matrix $\Psi_{X, t, k}$ maps the worker observables $X$ to utility in occupation $k$ in the current period. That is, the non-pecuniary or continuation value of each occupation $k$ will differ by workers' characteristics like age or education in practice. We further let idiosyncratic occupation valuations $v_{i, t, k}$ be mean zero and independent across individuals. They may be correlated across occupations for a given individual. Finally, notice already that only relative values $V_{i, t, k}$, and thus the parameters $\Psi_{X, t, k}$ compared to a chosen reference occupation, will be identifiable from workers' observed choices and wages.

With definition~\eqref{mc:eq:potutility} and utility maximization at hand, we can make the derivations corresponding to the baseline model. In particular, Equation~(B.6) of \citet[][Online Appendix]{BGS} holds in utility terms. Using an Approximation of the indifference point corresponding to Assumption~\ref{mc:ass:approx}
\begin{equation*}
	\frac{1}{2} \Big( (U_{i, t - 1, k(i, t - 1)} - U_{i, t - 1, k(i, t)}) + (U_{i, t, k(i, t - 1)} - U_{i, t, k(i, t)}) \Big)=0
\end{equation*}

we get:

\begin{equation}
	\begin{aligned}\Delta U_{i,t} & =\Delta w_{i,t} + \Delta V_{i,t} =U_{i,t,k(i,t)}-U_{i,t-1,k(i,t-1)}+\\
		& +\frac{1}{2}\Big((U_{i,t-1,k(i,t-1)}-U_{i,t-1,k(i,t)})+(U_{i,t,k(i,t-1)}-U_{i,t,k(i,t)})\Big)\\
		& =\frac{1}{2}(U_{i,t,k(i,t)}-U_{i,t-1,k(i,t)})+\frac{1}{2}(U_{i,t,k(i,t-1)}-U_{i,t-1,k(i,t-1)})\\
		& =\frac{1}{2}\big(\Delta{w}_{i,t,k(i,t)}+\Delta{V}_{i,t,k(i,t)}+\Delta{w}_{i,t,k(i,t-1)}+\Delta{V}_{i,t,k(i,t-1)}\big)
	\end{aligned}
	\label{eq:Result-1-utility}
\end{equation}

The intuition here is also parallel to Result~\eqref{mc:result:wagegrowth}: if a worker stays in his occupation, his realized utility gain is the change of his potential utility in that occupation. That is, $\Delta U_{i,t}=\Delta{U}_{i,t,k(i,t)}=\Delta{w}_{i,t,k(i,t)}+\Delta{V}_{i,t,k(i,t)}$ if $k(i, t)=k(i, t-1)$, which is not an approximation. If the worker switches ($k(i, t)\neq k(i, t-1)$), he obtains part of the origin occupation's utility gain (or loss) as well as part of the destination occupation's utility gain, set to half-half by the approximation (i.e., $\Delta U_{i,t}=\frac{1}{2}\Delta{U}_{i,t,k(i,t)}+\frac{1}{2}\Delta{U}_{i,t,k(i,t-1)}$). The results from Section~\ref{mc:sub:idiosyncratic-shocks} that the approximation error is negligible apply.

We want to solve equation~\eqref{eq:Result-1-utility} for $\Delta w_{i,t}$, which is observable in the data. Consider
\begin{equation*}
	\begin{aligned}\Delta V_{i,t} & =V_{i,t,k(i,t)}-V_{i,t-1,k(i,t-1)}\\
		& =(V_{i,t,k(i,t)}-V_{i,t-1,k(i,t)})+(V_{i,t-1,k(i,t)}-V_{i,t-1,k(i,t-1)})\\
		& =\Delta{V}_{i,t,k(i,t)}+(V_{i,t-1,k(i,t)}-V_{i,t-1,k(i,t-1)})
	\end{aligned}
\end{equation*}
and equivalently $\Delta V_{i,t} = \Delta{V}_{i,t,k(i,t-1)}+(V_{i,t,k(i,t)}-V_{i,t,k(i,t-1)})$. Defining an occupation's average amenity value over both periods, $t$ and $t-1$, as $\bar{V}_{i, t, k}\equiv\frac{1}{2}(V_{i, t-1, k}+V_{i, t, k})$ we have 
\begin{equation*}
	\begin{aligned}\Delta V_{i,t} & =\frac{1}{2}(\Delta{V}_{i,t,k(i,t)}+\Delta{V}_{i,t-1,k(i,t)})\\
		& +(\bar{V}_{i, t, k(i,t)}-\bar{V}_{i, t, k(i,t-1)})
	\end{aligned}
\end{equation*}

Inserting this into Equation~\eqref{eq:Result-1-utility}, the realized wage growth of individual worker $i$ in the generalized Roy model becomes:
\begin{equation}
  \label{mc:eq:wagegrowth-augm}
  \Delta w_{i,t} = \frac{1}{2}\big(\Delta{w}_{i,t,k(i,t)}+\Delta{w}_{i,t,k(i,t-1)}\big)-\frac{1}{2}\big(\bar{V}_{i, t, k(i,t)}-\bar{V}_{i, t, k(i,t-1)}\big)
\end{equation}
This result firstly has a purely pecuniary part as in Result~\ref{mc:result:wagegrowth} of Section~\ref{mc:sub:baseline-model}: if a worker stays in his occupation, his wage gain is the potential wage change (i.e., price growth and skill accumulation) in that occupation. If the worker switches, he obtains half of the origin's as well as half of the destination's potential wage change. Similar to the baseline model, Equation~\eqref{mc:eq:wagegrowth-augm} accommodates endogenous switches, which in this case may be due to changes in amenity/continuation values in addition to changes in potential wages.

The second summand on the right of Equation~\eqref{mc:eq:wagegrowth-augm} is then the intuitive extension of a purely pecuniary/static model: with optimal choices, a worker's observed wage growth is the change in the potential wage of his chosen occupations minus the utility gain (loss) from the behavioral response of switching occupations. That is, if a utility-optimizing worker chooses to switch occupations (i.e., from $k(i, t-1)$ to $k(i, t)\neq k(i, t-1)$), we observe lower wage growth than the change in relevant potential wages when he gains amenities or net present value of future earnings (i.e., $\bar{V}_{i, t, k(i,t)}>\bar{V}_{i, t, k(i,t-1)}$) via the move. Vice versa, we observe higher wage growth than the potential wage changes when he moves to a less desirable occupation in these respects ($\bar{V}_{i, t, k(i,t)}<\bar{V}_{i, t, k(i,t-1)}$).

Notice in Equation~\eqref{mc:eq:wagegrowth-augm} it is the \emph{average} non-pecuniary value over both periods $\bar{V}_{i, t, k}$ that the worker is moving into which matters for wage changes. For a switcher
\[
  \bar{V}_{i, t, k(i,t)}-\bar{V}_{i, t, k(i,t-1)}=\frac{1}{2}(V_{i, t, k(i,t)}-V_{i, t, k(i,t-1)})+\frac{1}{2}(V_{i, t-1, k(i,t)}-V_{i, t-1, k(i,t-1)}),
\]
conditional on wage gains associated with average choices, moving into the currently high-value occupation (i.e., $V_{i, t, k(i,t)}-V_{i, t, k(i,t-1)}>0$) is offset with lower wage growth. But also moving into a occupation that last period carried high value is associated with lower wage growth ($V_{i, t-1, k(i,t)}-V_{i, t-1, k(i,t-1)}>0$) because it implies that the worker was compensated last period for working in the low-value occupation, which now falls away with the switch. Both of these factors enter equally into the wage Equation~\eqref{mc:eq:wagegrowth-augm}. Hence one cannot distinguish them empirically and only identify the average value over the two periods. As is obvious from Equation~\eqref{mc:eq:wagegrowth-augm}, one can also not distinguish between the amenity and any continuation value considerations but only estimate a joint parameter $\bar{V}_{i, t, k}$. Finally, notice when the worker makes no switch, the value considerations do not come into play at all (i.e., $V_{i, t, k(i,t)}=V_{i, t, k(i,t-1)}$ and $V_{i, t-1, k(i,t)}=V_{i, t-1, k(i,t-1)}$) and the changing wage is just the changing skill price plus skill accumulation.

We now discuss empirical implementation for different versions of Equation~\eqref{mc:eq:amenity}. First, if non-pecuniary values are constant such that $\bar{V}_{i,k,}$ does not carry a time index, they will be simply incorporated in the skill accumulation parameters. That is, since $\Gamma_{a(i, t - 1),  k(i, t - 1),  k}$ in Equation~\eqref{mc:eq:estimation-main} is a fully interacted model of all task choice combinations and worker observables, it absorbs the term $\bar{V}_{i, t, k(i,t)}-\bar{V}_{i, t, k(i,t-1)}$. Because of this, also in the time-varying case, average $\bar{V}_{i, t, k}$ parameters can only be identified relative to their base period values. Our main estimation specification therefore already controls for general time-invariant non-pecuniary values as well as forward-looking considerations of occupation choice (with the interpretation of the parameter estimates $\Gamma_{a,  k,  k}$ adjusted accordingly).

If instead non-pecuniary values are time-varying, we first of all note again that only $\bar{V}_{i, t, k}$ relative to a reference occupation can be identified. The mechanical reason is that the $\Delta I_{k(i,t)}$ sum to zero over all $K$, and thus one of them has to be left out of the estimation due to multicollinearity. The economic intuition is that we can use choices and wages to identify relative utilities but not their levels. Other than that, it is straightforward to introduce a full set of task choice changes into estimation Equation~\eqref{mc:eq:estimation-main} and also interact them with worker characteristics. That is, $\bar{\Psi}_{X, t, k}$ in augmented regression
\begin{equation}
	\begin{aligned}
		\label{mc:eq:estimation-utility}
		\Delta w_{i,t} = & \;  \Delta{\pi}_{t, k(i, t - 1)} \cdot \frac{I_{k(i,t-1)}}{2} + \Delta{\pi}_{t, k(i,  t)} \cdot \frac{I_{k(i,t)}}{2} \\[1ex]
		& \; + \Gamma_{a(i, t - 1),  k(i, t - 1),  k(i, t - 1)} \cdot \frac{I_{a(i,t-1)} \cdot I_{k(i,t-1)}}{2}              \\[1ex]
		& \; + \Gamma_{a(i, t - 1),  k(i, t - 1), k(i, t)} \cdot \frac{I_{a(i,t-1)} \cdot I_{k(i,t-1)} \cdot I_{k(i,t)}}{2}     \\[1ex]
		& \; + \bar{\Psi}_{X(i, t-1), t, k(i,t-1)} \cdot {I_{a(i,t-1)} \cdot \Delta{I}_{k(i,t-1)}}     \\[1ex]	
		& \; + \bar{\Psi}_{X(i, t-1), t, k(i,t)} \cdot {I_{a(i,t-1)} \cdot \Delta{I}_{k(i,t)}} + \varepsilon_{i, t}    
	\end{aligned}
\end{equation}
identifies the average between time $t$ and $t-1$ amenity/continuation values in $k$ relative to a reference occupation and to the base period by age group.

The reason for why \eqref{mc:eq:estimation-utility} is identified, even with time-varying valuations, is the above-discussed fact that moving into current as well as past $V_{i, t, k}$ both matter equally for wage changes. Therefore, this contribution to wage growth is only via the changing sorting $\Delta{I}_{k(i,t)}$ into average non-pecuniary values, whereas the contribution to wage growth from changing skill prices and skill accumulation is only via the average sorting $\frac{I_{k(i,t)}}{2}$. The estimation method is thus robust to both changing potential wages and non-pecuniary values over time.

Finally, conditional on specification \eqref{mc:eq:estimation-utility}, an additional (average) error term $\bar{v}_{i, t, k}$ in \eqref{mc:eq:amenity} does not much affect the estimates, which to some extent parallels the limited confounding role of idiosyncratic skill shocks  \citep[see][for more detailed discussion of the idiosyncratic non-pecuniary error term]{B2018}. We estimate \eqref{mc:eq:estimation-utility} in Figure~\ref{appdx:fig:price-estimates-method-amenities}  and report very similar skill price estimates to the results in \citet{BGS}. For younger workers, the non-pecuniary values of Mgr-Prof-Tech, Sales-Office, and Srvc-Care have modestly declined compared to Prod-Op-Crafts over the sample period.

\subsection{Learning About Skills}
\label{mc:sub:learning}

In Section~\ref{mc:sec:theory} we have assumed that, aside from skill prices, all changes in individuals' wages over time are due to systematic skill accumulation and idiosyncratic skill shocks. In this section, we show that the model's interpretation can be widened to include imperfect information about skills and employer learning over time in addition to skill accumulation.\footnote{\citet{Groes2014} do it the other way around; they set up their theoretical model as employer learning but then clarify that it could alternatively be ``shocks to workers' ability'' \citet[p. 5][]{Groes2014}.}

Suppose that, as in the employer learning literature \citep[e.g.,][]{Altonji2001,Gibbons2005}, information about skills is imperfect. Each period an additional noisy signal of the worker's productivity arrives; employers form expectations about skills based on this as well as on all past observable information. Expectations are rational in the sense that employers' beliefs are correct on average. Information is symmetric, employers are competitive, all market participants are risk neutral, and a spot market for labor exists.

In this setup, workers' potential log wages in each occupation equal their expected productivity conditional on all available information:
\begin{equation}
  \label{mc:eq:log_wages_basic-1}
  w_{i, t, k}=\pi_{t,k}+E_{t}(s_{i, t, k})\;\;\forall\;\,k\in\{1,\ldots,K\},
\end{equation}
where $E_{t}$ indicates that we are conditioning on all the information available in $t$. We assume that workers maximize their log incomes by choosing the occupation in which they earn the highest wage. This yields a modified version of equation~\eqref{mc:eq:wagegrowth} for observed wage growth over time:
\begin{equation}\label{mc:eq:wagegrowth}
	\Delta w_{i, t} = \frac{1}{2}\big(\Delta{\pi}_{t, k(i, t)}+\Delta{E}_{t}({s}_{i, t, k(i, t)})\big) + \frac{1}{2}\big(\Delta{\pi}_{t, k(i, t - 1)}+\Delta{E}_{t}(s_{i, t, k(i, t - 1)})\big), 
\end{equation}
where $\Delta E_{t}(s_{i, t, k})\equiv E_{t}(s_{i, t, k})-E_{t-1}(s_{i, t-1, k})$ and the linearity in logs allows us to swap the first differencing, and expectations operators. The average skill accumulation remains the same as in equation~\eqref{mc:eq:accumulation} as $E_t\big[ \Delta s_{i, t, k}\; |\; a(i, t - 1), k(i, t - 1), k \big]=E\big[ \Delta s_{i, t, k}\; |\; a(i, t - 1), k(i, t - 1), k \big]$. The only differences lies in the interpretation of the innovations  $u_{i, t, k} = \Delta E_{t}(s_{i, t, k}) - \Gamma_{a(i, t - 1), k(i, t - 1), k}$, which now represent an update of employers' expectations about individual $i$'s occupation $k$ skill. These changes are immaterial for our estimation strategy; our results remain valid under a basic model of employer learning about skills as an alternative or in addition to systematic skill accumulation and idiosyncratic skill shocks.

\section{Fixed Effects as an Alternative Approach}
\label{mc:sec:fixed-effects-model}

In this section, we examine the occupation-specific fixed effects approach for estimating skill prices as an alternative to our method. We show that under a flexible model of skill accumulation, this approach requires controlling for workers' history of occupation-specific experience or, more feasibly, extending the fixed effects to being occupation-stint specific. As generally panel data based approaches, a base period or some other restriction on the skill accumulation are needed. With idiosyncratic skill shocks, an endogeneity bias emerges that is due to the fixed effects themselves. The results from the Monte Carlo simulations support our analytical arguments.

Several papers have used fixed effects approaches in order to address worker heterogeneity when estimating skill prices \citep[e.g.,][]{C2016,CE2020}.\footnote{In more broadly related settings, \citet{Combes2008} estimate city wage premia, taking into account sorting across locations. Analyzing variation over the business cycle, \citet{Solon1994} account for skill selection into the labor market market, while \citet{McLaughlin2001} examine skill selection across sectors.} To be specific, consider Cortes' time-varying model for the potential wage of individual $i$ in occupation $k$ at time $t$:
\begin{equation}
	\label{eq:appdx:CortesEq6}
	w_{i, t, k}=\pi_{t, k}+s_{i, t, k}=\pi_{t, k}+\Gamma_{X(i, t-1), k}+\eta_{k(i)}.
\end{equation}
The changing characteristics vector $X(i, t-1)$ can increase skills differentially with age or experience across occupations according to $\Gamma_{X(i, t-1), k}$. In addition, $\eta_{k(i)}$ are occupation-specific time-invariant skill levels, which will be introduced into the regression by individual-occupation specific fixed effects. \citet{C2016} and \citet{CE2020} interchangeably call these occupation- or sector-\emph{spell} fixed effects, which is why we instead use the term `stint' for a worker's self-contained stay (i.e., without switches in between) in a given occupation below. Consistent with \eqref{eq:appdx:CortesEq6}, Cortes' estimation equation (8) in our notation is:\footnote{Similar to us, \citet{C2016} uses ten year age bins in $X(i, t-1)$, allowing for the convexity of the life-cycle profile parallel to our Equation~\eqref{mc:eq:accumulation}.}
\begin{equation}
	\begin{aligned}
		\label{mc:eq:wageFE}
		w_{i,t}  & \; = {\pi}_{t, k(i,  t)} \cdot I_{k(i,t)} \\[1.5ex]
		& \; + \Gamma_{X(i, t - 1),  k(i, t)} \cdot I_{X(i,t-1)} \cdot I_{k(i,t)}           \\[1.5ex]
		& \; + \eta_{k(i)} \cdot I_{k(i,t)} + \varphi_{i,t}
	\end{aligned}
\end{equation}
where we have added the idiosyncratic error term $\varphi_{i,t}$. In the following, we examine under what conditions estimation of Equation~\eqref{mc:eq:wageFE} may then identify the correct skill prices.

\subsection{Systematic Skill Accumulation}
\label{mc:sub:fixed-effects-model-skill-acc}

We start by assuming that, as in \citet{C2016} or \citet{CE2020}, $\varphi_{i,t}$ is simply measurement error and thus not decision-relevant (exogenous mobility assumption of fixed effects approaches discussed in the main text). The skill accumulation becomes
\begin{equation}
	\label{mc:eq:accumulationFEdiff}
	\Delta s_{i, t, k} = \Gamma_{k(i, t - 1), k}
\end{equation}
where, compared to Equation~\eqref{mc:eq:accumulation}, we omit for now the $a(i, t-1)$ (or $X(i, t-1)$)-specificity of the skill accumulation function as another simplifying assumption and thus $\Gamma_{k(i, t - 1), k}$ is a scalar. Writing this out from when the worker joined the labor market at time $t_{i,0}$ gives
\begin{equation}
	\label{mc:eq:accumulationFE}	
	s_{i, t, k} = \eta_{k(i)} + \sum_{\tau = t_{i,0} + 1}^t   \Gamma_{k(i, \tau - 1),k(i, \tau)}
\end{equation}
for $t\geq t_{i,0}$ and $\eta_{k(i)}$ the initial skill endowments of $i$ in $k$ when he joins the labor market. Therefore, if we are willing to assume that skill accumulation occurs similarly in each occupation of origin ($\Gamma_{k(i, t - 1),k}=\Gamma_{k},\forall k(i, t - 1),k$), this simplifies to $s_{i, t, k}=\eta_{k(i)}+(t-t_{i,0})\cdot\Gamma_{k(i, t)}$ and Estimation \eqref{mc:eq:wageFE} identifies the correct skill prices, initial endowments, and skill accumulation parameters. In this case, $X(i, t - 1)=t-t_{i,0}$ represents labor market experience (proxied in \citet{C2016} by age dummies) in the estimation. Notice that this specification assumes that labor market experience is not occupation-specific, just that general experience is valued differently in different occupations. \citet{BGS} formally test and reject such a one-dimensional skill model.

\paragraph{The need for a base period or similar restriction:} We have argued in the main text that any approach using panel data needs a base period or other fundamental restriction of the skill accumulation function. This is the same in Estimation \eqref{mc:eq:wageFE} and easiest to see if we simplify it to its essence. First, noting that, because of the individual-occupation-specific fixed effects $\eta_{k(i)}$, the changing skill prices $\pi_{t, k}$ are fundamentally identified from wage growth of occupation stayers. Therefore, we can condition on the respective occupation $k$:
\begin{equation}
	\label{eq:appdx:CortesEq6-1}
	w_{i, t, k(i, t)}=\pi_{t, k(i, t)}+(t-t_{i,0})\Gamma_{k(i, t)}+\eta_{k(i)}+\varphi_{i,t}
\end{equation}
Second, fixed effects estimates are asymptotically equivalent to first differences (and exactly the same in finite samples if $T=2$). First-differencing gives the effective variation that the occupation-specific fixed effects approach identifies from:\footnote{We could make the same argument absorbing the fixed effects, i.e., $w_{i, t, k(i, t)}-\overline{w}_{k(i, t)}=\pi_{t, k(i, t)}-\overline{\pi}_{k(i, t)}+[(t-t_{i,0})-\overline{(t-t_{i,0})}]\Gamma_{k(i, t)}+\varphi_{i,t}-\overline{\varphi}_{i}$, but Equation~\eqref{eq:appdx:CortesEq6-1-1} seems clearer.}
\begin{equation}
	\label{eq:appdx:CortesEq6-1-1}
	\Delta w_{i, t, k(i, t)}=\Delta\pi_{t, k(i, t)}+\Gamma_{k(i, t)}+\Delta\varphi_{i,t}
\end{equation}
The first thing to note from Equation~\eqref{eq:appdx:CortesEq6-1-1} is that the levels of skill prices do not appear; rather only changes are identified \citep[][normalizes skill prices to zero in 1976]{C2016}. However, the parameters are still fundamentally non-identified without a further restriction since $\Delta\pi_{t, k}$ and $\Gamma_{k}$ are perfectly collinear. Either a base period where skill prices do not change (i.e., $\Delta\pi_{t, k}=0,\;\forall \; k \in \{1, \ldots K\}, t \in \{ 1, \ldots, T_{\text{base}}\}$) is needed, as we do in this paper, or a restriction on $\Gamma_{k}$ implicitly made. E.g., Stata does this automatically when we implement Estimation \eqref{mc:eq:wageFE} without base period in the Monte Carlo simulations, omitting the skill accumulation parameter for one of the age groups (i.e., setting it to zero). We prefer explicitly defining a base period instead.

\paragraph{A generalized skill accumulation specification:} Suppose we have used a base period where skill prices are indeed constant. Does Estimation \eqref{mc:eq:wageFE} then identify the parameters in the analysis period? The evidence in \citet{BGS} and various other papers strongly suggests that experience is occupation-specific.\footnote{For example, the skill accumulation estimates in \citet{BGS} indicate this as well as the fact that large wage differences between entrants and incumbents persist when controlling for general age or experience (Figure~A.1 in their Online Appendix). They also directly rejects the one-dimensional model in favor of multi-dimensional skills (changes).} 

A model that is aligned with the evidence hence allows for this, that is, for example allows for the fact that previous managerial experience imparts more managerial skills than previous experience in production jobs. Equation \eqref{mc:eq:accumulationFE} becomes $s_{i, t, k}=\eta_{k(i)}+\sum_{k(i,t-1)=1}^{K} exp_{k(i,t-1)}\cdot\Gamma_{k(i,t-1), k}$, where $exp_{k(i,t-1)}\equiv\sum_{\tau=t_{i,0}}^{t-1}I_{k(i,\tau)}$ is the worker's occupation $k'$ specific experience. Running regression \eqref{mc:eq:wageFE} gives an error term $\varphi_{i, t}=\sum_{k(i,t-1)=1}^{K} exp_{k(i,t-1)}\cdot\Gamma_{k(i,t-1), k(i,t)}-(t-t_{i,0})\cdot\Gamma_{k(i,t)}$ in that case which varies with the choice $I_{k(i,t)}$ and is thus systematically related to the regressors. This yields biased estimates even without any unobserved idiosyncratic skill shocks that lead to endogenous switching or staying in sectors.

The correct fixed effects regression for skill prices is instead
\begin{equation}
	\begin{aligned}
		\label{mc:eq:wageFEspecific}
		w_{i,t}  & \; = {\pi}_{t, k(i,  t)} \cdot I_{k(i,t)} \\[1.5ex]
		& \; + \sum_{k(i,t-1)=1}^{K} \Gamma_{k(i,t-1),k(i, t)} \cdot exp_{k(i,t-1)}  \cdot I_{k(i,t)} \\[1.5ex]
		& \; + \eta_{k(i)} \cdot I_{k(i,t)} + \varphi_{i,t}
	\end{aligned}
\end{equation}
that is, it controls for all previous occupation-specific experience separately. While this is conceptually possible to do, its practical implementation is difficult. It introduces many parameters to be estimated (even more when we realistically allow for occupation-specific skill accumulation to vary with age; e.g., see general skill accumulation Equation~\eqref{mc:eq:accumulation}) and it requires high-quality panel data in order to compute the full occupation- and age-specific work experience history of each individual. \citet{C2016} accounts for the fact that labor market experience is occupation-specific by introducing controls for occupation-specific tenure into regression Equation~\eqref{mc:eq:wageFE}. In order to deal with the growth in the number of parameters and the length of the employment history that is required for this approach, he assumes that tenure only affects the current job and that workers lose all of its effect once they switch.

We think that occupation-specific tenure is an especially powerful control when at the same time adding separate individual fixed effects for each occupation stint.\footnote{In the SIAB data of \citet{BGS}, 21\% of workers have multiple stints in an occupation during their career.} That is, to use $\eta_{k(i,t_{i,0,k})}$ which differs flexibly by in time of entry $t_{i,0,k}$ for each continuous period in $i$'s career during which he works in occupation $k$. Skill accumulation can then be only occupation-specific and straightforwardly interacted with observable characteristics such as age or education (again omitted for brevity):
\begin{equation}
	\begin{aligned}
		\label{mc:eq:wageFEspecific-stint}
		w_{i,t}  & \; = {\pi}_{t, k(i,  t)} \cdot I_{k(i,t)} \\[1.5ex]
		& \; + \Gamma_{k(i,t),k(i, t)} \cdot (t-t_{i,0,k}) \cdot I_{k(i,t)} \\[1.5ex]
		& \; + \eta_{k(i,t_{i,0,k})} \cdot I_{k(i,t)} + \varphi_{i,t}
	\end{aligned}
\end{equation}
Here $t-t_{i,0,k}$ is the number of years the individual has spent in this occupation stint at time $t$, which is effectively tenure (but interacted with age, which is omitted in Equation~\eqref{mc:eq:wageFEspecific-stint}) conditional on occupation-stint-specific fixed effects. This is in our view the best specification and close to \citet{C2016}'s arguably most flexible estimation specification. \citet{CE2020} also employ this specification throughout their analysis and we use it as an alternative estimation approach in the Monte Carlo simulations as well as in \citet{BGS}'s SIAB data below.

\subsection{Idiosyncratic Skill Shocks}

A substantive difference between Section~\ref{mc:sec:theory}'s proposed method and the fixed effects approach arises in the presence of idiosyncratic skill shocks and endogenous choice, which are indicated by \citet{BGS}'s cross-accumulation parameters of switchers  as well as higher-than-average skill shocks of occupation incumbents and stayers. We use a simplified analytical argument here.

With idiosyncratic skill shocks, the right-hand-side of skill change Equation~\eqref{mc:eq:accumulationFE} becomes:
\begin{equation}
	\label{mc:eq:accumulationFEidiosync}
	s_{i, t, k(i,t)} = \eta_{k(i,t_{i,0,k})} + (t-t_{i,0,k}) \cdot  \Gamma_{k(i,t),k(i,t)}+\sum_{\tau = t_{i,0,k} + 1}^t  u_{i,\tau,k(i,t)}
\end{equation}
The regression error in Equation~\eqref{mc:eq:wageFEspecific-stint}, $\varphi_{i,t}\equiv\sum_{\tau = t_{i,0,k} + 1}^t  u_{i,\tau,k(i,t)}$, now systematically depends on the full history of previous idiosyncratic skill shocks, which influence current choice (i.e., the regressor in Equation~\eqref{mc:eq:wageFEspecific-stint}). One might expect that the occupation-stint-specific controls in regression \eqref{mc:eq:wageFEspecific-stint} in principle address this problem, similar to the differenced approach \eqref{mc:eq:estimation-main}. But this is not the case.

To see the argument and the bias most clearly suppose for simplicity that all time-varying skill parameters are zero ($\Gamma_{k(i, t-1),k(i, t)}=0$, $\forall k(i, t-1),k(i, t)$). Suppose also that there are only two sectors, $k$ and a reference occupation $k'$, and consider first a two-period base period where we assume that $\pi_{t,  k}=\pi_{t, k'}=const$ for $t=1,\ldots,2$. In this case, simplified wage Equation~\eqref{mc:eq:wageFEspecific-stint} becomes
\begin{equation}
	\label{mc:eq:wageFEsimple}
	w_{i,t} = \eta_{k'(i,t_{i,0,k'})}+\tilde{\eta}_{k(i,t_{i,0,k})} {I}_{k(i, t)}+u_{k'(i,t_{i,0,k'})}+\tilde{u}_{k(i,t_{i,0,k})} {I}_{k(i, t)}\;\text{{for}}\;t=1,\ldots,2,
\end{equation}
where $\tilde{\eta}_{k(i,t_{i,0,k})}\equiv\eta_{k(i,t_{i,0,k})}-\eta_{k'(i,t_{i,0,k'})}$ and $\tilde{u}_{i, t, k}\equiv u_{k(i,t_{i,0,k})}-u_{k'(i,t_{i,0,k'})}$ are relative skill endowments and skill shocks in occupation $k$. The regression \eqref{mc:eq:wageFEsimple} is classically endogeneity-biased because the error term $\tilde{u}_{k(i,t_{i,0,k})} {I}_{k(i, t)}$ most likely positively correlates with the regressor $\tilde{\eta}_{k(i,t_{i,0,k})}{I}_{k(i, t)}$, even for the stayers in an occupation stint which we are identifying from. This will lead to an overestimation of $\tilde{\eta}_{k(i,t_{i,0,k})}$.

\section{Monte Carlo Evidence}
\label{mc:sec:simul}

In this section we provide Monte Carlo evidence for the performance of our and other estimation methods under various assumptions about the data generating process. We  describe the data generating process, which we attempt to keep reasonably close to our sample on the one hand while allowing us to evaluate the impact of key changes on the other hand. We also highlight some stylized facts in the data---for example on occupational switching and the distribution of period-by-period wage changes---that may help in judging what constitutes reasonable parameter values. We then discuss the actual simulation results in detail. Before showing the details, we start with a short summary:

\subsection{Summary}
\label{mc:sub:model_performance}

We test the limits of our estimation method in a broad range of Monte Carlo experiments, also exploring extensions of the underlying economic model. Furthermore, we compare the performance of our approach to an alternative that uses occupation-specific fixed effects pioneered by \citet{C2016}. We limit ourselves to a short description of the results, all details can be found in Section~\ref{mc:sec:simul} of the Appendix.

In the Monte Carlo simulations, we aim to create a fairly realistic setting. We draw a sample of occupations and wages at labor market entry from our SIAB dataset. The remaining potential wages are drawn from truncated distributions so that the observed initial choice is optimal within the model. The subsequent trajectories of wages in all occupations are simulated using our estimates for price changes and skill accumulation, varying the dispersion of the idiosyncratic shocks across experiments. We stick to the four broad occupation groups and draw $100 \times 50,\!000$ careers for each experiment. This balances the ability to summarize the results on the one hand and broadly resembles the effective size of detailed occupations in our application on the other hand. Section~\ref{mc:sub:key_facts_in_the_siab_data} of the Appendix reports on some dimensions of the actual data---occupational switchers, the distribution of wage innovations, and the evolution of wage inequality---that serve as a backdrop for judging what may constitute reasonable values for simulation inputs like, for example, the variance of skill shocks.

\paragraph{Correctly specified data generating process:} In Section~\ref{mc:sub:idiosyncratic-shocks}, we analyze the performance of our estimation method when the data generating process is the one described in Section~\ref{mc:sub:baseline-model}. A detailed verbal description is provided at the beginning of~\ref{mc:sub:idiosyncratic-shocks}; its four subsections contain tables and figures for varying specifications regarding the distribution of the idiosyncratic skill shocks. In order to judge the quality of the approximation~\eqref{mc:eq:wagegrowth}, we first shut these shocks off altogether. The only randomness in this experiment comes from the initial draws and from the evolving prices at the aggregate level. None of the $4 \times 100$ estimated lines is visually discernible from the respective truth; we thus conclude that the approximation of individual wage growth under optimal occupation choice in Equation~\eqref{mc:eq:wagegrowth} is unlikely to be causing a bias in our basic setting.

We then set the standard deviation of idiosyncratic skill shocks to half of the standard deviation of innovations to wages in the SIAB. This yields switching behavior, wage innovations, and an evolution of the wage structure very similar to those in that actual data; we thus term this distribution to have ``moderate shocks''. As predicted at the end of the previous section, the OLS estimates show a modest upward bias of stayers' skill accumulation coefficients, whereas the IV estimates are almost exactly on target.\footnote{Also as expected, the cross-accumulation parameters are generally upward-biased in the OLS; and in the IV with weak instruments, they are large in absolute values.} Both sets of skill price estimates track the evolution of their actual values very closely. Intuitively, mistakes we make with respect to the structural accumulation in the base period cancel out in the analysis period, i.e., in Equation~\eqref{mc:eq:estimation-main} for the OLS. This basic pattern holds true even when tripling the size of the shocks.\footnote{The descriptives on the resulting data in \ref{mc:sub:idiosyncratic-shocks-vlarge} show that tripling the shocks is clearly an extreme case. There is far more switching in all directions compared to the SIAB, wage growth is twice as high and more dispersed than in the data, and wage inequality is skyrocketing.} We overestimate skill accumulation, particularly when using OLS, but skill price estimates remain remarkably close to their targets. Finally, adding persistence to the skill shocks by means of an AR(1)-process does not alter these conclusions either.

\paragraph{Switching Costs:}One aspect that previous literature has emphasized are fixed costs of switching occupations \citep[e.g., ][]{Cortes2017,edin2019individual}. In our framework, the point of indifference between staying in an occupation and switching will now be determined by wages adjusted for switching costs. This means, however, that unadjusted wages of switchers will exhibit jumps at the indifference point, introducing a potential bias to our estimates. We work this case out theoretically in Section~\ref{mc:sec:model-extensions}; Section~\ref{mc:sub:costs_of_switching_occupations} presents Monte Carlo analyses examining the bias' importance. First, in a model without skill shocks and with moderate switching costs (5\% of annual wages), our approximation~\eqref{mc:eq:wagegrowth} continues to work well. OLS estimates recover skill prices and stayers' skill accumulation coefficients almost exactly in such a specification. As previously, we then add moderate and large skill shocks, paired with moderate and high (20\% of annual wages) switching costs. Our basic conclusions from the corresponding exercises without switching costs remain the same. We slightly overestimate the structural skill accumulation coefficients on the diagonal. As expected, the inertia generated by switching costs leads to a stronger overestimation of the off-diagonal elements of $\Gamma$. Nevertheless, skill prices are estimated with remarkable precision. We therefore expect our empirical results to be robust to switching costs of plausible magnitude.

\paragraph{Occupation-specific amenities and future values:} Another key extension of our approach is to the generalized Roy model, including non-pecuniary aspects or discounted future values of occupations in the worker's decision problem \citep[e.g., as in][]{Lee2006}.  Similar to the case with switching costs, workers who move to an occupation with lower (higher) non-pecuniary value will exhibit positive (negative) jumps in wages to compensate for the amenity difference. We show formally in Appendix \ref{mc:sec:model-extensions} that, if the non-pecuniary or future values are time-constant, the skill accumulation parameter $\hat{\Gamma}_{k', k}$ in our main specification will absorb them. If they are time-changing, the estimation equation~\eqref{mc:eq:estimation-main} has to be augmented and include regressors for occupation switches ($\Delta{I}_{k,i,t}$) on top of average occupation choices ($\bar{I}_{k,i,t}$) to control for (and estimate) the respective ``wage compensation''. Section~\ref{mc:subsub:amenities} of the Monte Carlo analyses examines such a case with rising amenities in one of the occupations, finding that the $\Delta{I}_{k,i,t}$ correction is indeed necessary but then we recover the skill prices and skill accumulation as well as before (plus the changing amenities or future values themselves).

\paragraph{Employer Learning:}We also show formally in the Appendix that what we have referred to as idiosyncratic skill shocks is observationally equivalent in our analysis to a basic model of employer learning about workers' skills \citep[e.g., as in][]{Altonji2001,Gibbons2005}. This is due to the fact that log-linearity allows us to write the model in terms of expected skills, which can evolve because of changes in actual skills (our formulation above) or because employers change their expectations about individuals' skills over time. The two interpretations are not mutually exclusive, of course.

\paragraph{Fixed effects estimation:}

Finally, we examine an alternative panel data approach for estimating skill prices due to \citet{C2016}, who uses individual $\times$ occupation specific fixed effects in order to control for skill selection. First, we show theoretically how to generalize \citeauthor{C2016}' estimation in order to flexibly control for a rich model of worker skill accumulation. We then implement this approach in the Monte Carlo simulations and find that it performs well in most cases. Exceptions are specifications with a lot of switching (i.e., a large number of occupations $K$ or large skill shocks), when the `exogenous mobility' assumption of fixed effects approaches discussed in  Appendix~\ref{mc:sec:fixed-effects-model} becomes quantitatively important. We conclude that the generalized version of \citeauthor{C2016}' method is a useful alternative when the goal is to estimate low-dimensional skill prices; it seems less suitable for applications that feature a large number of sectors as in our main estimation with 120 detailed occupations.

\subsection{Data Generating Process}

We generate panel datasets similar in structure to the actual SIAB data and set the data generating processes' parameters to the values we will eventually estimate. This allows us to keep some features disciplined by the data while varying components that appear critical to the model. We believe this is much more transparent to the reader than picking arbitrary distributions in a fully stylized setting.

We randomly draw the initial observations of 50,000 individuals from the SIAB as described in Section~A.1 of the Online Appendix to \citet{BGS}. The variables we use include initial wages, occupational choices, age (25--54) and year (1975--2010). We then use the parameters from their baseline estimation to decompose the initial wage of a worker into a price and a skill component:
\begin{equation}
  \label{mc:eq:appdx-mc:initial-skills}
  s_{k, i, t_{i, 0}} = w_{i, t_{i, 0}} - \pi_{k, t_{i, 0}} \quad \text{if} \quad i \in k,
\end{equation}
where $t_{i, 0} \in \{1975, \dots, 2010 \}$ is the year a worker is first observed.

As the Roy model implies that the initial choice must be optimal in the sense that $w_{k, i, t_{i, 0}} \ge w_{k', i, t_{i, 0}} \forall k'$, we have a natural bound for the skills a worker possesses in the remaining sectors.
\begin{equation}
  \label{mc:eq:appdx-mc:initial-latent-skills}
  s_{k', i, t_{i, 0}} \le s_{k, i, t_{i, 0}} + \pi_{k, t_{i, 0}} - \pi_{k', t_{i, 0}} \quad \text{if} \quad i \in k.
\end{equation}
We draw the initial skills separately and independently for every worker in the sample from a truncated normal distribution with the upper bound given by $s_{k, i, t_{i, 0}} + \pi_{k, t_{i, 0}} - \pi_{k', t_{i, 0}}$. We set the location parameter $\mu_{k, i, t_{i, 0}} = s_{k, i, t_{i, 0}} + \pi_{k, t_{i, 0}}$ and fix the scaling parameter $\sigma = 3$ across workers.

For the following years of a worker's career, we then simulate wage growth as the sum of systematic skill growth and price growth given by $\hat{\Gamma}_{k', k}, \Delta \hat{\pi}_{k, t}$. On top of that, we add idiosyncratic skill shocks depending on the specification and finally let workers choose their preferred sector based on comparative advantage (possibly including costs of switching and changing non monetary amenities). We repeat this until a worker's maximum age of 54 is reached or the sample period ends. We rerun the exercise for 100 Monte Carlo repetitions and estimate price and skill changes on each sample. We then compute the average price trends and skill accumulation function across repetitions. For computational reasons, we only use four occupations (i.e., the four broad occupation groups of the main text).
In the subsections from \ref{mc:sub:idiosyncratic-shocks} onward, we report the results by comparing estimated and true parameter values $\hat{\pi}_{k, t}, \pi_{k, t}$ for different Monte Carlo specifications.

\subsection{Key facts in the SIAB data}
\label{mc:sub:key_facts_in_the_siab_data}

We first document some facts in the SIAB data. While the goal of our Monte Carlo studies is not to replicate these facts it is useful to keep them in mind in order to judge what may be reasonable parameter values. For example, if a specification yields consistent parameter estimates but switching between occupations is shut down completely (as would happen, for example, if switching costs are high and idiosyncratic shocks are small), this would not constitute a very realistic exercise.

\begin{figure}[ht!]
  \caption{Descriptive statistics in the SIAB data}
  \label{mc:fig:descriptives-siab}
  \centering
  \begin{minipage}[t]{\panelwidth}
    \subcaption{Occupation entrants/incumbents in $t+1$}
    \label{mc:fig:descriptives-siab-switchers-entrants}
    \includegraphics[width=\textwidth]{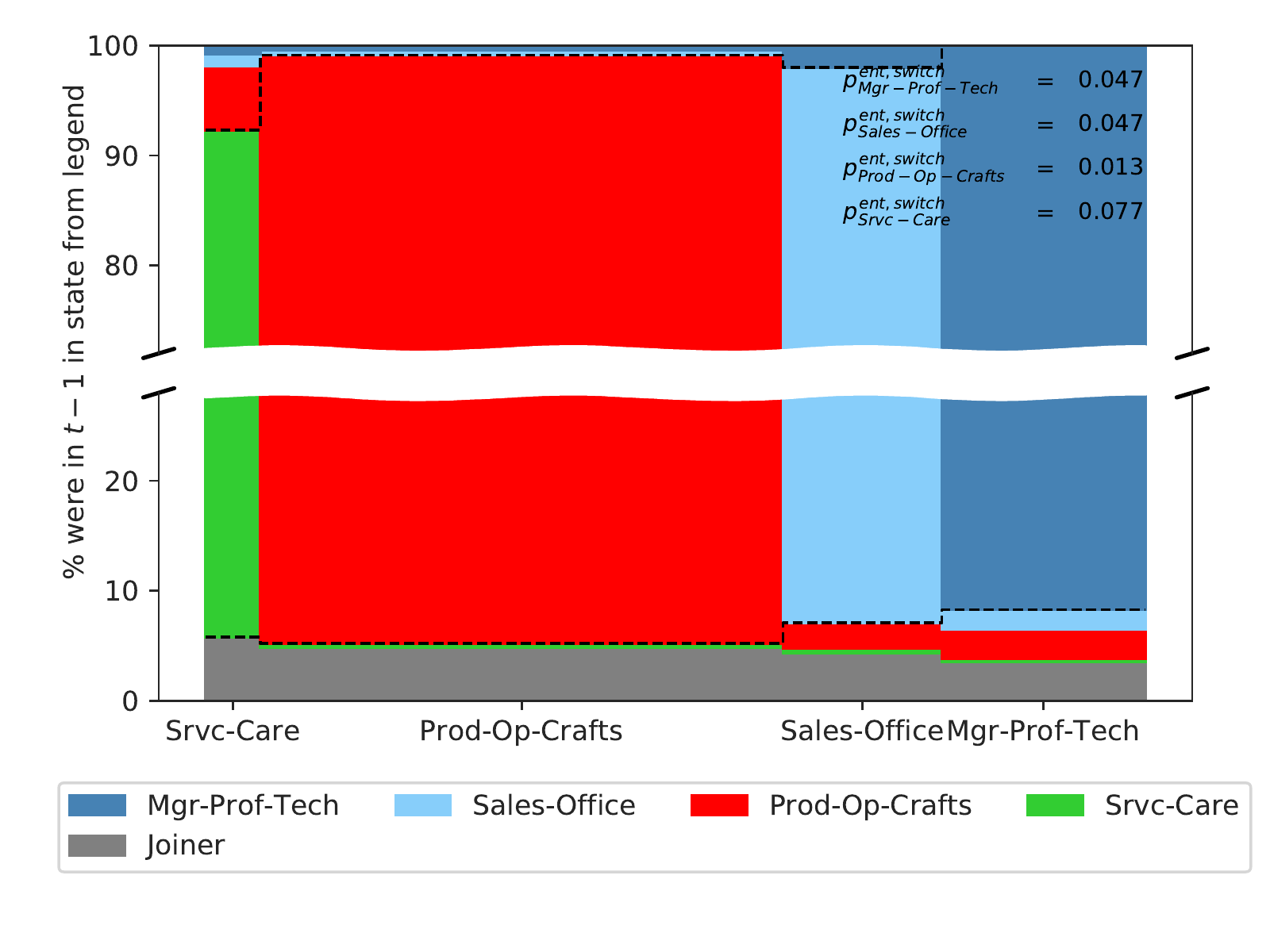}
  \end{minipage}
  \begin{minipage}[t]{\panelwidth}
    \subcaption{Occupation leavers/stayers in $t-1$}
    \label{mc:fig:descriptives-siab-switchers-leavers}
    \includegraphics[width=\textwidth]{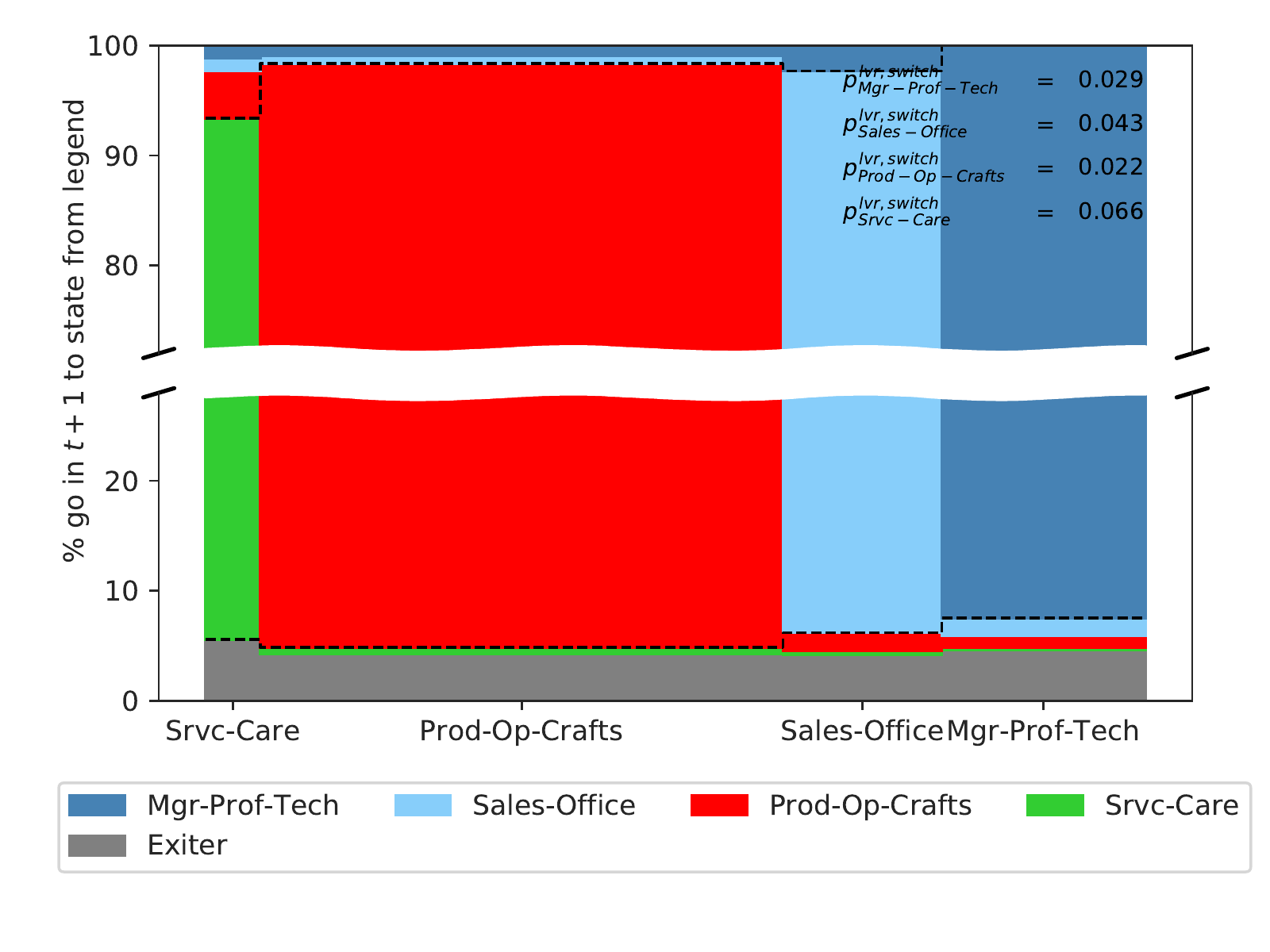}
  \end{minipage}
  \begin{minipage}[t]{\panelwidth}
    \subcaption{Distribution of annual wage growth}
    \label{mc:fig:descriptives-siab-wage-growth-distribution}
    \includegraphics[width=\textwidth]{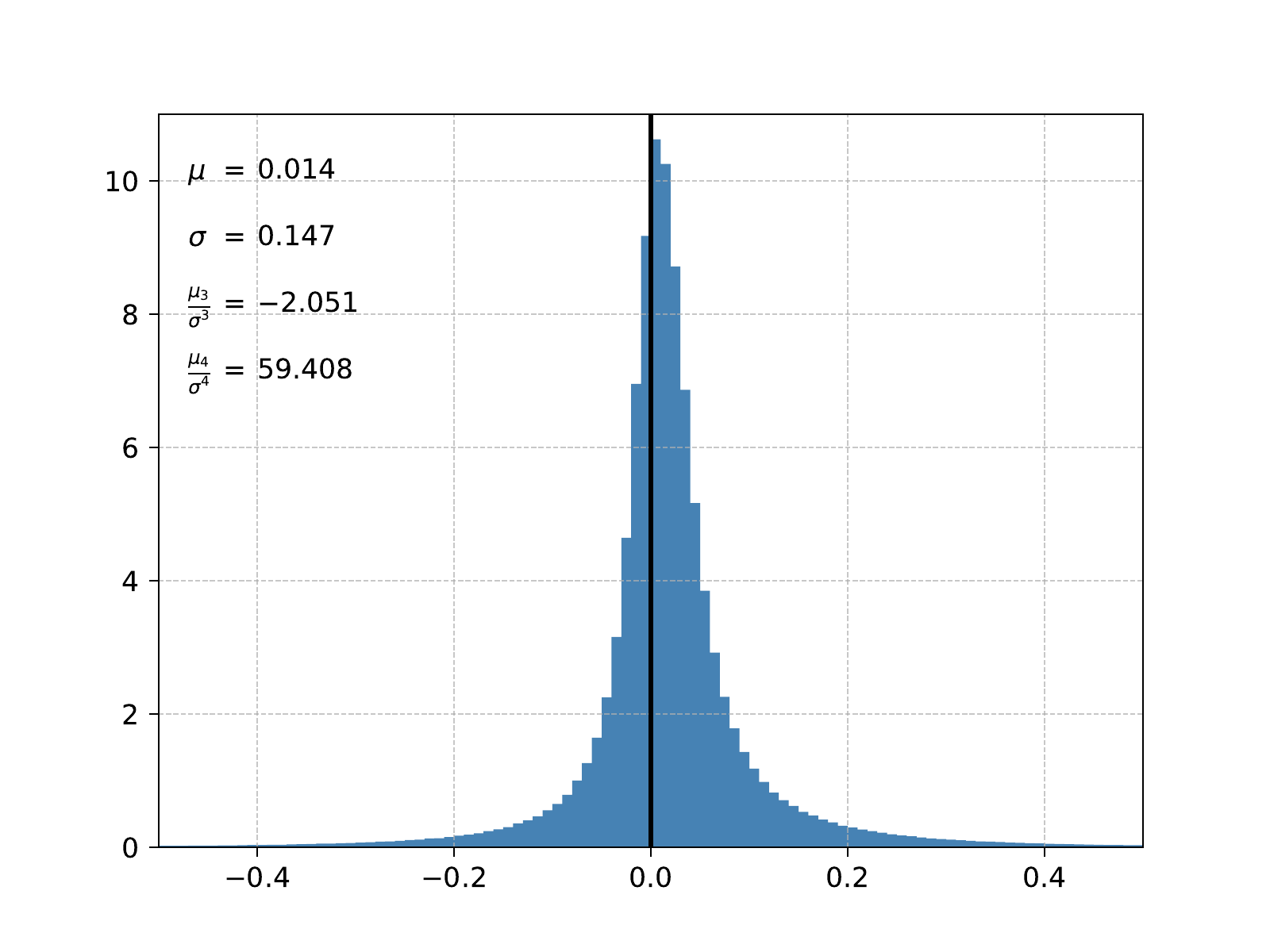}
  \end{minipage}
  \begin{minipage}[t]{\panelwidth}
    \subcaption{Evolution of the wage distribution}
    \label{mc:fig:descriptives-siab-evolution-wage-inequality}
    \vskip2ex
    \includegraphics[width=\textwidth]{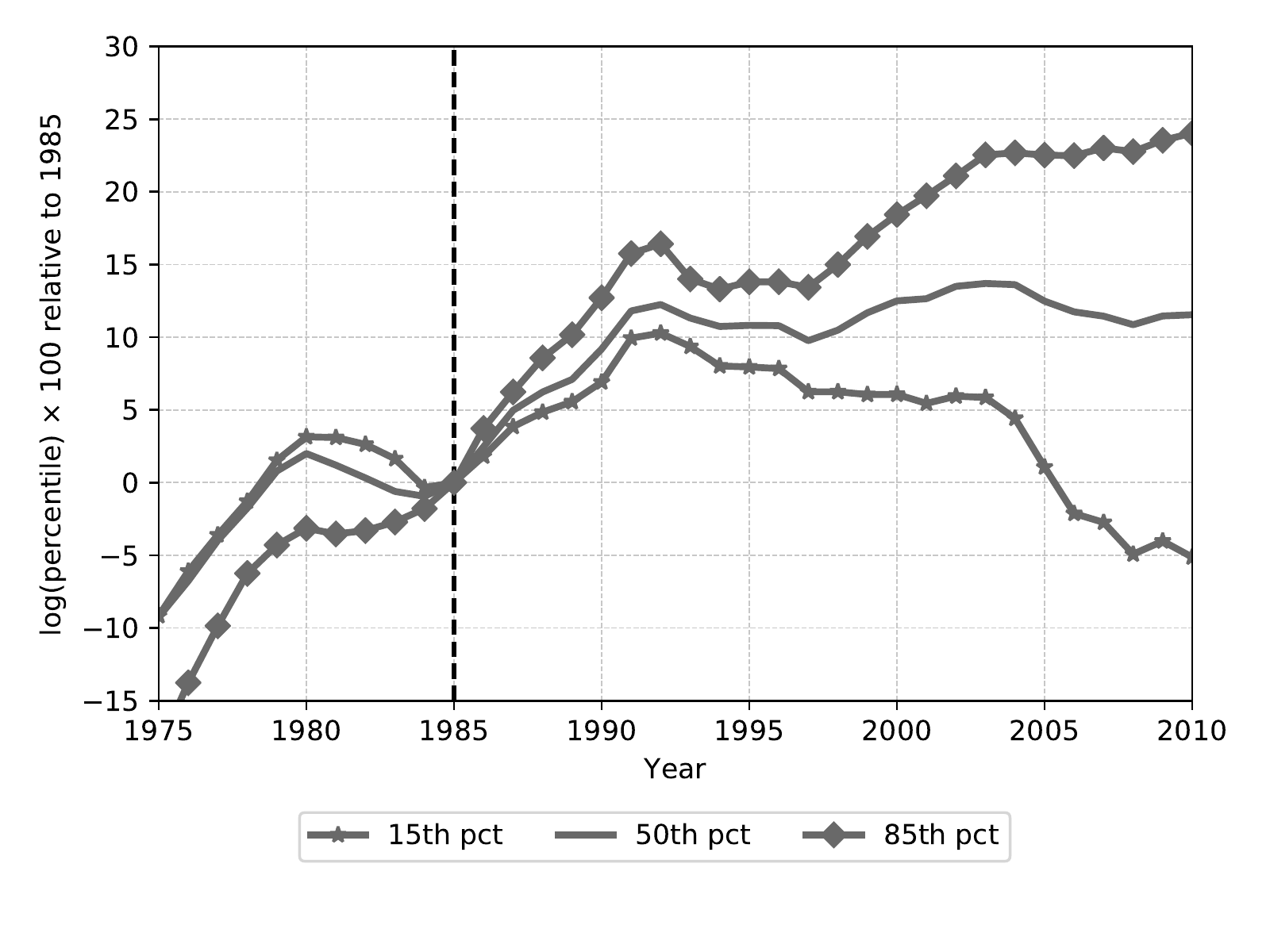}
  \end{minipage}
  \begin{minipage}{\textwidth}
    \scriptsize
    \emph{Notes:} \siabdescriptives Panels (a) and (b) use the specification with filled-up unemployment and out-of-the-labor-force spells (see Section~A.1.4 of the Online Appendix to \citet{BGS}) for better comparability later on, since unemployment or out-of-the-labor-force do not form a part of our Monte Carlo exercise.
  \end{minipage}
\end{figure}

Figure~\ref{mc:fig:descriptives-siab-switchers-entrants} shows the composition of incumbents and entrants into occupation groups from $t$ to $t+1$. The groups are ranked horizontally by their average wages over the years 1985--2010. They are ranked vertically also by average wages, with the incumbents in the middle and lower/higher earning occupations from which workers enter bottom/top, respectively. Figure~\ref{mc:fig:descriptives-siab-switchers-leavers} depicts the same thing for stayers and leavers from $t-1$ to $t$. Entrants into our sample at age 25 (joiners) and leavers at age 54 (exiters) are coded in gray at the very bottom. Figures~\ref{mc:fig:descriptives-siab-switchers-entrants} and \ref{mc:fig:descriptives-siab-switchers-leavers} use the specification with filled-up unemployment and out-of-the-labor-force spells \citep[see Section~A.1.4 of the Online Appendix to][]{BGS} for better comparability later on, since unemployment or out-of-the-labor-force do not form a part of our Monte Carlo exercise.

Both panels of Figure~\ref{mc:fig:descriptives-siab} show that a substantial amount of occupation entry and exit are at the ends of the sample age range. But there is also a large share of Srvc-Care workers who switch to Prod-Op-Crafts and even more who switch from Prod-Op-Crafts into Srvc-Care, which is consistent with a growth of Srvc-Card and decline of Prod-Op-Crafts over time. There are in addition substantial switches between Sales-Office and Mgr-Prof-Tech, among others.

The bottom panels of Figure~\ref{mc:fig:descriptives-siab} depict a histogram of workers' annual wage growth and repeat the distribution of key quantiles of the wage distribution for comparability.

\subsection{Baseline model}
\label{mc:sub:idiosyncratic-shocks}

This subsection shows the results for a Monte Carlo exercise where wage growth only stems from price growth or systematic skill growth but not from idiosyncratic shocks. The primary reason for this specification is to assess the quality of the approximation in Equation~\eqref{mc:eq:approx-in-assumption-1}.

\begin{figure}[ht!]
  \caption{Descriptives, no shocks}
  \label{mc:fig:descriptives-no-shocks}
  \centering
  \begin{minipage}[t]{\panelwidth}
    \subcaption{Occupation entrants/incumbents in $t+1$}
    \label{mc:fig:descriptives-no-shocks-switchers-entrants}
    \includegraphics[width=\textwidth]{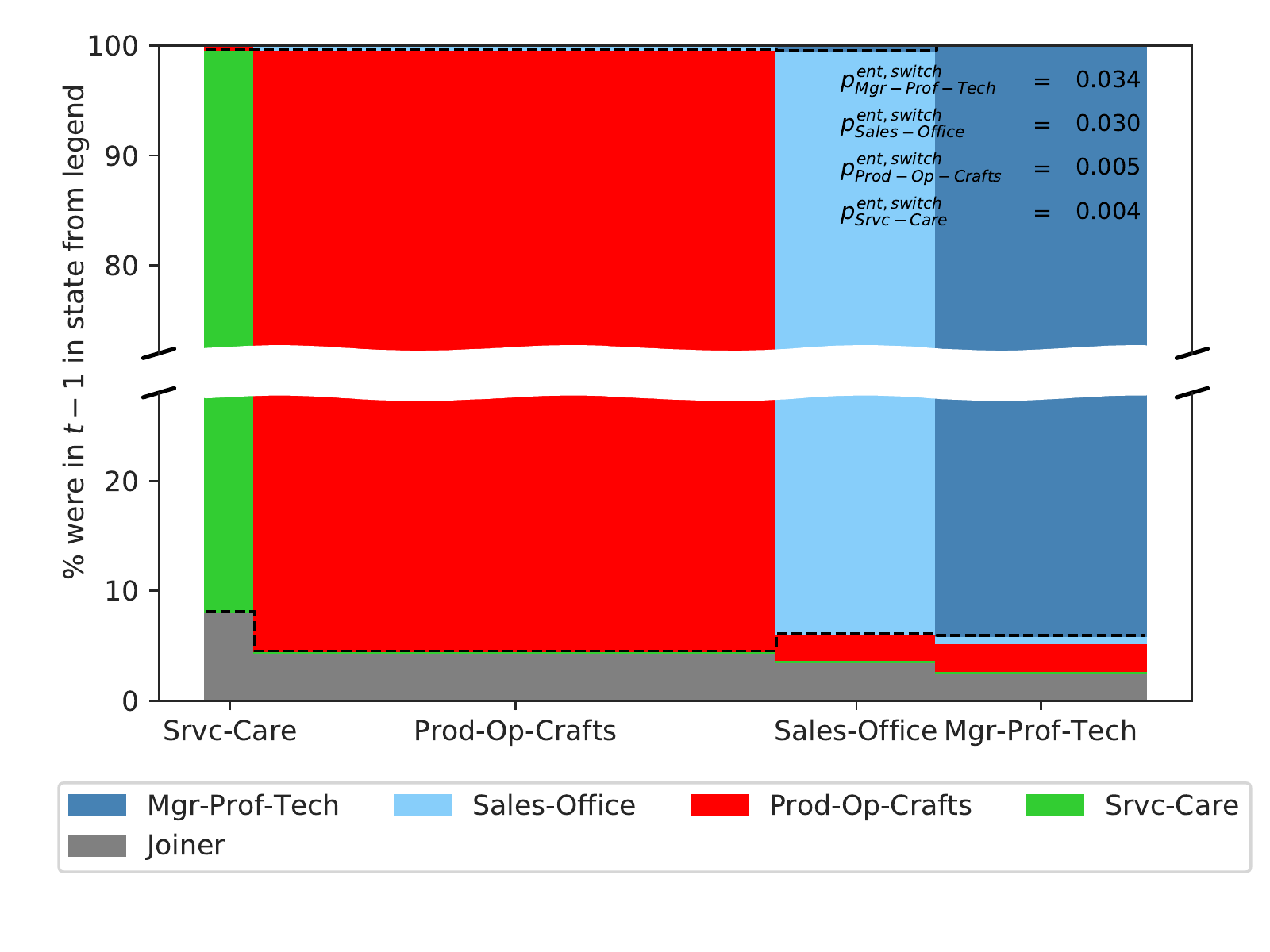}
  \end{minipage}
  \begin{minipage}[t]{\panelwidth}
    \subcaption{Occupation leavers/stayers in $t-1$}
    \label{mc:fig:descriptives-no-shocks-switchers-leavers}
    \includegraphics[width=\textwidth]{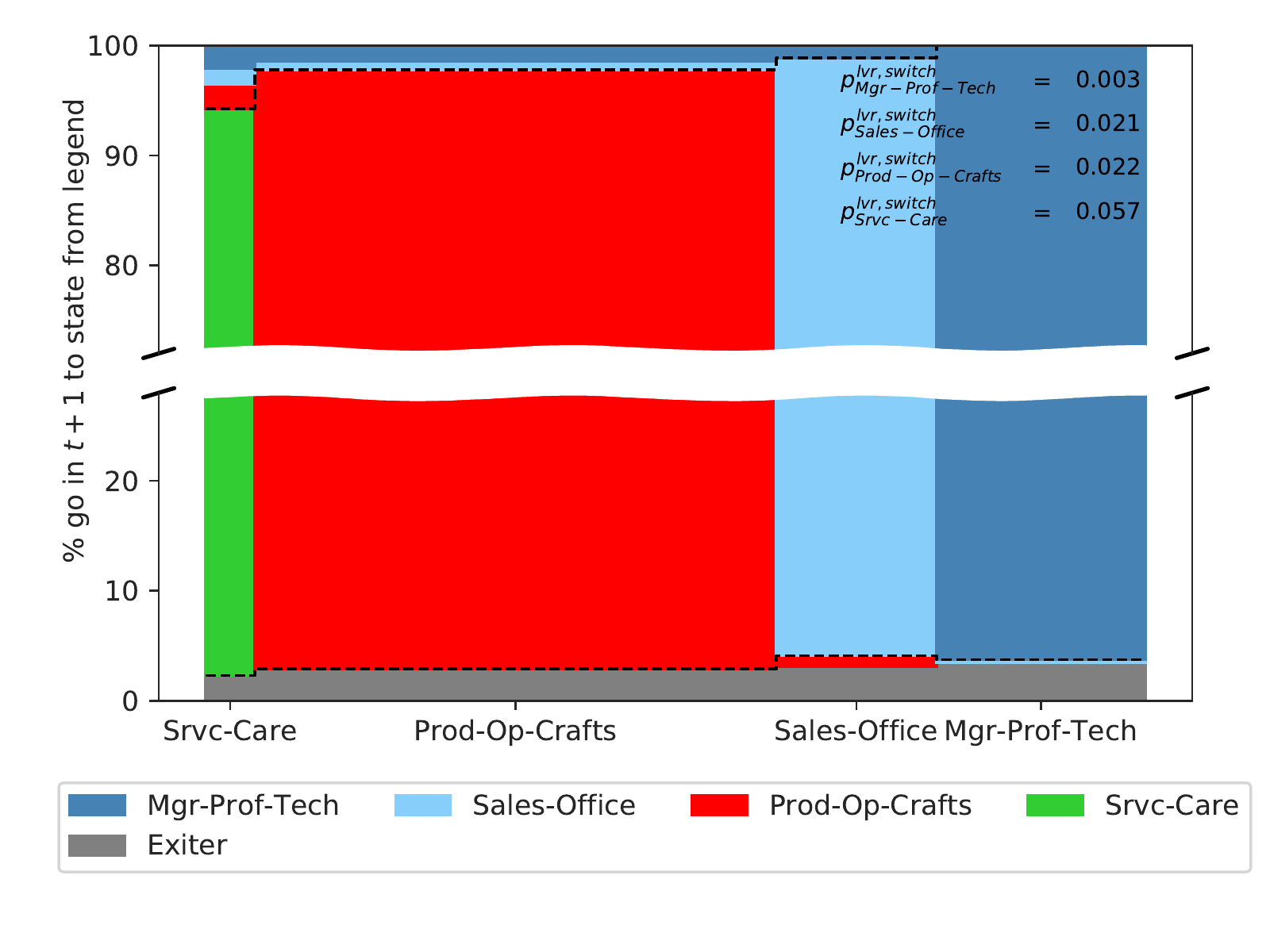}
  \end{minipage}
  \begin{minipage}[t]{\panelwidth}
    \label{mc:fig:descriptives-no-shocks-wage-growth-distribution}
    \subcaption{Distribution of annual wage growth}
    \includegraphics[width=\textwidth]{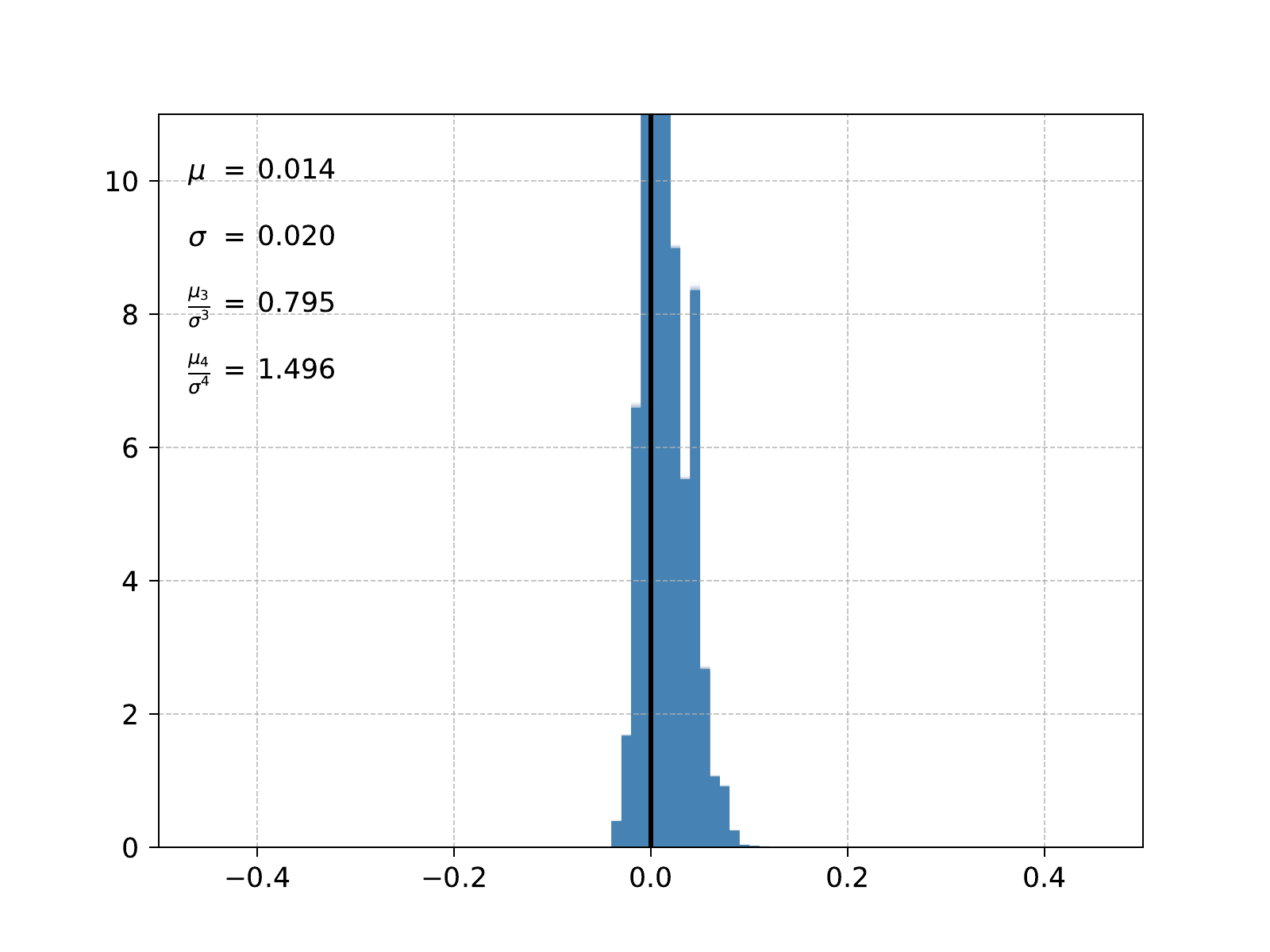}
  \end{minipage}
  \begin{minipage}[t]{\panelwidth}
    \label{mc:fig:descriptives-no-shocks-evolution-wage-inequality}
    \subcaption{Evolution of the wage distribution}
    \vskip2ex
    \includegraphics[width=\textwidth]{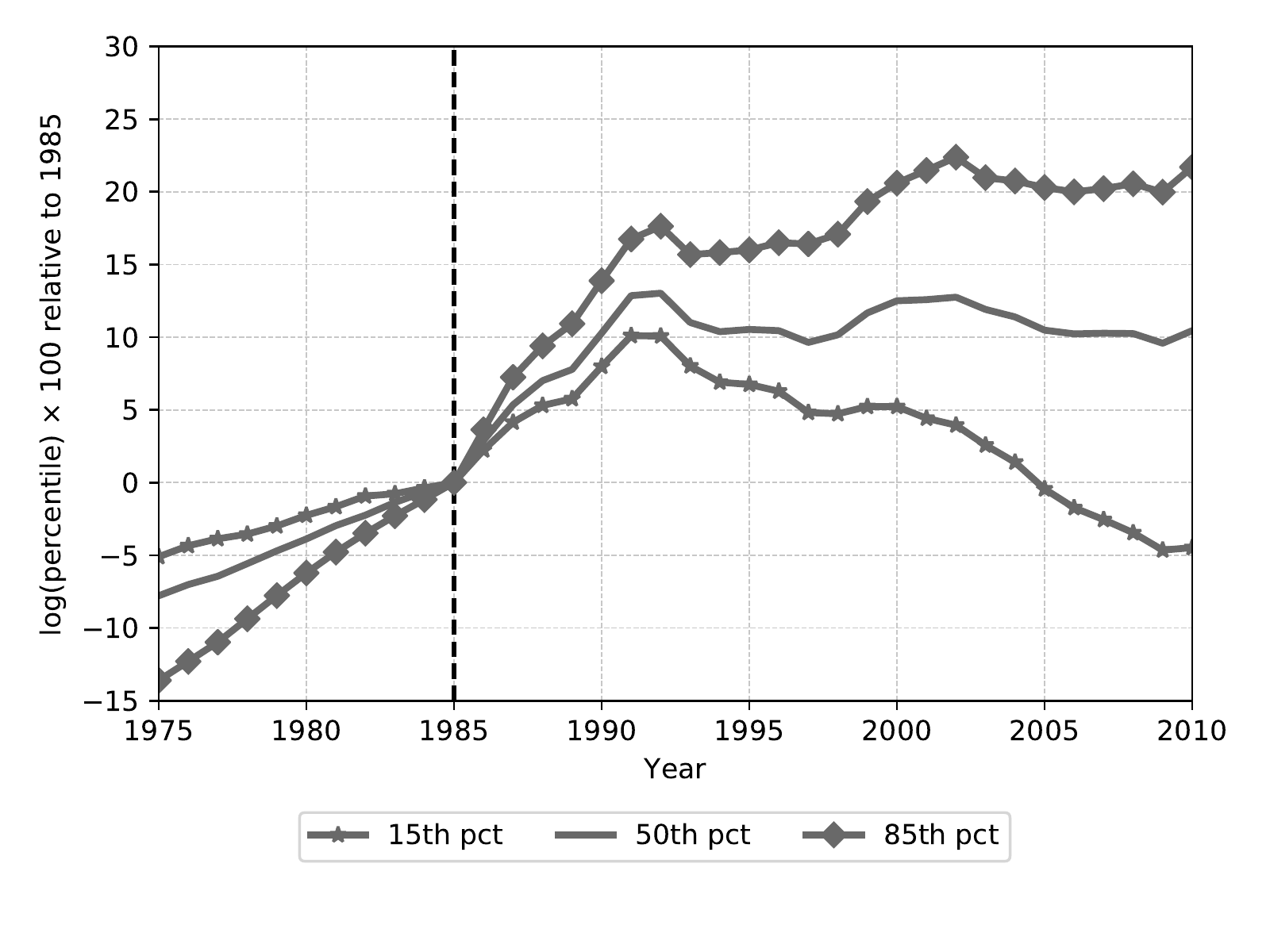}
  \end{minipage}
  \begin{minipage}{\textwidth}
    \scriptsize
    \emph{Notes:} \mcdescriptives
  \end{minipage}
\end{figure}

Despite the absence of shocks, Figure~\ref{mc:fig:descriptives-no-shocks} shows that there are a substantial number of switchers just because of prices and (cross-)skill accumulation. Unsurprisingly, other statistics do not match up well (for example, the standard deviation of annual wage growth is smaller by a factor of 7 than in Figure~\ref{mc:fig:descriptives-siab-wage-growth-distribution}), but that is not the point in this particular exercise.

\begin{figure}[ht!]
  \caption{Estimation results, no shocks}
  \label{mc:fig:estimation-no-shocks}
  \centering
  \begin{minipage}[t]{\panelwidth}
    \subcaption{Cumulative prices, saturated OLS}
    \label{mc:fig:estimation-ols-no-shocks-prices}
    \includegraphics[width=\textwidth]{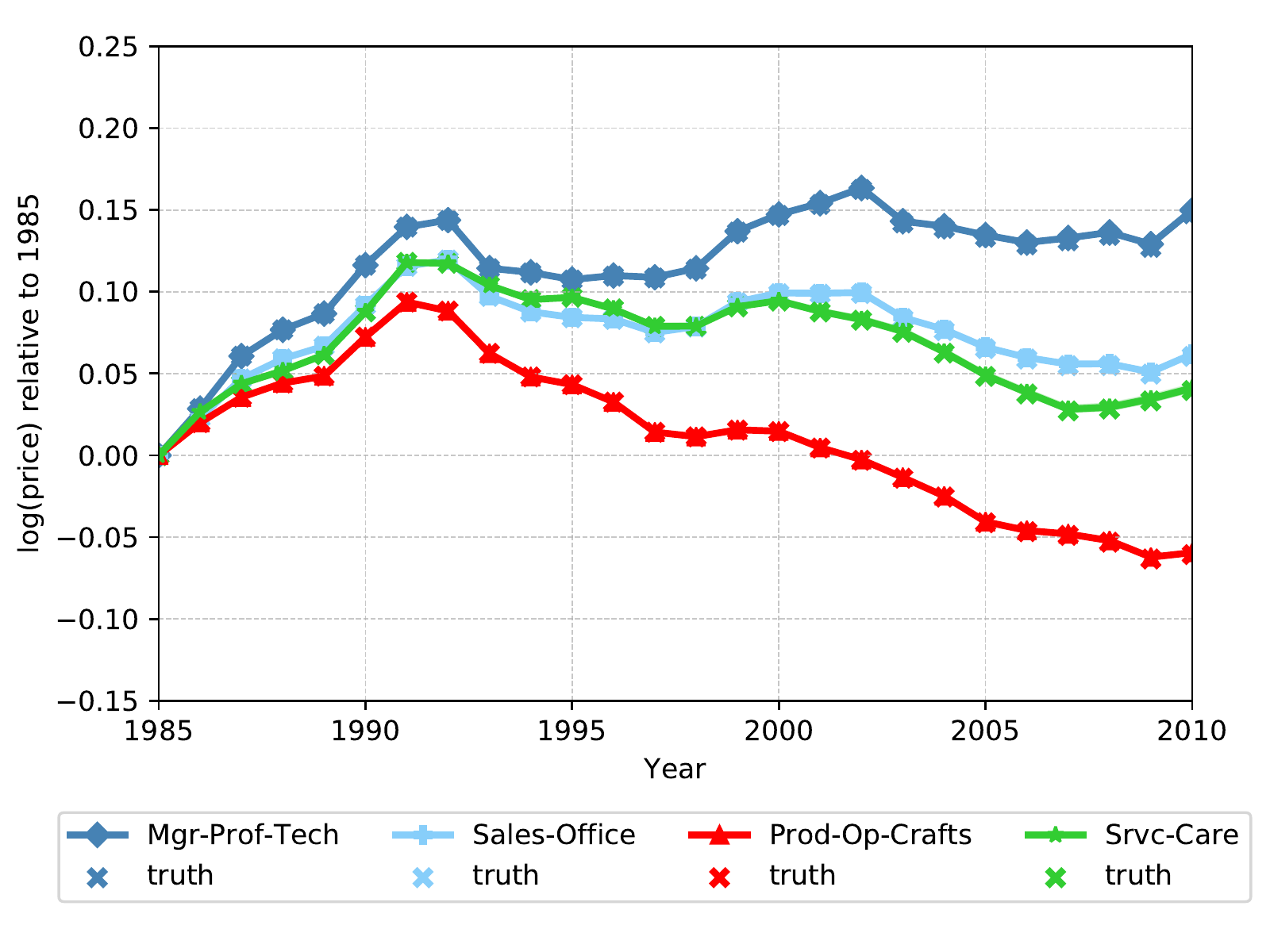}
  \end{minipage}
  \begin{minipage}[t]{\panelwidth}
    \subcaption{Skill accumulation, saturated OLS}
    \label{mc:fig:estimation-ols-no-shocks-skills}
    \includegraphics[width=\textwidth]{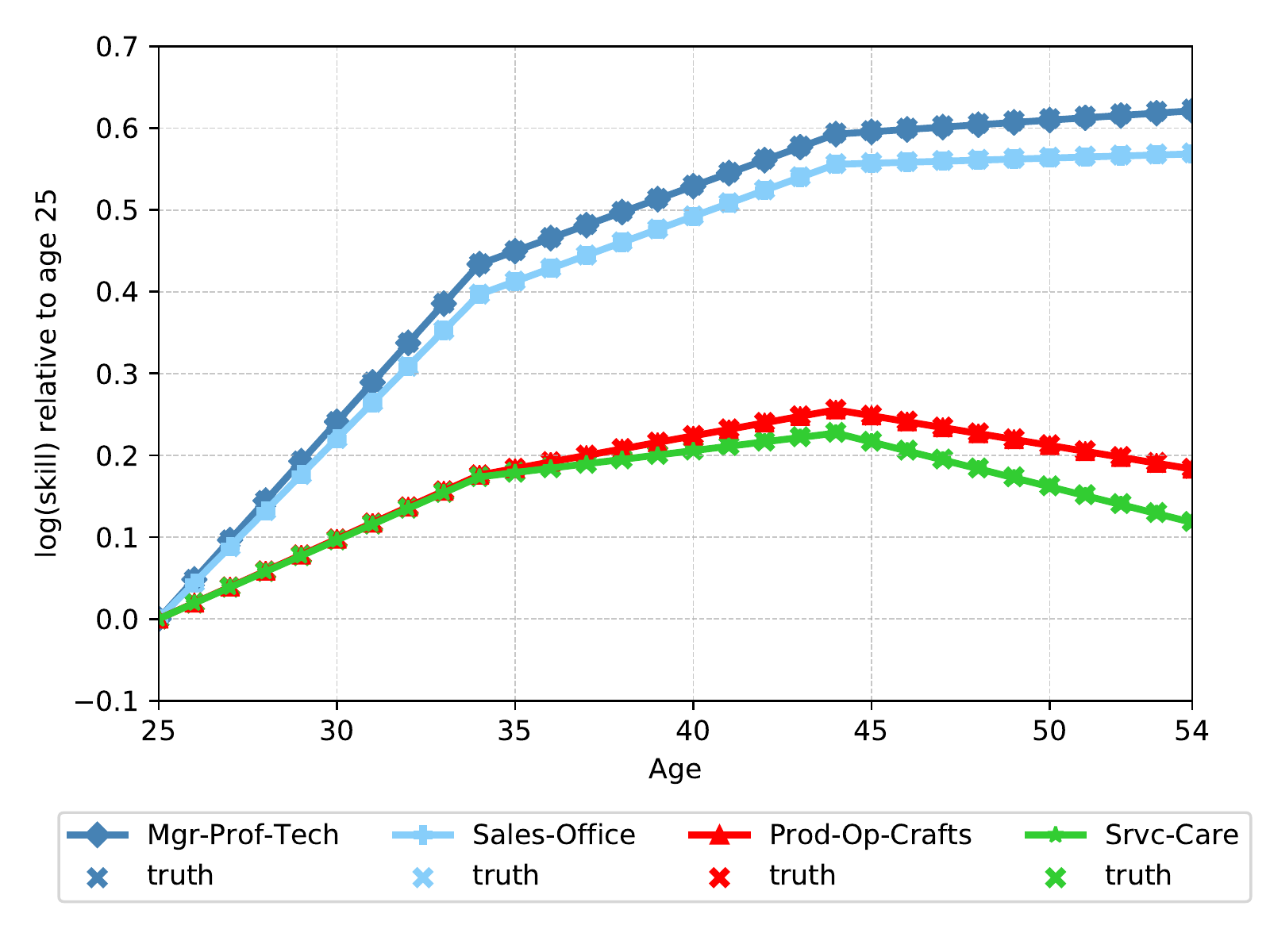}
  \end{minipage}
  \begin{minipage}{\textwidth}
    \scriptsize
    \emph{Notes:} \mcresults \norandomness {} \olsresultsmc
  \end{minipage}
\end{figure}

Figure~\ref{mc:fig:estimation-no-shocks} shows cumulative skill prices and skills in the four occupations, with estimates depicted in the solid lines and true parameter values as crosses. Clearly, the proposed method is able to estimate skill price and skills trends from observed wage changes. Note this is the case for all 100 experiments as no individual lines for specific experiments are to be seen. The only randomness in these cases comes from the initial distributions and it should be taken care of by our estimates if the approximation in Equation~\eqref{mc:eq:approx-in-assumption-1} works well. The graph certainly suggests that it does.

\begin{figure}[ht!]
  \caption{Descriptives, moderate shocks}
  \label{mc:fig:descriptives-moderate-shocks}
  \centering
  \begin{minipage}[t]{\panelwidth}
    \subcaption{Occupation entrants/incumbents in $t+1$}
    \label{mc:fig:descriptives-moderate-shocks-switchers-entrants}
    \includegraphics[width=\textwidth]{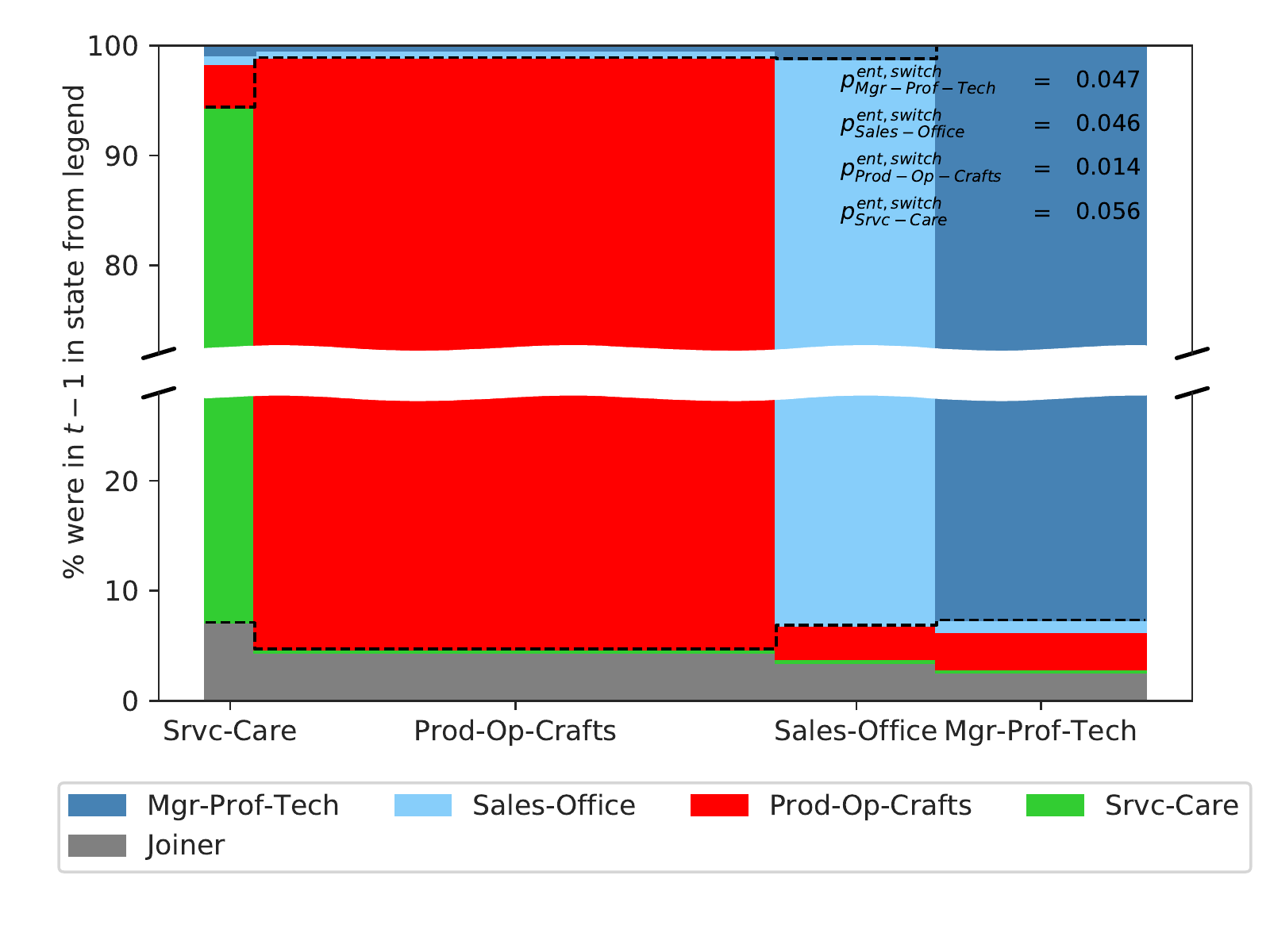}
  \end{minipage}
  \begin{minipage}[t]{\panelwidth}
    \subcaption{Occupation leavers/stayers in $t-1$}
    \label{mc:fig:descriptives-moderate-shocks-switchers-leavers}
    \includegraphics[width=\textwidth]{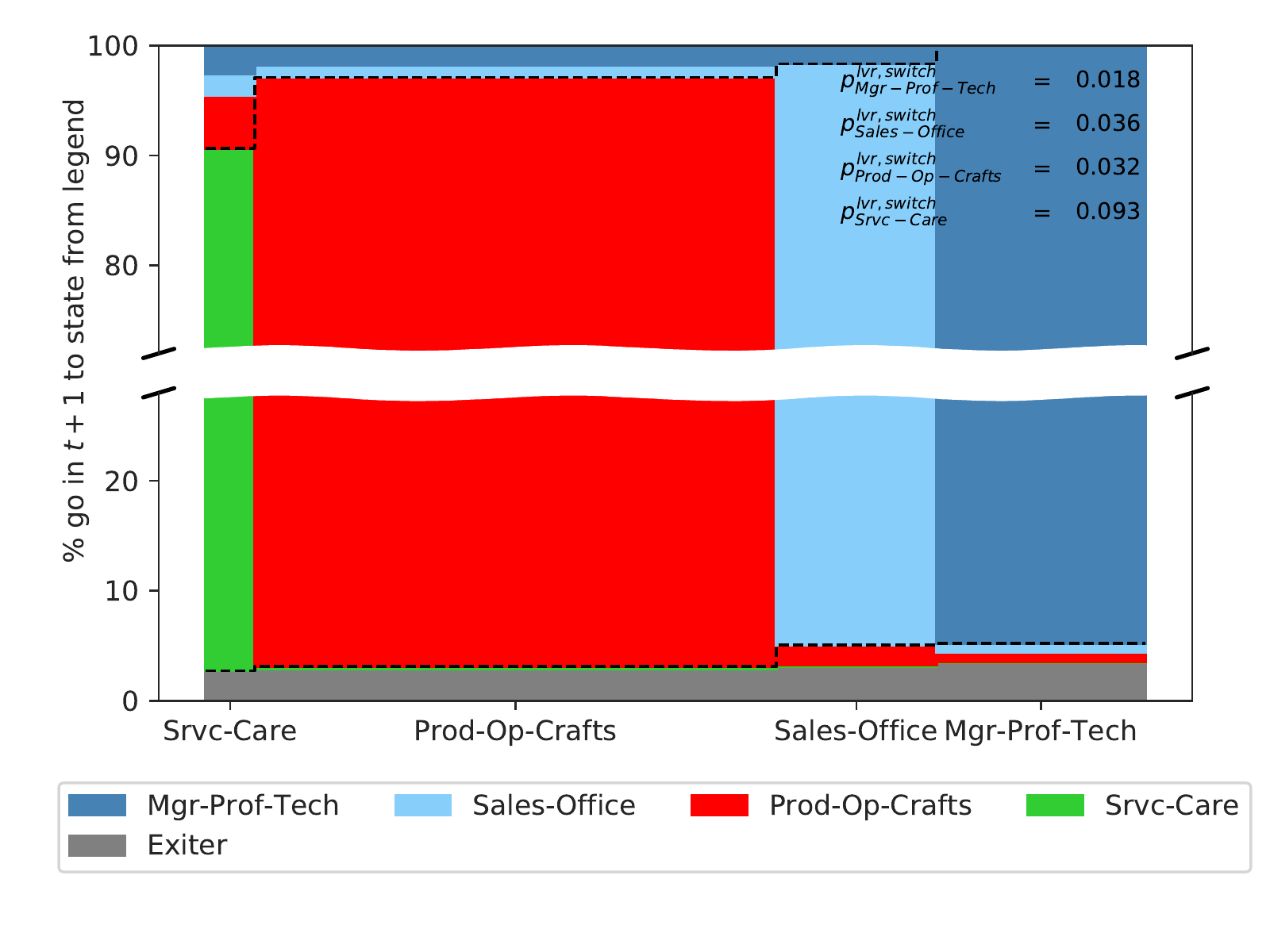}
  \end{minipage}
  \begin{minipage}[t]{\panelwidth}
    \subcaption{Distribution of annual wage growth}
    \label{mc:fig:descriptives-moderate-shocks-wage-growth-distribution}
    \includegraphics[width=\textwidth]{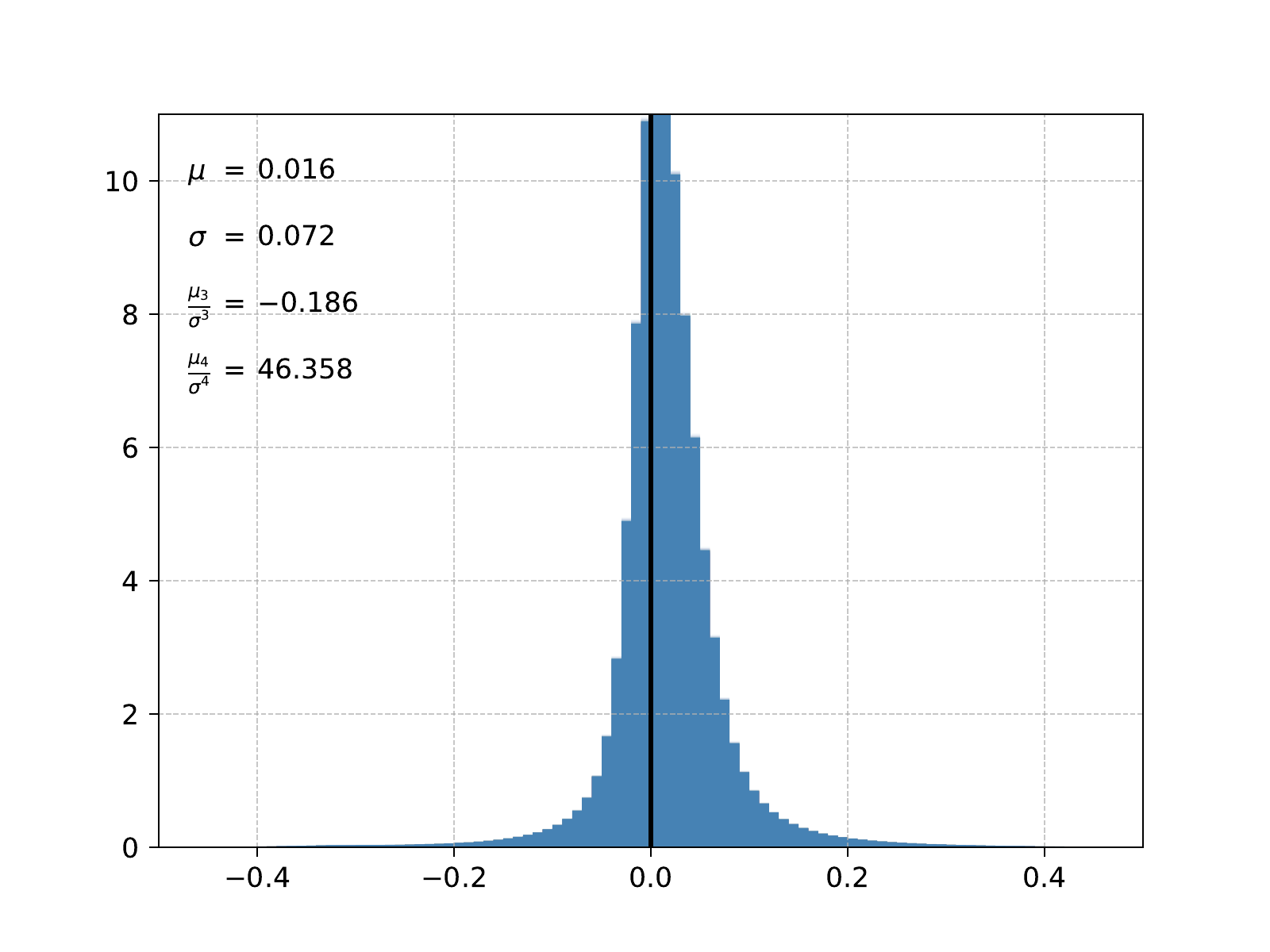}
  \end{minipage}
  \begin{minipage}[t]{\panelwidth}
    \subcaption{Evolution of the wage distribution}
    \label{mc:fig:descriptives-moderate-shocks-evolution-wage-inequality}
    \vskip2ex
    \includegraphics[width=\textwidth]{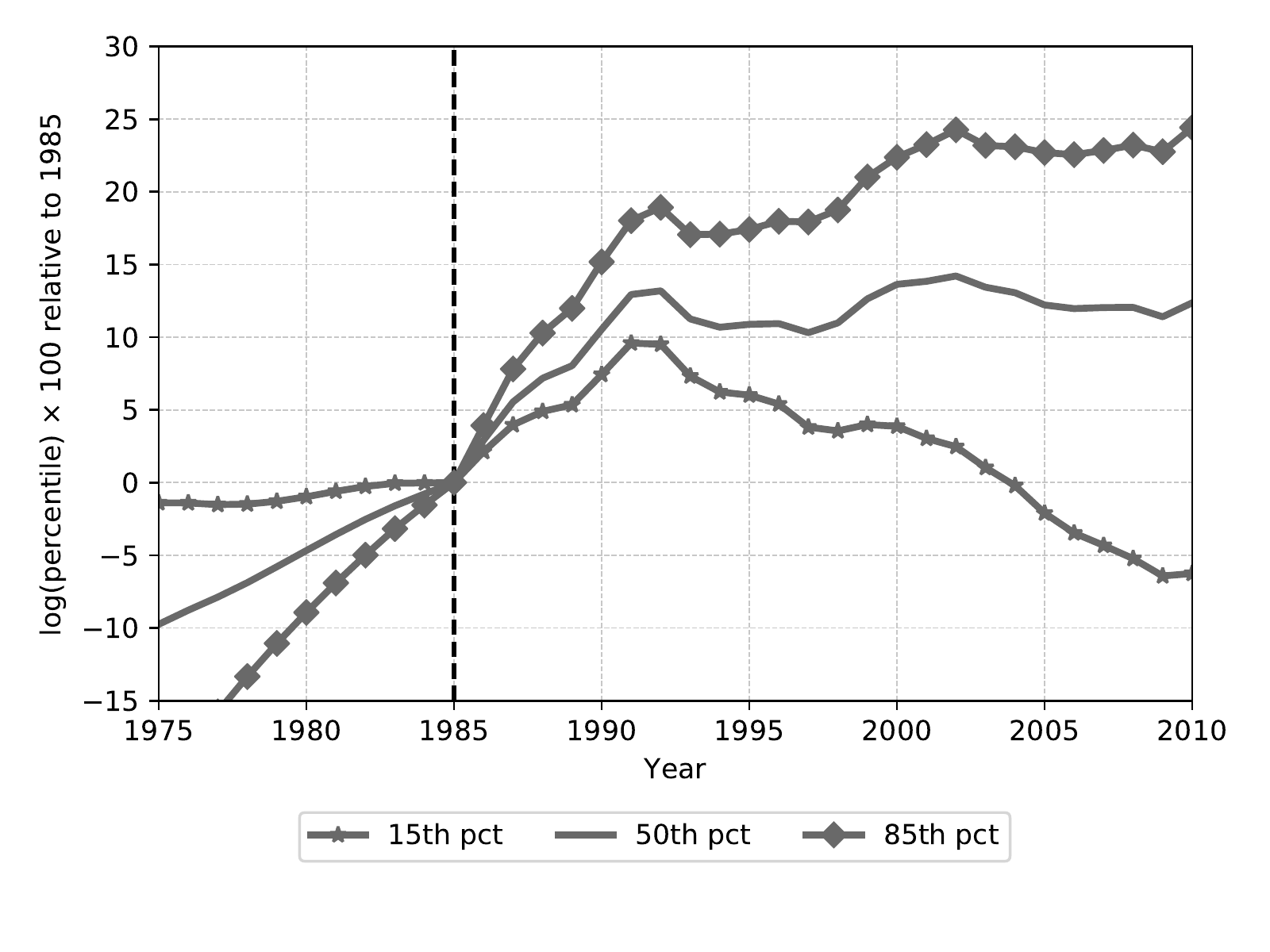}
  \end{minipage}
  \begin{minipage}{\textwidth}
    \scriptsize
    \emph{Notes:} \mcdescriptives {} \olsresultsmc \ivresults
  \end{minipage}
\end{figure}

Idiosyncratic skill shocks $u_{k, i, t}$ as in Equation~\eqref{mc:eq:accumulation} introduce an endogeneity bias to our estimates, which is rising with the standard deviation of the shocks. Figure~\ref{mc:fig:descriptives-moderate-shocks} shows information about the amount of switchers and the dispersion of log wage growth under a scenario where the dispersion of (normally distributed, center 0) skill shocks is $\sigma = 0.5 \cdot \sigma^{SIAB}_{\Delta log(w_{i, t})}$. Clearly, both the distribution of wage growth as well as the distribution of switchers is more comparable in Monte Carlo and SIAB data in this setting.

\begin{figure}[ht!]
  \caption{Estimation results, moderate shocks}
  \label{mc:fig:estimation-moderate-shocks}
  \centering
  \begin{minipage}[t]{\panelwidth}
    \subcaption{Cumulative prices, saturated OLS}
    \label{mc:fig:estimation-ols-moderate-shocks-prices}
    \includegraphics[width=\textwidth]{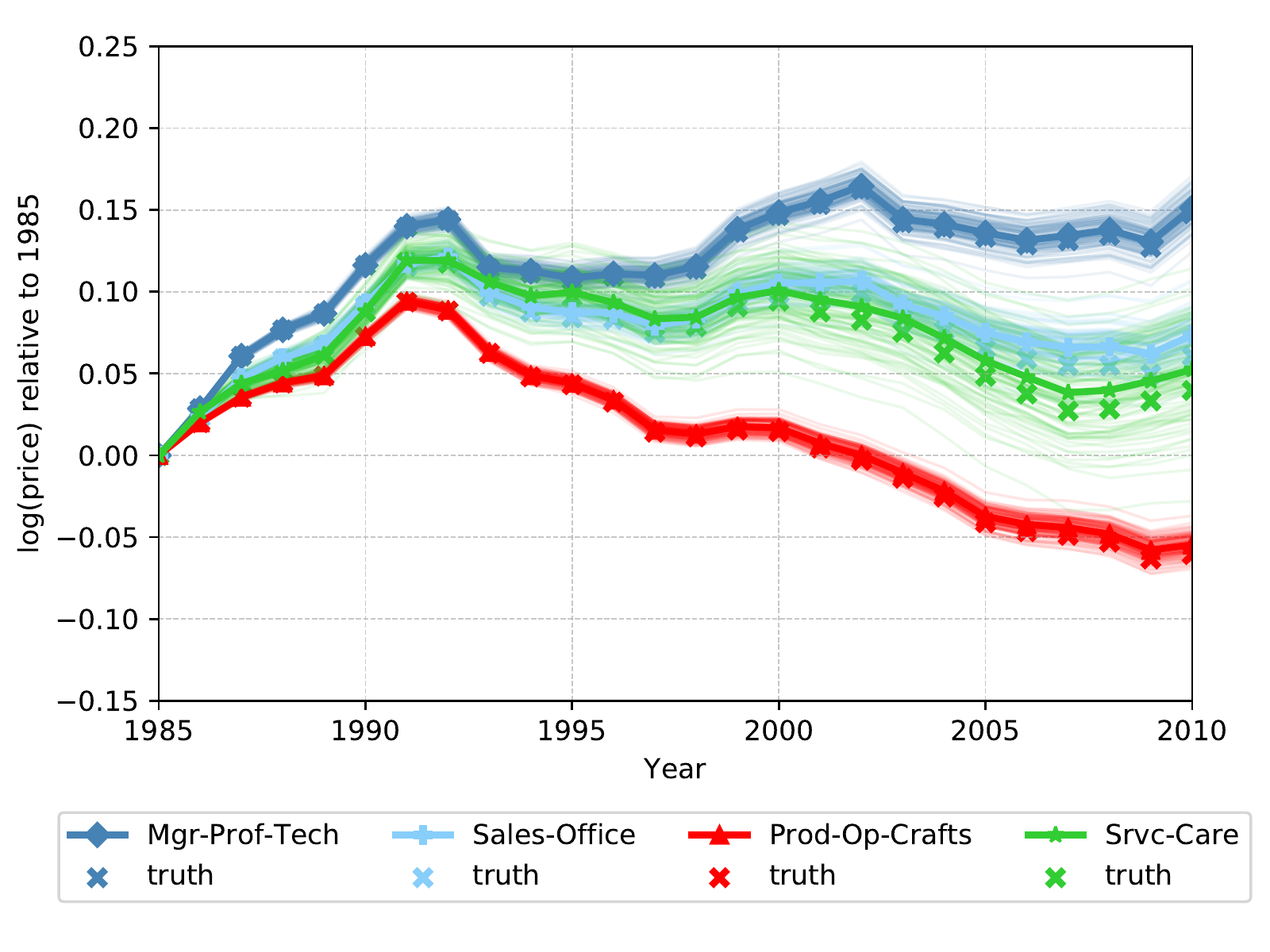}
  \end{minipage}
  \begin{minipage}[t]{\panelwidth}
    \subcaption{Skill accumulation, saturated OLS}
    \label{mc:fig:estimation-ols-moderate-shocks-skills}
    \includegraphics[width=\textwidth]{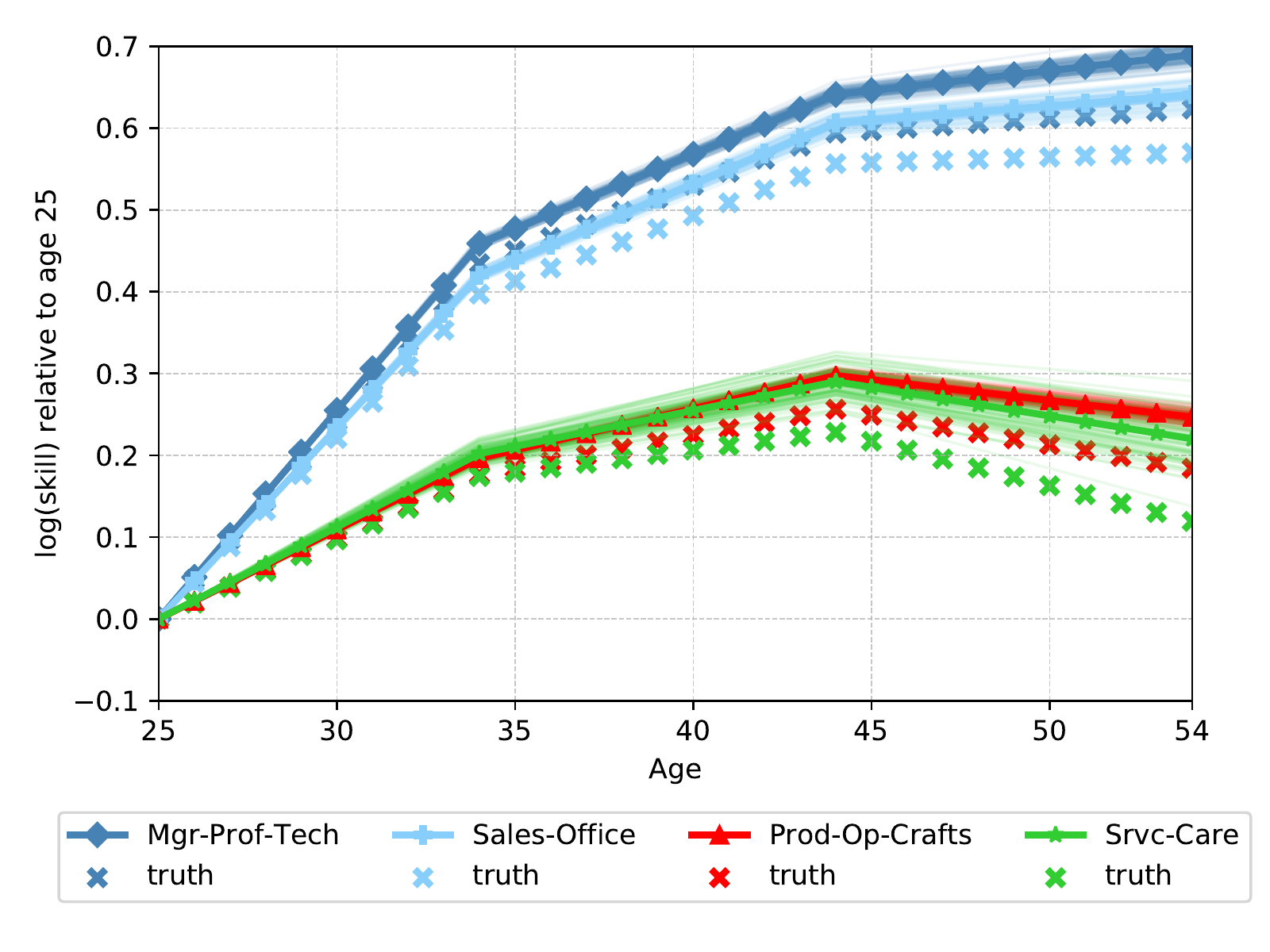}
  \end{minipage}
  \begin{minipage}[t]{\panelwidth}
    \subcaption{Cumulative prices, IV}
    \label{mc:fig:estimation-iv2-moderate-shocks-prices}
    \includegraphics[width=\textwidth]{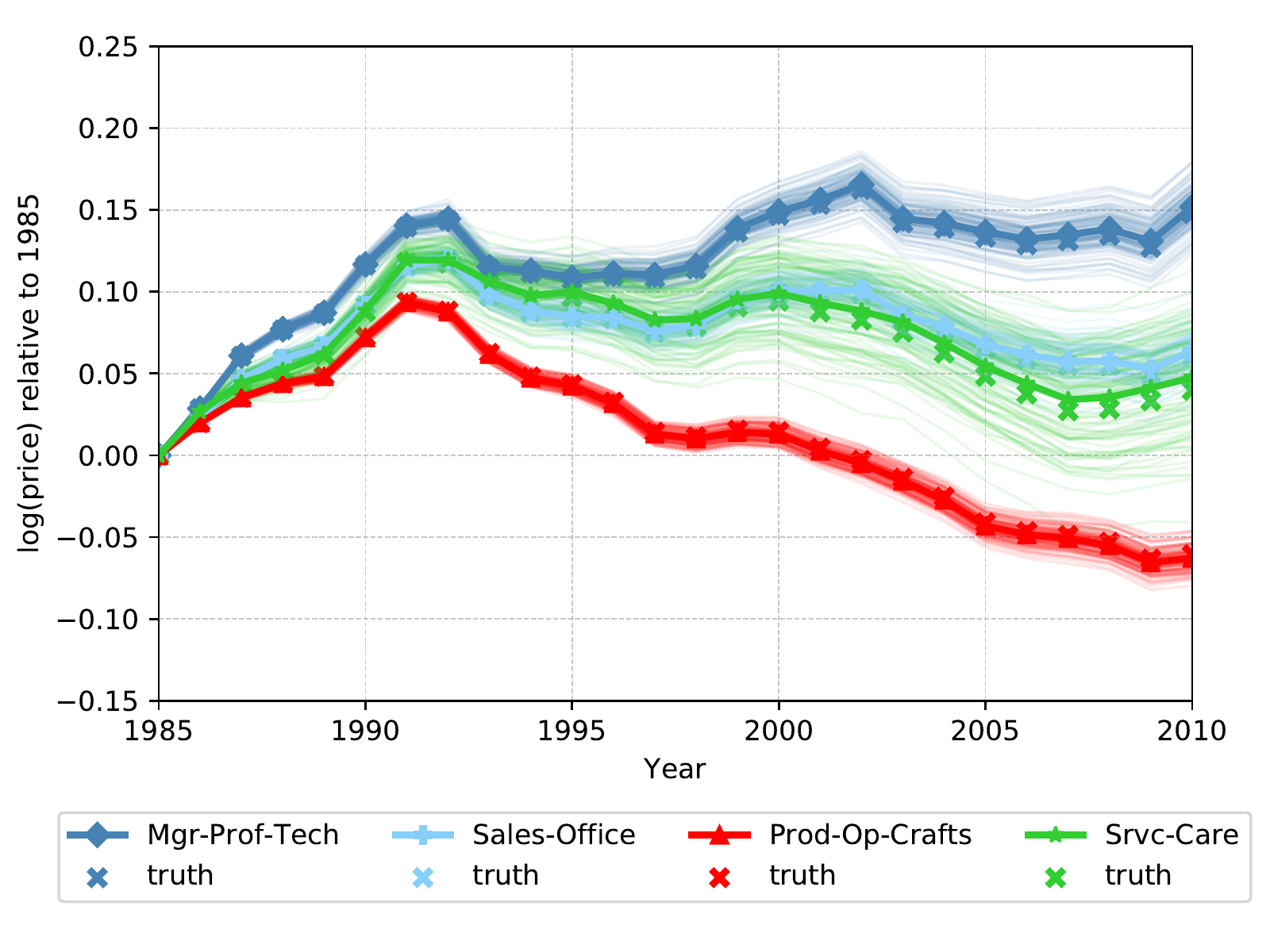}
  \end{minipage}
  \begin{minipage}[t]{\panelwidth}
    \subcaption{Skill accumulation, IV}
    \label{mc:fig:estimation-iv2-moderate-shocks-skills}
    \includegraphics[width=\textwidth]{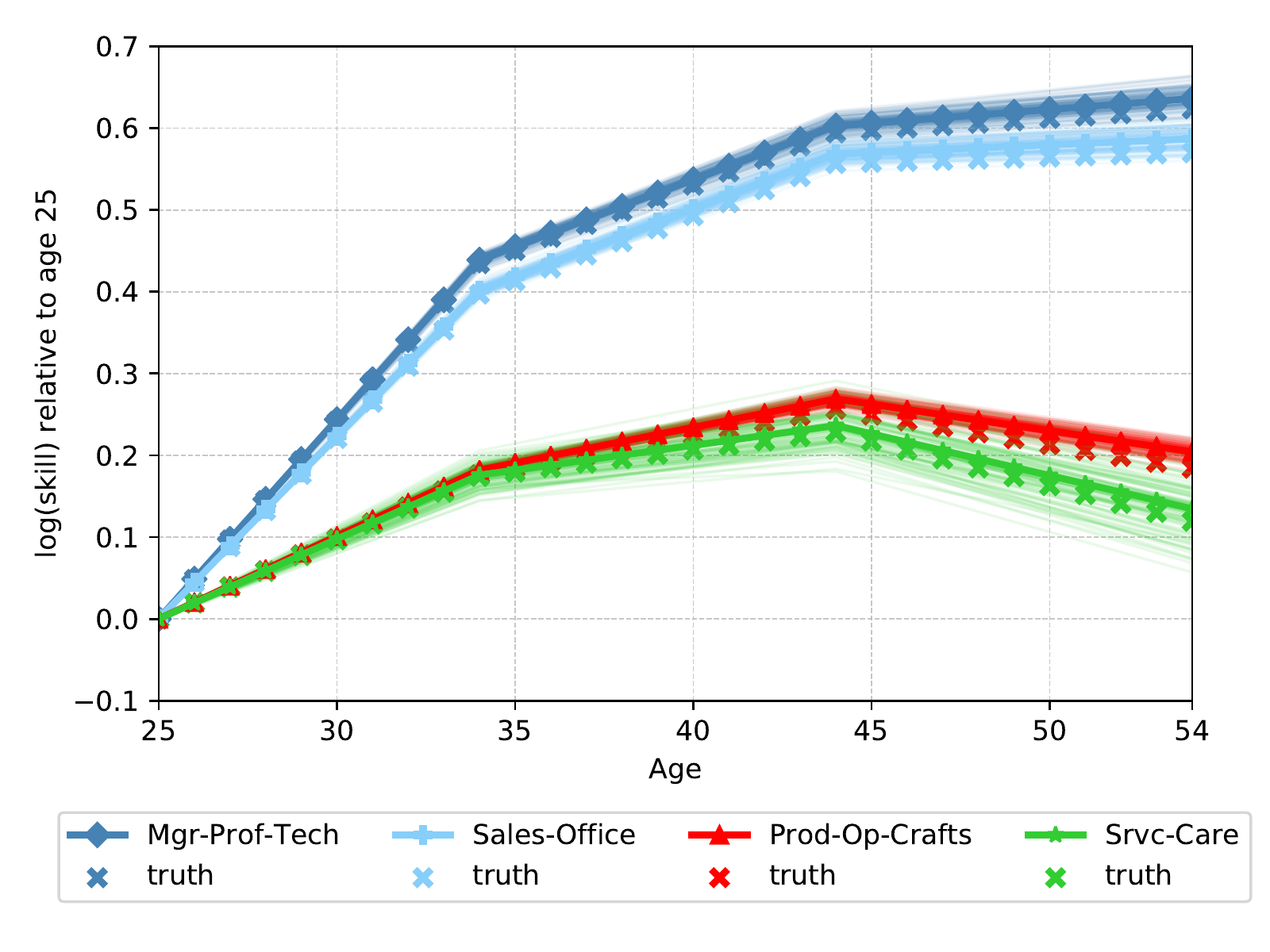}
  \end{minipage}
  \begin{minipage}{\textwidth}
    \scriptsize
    \emph{Notes:} \mcresults {} \olsresultsmc \ivresults
  \end{minipage}
\end{figure}

Figure~\ref{mc:fig:estimation-moderate-shocks} shows that, as predicted in Section~\ref{mc:sub:ols_estimation_strategy}, the OLS estimates contain modest upward bias of stayers' skill accumulation coefficients, whereas the IV estimates are almost exactly on target. Also as expected, the cross-accumulation parameters are generally upward-biased in the OLS (Table~\ref{mc:tab:estimation-ols-moderate-shocks-skill-acc-coeffs}); and in the IV with weak instruments, they are large in absolute values (Table~\ref{mc:tab:estimation-iv2-moderate-shocks-skill-acc-coeffs}). Both sets of skill price estimates track the evolution of their actual values very closely on average (individual experiments in the shimmering lines do modestly vary around the truth).

\begin{figure}[ht!]
  \caption{Descriptives, highly dispersed shocks}
  \label{mc:fig:descriptives-vlarge-shocks}
  \centering
  \begin{minipage}[t]{\panelwidth}
    \subcaption{Occupation entrants/incumbents in $t+1$}
    \label{mc:fig:descriptives-vlarge-shocks-switchers-entrants}
    \includegraphics[width=\textwidth]{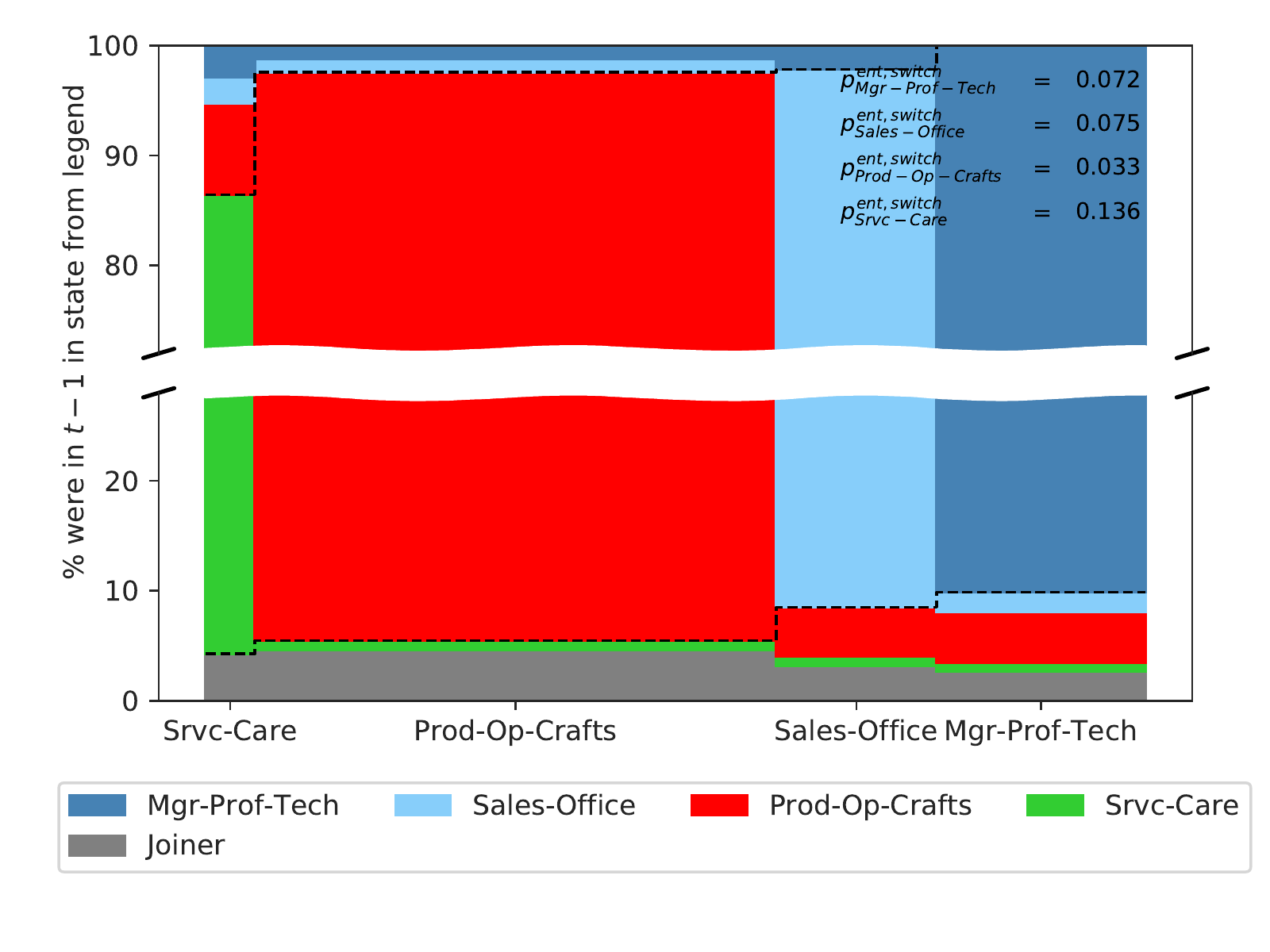}
  \end{minipage}
  \begin{minipage}[t]{\panelwidth}
    \subcaption{Occupation leavers/stayers in $t-1$}
    \label{mc:fig:descriptives-vlarge-shocks-switchers-leavers}
    \includegraphics[width=\textwidth]{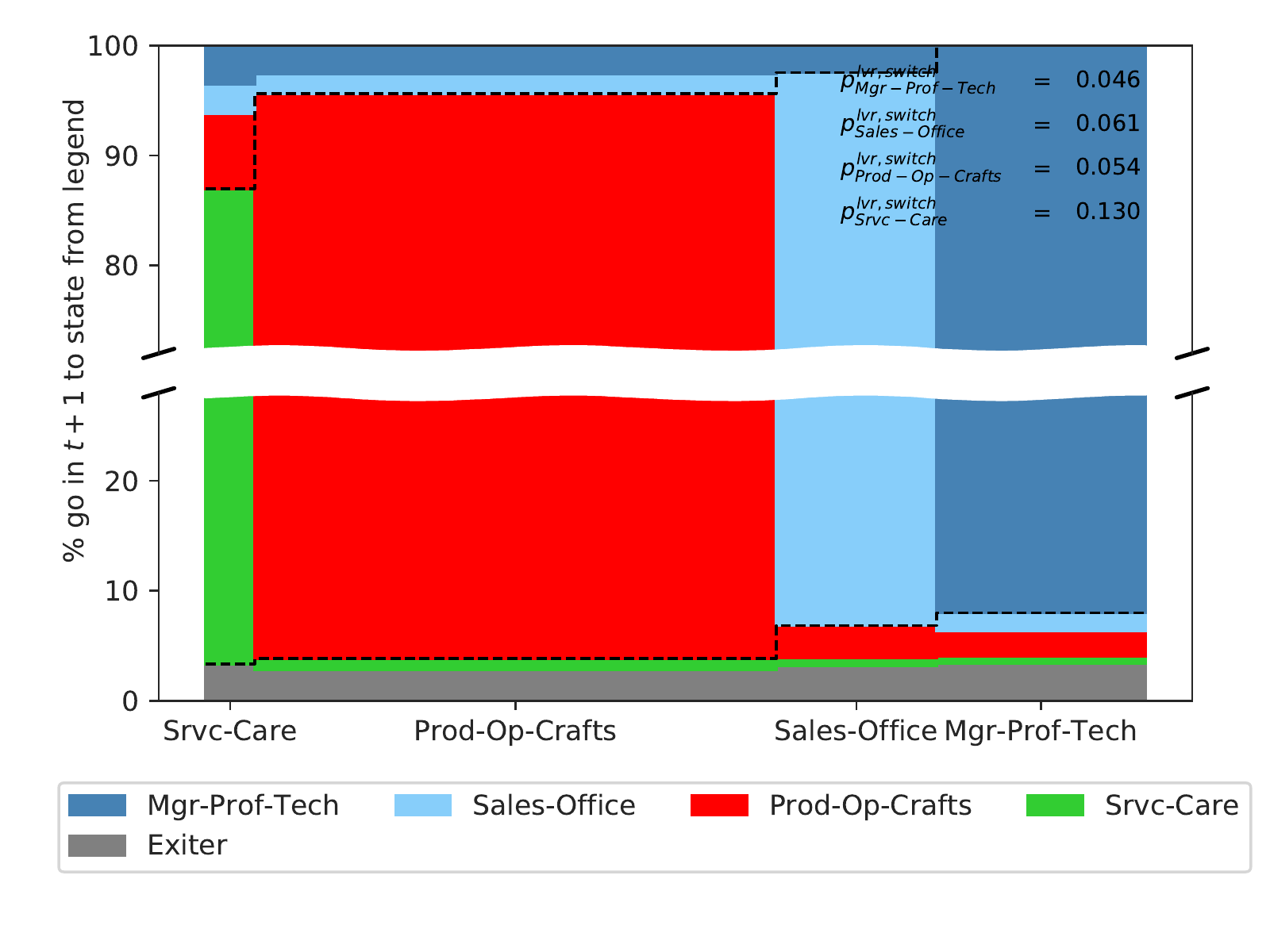}
  \end{minipage}
  \begin{minipage}[t]{\panelwidth}
    \subcaption{Distribution of annual wage growth}
    \label{mc:fig:descriptives-vlarge-shocks-wage-growth-distribution}
    \includegraphics[width=\textwidth]{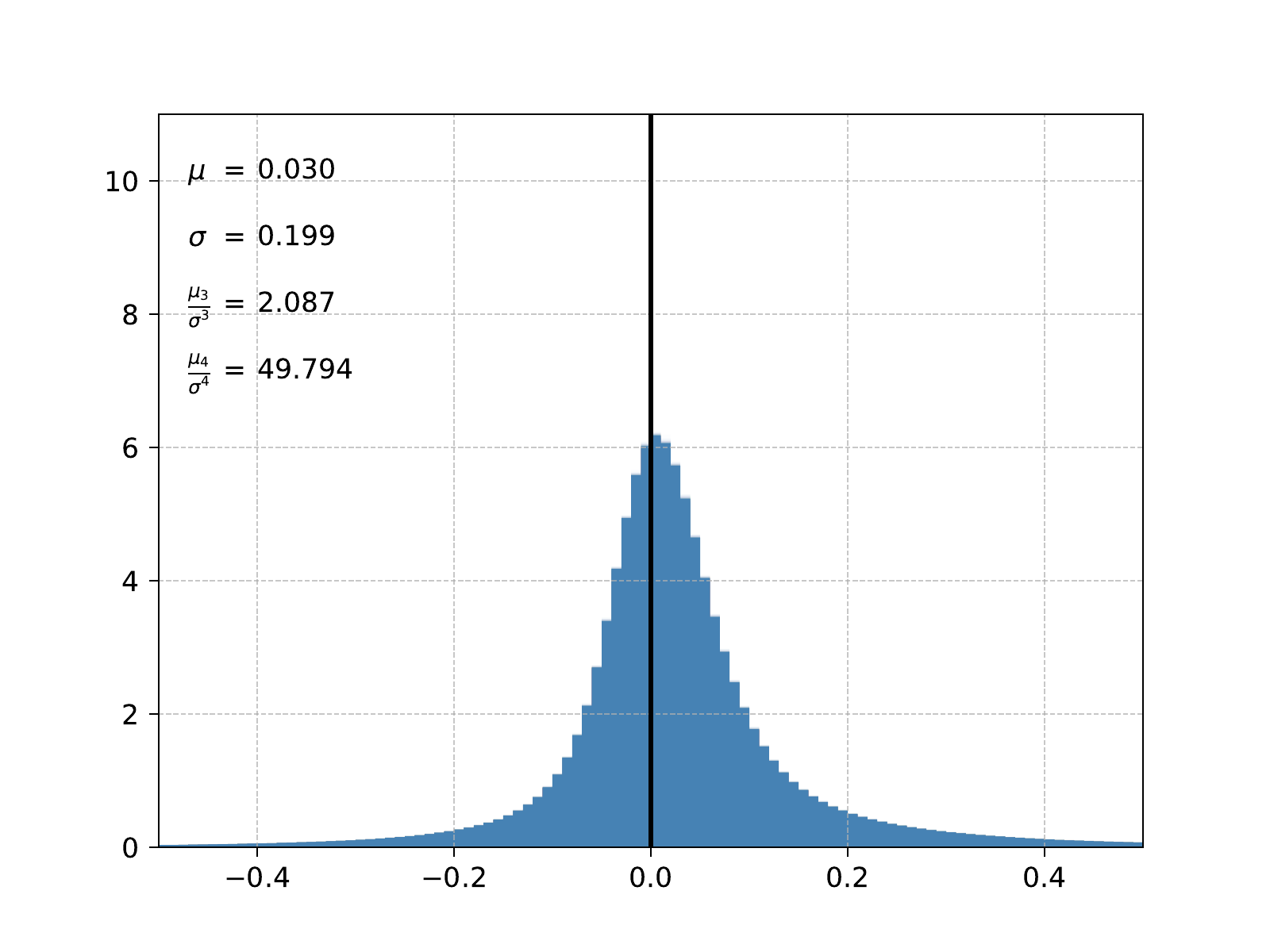}
  \end{minipage}
  \begin{minipage}[t]{\panelwidth}
    \subcaption{Evolution of the wage distribution}
    \label{mc:fig:descriptives-vlarge-shocks-evolution-wage-inequality}
    \vskip2ex
    \includegraphics[width=\textwidth]{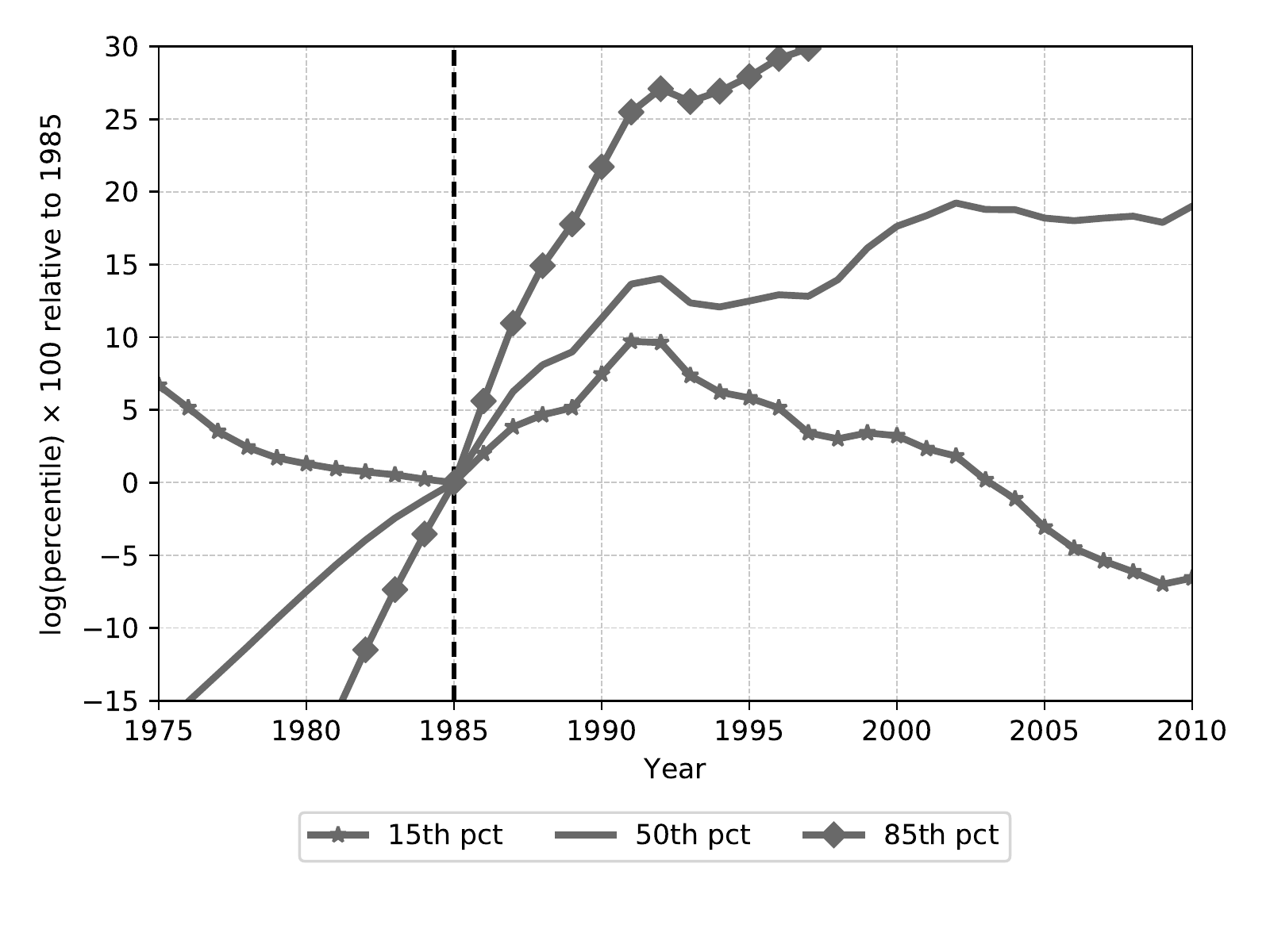}
  \end{minipage}
  \begin{minipage}{\textwidth}
    \scriptsize
    \emph{Notes:} \mcdescriptives
  \end{minipage}
\end{figure}

Next, we increase the standard deviation of $u_{k, i, t}$ to 1.5 times the standard deviation of log wage growth observed in the SIAB data. Figure~\ref{mc:fig:descriptives-vlarge-shocks} shows that he number of switches in the Monte Carlo sample increase a lot and it is now much larger than in the actual SIAB data. Additionally, the simulated distribution of log wage growth and overall wage inequality are much more dispersed compared to the observed distributions. So this is clearly an extreme setting, which we create to see whether the bias due to skill shocks can in some instances become substantial. The OLS estimates in Figure~\ref{mc:fig:estimation-vlarge-shocks} show that the bias in the skill accumulation parameters indeed becomes large. However, skill price changes are only slightly downward biased and not far off their targets (the bias is a bit larger for Mgr-Prof-Tech).

In Panels~\ref{mc:fig:estimation-iv2-vlarge-shocks-prices} and \ref{mc:fig:estimation-iv2-vlarge-shocks-skills}, we implement the instrumental variables strategy that was outlined in the main text in order to deal with the remaining bias due to idiosyncratic skill shocks. It turns out that IV does indeed estimate the correct skill price changes even in this extreme case. The skill accumulation parameters are still upward biased, but less so than in the OLS. Therefore, the instrumental variables estimation is robust even to rather extreme idiosyncratic skill shocks. It seems that implementing the IV will be helpful in practice to check whether the (main) OLS estimates might be biased because of such large shocks.

Finally, we make skill shocks persistent by introducing autocorrelation between shocks in $t$ and $t - x$ as $u_{k, i, t} = 0.3 \cdot u_{k, i, t - 1} + \eta_{k, i , t}$. We calibrate $\sigma = 0.5 \cdot \sigma^{SIAB}_{\Delta log(w_{i, t})}$.\footnote{The evidence on the existence and importance of correlated wage shocks seems to be mixed \citep[e.g., see][and the references therein]{Gibbons1999}.} When workers switch because of a positive skill shocks, they move into an occupation where previous skill shocks tend to be positive as well due to the autocorrelation coefficient $\rho$. The results are displayed in Figure~\ref{mc:fig:estimation-persistent-shocks}. Again, any bias in the skill price estimates for both the OLS and IV is minor, while the structural skill accumulation parameters are unsurprisingly quite off.

\subsubsection{No Idiosyncratic Skill Shocks}
\label{mc:sub:no-shocks}

\begin{table}[h!]
  \centering
  \caption{Parameters}
  \label{mc:tab:no-shocks}
  \begin{minipage}[t]{\linewidth}
    \centering
    \begin{tabular}{lp{8cm}} \toprule 
Parameter & Value \tabularnewline
\midrule
$N$ & 50000 \tabularnewline
Repetitions & 100 \tabularnewline
Skill shocks in $k$ & uniform, $(\mu, \sigma_k) = (0, 0 \cdot \sigma^{SIAB}_{\Delta \log(w_i)})$ \tabularnewline
Stayers accumulation $\gamma_{{k, k, a}}, k' = k $ & $\hat{{\gamma}}^{{SIAB}}_{{k, k, a}} $  \tabularnewline
Cross accumulation $\gamma_{k', k, a}, k' \ne k $ & $\frac{1}{3} \hat{\gamma}^{SIAB}_{k', k, a}$ \tabularnewline
$\rho$ in $\varepsilon_{i, t} = \rho \varepsilon_{i, t-1} + v_{i, t}$ & 0 \tabularnewline
Switching costs $c$ & $ 0 $ \tabularnewline
Amenity trends, $t = 1985,...,2010$ & $  [\Delta \Psi_{k, t}]_{k = 1,...,4} = [0, 0, 0, 0] $ \tabularnewline
\bottomrule
\end{tabular}
  \end{minipage}
\end{table}

\clearpage

\clearpage

\subsubsection{Moderately Dispersed Shocks}
\label{mc:sub:idiosyncratic-shocks-moderate}

\begin{table}[ht!]
  \centering
  \caption{Parameters}
  \label{mc:tab:moderate-shocks}
  \begin{minipage}[t]{\linewidth}
    \centering
    \begin{tabular}{lp{8cm}} \toprule 
Parameter & Value \tabularnewline
\midrule
$N$ & 50000 \tabularnewline
Repetitions & 100 \tabularnewline
Skill shocks in $k$ & observed wage growth distribution, $(\mu, \sigma_k) = (0, 0.5 \cdot \sigma^{SIAB}_{\Delta \log(w_i)})$ \tabularnewline
Stayers accumulation $\gamma_{{k, k, a}}, k' = k $ & $\hat{{\gamma}}^{{SIAB}}_{{k, k, a}} $  \tabularnewline
Cross accumulation $\gamma_{k', k, a}, k' \ne k $ & $\frac{1}{3} \hat{\gamma}^{SIAB}_{k', k, a}$ \tabularnewline
$\rho$ in $\varepsilon_{i, t} = \rho \varepsilon_{i, t-1} + v_{i, t}$ & 0 \tabularnewline
Switching costs $c$ & $ 0 $ \tabularnewline
Amenity trends, $t = 1985,...,2010$ & $  [\Delta \Psi_{k, t}]_{k = 1,...,4} = [0, 0, 0, 0] $ \tabularnewline
\bottomrule
\end{tabular}
  \end{minipage}
\end{table}

\clearpage

\begin{table}[p]
  \centering
  \caption{True and estimated skill accumulation parameters, saturated OLS}
  \label{mc:tab:estimation-ols-moderate-shocks-skill-acc-coeffs}
  \begin{minipage}[t]{\linewidth}
    \centering
    \begin{tabular}{lllrrrrrr}
\toprule & & \multicolumn{7}{c}{Age group} \\ \cmidrule{4-9}
           &           &                   & \multicolumn{2}{l}{[25, 34]} & \multicolumn{2}{l}{[35, 44]} & \multicolumn{2}{l}{[45, 54]} \\
           Previous sector &           Current sector &                   & $\hat{\gamma}_{k', k, a}$ & $\gamma^{true}_{k', k, a}$ & $\hat{\gamma}_{k', k, a}$ & $\gamma^{true}_{k', k, a}$ & $\hat{\gamma}_{k', k, a}$ & $\gamma^{true}_{k', k, a}$ \\
\midrule
Mgr-Prof-Tech  & Mgr-Prof-Tech & $\gamma$ &                     0.051 &                      0.048 &                     0.018 &                      0.016 &                     0.005 &                      0.003 \\
           &           & $\sigma_{\gamma}$ &                     0.000 &                            &                     0.000 &                            &                     0.000 &                            \\
           & Sales-Office & $\gamma$ &                     0.069 &                      0.063 &                     0.002 &                      0.009 &                    -0.045 &                     -0.010 \\
           &           & $\sigma_{\gamma}$ &                     0.009 &                            &                     0.012 &                            &                     0.021 &                            \\
           & Prod-Op-Crafts & $\gamma$ &                     0.031 &                      0.023 &                    -0.032 &                     -0.011 &                    -0.050 &                     -0.022 \\
           &           & $\sigma_{\gamma}$ &                     0.014 &                            &                     0.014 &                            &                     0.017 &                            \\
           & Srvc-Care & $\gamma$ &                    -0.093 &                     -0.008 &                    -0.131 &                     -0.036 &                    -0.031 &                     -0.004 \\
           &           & $\sigma_{\gamma}$ &                     0.061 &                            &                     0.069 &                            &                     0.034 &                            \\ \midrule
Sales-Office  & Mgr-Prof-Tech & $\gamma$ &                     0.098 &                      0.088 &                     0.030 &                      0.027 &                     0.003 &                      0.009 \\
           &           & $\sigma_{\gamma}$ &                     0.006 &                            &                     0.009 &                            &                     0.013 &                            \\
           & Sales-Office & $\gamma$ &                     0.047 &                      0.044 &                     0.019 &                      0.016 &                     0.004 &                      0.001 \\
           &           & $\sigma_{\gamma}$ &                     0.000 &                            &                     0.000 &                            &                     0.000 &                            \\
           & Prod-Op-Crafts & $\gamma$ &                     0.070 &                      0.056 &                     0.032 &                      0.019 &                    -0.008 &                     -0.008 \\
           &           & $\sigma_{\gamma}$ &                     0.007 &                            &                     0.007 &                            &                     0.012 &                            \\
           & Srvc-Care & $\gamma$ &                    -0.042 &                      0.010 &                    -0.125 &                     -0.034 &                    -0.083 &                     -0.024 \\
           &           & $\sigma_{\gamma}$ &                     0.042 &                            &                     0.067 &                            &                     0.065 &                            \\ \midrule
Prod-Op-Crafts  & Mgr-Prof-Tech & $\gamma$ &                     0.086 &                      0.075 &                     0.035 &                      0.042 &                     0.012 &                      0.021 \\
           &           & $\sigma_{\gamma}$ &                     0.004 &                            &                     0.005 &                            &                     0.006 &                            \\
           & Sales-Office & $\gamma$ &                     0.037 &                      0.036 &                     0.017 &                      0.022 &                    -0.014 &                      0.000 \\
           &           & $\sigma_{\gamma}$ &                     0.006 &                            &                     0.006 &                            &                     0.009 &                            \\
           & Prod-Op-Crafts & $\gamma$ &                     0.022 &                      0.020 &                     0.010 &                      0.008 &                    -0.005 &                     -0.007 \\
           &           & $\sigma_{\gamma}$ &                     0.000 &                            &                     0.000 &                            &                     0.000 &                            \\
           & Srvc-Care & $\gamma$ &                    -0.056 &                     -0.017 &                    -0.076 &                     -0.014 &                    -0.038 &                     -0.009 \\
           &           & $\sigma_{\gamma}$ &                     0.022 &                            &                     0.025 &                            &                     0.016 &                            \\ \midrule
Srvc-Care  & Mgr-Prof-Tech & $\gamma$ &                     0.107 &                      0.099 &                     0.063 &                      0.063 &                     0.047 &                      0.041 \\
           &           & $\sigma_{\gamma}$ &                     0.011 &                            &                     0.014 &                            &                     0.014 &                            \\
           & Sales-Office & $\gamma$ &                     0.092 &                      0.090 &                     0.048 &                      0.048 &                     0.014 &                      0.015 \\
           &           & $\sigma_{\gamma}$ &                     0.012 &                            &                     0.016 &                            &                     0.026 &                            \\
           & Prod-Op-Crafts & $\gamma$ &                     0.119 &                      0.106 &                     0.085 &                      0.075 &                     0.055 &                      0.037 \\
           &           & $\sigma_{\gamma}$ &                     0.007 &                            &                     0.008 &                            &                     0.008 &                            \\
           & Srvc-Care & $\gamma$ &                     0.023 &                      0.019 &                     0.009 &                      0.005 &                    -0.007 &                     -0.011 \\
           &           & $\sigma_{\gamma}$ &                     0.001 &                            &                     0.001 &                            &                     0.001 &                            \\
\bottomrule
\end{tabular}

  \end{minipage}
  \begin{minipage}{\linewidth}
    \scriptsize
    \emph{Notes:} \crossgammas \broadgroupsmc {} \olsresultsmc
  \end{minipage}
\end{table}

\clearpage

\begin{table}[p]
  \centering
  \caption{True and estimated skill accumulation parameters, IV}
  \label{mc:tab:estimation-iv2-moderate-shocks-skill-acc-coeffs}
  \begin{minipage}[t]{\linewidth}
    \centering
    \begin{tabular}{lllrrrrrr}
\toprule & & \multicolumn{7}{c}{Age group} \\ \cmidrule{4-9}
           &           &                   & \multicolumn{2}{l}{[25, 34]} & \multicolumn{2}{l}{[35, 44]} & \multicolumn{2}{l}{[45, 54]} \\
           Previous sector &           Current sector &                   & $\hat{\gamma}_{k', k, a}$ & $\gamma^{true}_{k', k, a}$ & $\hat{\gamma}_{k', k, a}$ & $\gamma^{true}_{k', k, a}$ & $\hat{\gamma}_{k', k, a}$ & $\gamma^{true}_{k', k, a}$ \\
\midrule
Mgr-Prof-Tech  & Mgr-Prof-Tech & $\gamma$ &                     0.049 &                      0.048 &                     0.016 &                      0.016 &                     0.003 &                      0.003 \\
           &           & $\sigma_{\gamma}$ &                     0.001 &                            &                     0.000 &                            &                     0.000 &                            \\
           & Sales-Office & $\gamma$ &                     0.173 &                      0.063 &                     0.162 &                      0.009 &                     0.152 &                     -0.010 \\
           &           & $\sigma_{\gamma}$ &                     0.016 &                            &                     0.024 &                            &                     0.038 &                            \\
           & Prod-Op-Crafts & $\gamma$ &                     0.249 &                      0.023 &                     0.210 &                     -0.011 &                     0.192 &                     -0.022 \\
           &           & $\sigma_{\gamma}$ &                     0.030 &                            &                     0.027 &                            &                     0.030 &                            \\
           & Srvc-Care & $\gamma$ &                     0.283 &                     -0.008 &                     0.253 &                     -0.036 &                     0.152 &                     -0.004 \\
           &           & $\sigma_{\gamma}$ &                     0.141 &                            &                     0.134 &                            &                     0.054 &                            \\ \midrule
Sales-Office  & Mgr-Prof-Tech & $\gamma$ &                     0.164 &                      0.088 &                     0.132 &                      0.027 &                     0.123 &                      0.009 \\
           &           & $\sigma_{\gamma}$ &                     0.010 &                            &                     0.017 &                            &                     0.029 &                            \\
           & Sales-Office & $\gamma$ &                     0.045 &                      0.044 &                     0.017 &                      0.016 &                     0.002 &                      0.001 \\
           &           & $\sigma_{\gamma}$ &                     0.001 &                            &                     0.001 &                            &                     0.001 &                            \\
           & Prod-Op-Crafts & $\gamma$ &                     0.171 &                      0.056 &                     0.152 &                      0.019 &                     0.154 &                     -0.008 \\
           &           & $\sigma_{\gamma}$ &                     0.014 &                            &                     0.015 &                            &                     0.025 &                            \\
           & Srvc-Care & $\gamma$ &                     0.228 &                      0.010 &                     0.237 &                     -0.034 &                     0.169 &                     -0.024 \\
           &           & $\sigma_{\gamma}$ &                     0.095 &                            &                     0.134 &                            &                     0.109 &                            \\ \midrule
Prod-Op-Crafts  & Mgr-Prof-Tech & $\gamma$ &                     0.150 &                      0.075 &                     0.125 &                      0.042 &                     0.105 &                      0.021 \\
           &           & $\sigma_{\gamma}$ &                     0.013 &                            &                     0.013 &                            &                     0.015 &                            \\
           & Sales-Office & $\gamma$ &                     0.133 &                      0.036 &                     0.122 &                      0.022 &                     0.109 &                      0.000 \\
           &           & $\sigma_{\gamma}$ &                     0.013 &                            &                     0.011 &                            &                     0.019 &                            \\
           & Prod-Op-Crafts & $\gamma$ &                     0.020 &                      0.020 &                     0.009 &                      0.008 &                    -0.006 &                     -0.007 \\
           &           & $\sigma_{\gamma}$ &                     0.000 &                            &                     0.000 &                            &                     0.000 &                            \\
           & Srvc-Care & $\gamma$ &                     0.222 &                     -0.017 &                     0.180 &                     -0.014 &                     0.119 &                     -0.009 \\
           &           & $\sigma_{\gamma}$ &                     0.071 &                            &                     0.049 &                            &                     0.028 &                            \\ \midrule
Srvc-Care  & Mgr-Prof-Tech & $\gamma$ &                     0.171 &                      0.099 &                     0.134 &                      0.063 &                     0.108 &                      0.041 \\
           &           & $\sigma_{\gamma}$ &                     0.041 &                            &                     0.041 &                            &                     0.025 &                            \\
           & Sales-Office & $\gamma$ &                     0.159 &                      0.090 &                     0.124 &                      0.048 &                     0.100 &                      0.015 \\
           &           & $\sigma_{\gamma}$ &                     0.029 &                            &                     0.047 &                            &                     0.061 &                            \\
           & Prod-Op-Crafts & $\gamma$ &                     0.180 &                      0.106 &                     0.147 &                      0.075 &                     0.113 &                      0.037 \\
           &           & $\sigma_{\gamma}$ &                     0.018 &                            &                     0.018 &                            &                     0.016 &                            \\
           & Srvc-Care & $\gamma$ &                     0.020 &                      0.019 &                     0.006 &                      0.005 &                    -0.010 &                     -0.011 \\
           &           & $\sigma_{\gamma}$ &                     0.002 &                            &                     0.002 &                            &                     0.001 &                            \\
\bottomrule
\end{tabular}

  \end{minipage}
  \begin{minipage}{\linewidth}
    \scriptsize
    \emph{Notes:} \crossgammas \broadgroupsmc \ivresults
  \end{minipage}
\end{table}

\clearpage

\subsubsection{Highly Dispersed Shocks}
\label{mc:sub:idiosyncratic-shocks-vlarge}

\begin{table}[ht!]
  \centering
  \caption{Parameters}
  \label{mc:tab:vlarge-shocks}
  \begin{minipage}[t]{\linewidth}
    \centering
    \begin{tabular}{lp{8cm}} \toprule 
Parameter & Value \tabularnewline
\midrule
$N$ & 50000 \tabularnewline
Repetitions & 100 \tabularnewline
Skill shocks in $k$ & observed wage growth distribution, $(\mu, \sigma_k) = (0, 1.5 \cdot \sigma^{SIAB}_{\Delta \log(w_i)})$ \tabularnewline
Stayers accumulation $\gamma_{{k, k, a}}, k' = k $ & $\hat{{\gamma}}^{{SIAB}}_{{k, k, a}} $  \tabularnewline
Cross accumulation $\gamma_{k', k, a}, k' \ne k $ & $\frac{1}{3} \hat{\gamma}^{SIAB}_{k', k, a}$ \tabularnewline
$\rho$ in $\varepsilon_{i, t} = \rho \varepsilon_{i, t-1} + v_{i, t}$ & 0 \tabularnewline
Switching costs $c$ & $ 0 $ \tabularnewline
Amenity trends, $t = 1985,...,2010$ & $  [\Delta \Psi_{k, t}]_{k = 1,...,4} = [0, 0, 0, 0] $ \tabularnewline
\bottomrule
\end{tabular}
  \end{minipage}
\end{table}

\begin{figure}[ht!]
  \caption{Estimation results, highly dispersed shocks}
  \label{mc:fig:estimation-vlarge-shocks}
  \centering
  \begin{minipage}[t]{\panelwidth}
    \subcaption{Cumulative prices, saturated OLS}
    \label{mc:fig:estimation-ols-vlarge-shocks-prices}
    \includegraphics[width=\textwidth]{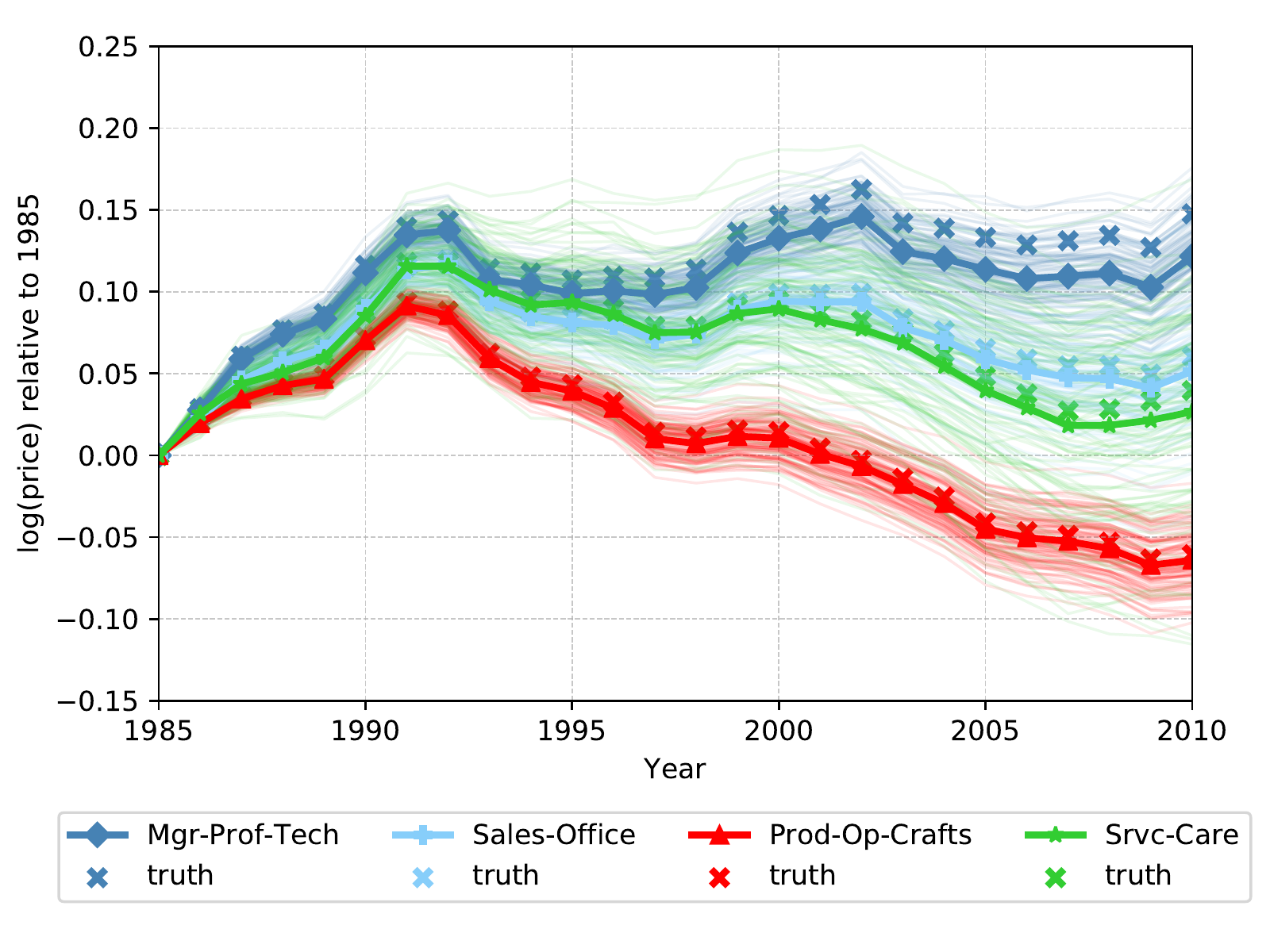}
  \end{minipage}
  \begin{minipage}[t]{\panelwidth}
    \subcaption{Skill accumulation, saturated OLS}
    \label{mc:fig:estimation-ols-vlarge-shocks-skills}
    \includegraphics[width=\textwidth]{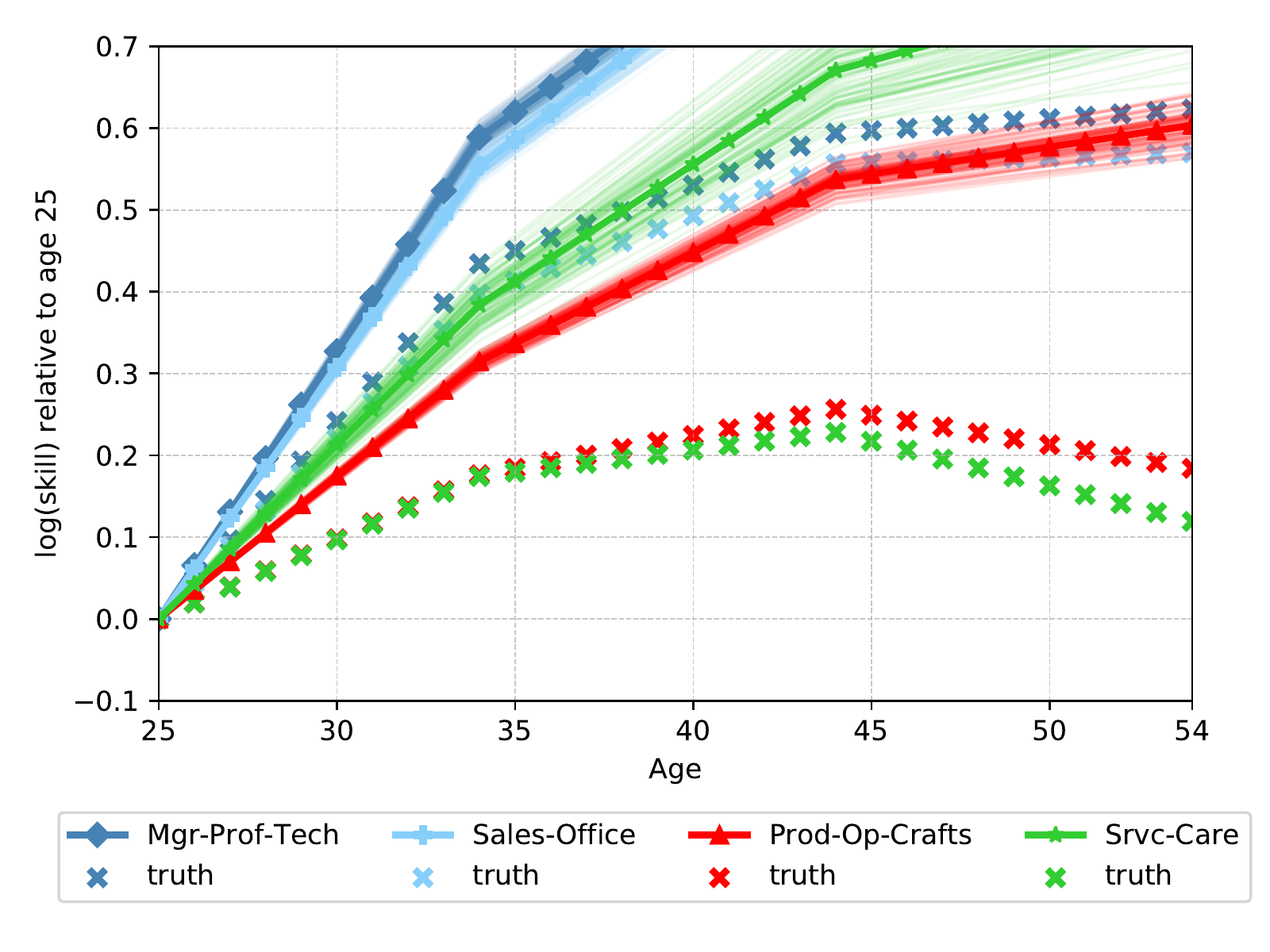}
  \end{minipage}
  \begin{minipage}[t]{\panelwidth}
    \subcaption{Cumulative prices, IV}
    \label{mc:fig:estimation-iv2-vlarge-shocks-prices}
    \includegraphics[width=\textwidth]{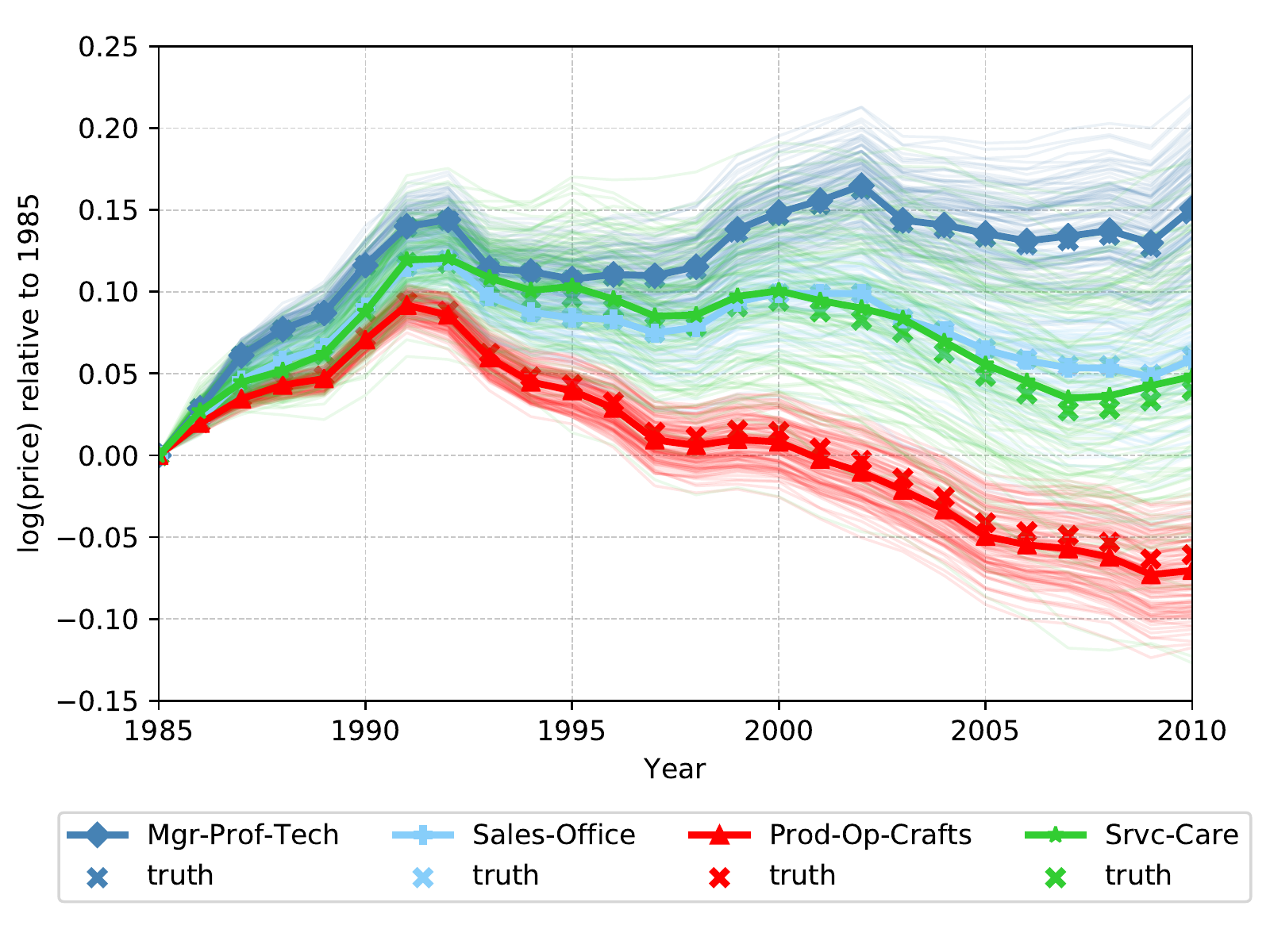}
  \end{minipage}
  \begin{minipage}[t]{\panelwidth}
    \subcaption{Skill accumulation, IV}
    \label{mc:fig:estimation-iv2-vlarge-shocks-skills}
    \includegraphics[width=\textwidth]{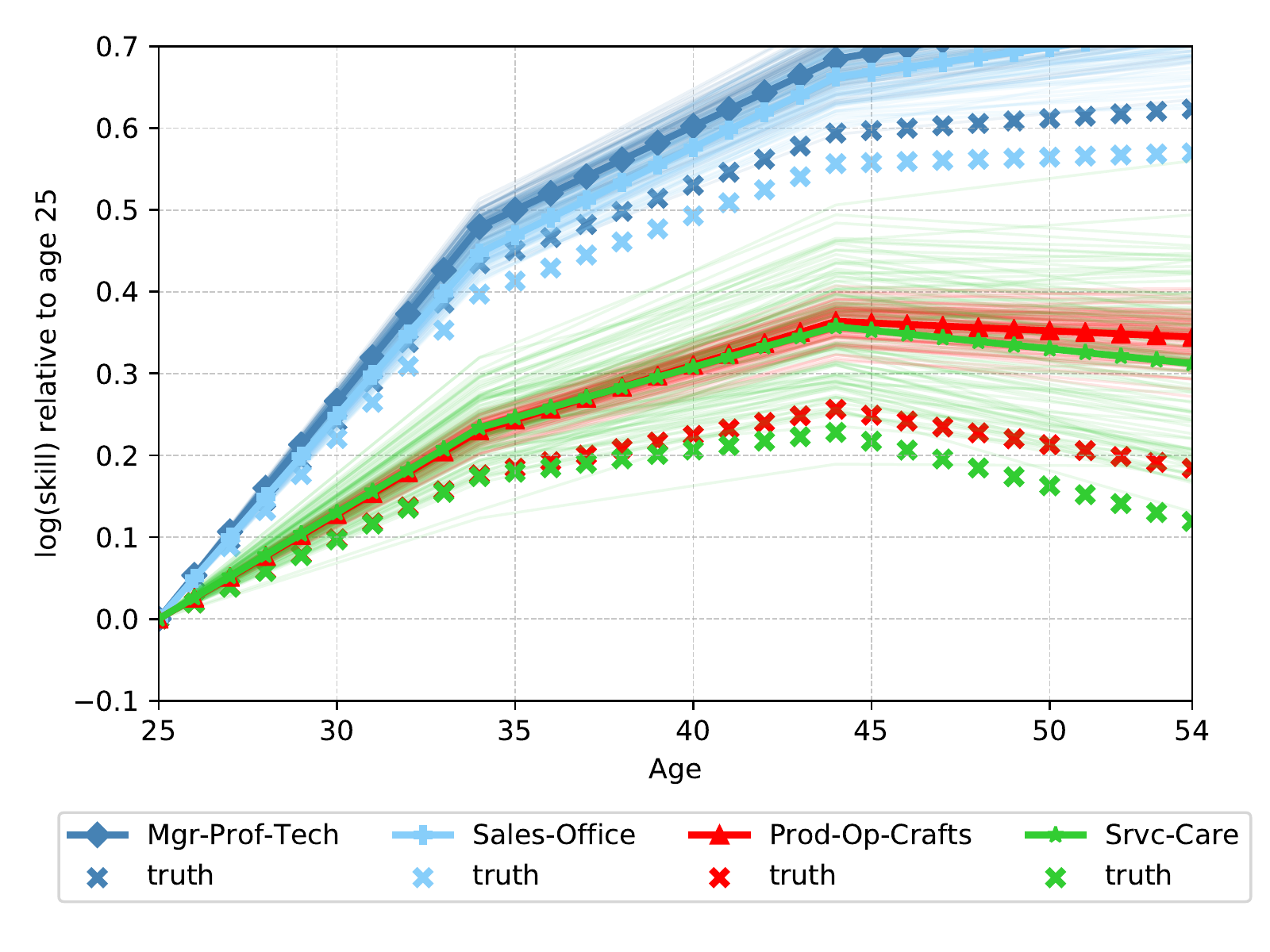}
  \end{minipage}
  \begin{minipage}{\textwidth}
    \scriptsize
    \emph{Notes:} \mcresults {} \olsresultsmc \ivresults
  \end{minipage}
\end{figure}

\clearpage

\begin{table}[h!]
  \centering
  \caption{True and estimated skill accumulation parameters, saturated OLS}
  \label{mc:tab:estimation-ols-vlarge-shocks-skill-acc-coeffs}
  \begin{minipage}[t]{\linewidth}
    \centering
    \begin{tabular}{lllrrrrrr}
\toprule & & \multicolumn{7}{c}{Age group} \\ \cmidrule{4-9}
           &           &                   & \multicolumn{2}{l}{[25, 34]} & \multicolumn{2}{l}{[35, 44]} & \multicolumn{2}{l}{[45, 54]} \\
           Previous sector &           Current sector &                   & $\hat{\gamma}_{k', k, a}$ & $\gamma^{true}_{k', k, a}$ & $\hat{\gamma}_{k', k, a}$ & $\gamma^{true}_{k', k, a}$ & $\hat{\gamma}_{k', k, a}$ & $\gamma^{true}_{k', k, a}$ \\
\midrule
Mgr-Prof-Tech  & Mgr-Prof-Tech & $\gamma$ &                     0.065 &                      0.048 &                     0.031 &                      0.016 &                     0.017 &                      0.003 \\
           &           & $\sigma_{\gamma}$ &                     0.001 &                            &                     0.001 &                            &                     0.001 &                            \\
           & Sales-Office & $\gamma$ &                     0.113 &                      0.063 &                     0.037 &                      0.009 &                    -0.001 &                     -0.010 \\
           &           & $\sigma_{\gamma}$ &                     0.022 &                            &                     0.023 &                            &                     0.031 &                            \\
           & Prod-Op-Crafts & $\gamma$ &                     0.097 &                      0.023 &                     0.030 &                     -0.011 &                     0.002 &                     -0.022 \\
           &           & $\sigma_{\gamma}$ &                     0.020 &                            &                     0.019 &                            &                     0.023 &                            \\
           & Srvc-Care & $\gamma$ &                     0.038 &                     -0.008 &                    -0.008 &                     -0.036 &                     0.010 &                     -0.004 \\
           &           & $\sigma_{\gamma}$ &                     0.052 &                            &                     0.053 &                            &                     0.043 &                            \\ \midrule
Sales-Office  & Mgr-Prof-Tech & $\gamma$ &                     0.130 &                      0.088 &                     0.069 &                      0.027 &                     0.038 &                      0.009 \\
           &           & $\sigma_{\gamma}$ &                     0.018 &                            &                     0.020 &                            &                     0.026 &                            \\
           & Sales-Office & $\gamma$ &                     0.062 &                      0.044 &                     0.032 &                      0.016 &                     0.016 &                      0.001 \\
           &           & $\sigma_{\gamma}$ &                     0.001 &                            &                     0.001 &                            &                     0.001 &                            \\
           & Prod-Op-Crafts & $\gamma$ &                     0.116 &                      0.056 &                     0.073 &                      0.019 &                     0.032 &                     -0.008 \\
           &           & $\sigma_{\gamma}$ &                     0.018 &                            &                     0.017 &                            &                     0.023 &                            \\
           & Srvc-Care & $\gamma$ &                     0.058 &                      0.010 &                     0.010 &                     -0.034 &                     0.021 &                     -0.024 \\
           &           & $\sigma_{\gamma}$ &                     0.044 &                            &                     0.056 &                            &                     0.060 &                            \\ \midrule
Prod-Op-Crafts  & Mgr-Prof-Tech & $\gamma$ &                     0.103 &                      0.075 &                     0.052 &                      0.042 &                     0.026 &                      0.021 \\
           &           & $\sigma_{\gamma}$ &                     0.011 &                            &                     0.013 &                            &                     0.016 &                            \\
           & Sales-Office & $\gamma$ &                     0.062 &                      0.036 &                     0.036 &                      0.022 &                     0.008 &                      0.000 \\
           &           & $\sigma_{\gamma}$ &                     0.015 &                            &                     0.016 &                            &                     0.021 &                            \\
           & Prod-Op-Crafts & $\gamma$ &                     0.035 &                      0.020 &                     0.022 &                      0.008 &                     0.007 &                     -0.007 \\
           &           & $\sigma_{\gamma}$ &                     0.001 &                            &                     0.001 &                            &                     0.001 &                            \\
           & Srvc-Care & $\gamma$ &                     0.008 &                     -0.017 &                     0.000 &                     -0.014 &                     0.002 &                     -0.009 \\
           &           & $\sigma_{\gamma}$ &                     0.026 &                            &                     0.027 &                            &                     0.026 &                            \\ \midrule
Srvc-Care  & Mgr-Prof-Tech & $\gamma$ &                     0.154 &                      0.099 &                     0.121 &                      0.063 &                     0.107 &                      0.041 \\
           &           & $\sigma_{\gamma}$ &                     0.025 &                            &                     0.027 &                            &                     0.028 &                            \\
           & Sales-Office & $\gamma$ &                     0.143 &                      0.090 &                     0.116 &                      0.048 &                     0.094 &                      0.015 \\
           &           & $\sigma_{\gamma}$ &                     0.027 &                            &                     0.033 &                            &                     0.043 &                            \\
           & Prod-Op-Crafts & $\gamma$ &                     0.159 &                      0.106 &                     0.129 &                      0.075 &                     0.101 &                      0.037 \\
           &           & $\sigma_{\gamma}$ &                     0.016 &                            &                     0.016 &                            &                     0.019 &                            \\
           & Srvc-Care & $\gamma$ &                     0.043 &                      0.019 &                     0.029 &                      0.005 &                     0.012 &                     -0.011 \\
           &           & $\sigma_{\gamma}$ &                     0.002 &                            &                     0.002 &                            &                     0.002 &                            \\
\bottomrule
\end{tabular}

  \end{minipage}
  \begin{minipage}{\textwidth}
    \scriptsize
    \emph{Notes:} \crossgammas \broadgroupsmc {} \olsresultsmc
  \end{minipage}
\end{table}

\begin{table}[h!]
  \centering
  \caption{True and estimated skill accumulation parameters, IV}
  \label{mc:tab:estimation-iv2-vlarge-shocks-skill-acc-coeffs}
  \begin{minipage}[t]{\linewidth}
    \centering
    \begin{tabular}{lllrrrrrr}
\toprule & & \multicolumn{7}{c}{Age group} \\ \cmidrule{4-9}
           &           &                   & \multicolumn{2}{l}{[25, 34]} & \multicolumn{2}{l}{[35, 44]} & \multicolumn{2}{l}{[45, 54]} \\
           Previous sector &           Current sector &                   & $\hat{\gamma}_{k', k, a}$ & $\gamma^{true}_{k', k, a}$ & $\hat{\gamma}_{k', k, a}$ & $\gamma^{true}_{k', k, a}$ & $\hat{\gamma}_{k', k, a}$ & $\gamma^{true}_{k', k, a}$ \\
\midrule
Mgr-Prof-Tech  & Mgr-Prof-Tech & $\gamma$ &                     0.053 &                      0.048 &                     0.020 &                      0.016 &                     0.007 &                      0.003 \\
           &           & $\sigma_{\gamma}$ &                     0.002 &                            &                     0.002 &                            &                     0.002 &                            \\
           & Sales-Office & $\gamma$ &                     0.406 &                      0.063 &                     0.405 &                      0.009 &                     0.403 &                     -0.010 \\
           &           & $\sigma_{\gamma}$ &                     0.050 &                            &                     0.052 &                            &                     0.068 &                            \\
           & Prod-Op-Crafts & $\gamma$ &                     0.477 &                      0.023 &                     0.442 &                     -0.011 &                     0.437 &                     -0.022 \\
           &           & $\sigma_{\gamma}$ &                     0.049 &                            &                     0.044 &                            &                     0.051 &                            \\
           & Srvc-Care & $\gamma$ &                     0.515 &                     -0.008 &                     0.488 &                     -0.036 &                     0.385 &                     -0.004 \\
           &           & $\sigma_{\gamma}$ &                     0.143 &                            &                     0.120 &                            &                     0.083 &                            \\ \midrule
Sales-Office  & Mgr-Prof-Tech & $\gamma$ &                     0.378 &                      0.088 &                     0.371 &                      0.027 &                     0.359 &                      0.009 \\
           &           & $\sigma_{\gamma}$ &                     0.045 &                            &                     0.048 &                            &                     0.062 &                            \\
           & Sales-Office & $\gamma$ &                     0.050 &                      0.044 &                     0.021 &                      0.016 &                     0.007 &                      0.001 \\
           &           & $\sigma_{\gamma}$ &                     0.002 &                            &                     0.002 &                            &                     0.002 &                            \\
           & Prod-Op-Crafts & $\gamma$ &                     0.417 &                      0.056 &                     0.396 &                      0.019 &                     0.393 &                     -0.008 \\
           &           & $\sigma_{\gamma}$ &                     0.043 &                            &                     0.041 &                            &                     0.054 &                            \\
           & Srvc-Care & $\gamma$ &                     0.470 &                      0.010 &                     0.462 &                     -0.034 &                     0.397 &                     -0.024 \\
           &           & $\sigma_{\gamma}$ &                     0.119 &                            &                     0.126 &                            &                     0.130 &                            \\ \midrule
Prod-Op-Crafts  & Mgr-Prof-Tech & $\gamma$ &                     0.353 &                      0.075 &                     0.351 &                      0.042 &                     0.339 &                      0.021 \\
           &           & $\sigma_{\gamma}$ &                     0.035 &                            &                     0.034 &                            &                     0.041 &                            \\
           & Sales-Office & $\gamma$ &                     0.376 &                      0.036 &                     0.368 &                      0.022 &                     0.362 &                      0.000 \\
           &           & $\sigma_{\gamma}$ &                     0.044 &                            &                     0.039 &                            &                     0.049 &                            \\
           & Prod-Op-Crafts & $\gamma$ &                     0.025 &                      0.020 &                     0.013 &                      0.008 &                    -0.002 &                     -0.007 \\
           &           & $\sigma_{\gamma}$ &                     0.001 &                            &                     0.001 &                            &                     0.001 &                            \\
           & Srvc-Care & $\gamma$ &                     0.435 &                     -0.017 &                     0.414 &                     -0.014 &                     0.371 &                     -0.009 \\
           &           & $\sigma_{\gamma}$ &                     0.076 &                            &                     0.060 &                            &                     0.056 &                            \\ \midrule
Srvc-Care  & Mgr-Prof-Tech & $\gamma$ &                     0.363 &                      0.099 &                     0.356 &                      0.063 &                     0.325 &                      0.041 \\
           &           & $\sigma_{\gamma}$ &                     0.080 &                            &                     0.078 &                            &                     0.068 &                            \\
           & Sales-Office & $\gamma$ &                     0.374 &                      0.090 &                     0.360 &                      0.048 &                     0.339 &                      0.015 \\
           &           & $\sigma_{\gamma}$ &                     0.076 &                            &                     0.092 &                            &                     0.114 &                            \\
           & Prod-Op-Crafts & $\gamma$ &                     0.369 &                      0.106 &                     0.351 &                      0.075 &                     0.324 &                      0.037 \\
           &           & $\sigma_{\gamma}$ &                     0.045 &                            &                     0.045 &                            &                     0.051 &                            \\
           & Srvc-Care & $\gamma$ &                     0.026 &                      0.019 &                     0.012 &                      0.005 &                    -0.004 &                     -0.011 \\
           &           & $\sigma_{\gamma}$ &                     0.004 &                            &                     0.004 &                            &                     0.004 &                            \\
\bottomrule
\end{tabular}

  \end{minipage}
  \begin{minipage}{\textwidth}
    \scriptsize
    \emph{Notes:} \crossgammas \broadgroupsmc \ivresults
  \end{minipage}
\end{table}

\clearpage

\subsubsection{Persistent Shocks}
\label{mc:sub:idiosyncratic-shocks-persistent}

\begin{table}[ht!]
  \centering
  \caption{Parameters}
  \label{mc:tab:persistent-shocks}
  \begin{minipage}[t]{\linewidth}
    \centering
    \begin{tabular}{lp{8cm}} \toprule 
Parameter & Value \tabularnewline
\midrule
$N$ & 50000 \tabularnewline
Repetitions & 100 \tabularnewline
Skill shocks in $k$ & observed wage growth distribution, $(\mu, \sigma_k) = (0, 0.5 \cdot \sigma^{SIAB}_{\Delta \log(w_i)})$ \tabularnewline
Stayers accumulation $\gamma_{{k, k, a}}, k' = k $ & $\hat{{\gamma}}^{{SIAB}}_{{k, k, a}} $  \tabularnewline
Cross accumulation $\gamma_{k', k, a}, k' \ne k $ & $\frac{1}{3} \hat{\gamma}^{SIAB}_{k', k, a}$ \tabularnewline
$\rho$ in $\varepsilon_{i, t} = \rho \varepsilon_{i, t-1} + v_{i, t}$ & 0.3 \tabularnewline
Switching costs $c$ & $ 0 $ \tabularnewline
Amenity trends, $t = 1985,...,2010$ & $  [\Delta \Psi_{k, t}]_{k = 1,...,4} = [0, 0, 0, 0] $ \tabularnewline
\bottomrule
\end{tabular}
  \end{minipage}
\end{table}

\begin{figure}[ht!]
  \caption{Descriptives, persistent shocks}
  \label{mc:fig:descriptives-persistent-shocks}
  \centering
  \begin{minipage}[t]{\panelwidth}
    \subcaption{Occupation entrants/incumbents in $t+1$}
    \label{mc:fig:descriptives-persistent-shocks-switchers-entrants}
    \includegraphics[width=\textwidth]{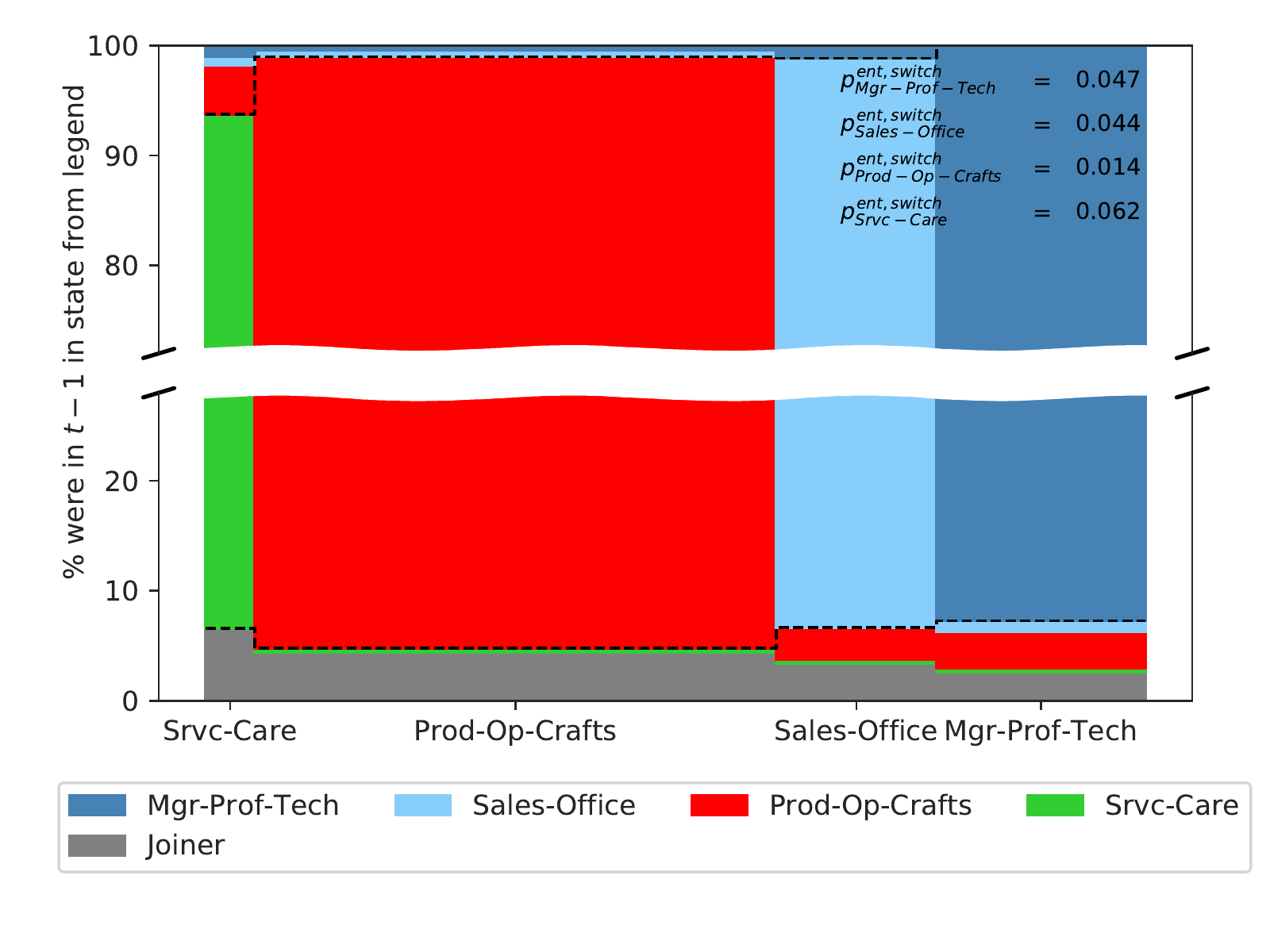}
  \end{minipage}
  \begin{minipage}[t]{\panelwidth}
    \subcaption{Occupation leavers/stayers in $t-1$}
    \label{mc:fig:descriptives-persistent-shocks-switchers-leavers}
    \includegraphics[width=\textwidth]{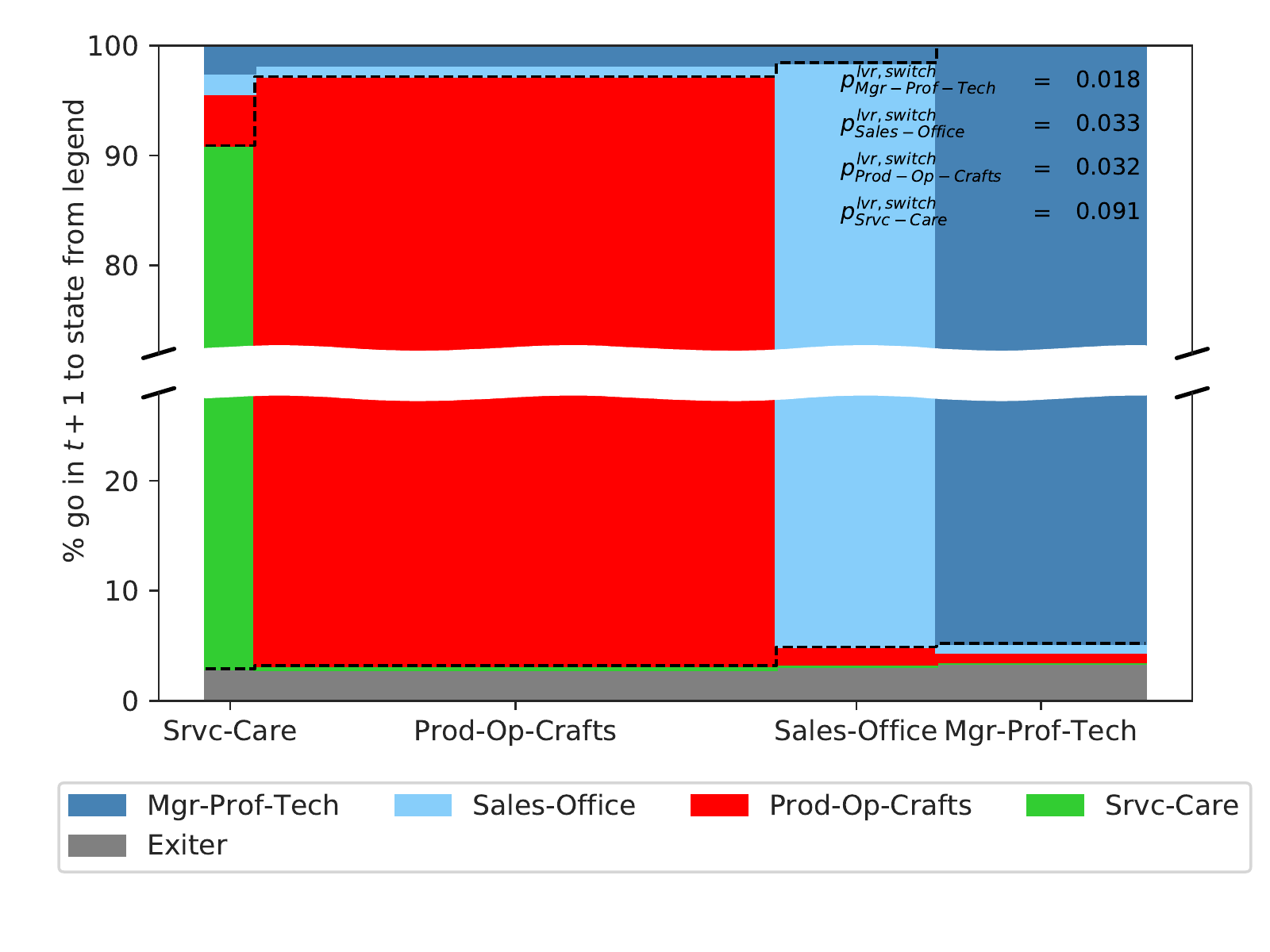}
  \end{minipage}
  \begin{minipage}[t]{\panelwidth}
    \subcaption{Distribution of annual wage growth}
    \label{mc:fig:descriptives-persistent-shocks-wage-growth-distribution}
    \includegraphics[width=\textwidth]{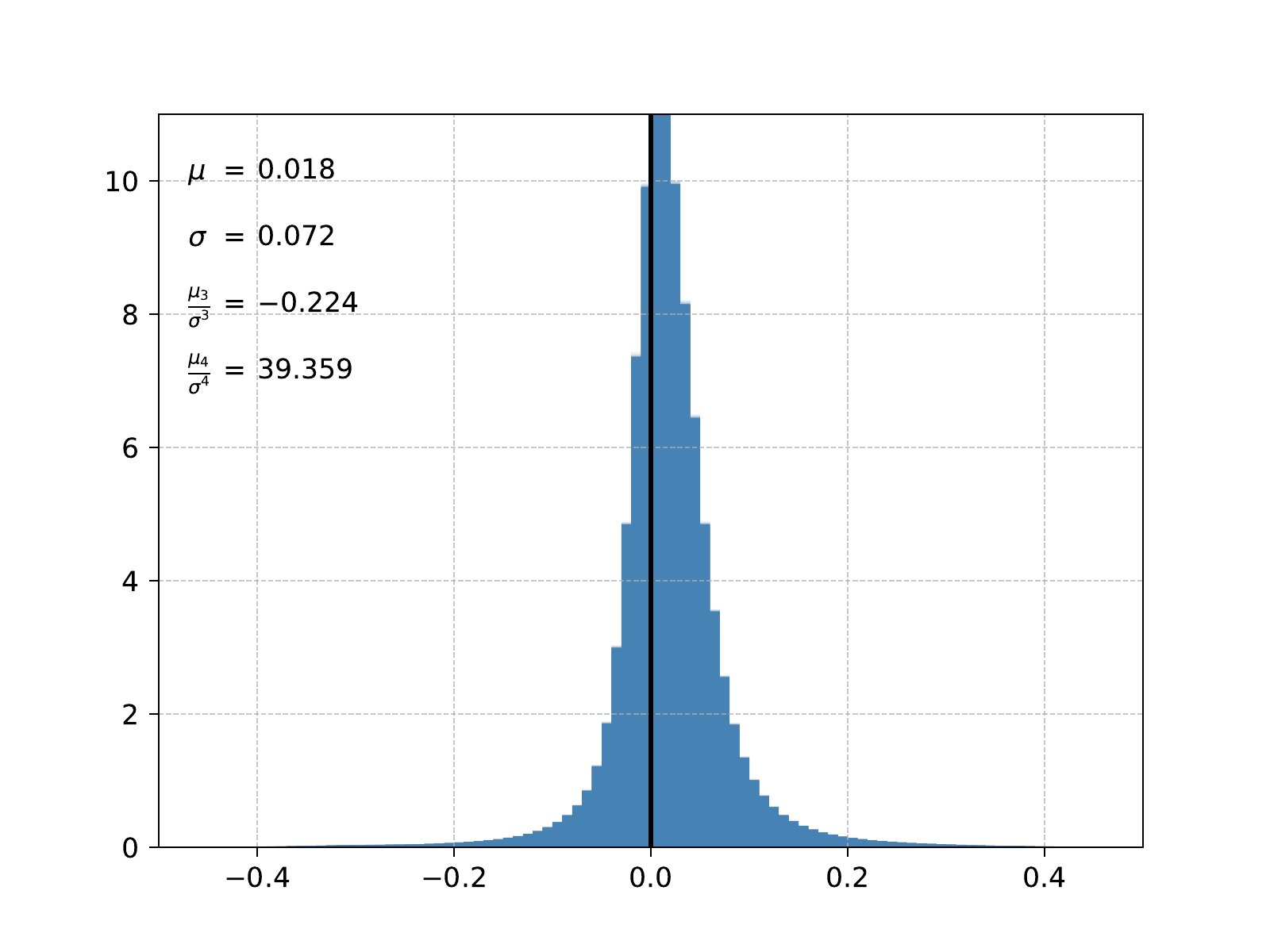}
  \end{minipage}
  \begin{minipage}[t]{\panelwidth}
    \subcaption{Evolution of the wage distribution}
    \label{mc:fig:descriptives-peristent-shocks-evolution-wage-inequality}
    \vskip2ex
    \includegraphics[width=\textwidth]{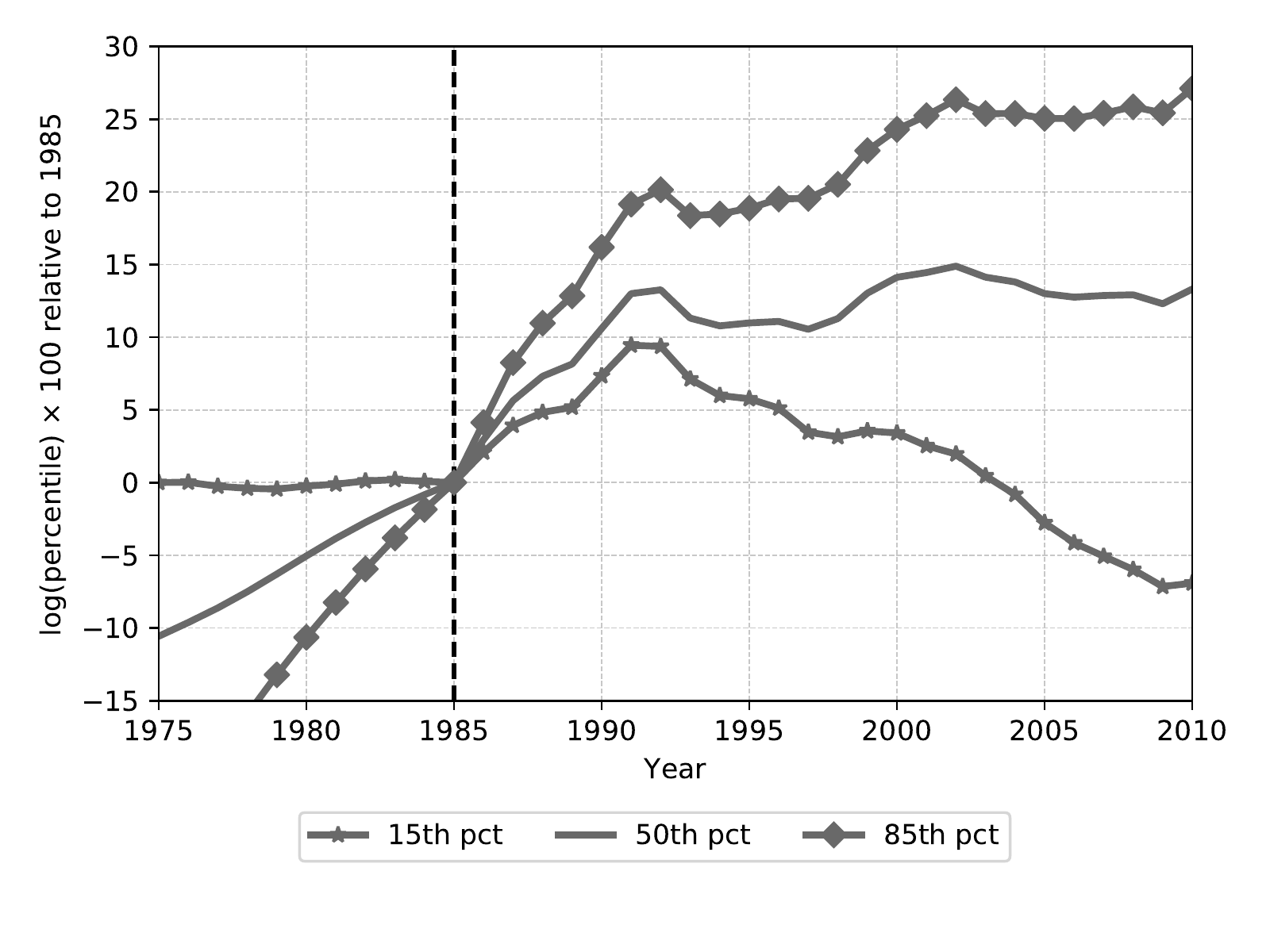}
  \end{minipage}
  \begin{minipage}{\textwidth}
    \scriptsize
    \emph{Notes:} \mcdescriptives
  \end{minipage}
\end{figure}

\begin{figure}[ht!]
  \caption{Estimation results, persistent shocks}
  \label{mc:fig:estimation-persistent-shocks}
  \centering
  \begin{minipage}[t]{\panelwidth}
    \subcaption{Cumulative prices, saturated OLS}
    \label{mc:fig:estimation-ols-persistent-shocks-prices}
    \includegraphics[width=\textwidth]{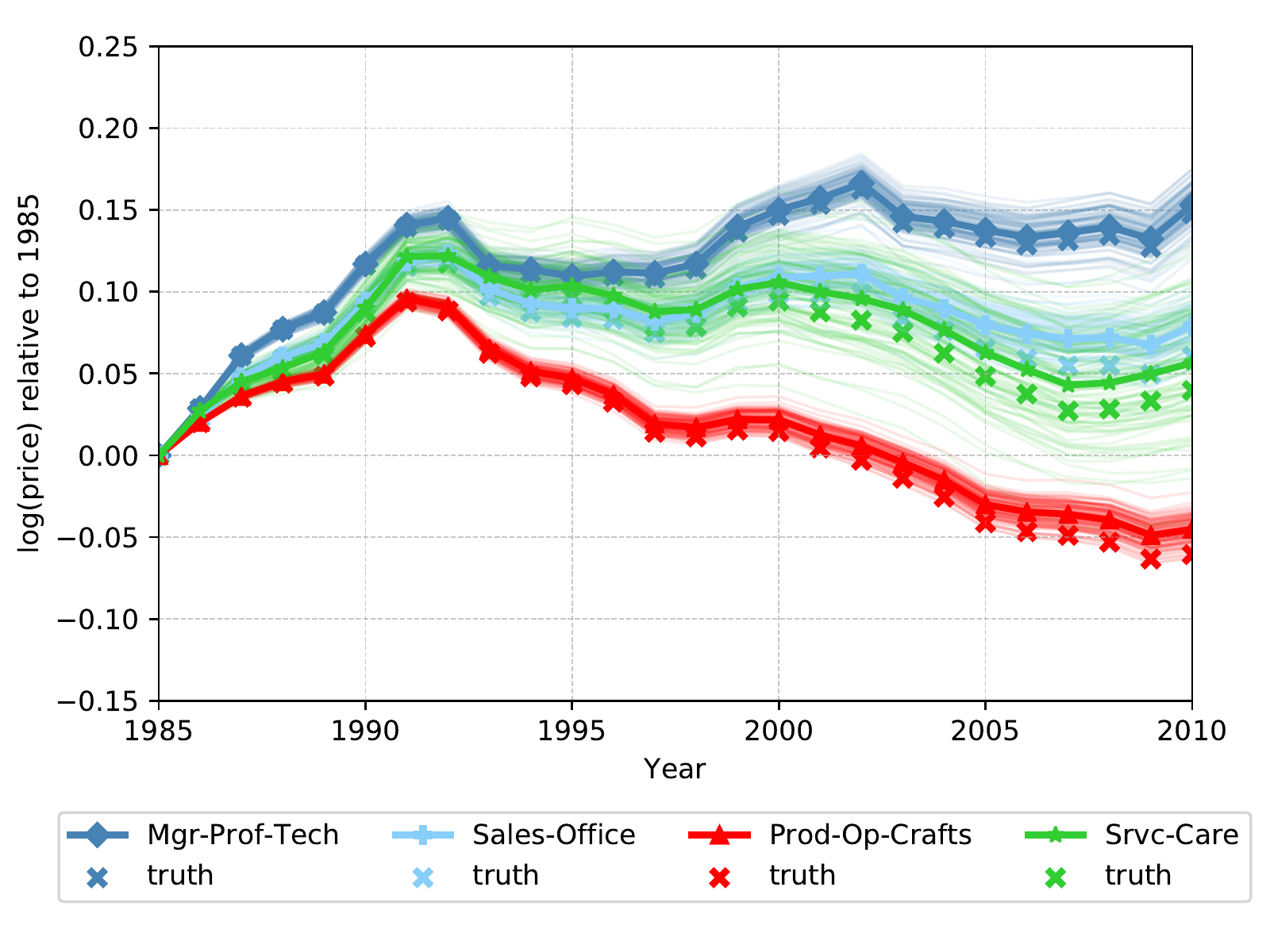}
  \end{minipage}
  \begin{minipage}[t]{\panelwidth}
    \subcaption{Skill accumulation, saturated OLS}
    \label{mc:fig:estimation-ols-persistent-shocks-skills}
    \includegraphics[width=\textwidth]{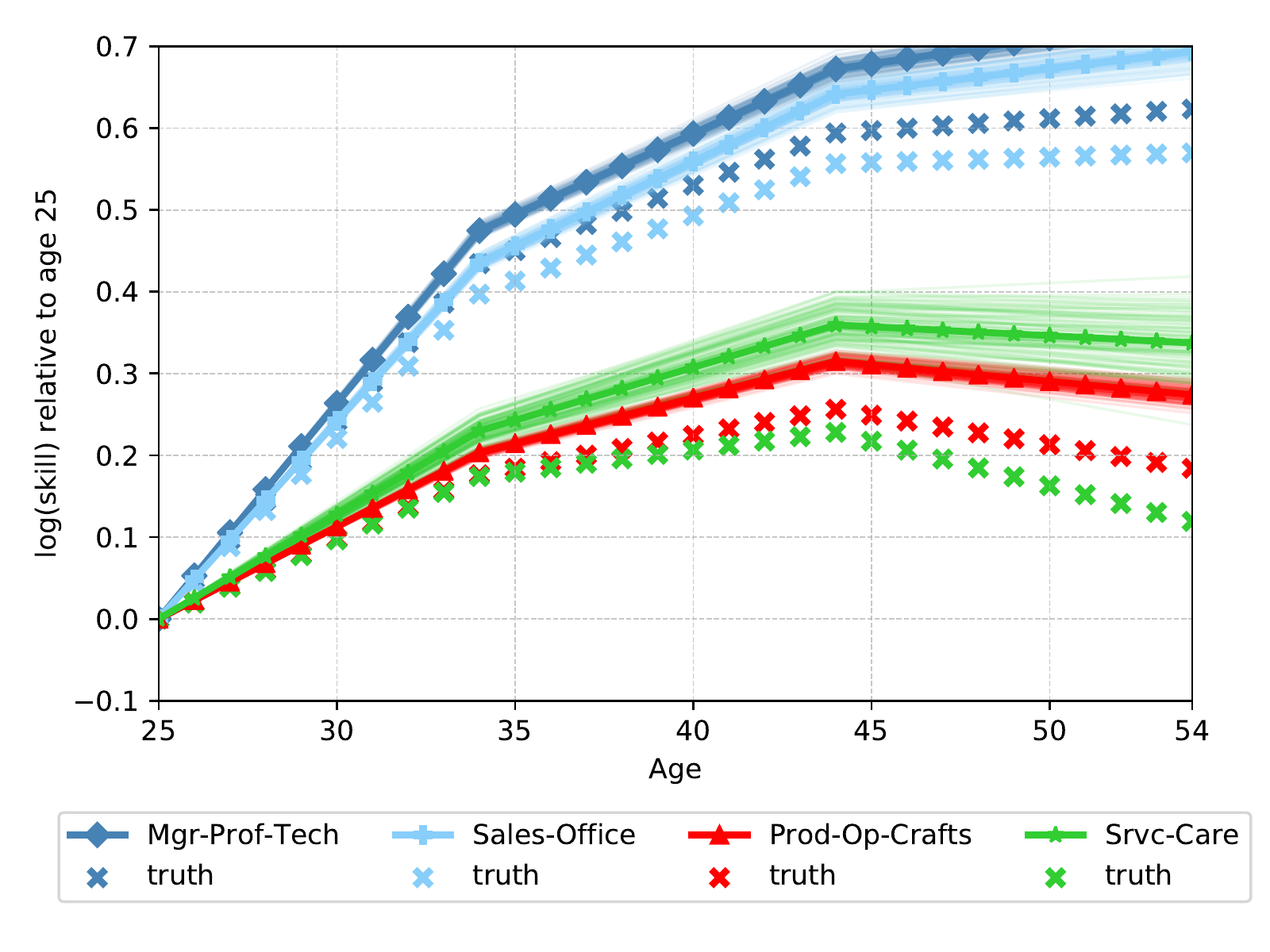}
  \end{minipage}
  \begin{minipage}[t]{\panelwidth}
    \subcaption{Cumulative prices, IV}
    \label{mc:fig:estimation-iv2-persistent-shocks-prices}
    \includegraphics[width=\textwidth]{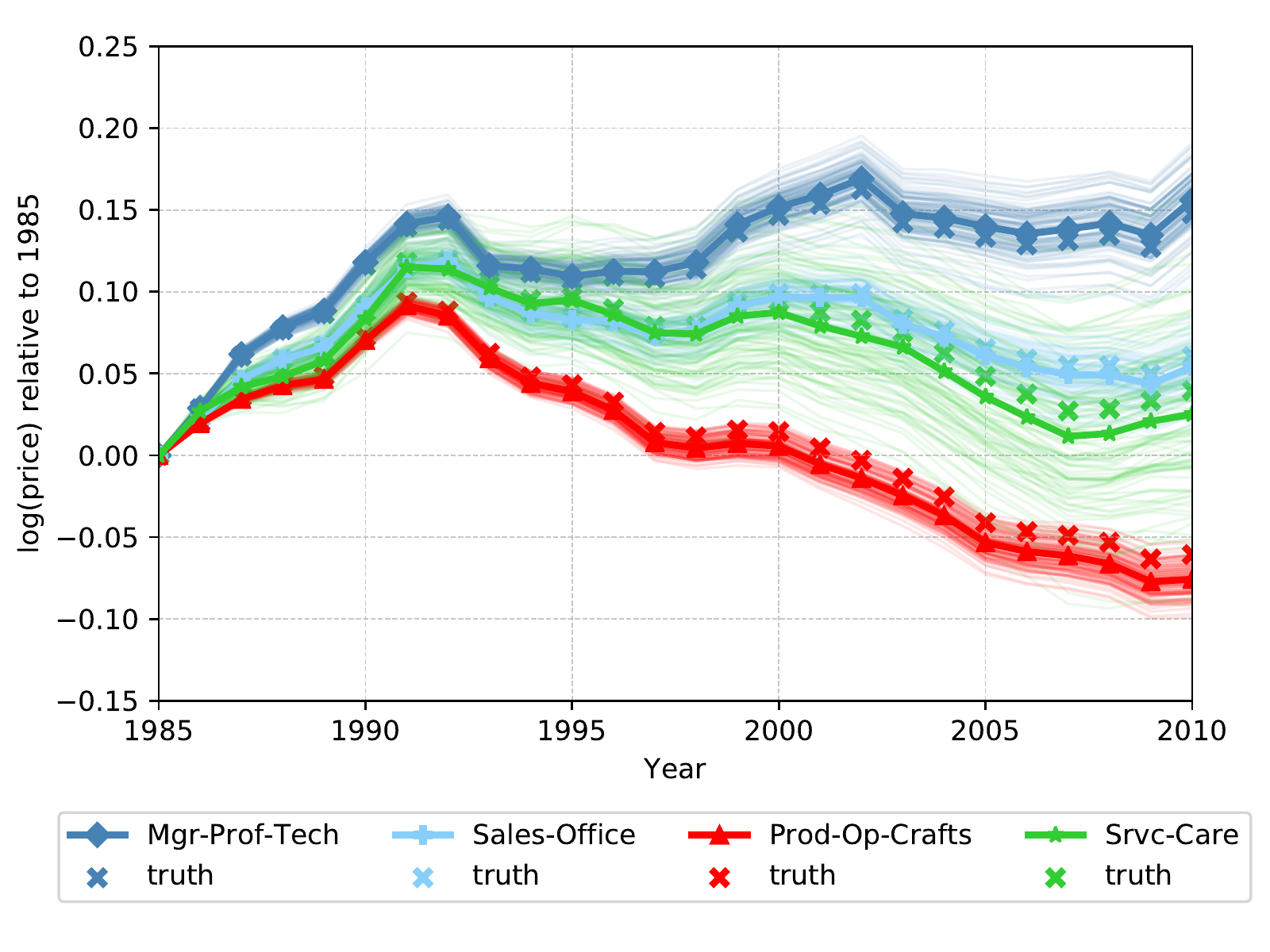}
  \end{minipage}
  \begin{minipage}[t]{\panelwidth}
    \subcaption{Skill accumulation, IV}
    \label{mc:fig:estimation-iv2-persistent-shocks-skills}
    \includegraphics[width=\textwidth]{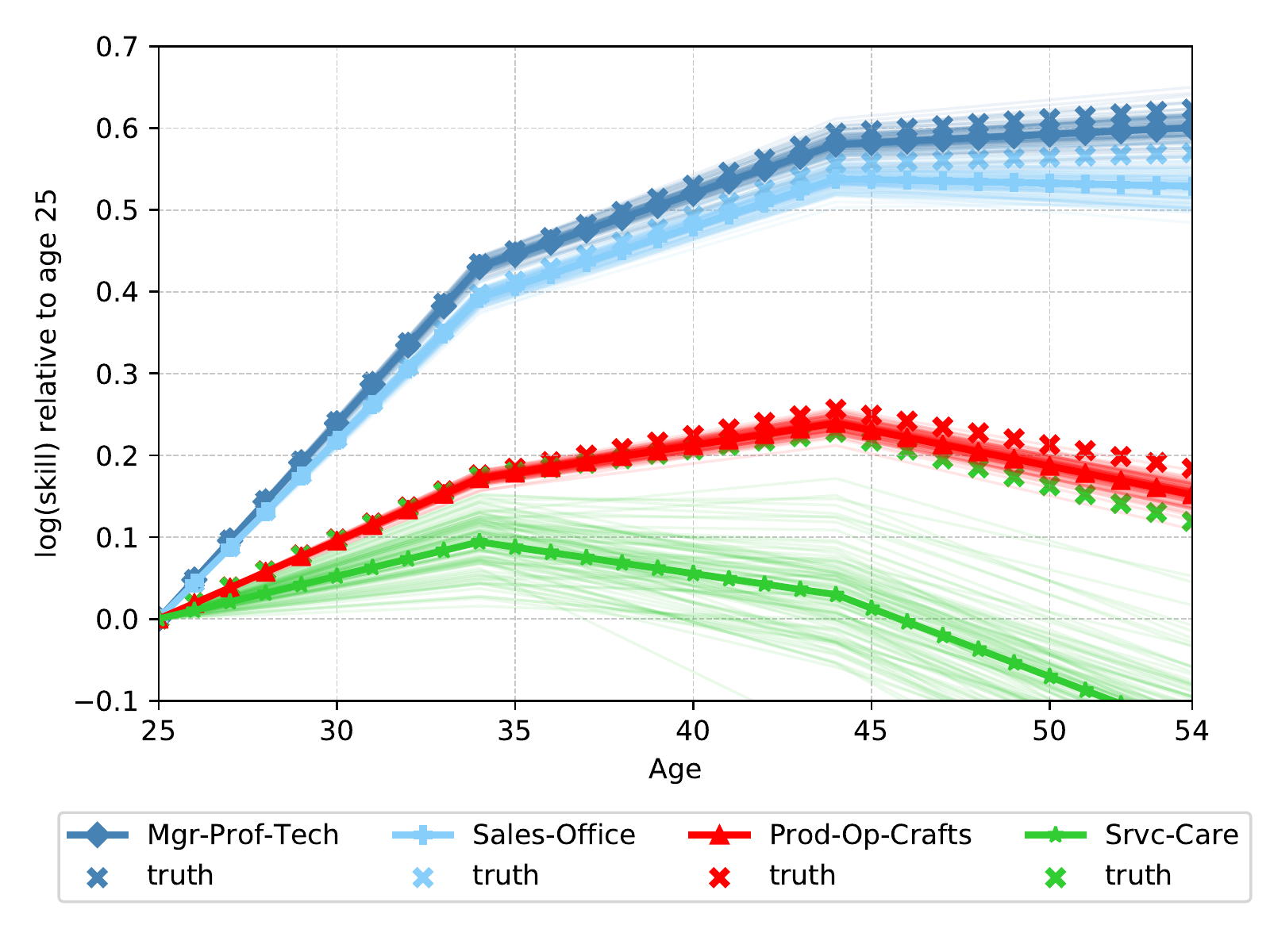}
  \end{minipage}
  \begin{minipage}{\textwidth}
    \scriptsize
    \emph{Notes:} \mcresults {} \olsresultsmc \ivresults
  \end{minipage}
\end{figure}

\clearpage

\subsection{Switching Costs}
\label{mc:sub:switching-costs}

\begin{figure}[ht!]
  \caption{Descriptives, moderate switch costs, no shocks}
  \label{mc:fig:descriptives-switch-costs-no-shocks}
  \centering
  \begin{minipage}[t]{\panelwidth}
    \subcaption{Occupation entrants/incumbents in $t+1$}
    \label{mc:fig:descriptives-switch-costs-no-shocks-switchers-entrants}
    \includegraphics[width=\textwidth]{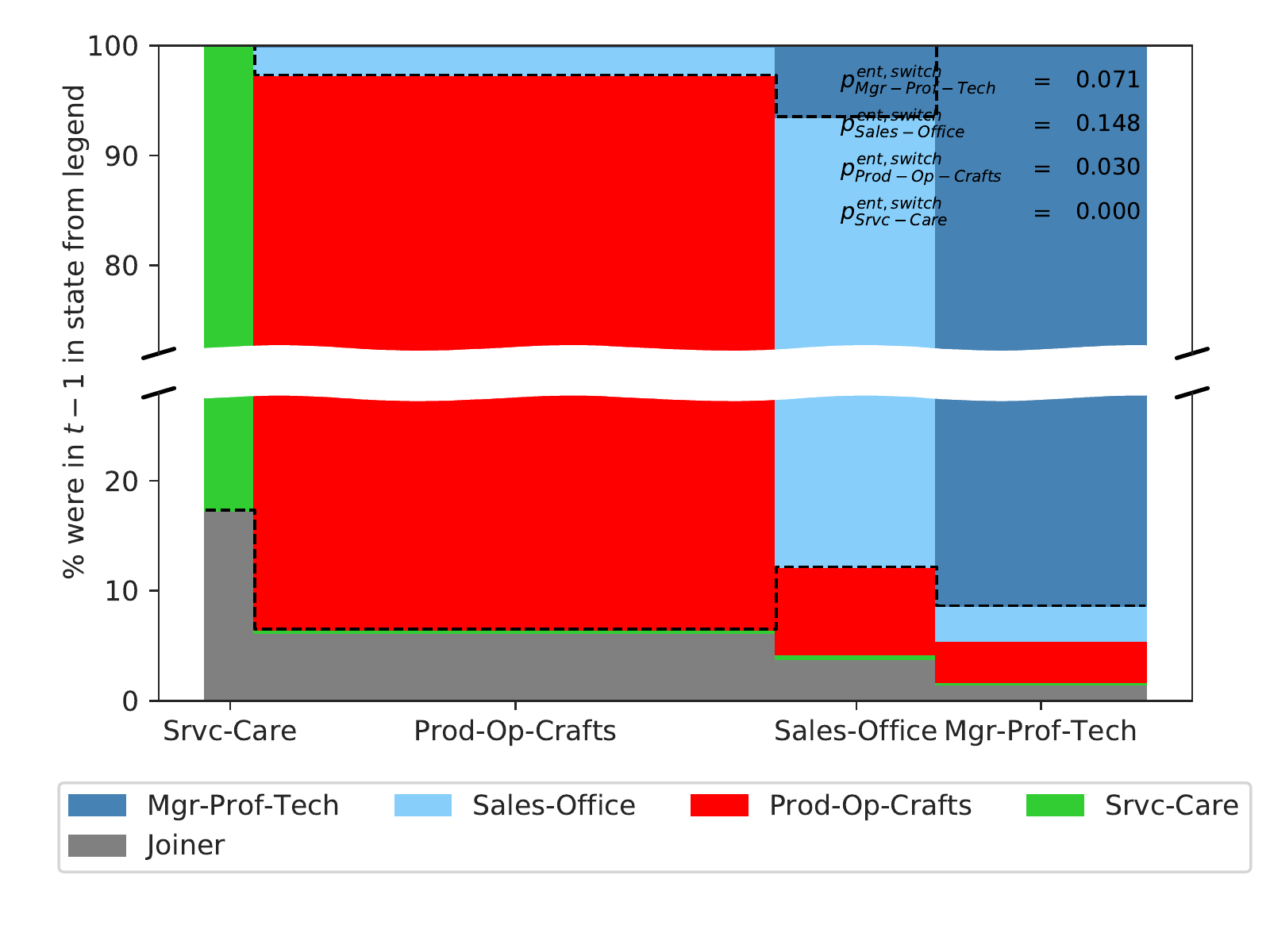}
  \end{minipage}
  \begin{minipage}[t]{\panelwidth}
    \subcaption{Occupation leavers/stayers in $t-1$}
    \label{mc:fig:descriptives-switch-costs-no-shocks-switchers-leavers}
    \includegraphics[width=\textwidth]{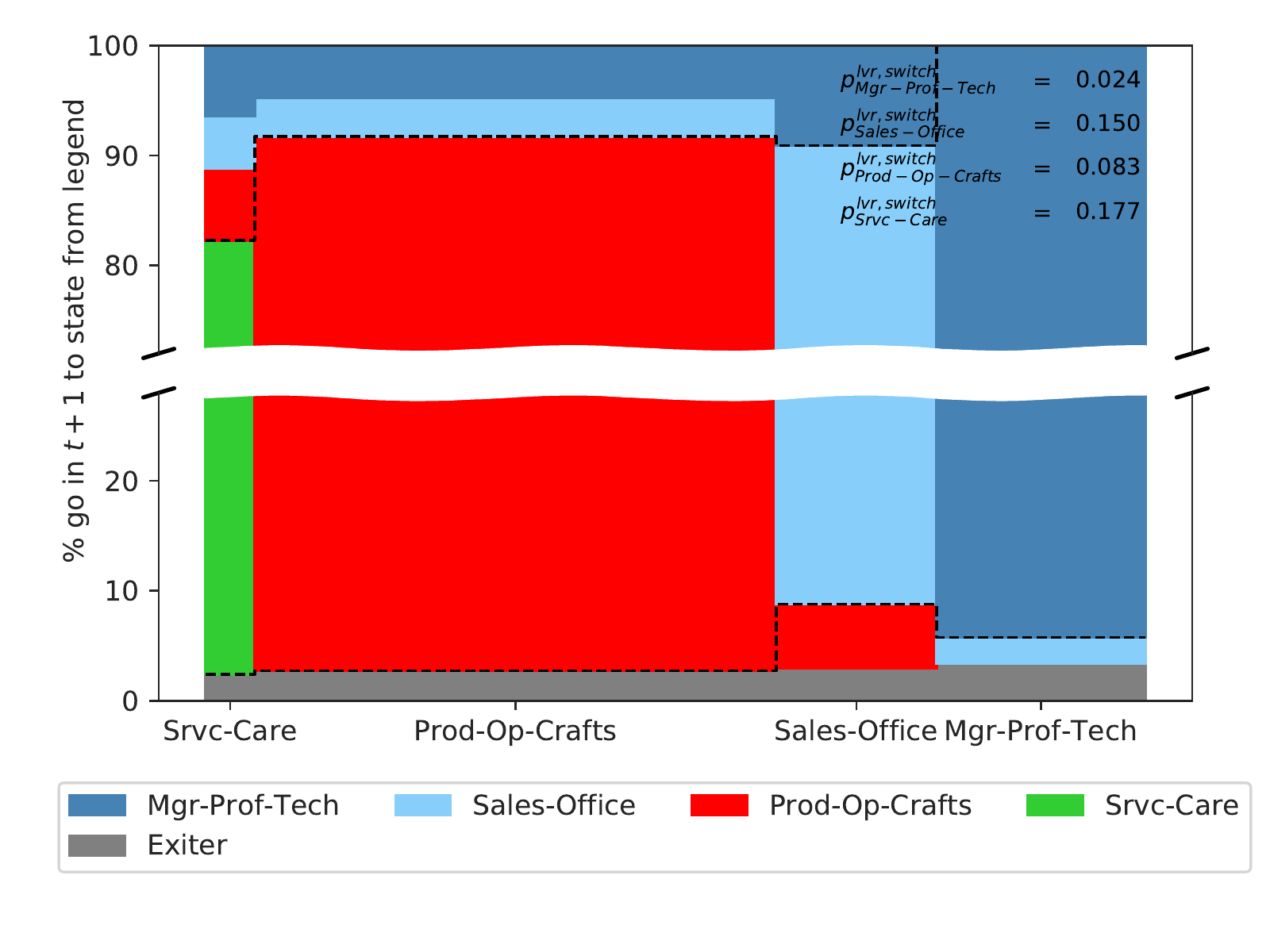}
  \end{minipage}
  \begin{minipage}[t]{\panelwidth}
    \subcaption{Distribution of annual wage growth}
    \label{mc:fig:descriptives-switch-costs-no-shocks-wage-growth-distribution}
    \includegraphics[width=\textwidth]{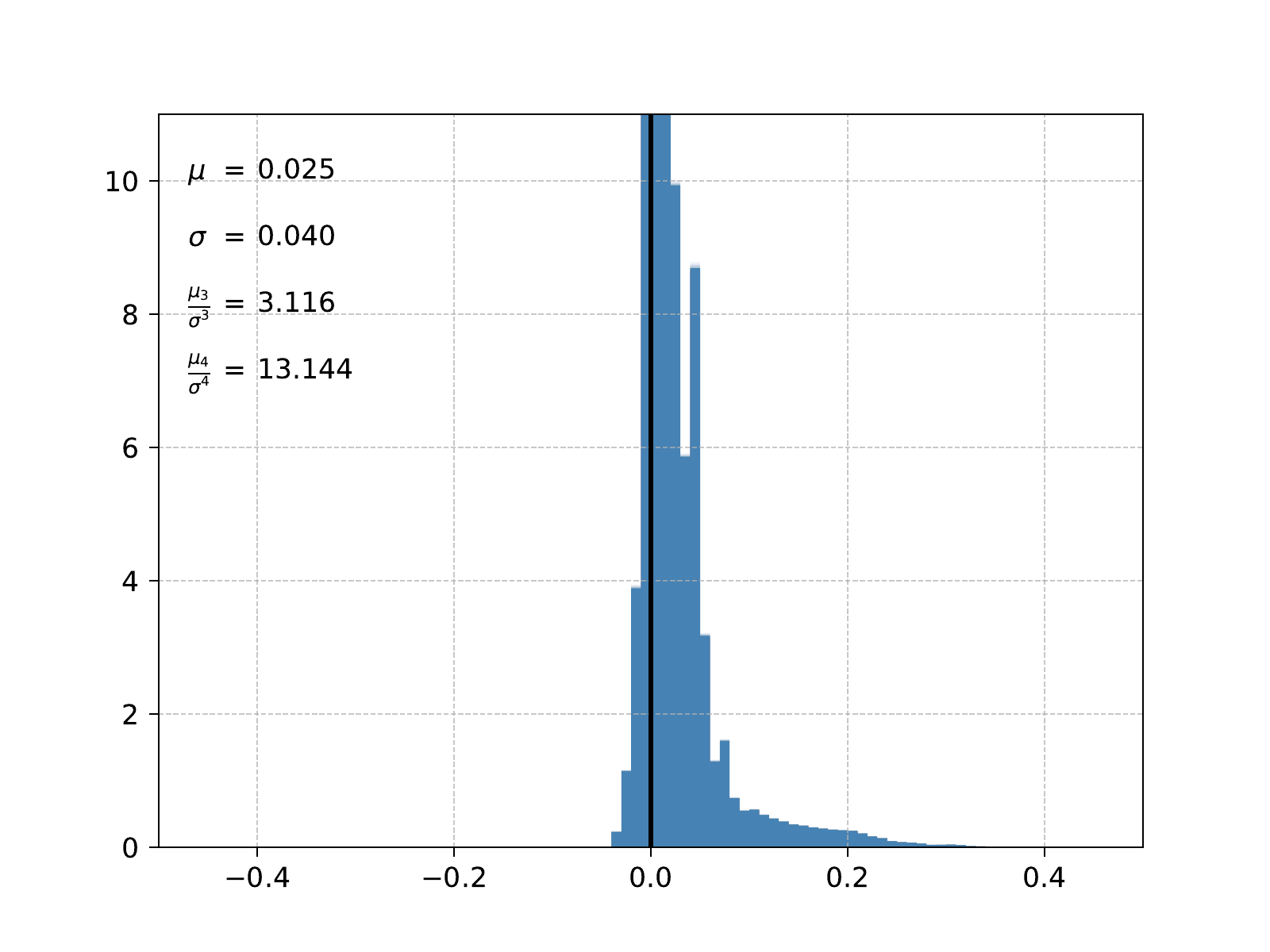}
  \end{minipage}
  \begin{minipage}[t]{\panelwidth}
    \subcaption{Evolution of the wage distribution}
    \label{mc:fig:descriptives-switch-costs-no-shocks-evolution-wage-inequality}
    \vskip2ex
    \includegraphics[width=\textwidth]{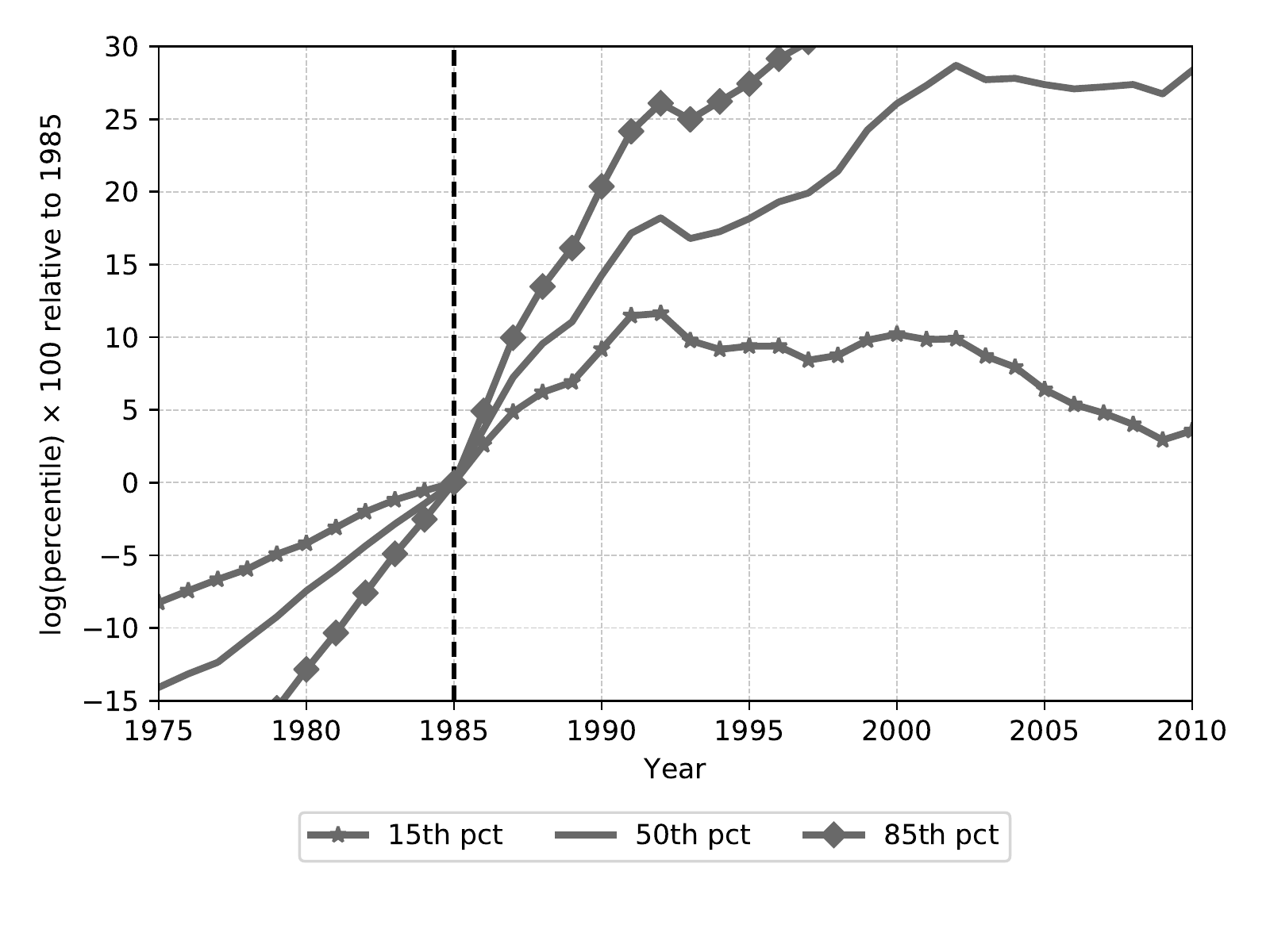}
  \end{minipage}
  \begin{minipage}{\textwidth}
    \scriptsize
    \emph{Notes:} \mcdescriptives
  \end{minipage}
\end{figure}

\begin{figure}[ht!]
  \caption{Estimation results, moderate switch costs, no shocks}
  \label{mc:fig:estimation-switch-costs-no-shocks}
  \centering
  \begin{minipage}[t]{\panelwidth}
    \subcaption{Cumulative prices, saturated OLS}
    \label{mc:fig:estimation-ols-switch-costs-no-shocks-prices}
    \includegraphics[width=\textwidth]{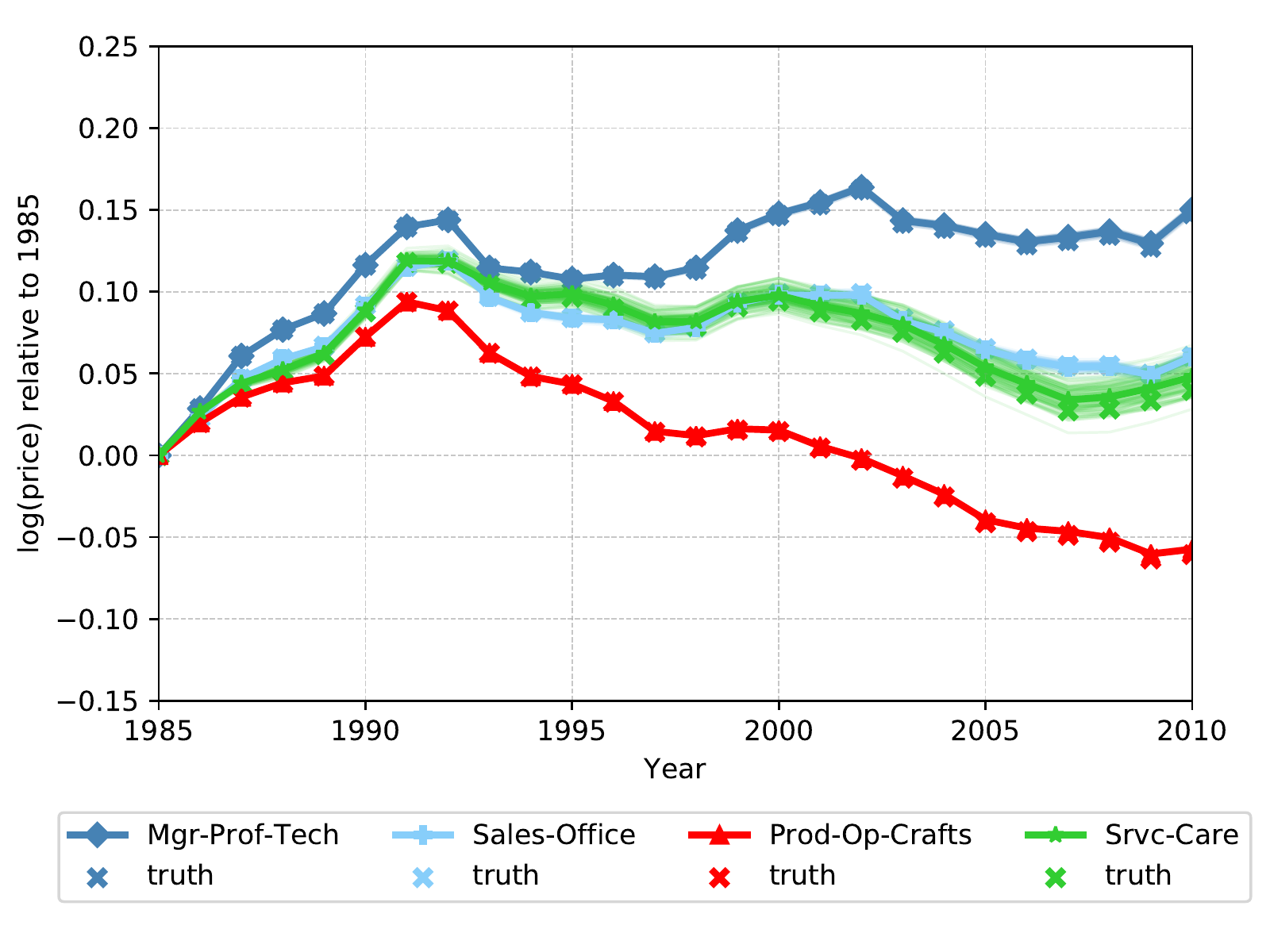}
  \end{minipage}
  \begin{minipage}[t]{\panelwidth}
    \subcaption{Skill accumulation, saturated OLS}
    \label{mc:fig:estimation-ols-switch-costs-no-shocks-skills}
    \includegraphics[width=\textwidth]{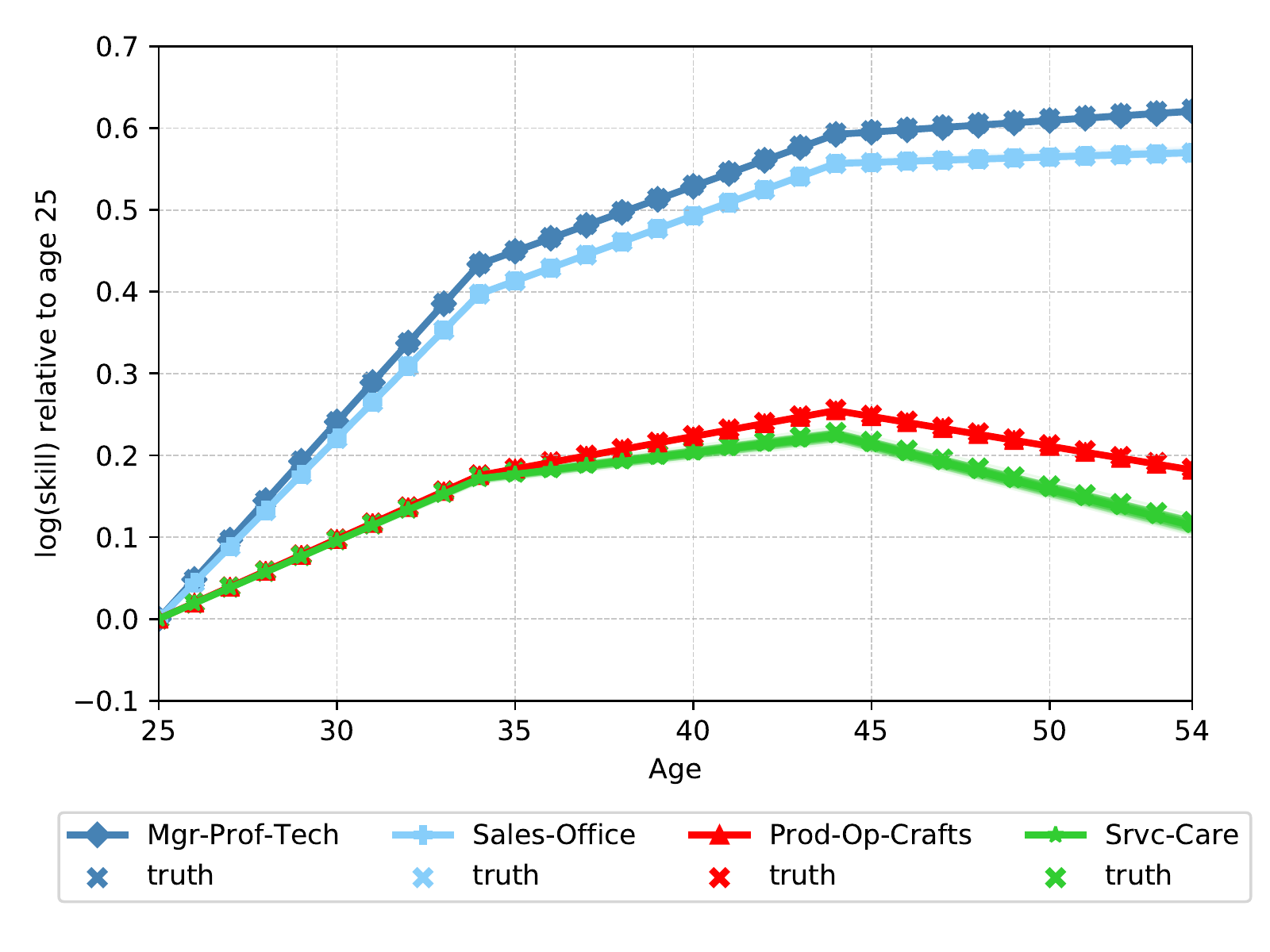}
  \end{minipage}
  \begin{minipage}{\textwidth}
    \scriptsize
    \emph{Notes:} \mcresults {} \olsresultsmc
  \end{minipage}
\end{figure}

In this section, we make the decision problem depend on non wage switching costs as a fraction of log wages. We assume that every worker has to pay a (psychic) utility cost when wanting to switch so that the potential utility amounts to $U = w$ when not switching and $U = (1 - c) w$ when switching. We start with switching costs of $c = 0.075$. First, in a model without skill shocks and with switching costs, our approximation~\eqref{mc:eq:approx-in-assumption-1} continues to work well (Figures~\ref{mc:fig:descriptives-switch-costs-no-shocks} and \ref{mc:fig:estimation-switch-costs-no-shocks}). We have trebled the size of cross-accumulation parameters in the data generating process here (see Table~\ref{mc:tab:switch-costs-no-shocks}), since otherwise we would hardly observe any switches without skill shocks and given the switching costs.

\begin{figure}[ht!]
  \caption{Descriptives, moderate switching costs and moderate shocks}
  \label{mc:fig:descriptives-moderate-switch-costs}
  \centering
  \begin{minipage}[t]{\panelwidth}
    \subcaption{Occupation entrants/incumbents in $t+1$}
    \label{mc:fig:descriptives-moderate-switch-costs-switchers-entrants}
    \includegraphics[width=\textwidth]{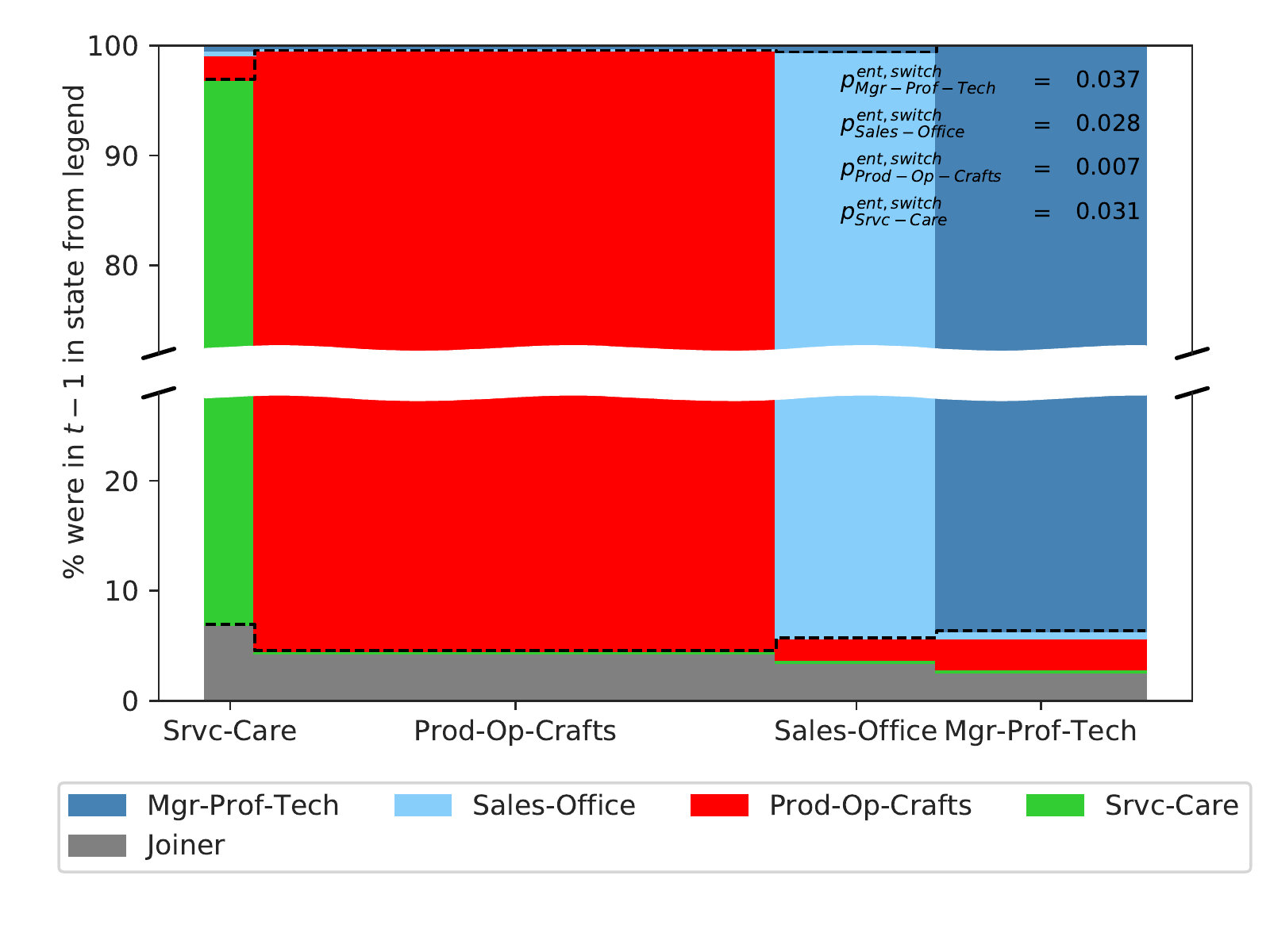}
  \end{minipage}
  \begin{minipage}[t]{\panelwidth}
    \subcaption{Occupation leavers/stayers in $t-1$}
    \label{mc:fig:descriptives-moderate-switch-costs-switchers-leavers}
    \includegraphics[width=\textwidth]{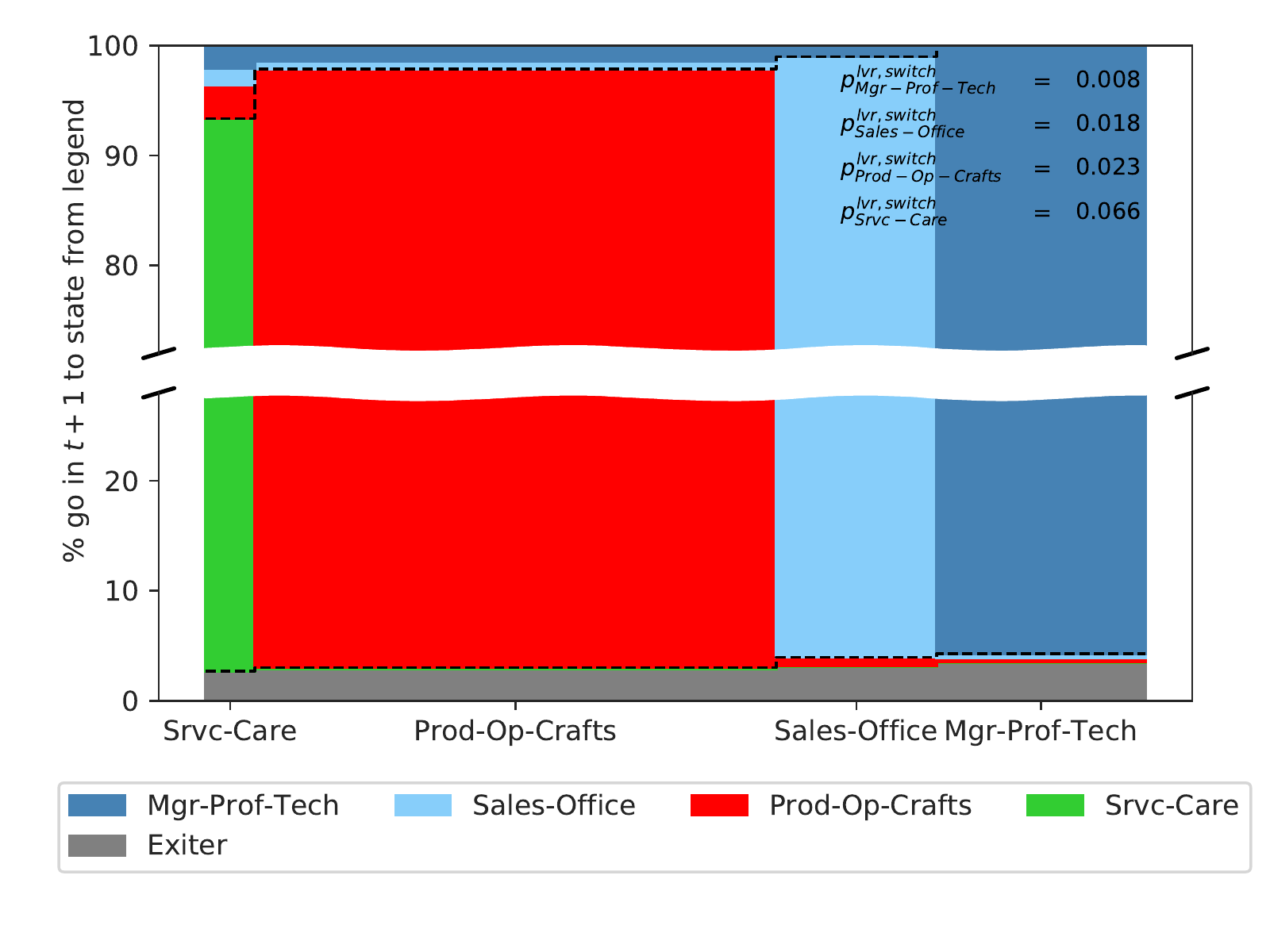}
  \end{minipage}
  \begin{minipage}[t]{\panelwidth}
    \subcaption{Distribution of annual wage growth}
    \label{mc:fig:descriptives-moderate-switch-costs-wage-growth-distribution}
    \includegraphics[width=\textwidth]{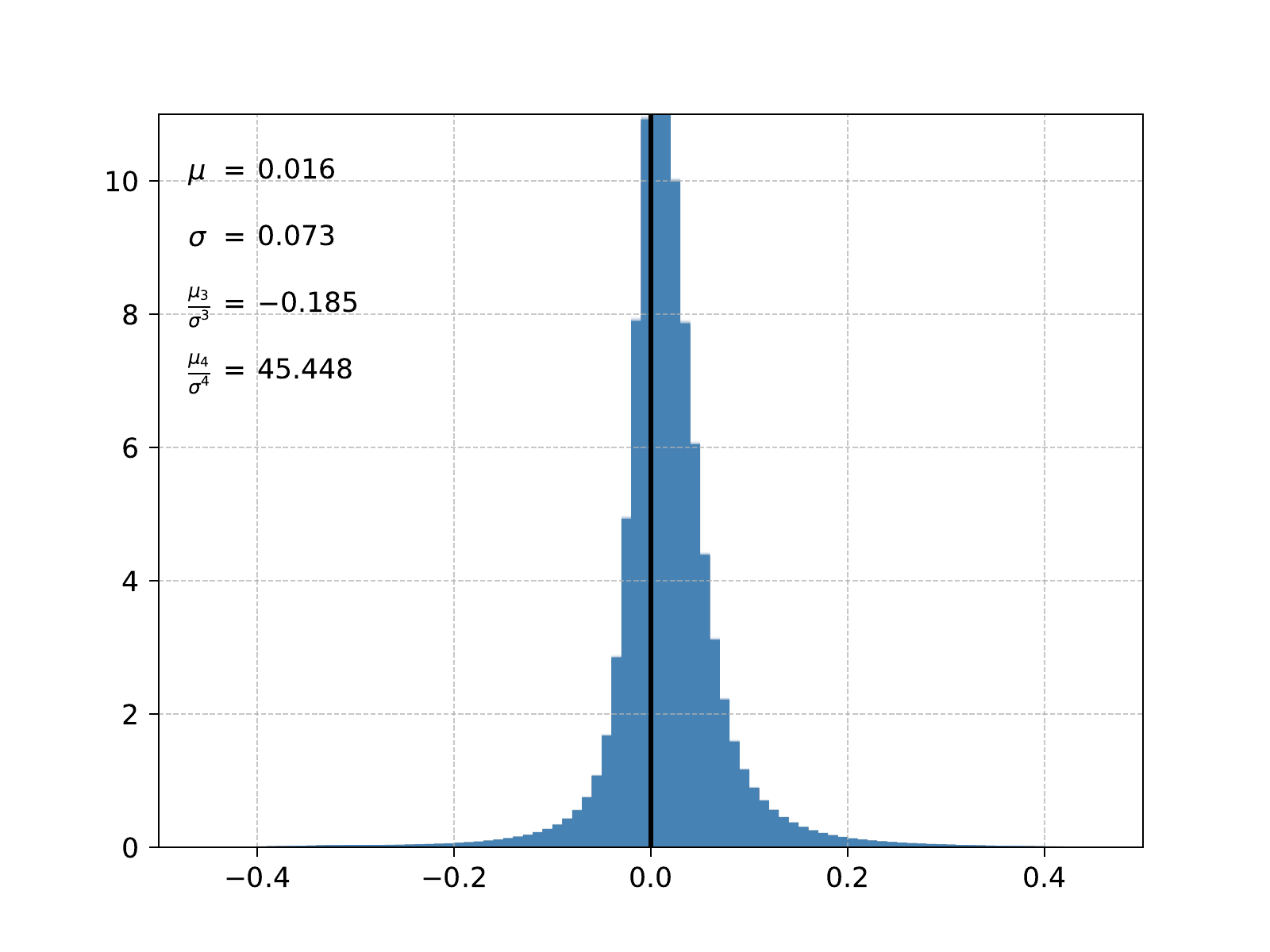}
  \end{minipage}
  \begin{minipage}[t]{\panelwidth}
    \subcaption{Evolution of the wage distribution}
    \label{mc:fig:descriptives-moderate-switch-costs-evolution-wage-inequality}
    \vskip2ex
    \includegraphics[width=\textwidth]{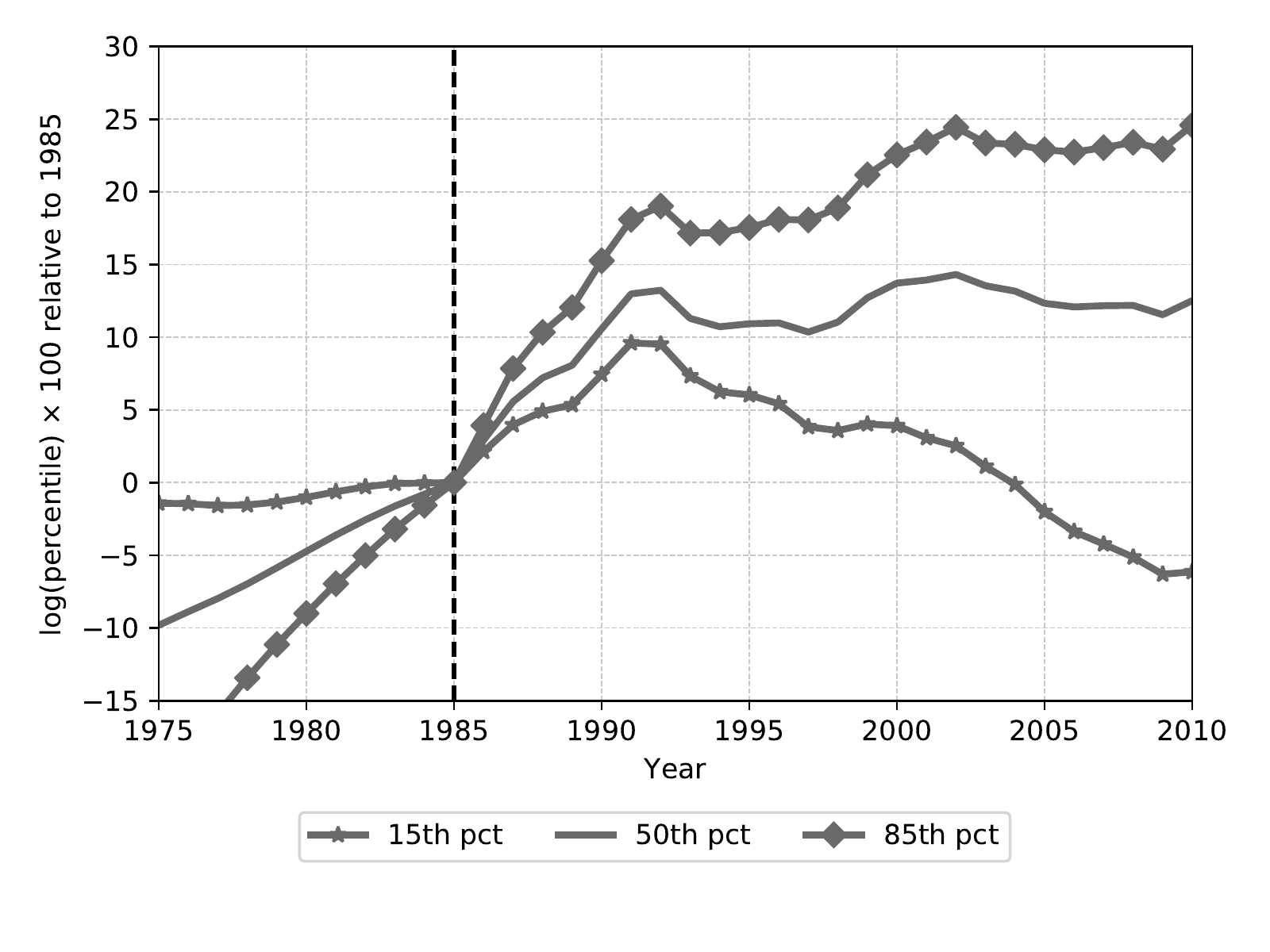}
  \end{minipage}
  \begin{minipage}{\textwidth}
    \scriptsize
    \emph{Notes:} \mcdescriptives
  \end{minipage}
\end{figure}

\begin{figure}[ht!]
  \caption{Estimation results, moderate switching costs and moderate shocks}
  \label{mc:fig:estimation-moderate-switch-costs}
  \centering
  \begin{minipage}[t]{\panelwidth}
    \subcaption{Cumulative prices, saturated OLS}
    \label{mc:fig:estimation-ols-moderate-switch-costs-prices}
    \includegraphics[width=\textwidth]{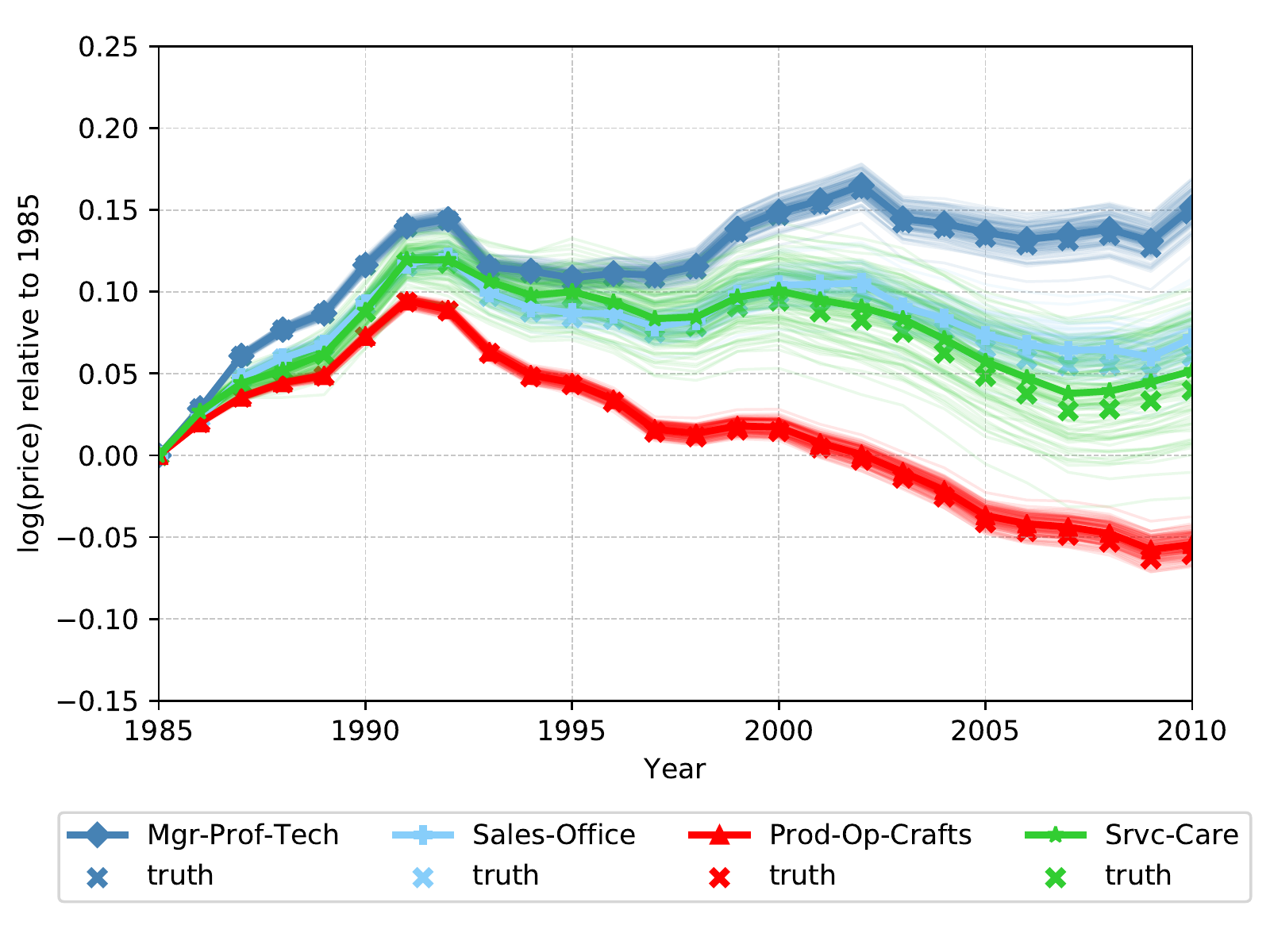}
  \end{minipage}
  \begin{minipage}[t]{\panelwidth}
    \subcaption{Skill accumulation, saturated OLS}
    \label{mc:fig:estimation-ols-moderate-switch-costs-skills}
    \includegraphics[width=\textwidth]{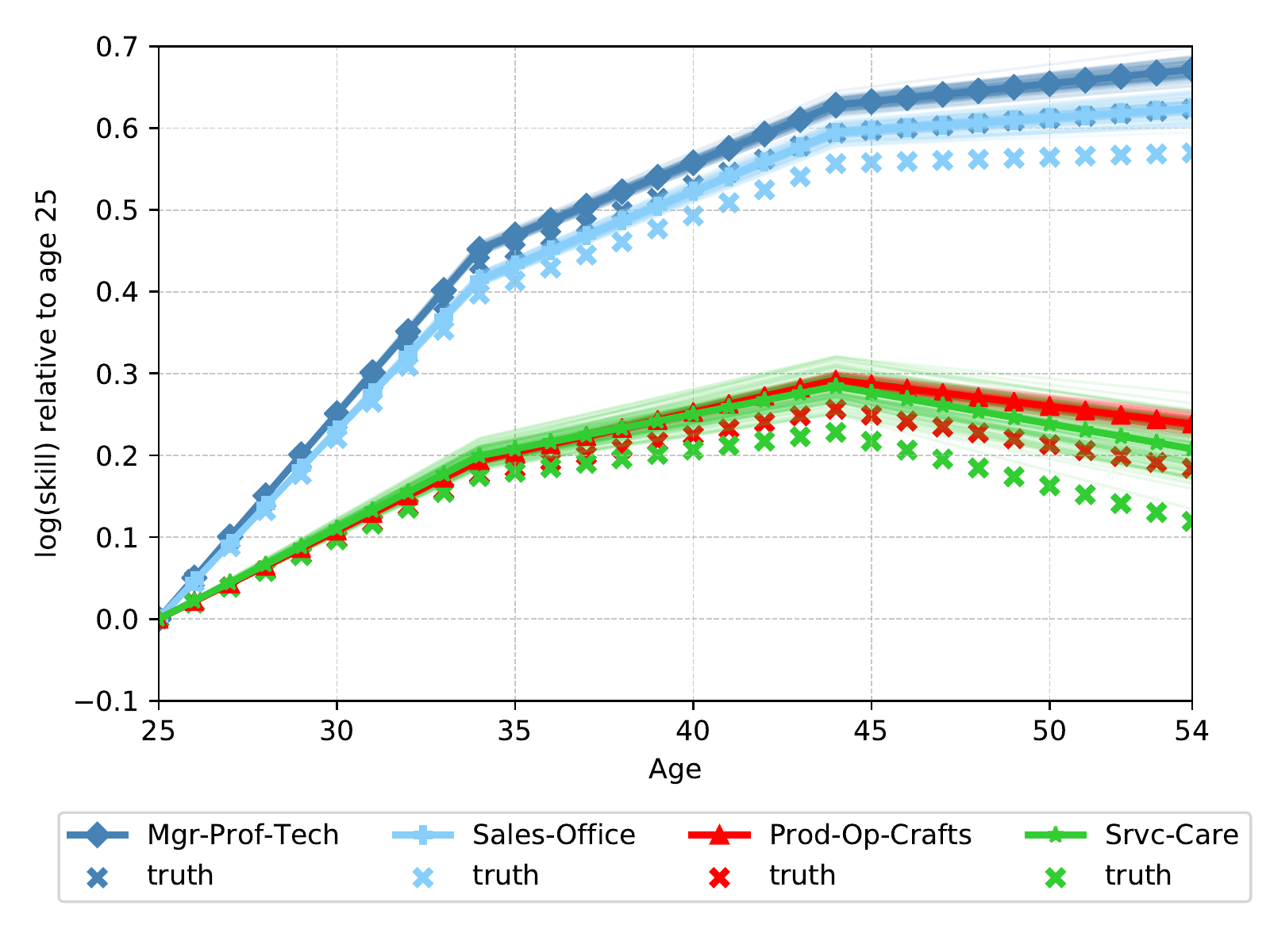}
  \end{minipage}
  \begin{minipage}[t]{\panelwidth}
    \subcaption{Cumulative prices, IV}
    \label{mc:fig:estimation-iv2-moderate-switch-costs-prices}
    \includegraphics[width=\textwidth]{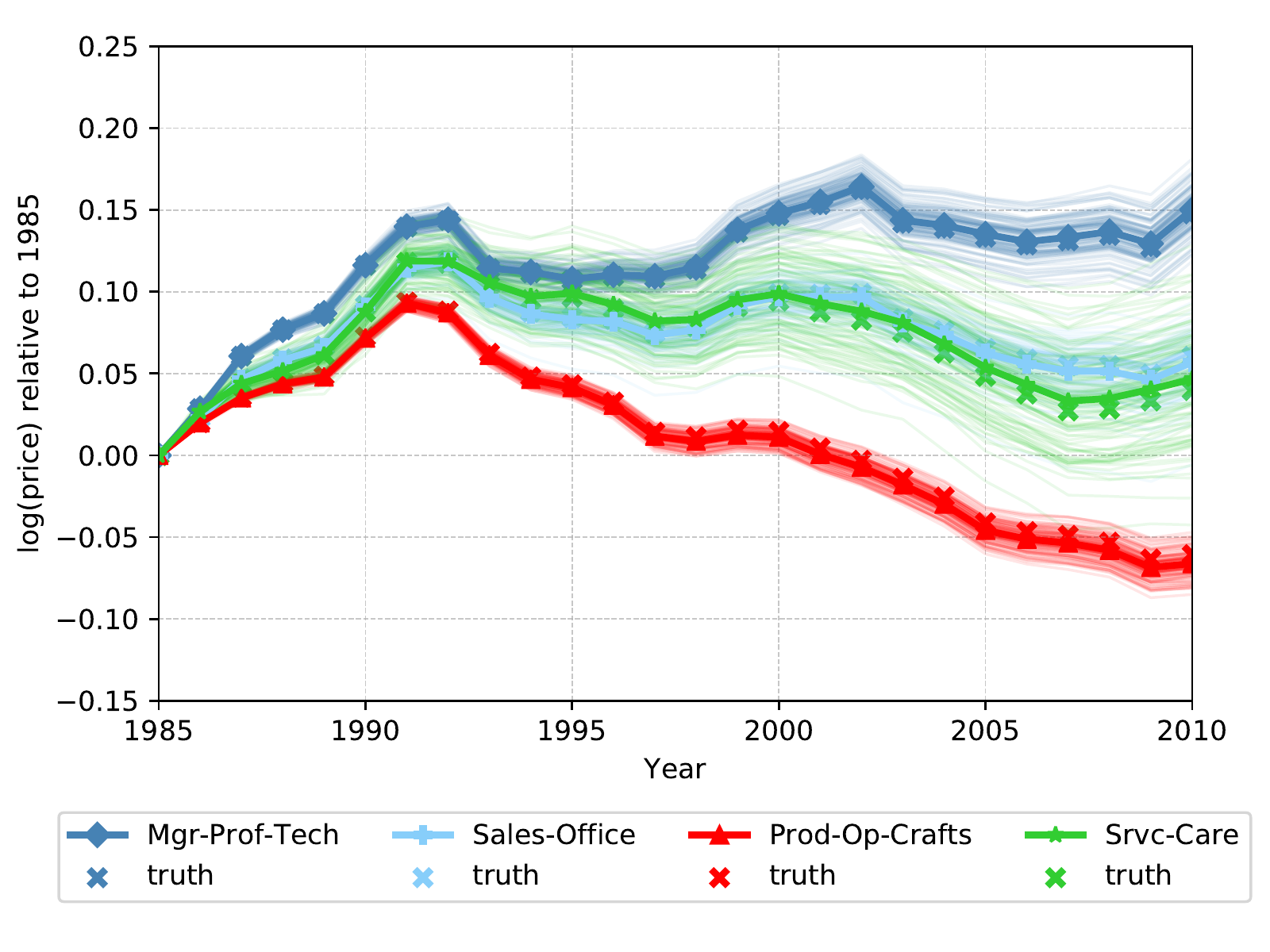}
  \end{minipage}
  \begin{minipage}[t]{\panelwidth}
    \subcaption{Skill accumulation, IV}
    \label{mc:fig:estimation-iv2-moderate-switch-costs-skills}
    \includegraphics[width=\textwidth]{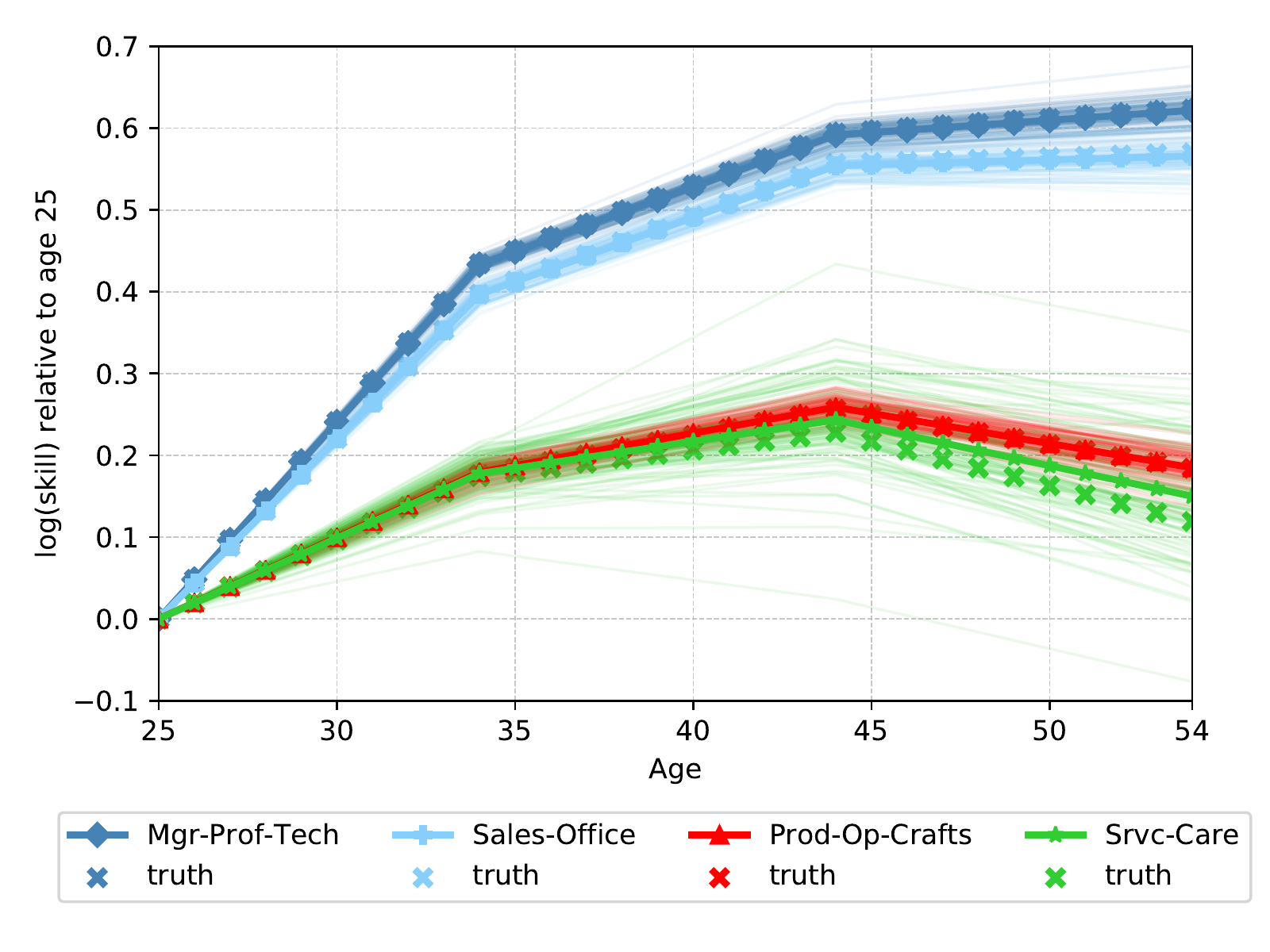}
  \end{minipage}
  \begin{minipage}{\textwidth}
    \scriptsize
    \emph{Notes:} \mcresults {} \olsresultsmc \ivresults
  \end{minipage}
\end{figure}

The switching cost also reduces the number of switchers under moderate skill shocks as depicted in Figure~\ref{mc:fig:descriptives-moderate-switch-costs}. Figure~\ref{mc:fig:estimation-moderate-switch-costs} shows that it does not bias our results. In fact, as skill prices are identified well from wage growth of stayers, switching costs make any bias of the skill prices less severe. Both OLS and IV therefore work for this scenario with moderate shocks and moderate switching costs, which we deem a rather realistic one.

An additional reason for why we think that the scenario with moderate skill shocks and moderate switching costs is a sensible benchmark are the estimated cross-accumulation parameters. This is a bit subtle: Note that OLS's upward-bias of the cross-accumulation parameters (${\Gamma}_{a(i,t-1), k(i,t-1),k(i,t)}$ for $k(i,t)\neq k(i,t-1)$) is exacerbated here because even larger idiosyncratic skill shocks are required to overcome the switch costs (Table~\ref{mc:tab:estimation-ols-moderate-switch-costs-skill-acc-coeffs}). The IV's weak instrument problem for the switchers is also more severe (Table~\ref{mc:tab:estimation-iv2-moderate-switch-costs-skill-acc-coeffs}). The data generating process of the Monte Carlos used a third of the cross-accumulation values estimated in the SIAB (see Table \ref{mc:tab:moderate-switch-costs}). The estimates based on the simulations with moderate shocks and moderate switching costs re-create these values, i.e., they overstate the target by a factor of three on average. Therefore, this scenario approximately ``replicates its own bias'' in the estimation of the cross-accumulation parameters. That is, what we pick as data generating process and the estimates that we receive are consistent with one another.

Finally, we increase the switching costs to $c = 0.2$ and the standard deviation of skill shocks to 1.5 times the standard deviation of log wage growth in the SIAB. Once again, the estimates of the skill prices, and especially in the IV, are quite close to their true values.

\clearpage

\subsubsection{Benchmark: Moderate Switching Costs, No Shocks}
\label{mc:sub:switching-no-shocks}

\begin{table}[ht!]
  \centering
  \caption{Parameters}
  \label{mc:tab:switch-costs-no-shocks}
  \begin{minipage}[t]{\linewidth}
    \centering
    \begin{tabular}{lp{8cm}} \toprule 
Parameter & Value \tabularnewline
\midrule
$N$ & 50000 \tabularnewline
Repetitions & 100 \tabularnewline
Skill shocks in $k$ & uniform, $(\mu, \sigma_k) = (0, 0 \cdot \sigma^{SIAB}_{\Delta \log(w_i)})$ \tabularnewline
Stayers accumulation $\gamma_{{k, k, a}}, k' = k $ & $\hat{{\gamma}}^{{SIAB}}_{{k, k, a}} $  \tabularnewline
Cross accumulation $\gamma_{k', k, a}, k' \ne k $ & $\frac{1}{1} \hat{\gamma}^{SIAB}_{k', k, a}$ \tabularnewline
$\rho$ in $\varepsilon_{i, t} = \rho \varepsilon_{i, t-1} + v_{i, t}$ & 0 \tabularnewline
Switching costs $c$ & $ 0.05 $ \tabularnewline
Amenity trends, $t = 1985,...,2010$ & $  [\Delta \Psi_{k, t}]_{k = 1,...,4} = [0, 0, 0, 0] $ \tabularnewline
\bottomrule
\end{tabular}
  \end{minipage}
\end{table}

\clearpage

\subsubsection{Moderate Switching Costs, Moderate Shocks}
\label{mc:sub:switching-moderate}

\begin{table}[ht!]
  \centering
  \caption{Parameters}
  \label{mc:tab:moderate-switch-costs}
  \begin{minipage}[t]{\linewidth}
    \centering
    \begin{tabular}{lp{8cm}} \toprule 
Parameter & Value \tabularnewline
\midrule
$N$ & 50000 \tabularnewline
Repetitions & 100 \tabularnewline
Skill shocks in $k$ & observed wage growth distribution, $(\mu, \sigma_k) = (0, 0.5 \cdot \sigma^{SIAB}_{\Delta \log(w_i)})$ \tabularnewline
Stayers accumulation $\gamma_{{k, k, a}}, k' = k $ & $\hat{{\gamma}}^{{SIAB}}_{{k, k, a}} $  \tabularnewline
Cross accumulation $\gamma_{k', k, a}, k' \ne k $ & $\frac{1}{3} \hat{\gamma}^{SIAB}_{k', k, a}$ \tabularnewline
$\rho$ in $\varepsilon_{i, t} = \rho \varepsilon_{i, t-1} + v_{i, t}$ & 0 \tabularnewline
Switching costs $c$ & $ 0.05 $ \tabularnewline
Amenity trends, $t = 1985,...,2010$ & $  [\Delta \Psi_{k, t}]_{k = 1,...,4} = [0, 0, 0, 0] $ \tabularnewline
\bottomrule
\end{tabular}
  \end{minipage}
\end{table}

\clearpage

\begin{table}[h!]
  \centering
  \caption{True and estimated skill accumulation parameters, saturated OLS}
  \label{mc:tab:estimation-ols-moderate-switch-costs-skill-acc-coeffs}
  \begin{minipage}[t]{\linewidth}
    \centering
    \begin{tabular}{lllrrrrrr}
\toprule & & \multicolumn{7}{c}{Age group} \\ \cmidrule{4-9}
           &           &                   & \multicolumn{2}{l}{[25, 34]} & \multicolumn{2}{l}{[35, 44]} & \multicolumn{2}{l}{[45, 54]} \\
           Previous sector &           Current sector &                   & $\hat{\gamma}_{k', k, a}$ & $\gamma^{true}_{k', k, a}$ & $\hat{\gamma}_{k', k, a}$ & $\gamma^{true}_{k', k, a}$ & $\hat{\gamma}_{k', k, a}$ & $\gamma^{true}_{k', k, a}$ \\
\midrule
Mgr-Prof-Tech  & Mgr-Prof-Tech & $\gamma$ &                     0.050 &                      0.048 &                     0.018 &                      0.016 &                     0.004 &                      0.003 \\
           &           & $\sigma_{\gamma}$ &                     0.000 &                            &                     0.000 &                            &                     0.000 &                            \\
           & Sales-Office & $\gamma$ &                     0.126 &                      0.063 &                     0.045 &                      0.009 &                    -0.021 &                     -0.010 \\
           &           & $\sigma_{\gamma}$ &                     0.018 &                            &                     0.022 &                            &                     0.036 &                            \\
           & Prod-Op-Crafts & $\gamma$ &                     0.027 &                      0.023 &                    -0.047 &                     -0.011 &                    -0.069 &                     -0.022 \\
           &           & $\sigma_{\gamma}$ &                     0.031 &                            &                     0.028 &                            &                     0.034 &                            \\
           & Srvc-Care & $\gamma$ &                    -0.121 &                     -0.008 &                    -0.160 &                     -0.036 &                    -0.007 &                     -0.004 \\
           &           & $\sigma_{\gamma}$ &                     0.095 &                            &                     0.107 &                            &                     0.063 &                            \\ \midrule
Sales-Office  & Mgr-Prof-Tech & $\gamma$ &                     0.176 &                      0.088 &                     0.105 &                      0.027 &                     0.078 &                      0.009 \\
           &           & $\sigma_{\gamma}$ &                     0.009 &                            &                     0.014 &                            &                     0.019 &                            \\
           & Sales-Office & $\gamma$ &                     0.046 &                      0.044 &                     0.018 &                      0.016 &                     0.003 &                      0.001 \\
           &           & $\sigma_{\gamma}$ &                     0.000 &                            &                     0.000 &                            &                     0.000 &                            \\
           & Prod-Op-Crafts & $\gamma$ &                     0.122 &                      0.056 &                     0.087 &                      0.019 &                     0.029 &                     -0.008 \\
           &           & $\sigma_{\gamma}$ &                     0.018 &                            &                     0.016 &                            &                     0.025 &                            \\
           & Srvc-Care & $\gamma$ &                    -0.041 &                      0.010 &                    -0.146 &                     -0.034 &                    -0.094 &                     -0.024 \\
           &           & $\sigma_{\gamma}$ &                     0.072 &                            &                     0.105 &                            &                     0.112 &                            \\ \midrule
Prod-Op-Crafts  & Mgr-Prof-Tech & $\gamma$ &                     0.173 &                      0.075 &                     0.127 &                      0.042 &                     0.103 &                      0.021 \\
           &           & $\sigma_{\gamma}$ &                     0.004 &                            &                     0.006 &                            &                     0.007 &                            \\
           & Sales-Office & $\gamma$ &                     0.110 &                      0.036 &                     0.093 &                      0.022 &                     0.062 &                      0.000 \\
           &           & $\sigma_{\gamma}$ &                     0.010 &                            &                     0.010 &                            &                     0.013 &                            \\
           & Prod-Op-Crafts & $\gamma$ &                     0.022 &                      0.020 &                     0.010 &                      0.008 &                    -0.005 &                     -0.007 \\
           &           & $\sigma_{\gamma}$ &                     0.000 &                            &                     0.000 &                            &                     0.000 &                            \\
           & Srvc-Care & $\gamma$ &                    -0.038 &                     -0.017 &                    -0.056 &                     -0.014 &                     0.008 &                     -0.009 \\
           &           & $\sigma_{\gamma}$ &                     0.035 &                            &                     0.040 &                            &                     0.027 &                            \\ \midrule
Srvc-Care  & Mgr-Prof-Tech & $\gamma$ &                     0.194 &                      0.099 &                     0.159 &                      0.063 &                     0.134 &                      0.041 \\
           &           & $\sigma_{\gamma}$ &                     0.012 &                            &                     0.015 &                            &                     0.018 &                            \\
           & Sales-Office & $\gamma$ &                     0.181 &                      0.090 &                     0.141 &                      0.048 &                     0.100 &                      0.015 \\
           &           & $\sigma_{\gamma}$ &                     0.014 &                            &                     0.020 &                            &                     0.034 &                            \\
           & Prod-Op-Crafts & $\gamma$ &                     0.201 &                      0.106 &                     0.174 &                      0.075 &                     0.136 &                      0.037 \\
           &           & $\sigma_{\gamma}$ &                     0.010 &                            &                     0.011 &                            &                     0.014 &                            \\
           & Srvc-Care & $\gamma$ &                     0.022 &                      0.019 &                     0.008 &                      0.005 &                    -0.008 &                     -0.011 \\
           &           & $\sigma_{\gamma}$ &                     0.001 &                            &                     0.001 &                            &                     0.001 &                            \\
\bottomrule
\end{tabular}

  \end{minipage}
  \begin{minipage}{\textwidth}
    \scriptsize
    \emph{Notes:} \crossgammas \broadgroupsmc {} \olsresultsmc
  \end{minipage}
\end{table}

\begin{table}[h!]
  \centering
  \caption{True and estimated skill accumulation parameters, IV}
  \label{mc:tab:estimation-iv2-moderate-switch-costs-skill-acc-coeffs}
  \begin{minipage}[t]{\linewidth}
    \centering
    \begin{tabular}{lllrrrrrr}
\toprule & & \multicolumn{7}{c}{Age group} \\ \cmidrule{4-9}
           &           &                   & \multicolumn{2}{l}{[25, 34]} & \multicolumn{2}{l}{[35, 44]} & \multicolumn{2}{l}{[45, 54]} \\
           Previous sector &           Current sector &                   & $\hat{\gamma}_{k', k, a}$ & $\gamma^{true}_{k', k, a}$ & $\hat{\gamma}_{k', k, a}$ & $\gamma^{true}_{k', k, a}$ & $\hat{\gamma}_{k', k, a}$ & $\gamma^{true}_{k', k, a}$ \\
\midrule
Mgr-Prof-Tech  & Mgr-Prof-Tech & $\gamma$ &                     0.048 &                      0.048 &                     0.016 &                      0.016 &                     0.003 &                      0.003 \\
           &           & $\sigma_{\gamma}$ &                     0.001 &                            &                     0.001 &                            &                     0.001 &                            \\
           & Sales-Office & $\gamma$ &                     0.377 &                      0.063 &                     0.393 &                      0.009 &                     0.369 &                     -0.010 \\
           &           & $\sigma_{\gamma}$ &                     0.102 &                            &                     0.147 &                            &                     0.223 &                            \\
           & Prod-Op-Crafts & $\gamma$ &                     0.496 &                      0.023 &                     0.432 &                     -0.011 &                     0.422 &                     -0.022 \\
           &           & $\sigma_{\gamma}$ &                     0.123 &                            &                     0.111 &                            &                     0.129 &                            \\
           & Srvc-Care & $\gamma$ &                     0.488 &                     -0.008 &                     0.506 &                     -0.036 &                     0.454 &                     -0.004 \\
           &           & $\sigma_{\gamma}$ &                     0.528 &                            &                     0.496 &                            &                     0.333 &                            \\ \midrule
Sales-Office  & Mgr-Prof-Tech & $\gamma$ &                     0.301 &                      0.088 &                     0.324 &                      0.027 &                     0.327 &                      0.009 \\
           &           & $\sigma_{\gamma}$ &                     0.062 &                            &                     0.131 &                            &                     0.250 &                            \\
           & Sales-Office & $\gamma$ &                     0.044 &                      0.044 &                     0.016 &                      0.016 &                     0.001 &                      0.001 \\
           &           & $\sigma_{\gamma}$ &                     0.001 &                            &                     0.001 &                            &                     0.001 &                            \\
           & Prod-Op-Crafts & $\gamma$ &                     0.386 &                      0.056 &                     0.387 &                      0.019 &                     0.383 &                     -0.008 \\
           &           & $\sigma_{\gamma}$ &                     0.095 &                            &                     0.111 &                            &                     0.163 &                            \\
           & Srvc-Care & $\gamma$ &                     0.513 &                      0.010 &                     0.476 &                     -0.034 &                     0.100 &                     -0.024 \\
           &           & $\sigma_{\gamma}$ &                     0.472 &                            &                     0.530 &                            &                     0.838 &                            \\ \midrule
Prod-Op-Crafts  & Mgr-Prof-Tech & $\gamma$ &                     0.247 &                      0.075 &                     0.288 &                      0.042 &                     0.274 &                      0.021 \\
           &           & $\sigma_{\gamma}$ &                     0.065 &                            &                     0.090 &                            &                     0.126 &                            \\
           & Sales-Office & $\gamma$ &                     0.314 &                      0.036 &                     0.316 &                      0.022 &                     0.311 &                      0.000 \\
           &           & $\sigma_{\gamma}$ &                     0.131 &                            &                     0.098 &                            &                     0.163 &                            \\
           & Prod-Op-Crafts & $\gamma$ &                     0.020 &                      0.020 &                     0.008 &                      0.008 &                    -0.007 &                     -0.007 \\
           &           & $\sigma_{\gamma}$ &                     0.001 &                            &                     0.001 &                            &                     0.001 &                            \\
           & Srvc-Care & $\gamma$ &                     0.416 &                     -0.017 &                     0.434 &                     -0.014 &                     0.365 &                     -0.009 \\
           &           & $\sigma_{\gamma}$ &                     0.285 &                            &                     0.236 &                            &                     0.188 &                            \\ \midrule
Srvc-Care  & Mgr-Prof-Tech & $\gamma$ &                     0.285 &                      0.099 &                     0.214 &                      0.063 &                     0.201 &                      0.041 \\
           &           & $\sigma_{\gamma}$ &                     0.132 &                            &                     0.171 &                            &                     0.139 &                            \\
           & Sales-Office & $\gamma$ &                     0.273 &                      0.090 &                     0.155 &                      0.048 &                     0.042 &                      0.015 \\
           &           & $\sigma_{\gamma}$ &                     0.104 &                            &                     0.248 &                            &                     0.424 &                            \\
           & Prod-Op-Crafts & $\gamma$ &                     0.279 &                      0.106 &                     0.235 &                      0.075 &                     0.224 &                      0.037 \\
           &           & $\sigma_{\gamma}$ &                     0.037 &                            &                     0.059 &                            &                     0.089 &                            \\
           & Srvc-Care & $\gamma$ &                     0.019 &                      0.019 &                     0.007 &                      0.005 &                    -0.009 &                     -0.011 \\
           &           & $\sigma_{\gamma}$ &                     0.003 &                            &                     0.004 &                            &                     0.004 &                            \\
\bottomrule
\end{tabular}

  \end{minipage}
  \begin{minipage}{\textwidth}
    \scriptsize
    \emph{Notes:} \crossgammas \broadgroupsmc \ivresults
  \end{minipage}
\end{table}

\clearpage

\subsubsection{High Switching Costs, Highly Dispersed Shocks}
\label{mc:sub:switching-vlarge}

\begin{table}[ht!]
  \centering
  \caption{Parameters}
  \label{mc:tab:high-switch-costs}
  \begin{minipage}[t]{\linewidth}
    \centering
    \begin{tabular}{lp{8cm}} \toprule 
Parameter & Value \tabularnewline
\midrule
$N$ & 50000 \tabularnewline
Repetitions & 100 \tabularnewline
Skill shocks in $k$ & observed wage growth distribution, $(\mu, \sigma_k) = (0, 1.5 \cdot \sigma^{SIAB}_{\Delta \log(w_i)})$ \tabularnewline
Stayers accumulation $\gamma_{{k, k, a}}, k' = k $ & $\hat{{\gamma}}^{{SIAB}}_{{k, k, a}} $  \tabularnewline
Cross accumulation $\gamma_{k', k, a}, k' \ne k $ & $\frac{1}{3} \hat{\gamma}^{SIAB}_{k', k, a}$ \tabularnewline
$\rho$ in $\varepsilon_{i, t} = \rho \varepsilon_{i, t-1} + v_{i, t}$ & 0 \tabularnewline
Switching costs $c$ & $ 0.2 $ \tabularnewline
Amenity trends, $t = 1985,...,2010$ & $  [\Delta \Psi_{k, t}]_{k = 1,...,4} = [0, 0, 0, 0] $ \tabularnewline
\bottomrule
\end{tabular}
  \end{minipage}
\end{table}

\begin{figure}[ht!]
  \caption{Descriptives, high switching costs and highly dispersed shocks}
  \label{mc:fig:descriptives-high-switch-costs}
  \centering
  \begin{minipage}[t]{\panelwidth}
    \subcaption{Occupation entrants/incumbents in $t+1$}
    \label{mc:fig:descriptives-high-switch-costs-switchers-entrants}
    \includegraphics[width=\textwidth]{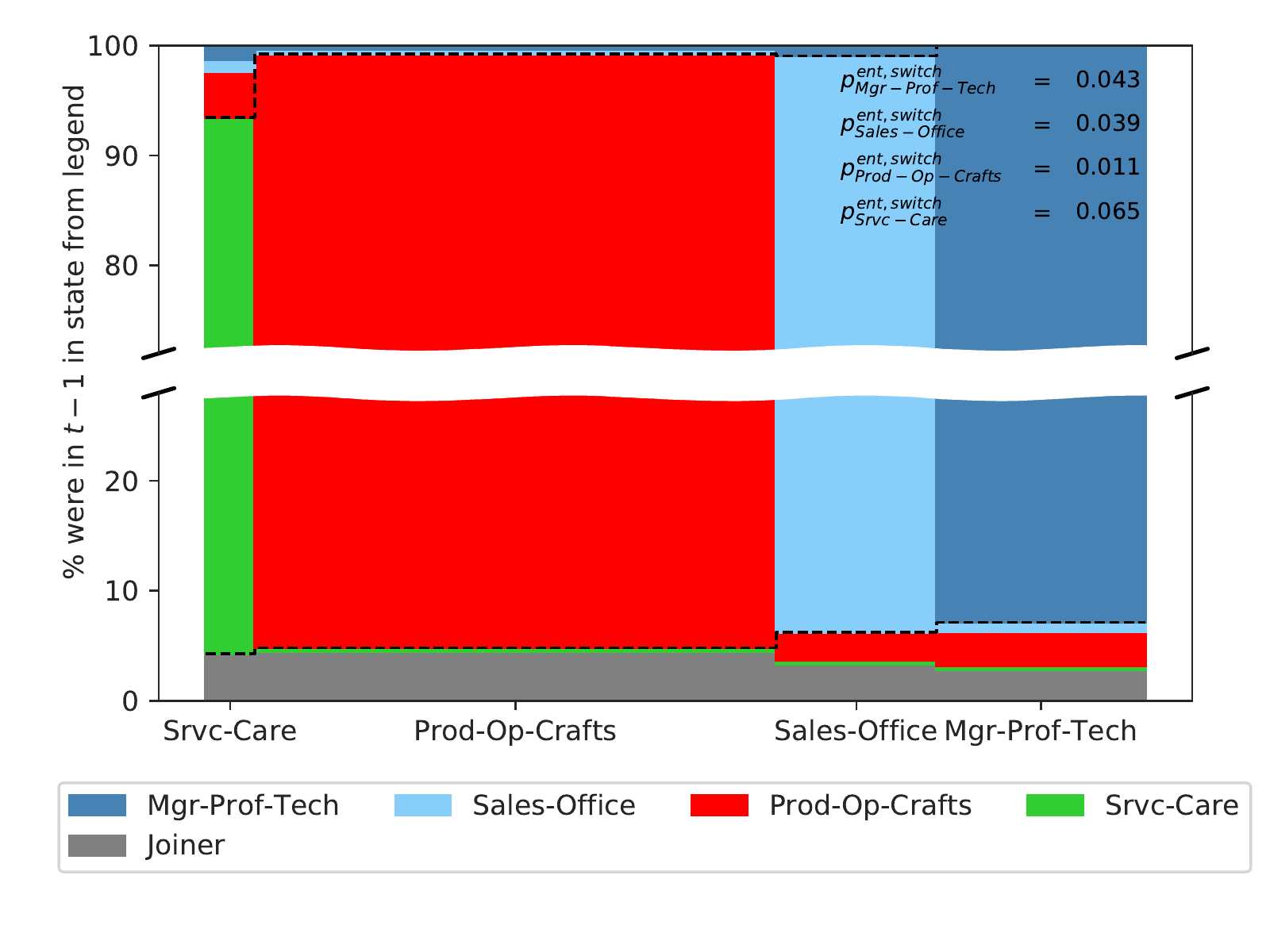}
  \end{minipage}
  \begin{minipage}[t]{\panelwidth}
    \subcaption{Occupation leavers/stayers in $t-1$}
    \label{mc:fig:descriptives-high-switch-costs-switchers-leavers}
    \includegraphics[width=\textwidth]{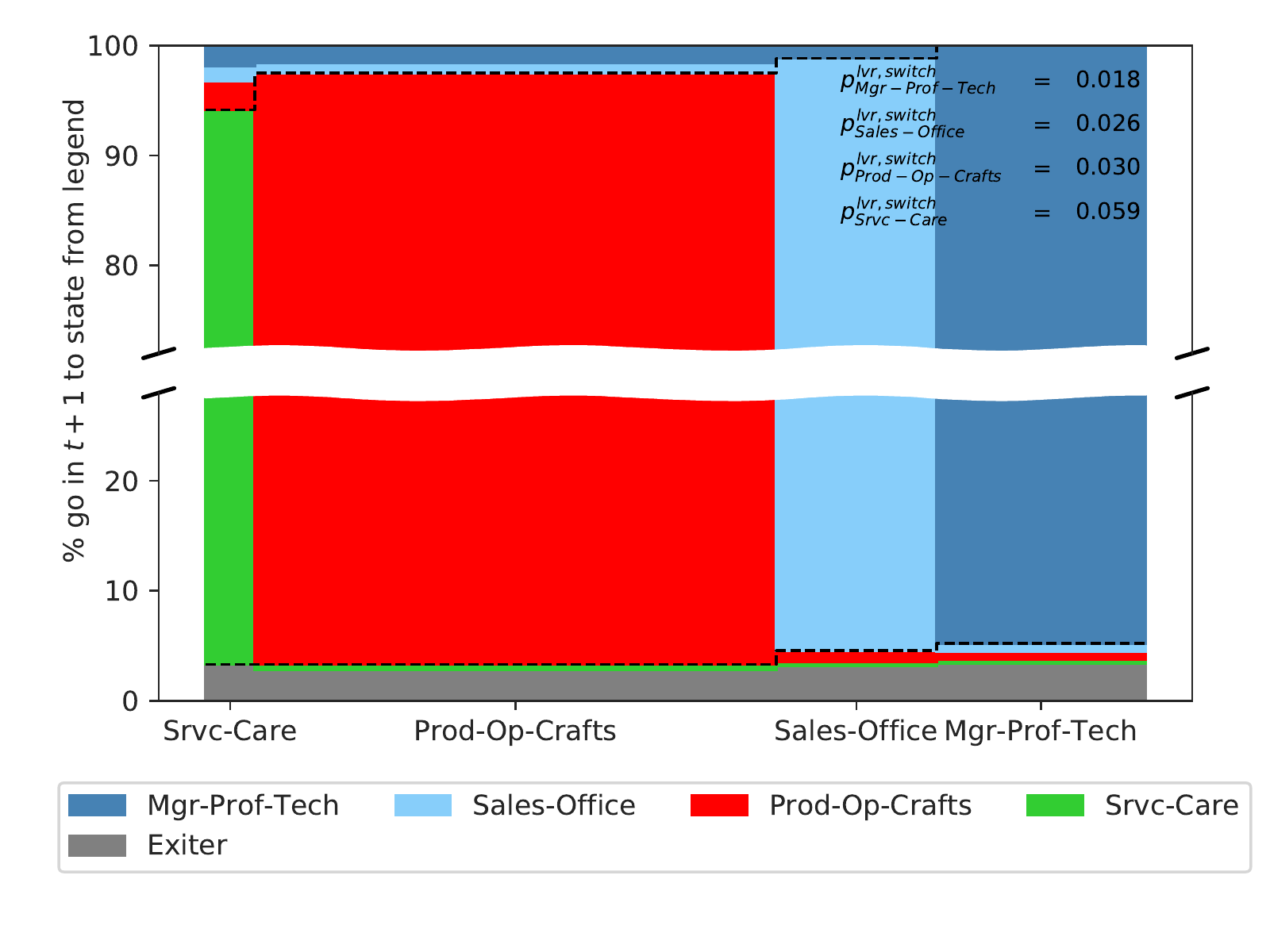}
  \end{minipage}
  \begin{minipage}[t]{\panelwidth}
    \subcaption{Distribution of annual wage growth}
    \label{mc:fig:descriptives-high-switch-costs-wage-growth-distribution}
    \includegraphics[width=\textwidth]{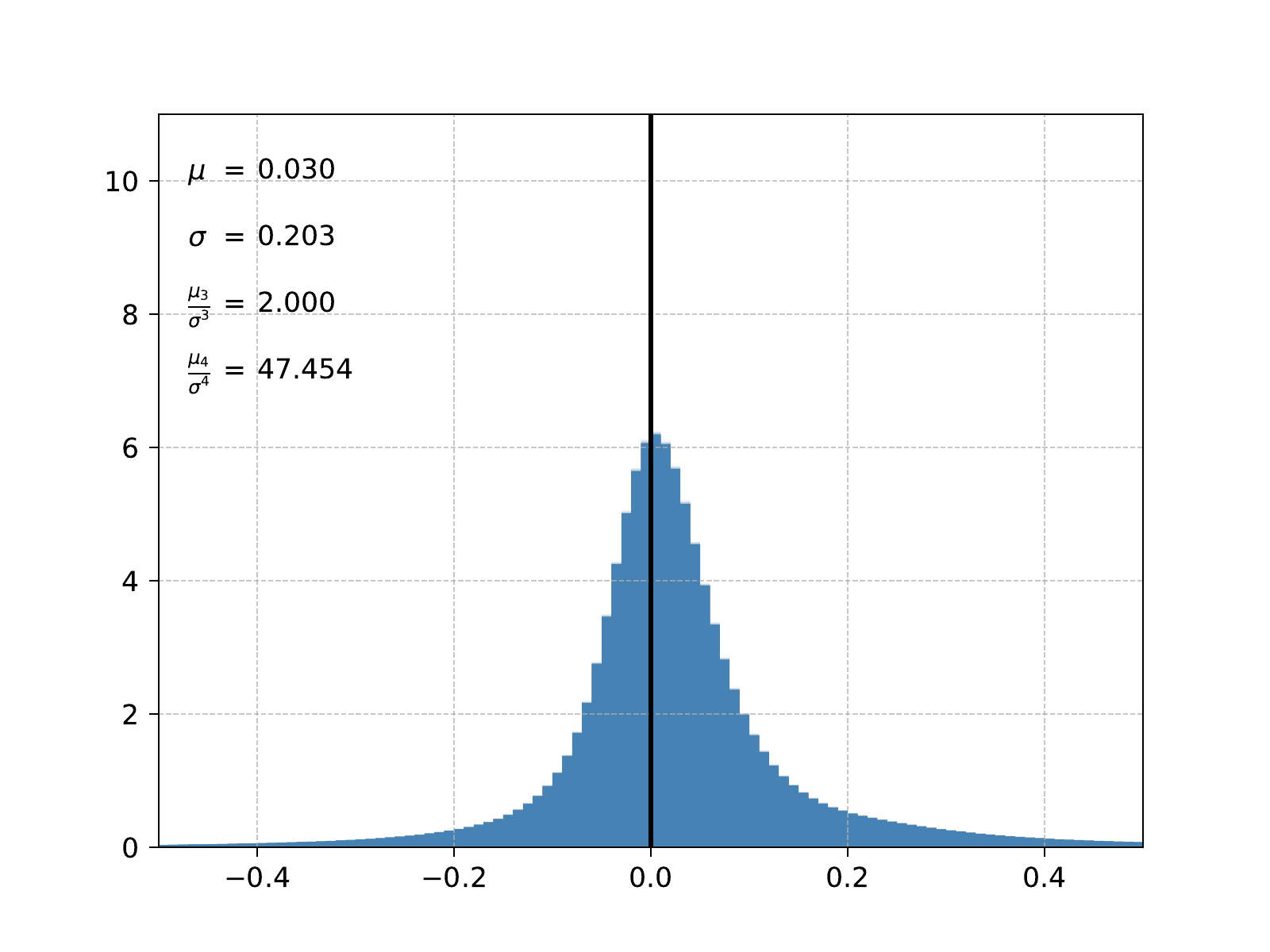}
  \end{minipage}
  \begin{minipage}[t]{\panelwidth}
    \subcaption{Evolution of the wage distribution}
    \label{mc:fig:descriptives-high-switch-costs-evolution-wage-inequality}
    \vskip2ex
    \includegraphics[width=\textwidth]{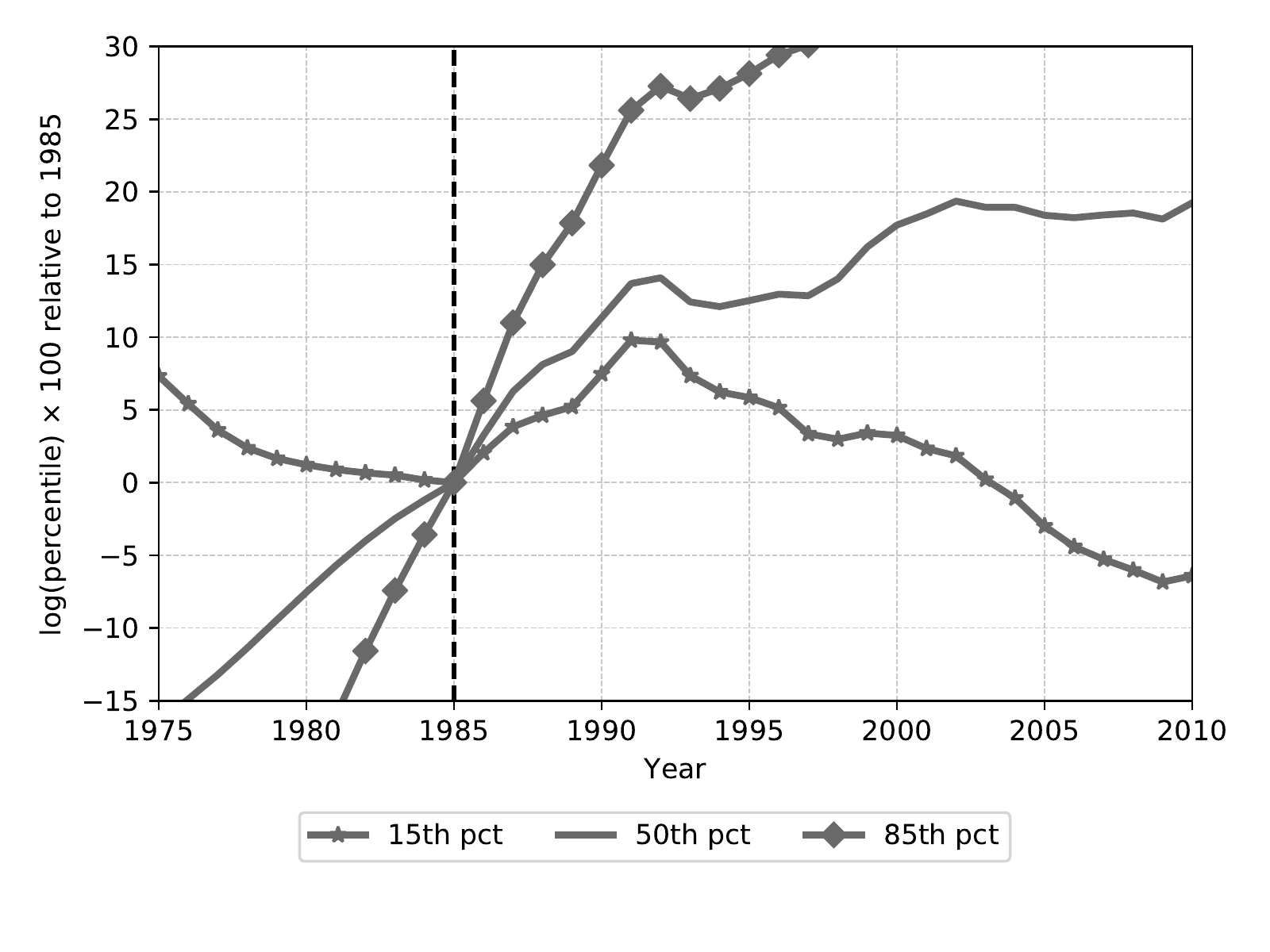}
  \end{minipage}
  \begin{minipage}{\textwidth}
    \scriptsize
    \emph{Notes:} \mcdescriptives
  \end{minipage}
\end{figure}

\begin{figure}[ht!]
  \caption{Estimation results, high switching costs and highly dispersed shocks}
  \label{mc:fig:estimation-high-switch-costs}
  \centering
  \begin{minipage}[t]{\panelwidth}
    \subcaption{Cumulative prices, saturated OLS}
    \label{mc:fig:estimation-ols-high-switch-costs-prices}
    \includegraphics[width=\textwidth]{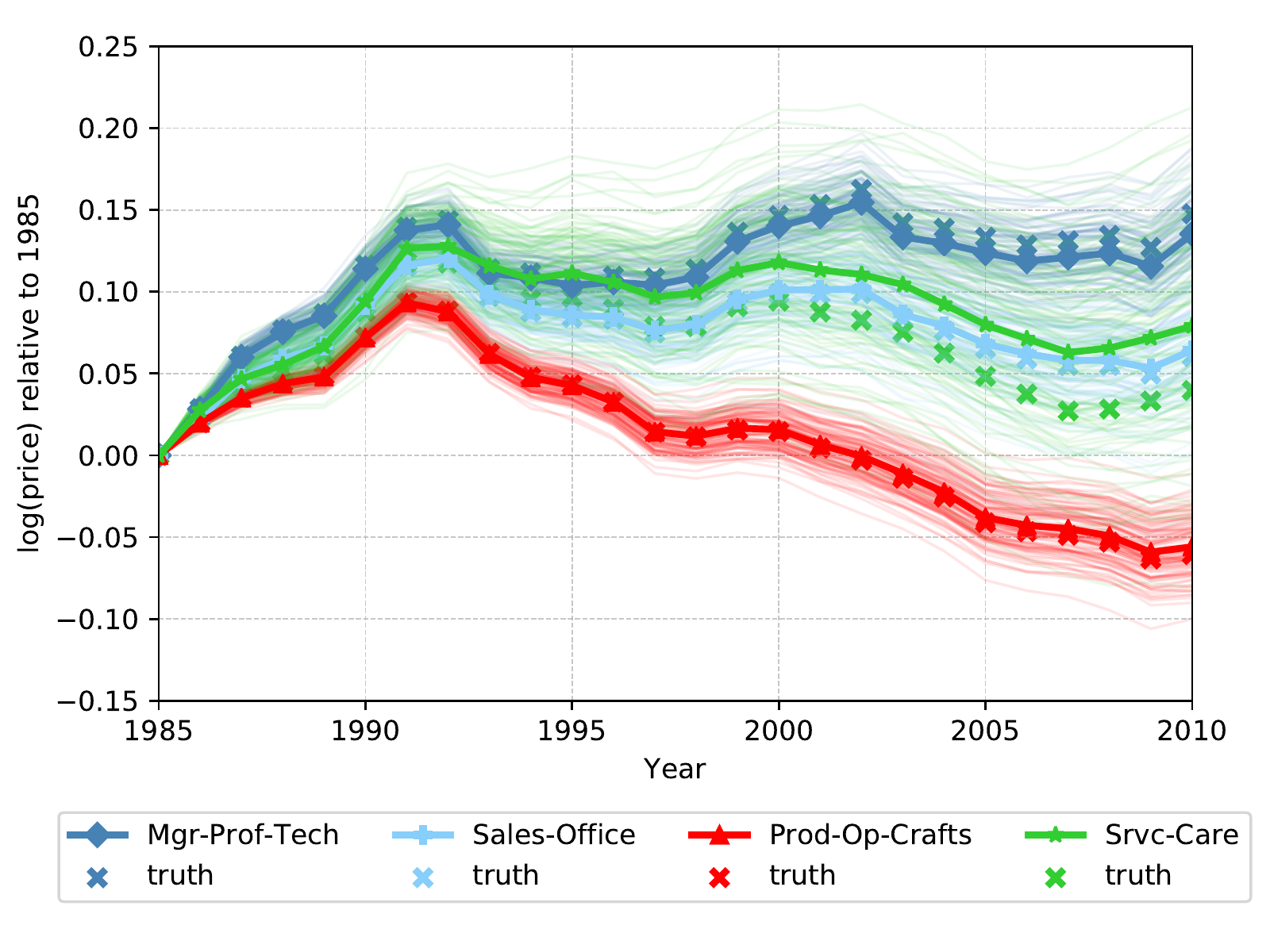}
  \end{minipage}
  \begin{minipage}[t]{\panelwidth}
    \subcaption{Skill accumulation, saturated OLS}
    \label{mc:fig:estimation-ols-high-switch-costs-skills}
    \includegraphics[width=\textwidth]{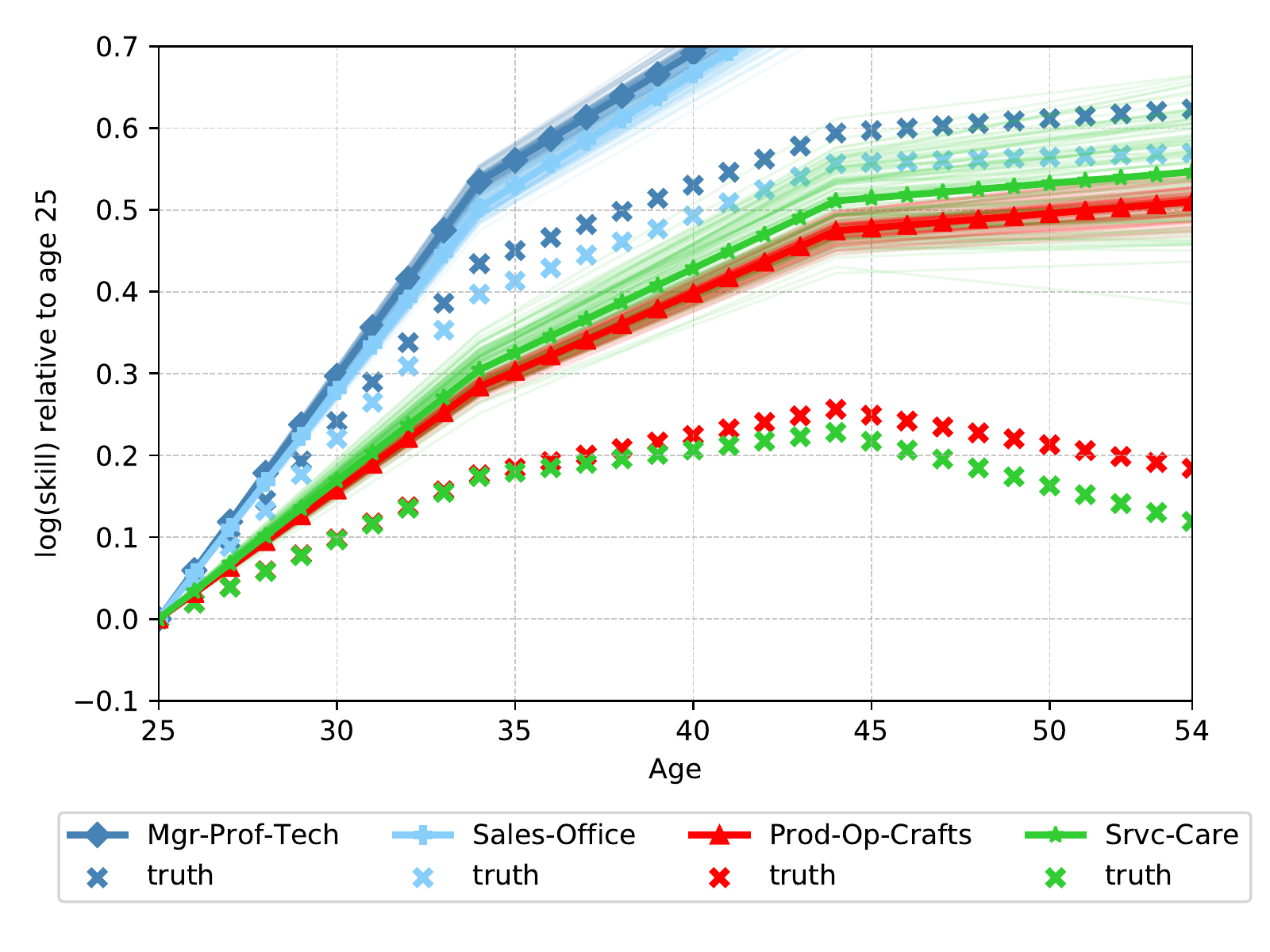}
  \end{minipage}
  \begin{minipage}[t]{\panelwidth}
    \subcaption{Cumulative prices, IV}
    \label{mc:fig:estimation-iv2-high-switch-costs-prices}
    \includegraphics[width=\textwidth]{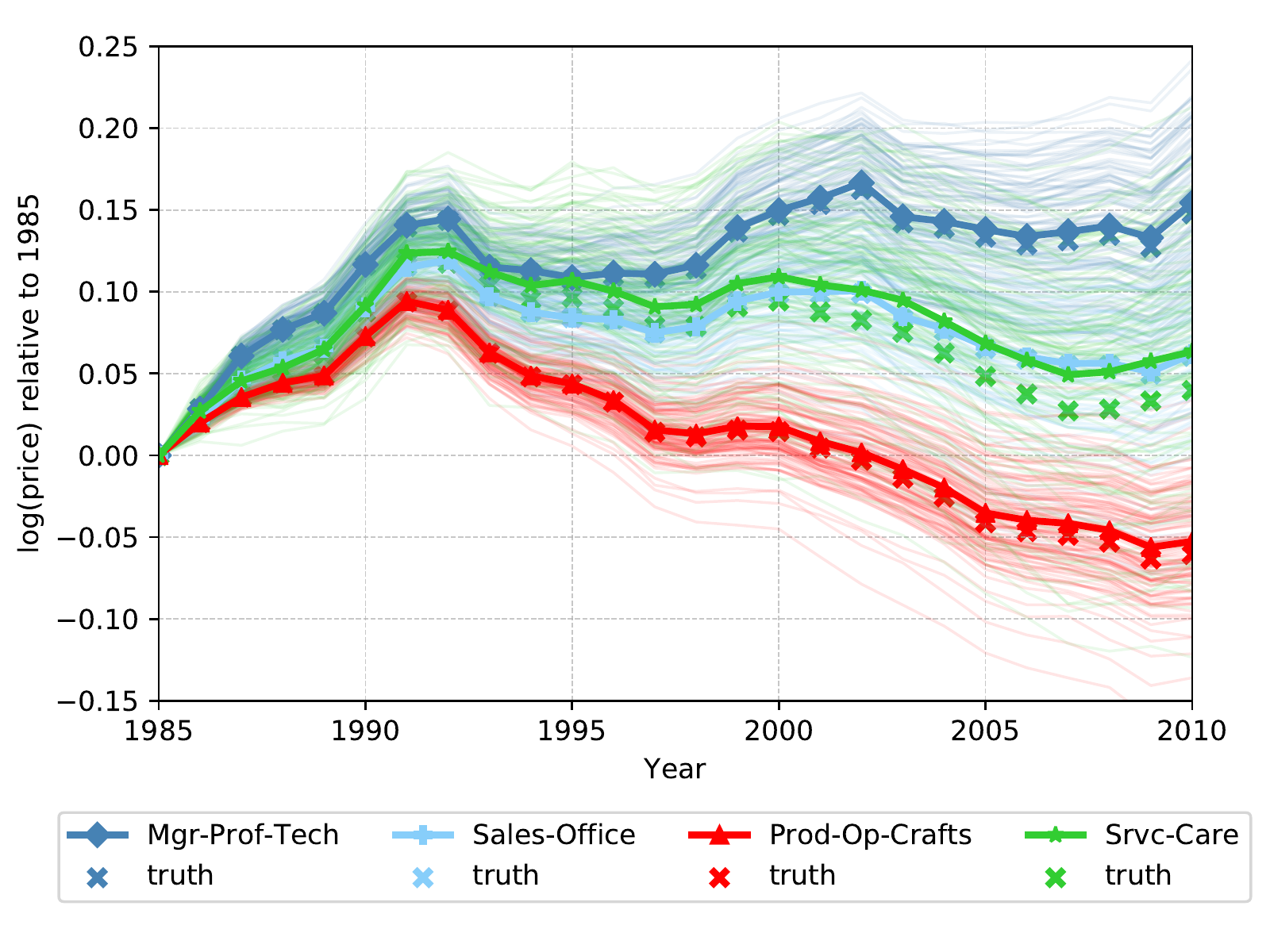}
  \end{minipage}
  \begin{minipage}[t]{\panelwidth}
    \subcaption{Skill accumulation, IV}
    \label{mc:fig:estimation-iv2-high-switch-costs-skills}
    \includegraphics[width=\textwidth]{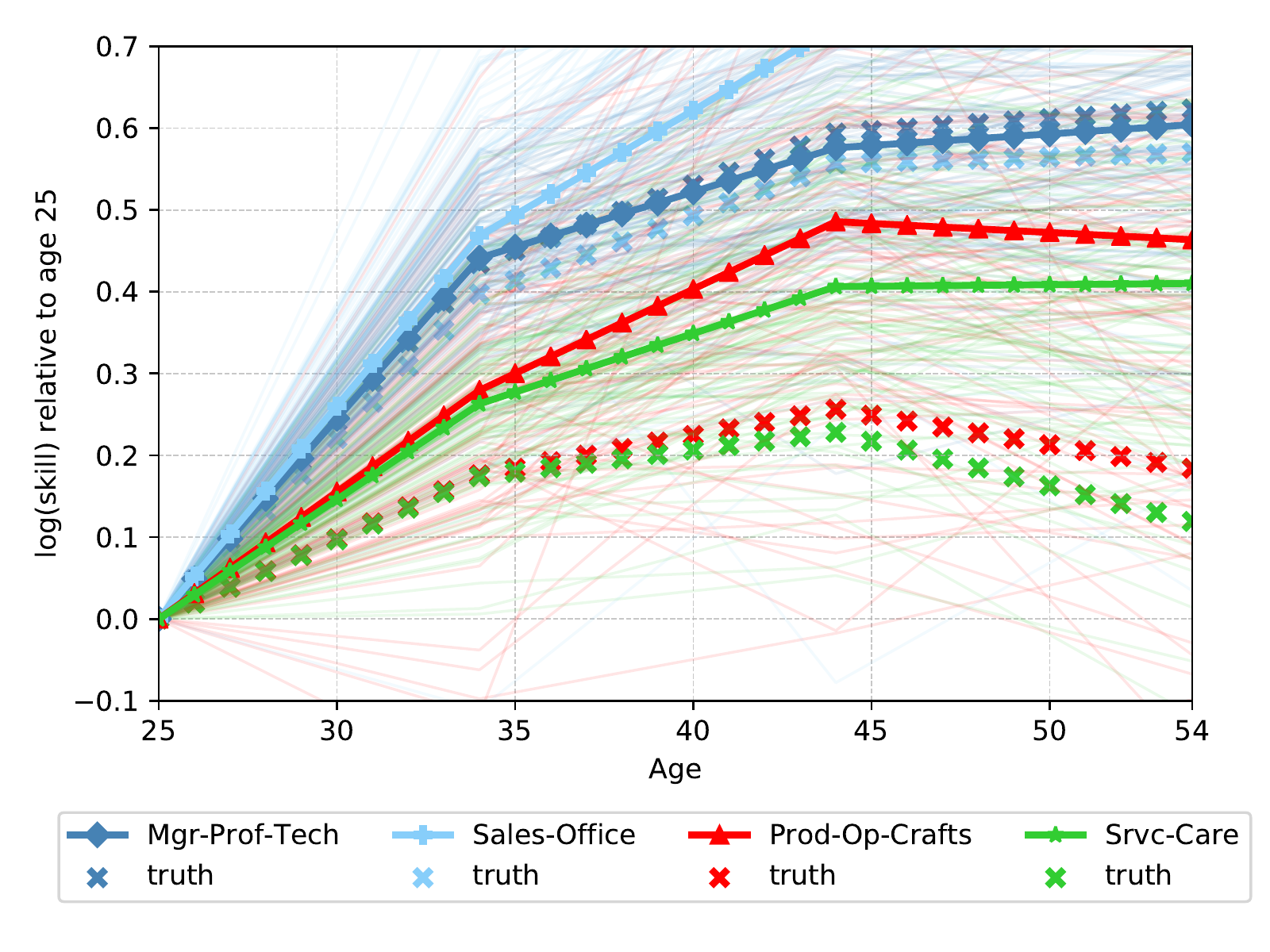}
  \end{minipage}
  \begin{minipage}{\textwidth}
    \scriptsize
    \emph{Notes:} \mcresults {} \olsresultsmc \ivresults
  \end{minipage}
\end{figure}

\clearpage

\subsection{Changing Amenities}
\label{mc:subsub:amenities}

\begin{figure}[ht!]
  \caption{Descriptives, trends in amenities}
  \label{mc:fig:descriptives-trends-amenities}
  \centering
  \begin{minipage}[t]{\panelwidth}
    \subcaption{Occupation entrants/incumbents in $t+1$}
    \label{mc:fig:descriptives-trends-amenities-switchers-entrants}
    \includegraphics[width=\textwidth]{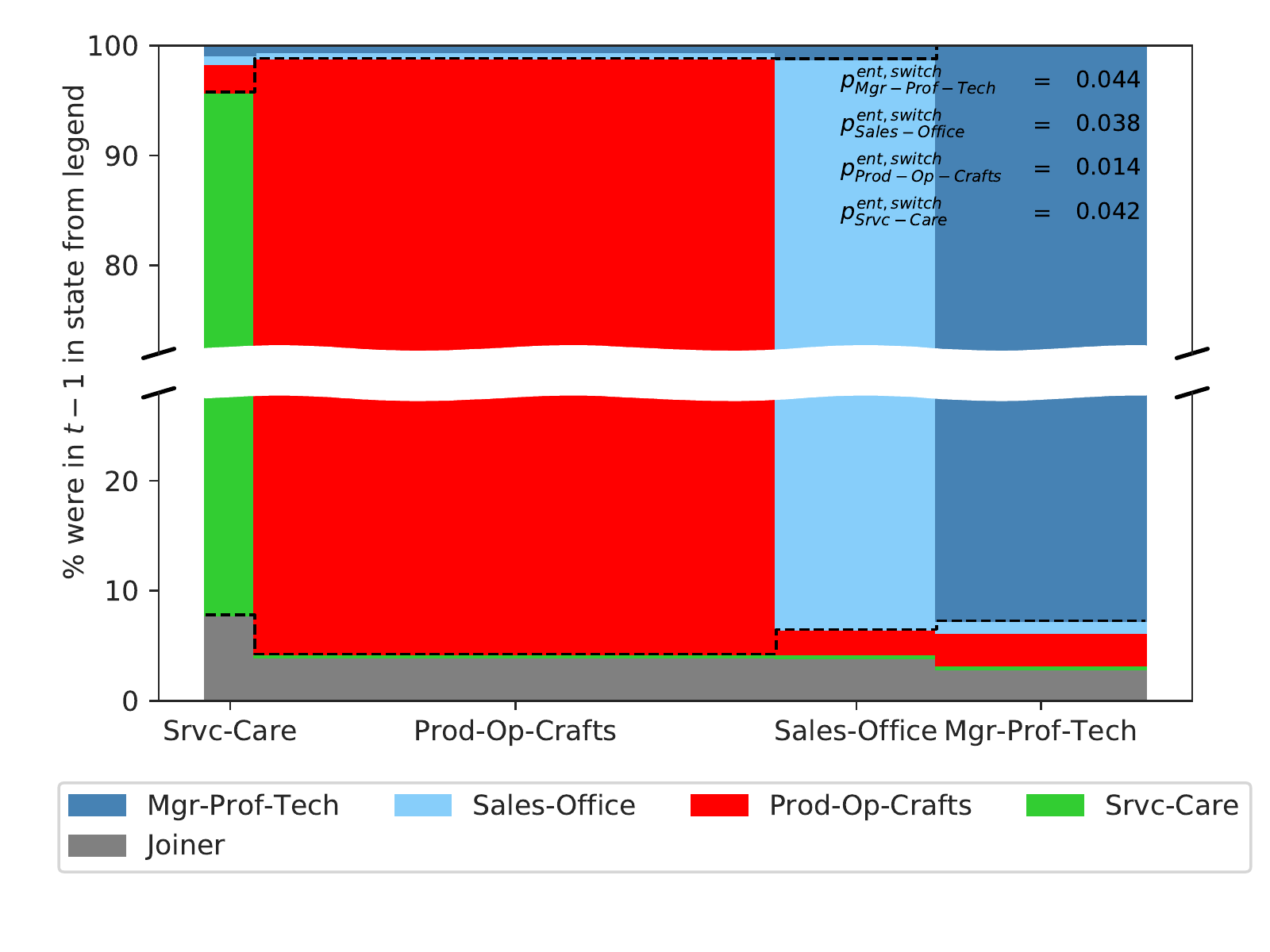}
  \end{minipage}
  \begin{minipage}[t]{\panelwidth}
    \subcaption{Occupation leavers/stayers in $t-1$}
    \label{mc:fig:descriptives-trends-amenities-switchers-leavers}
    \includegraphics[width=\textwidth]{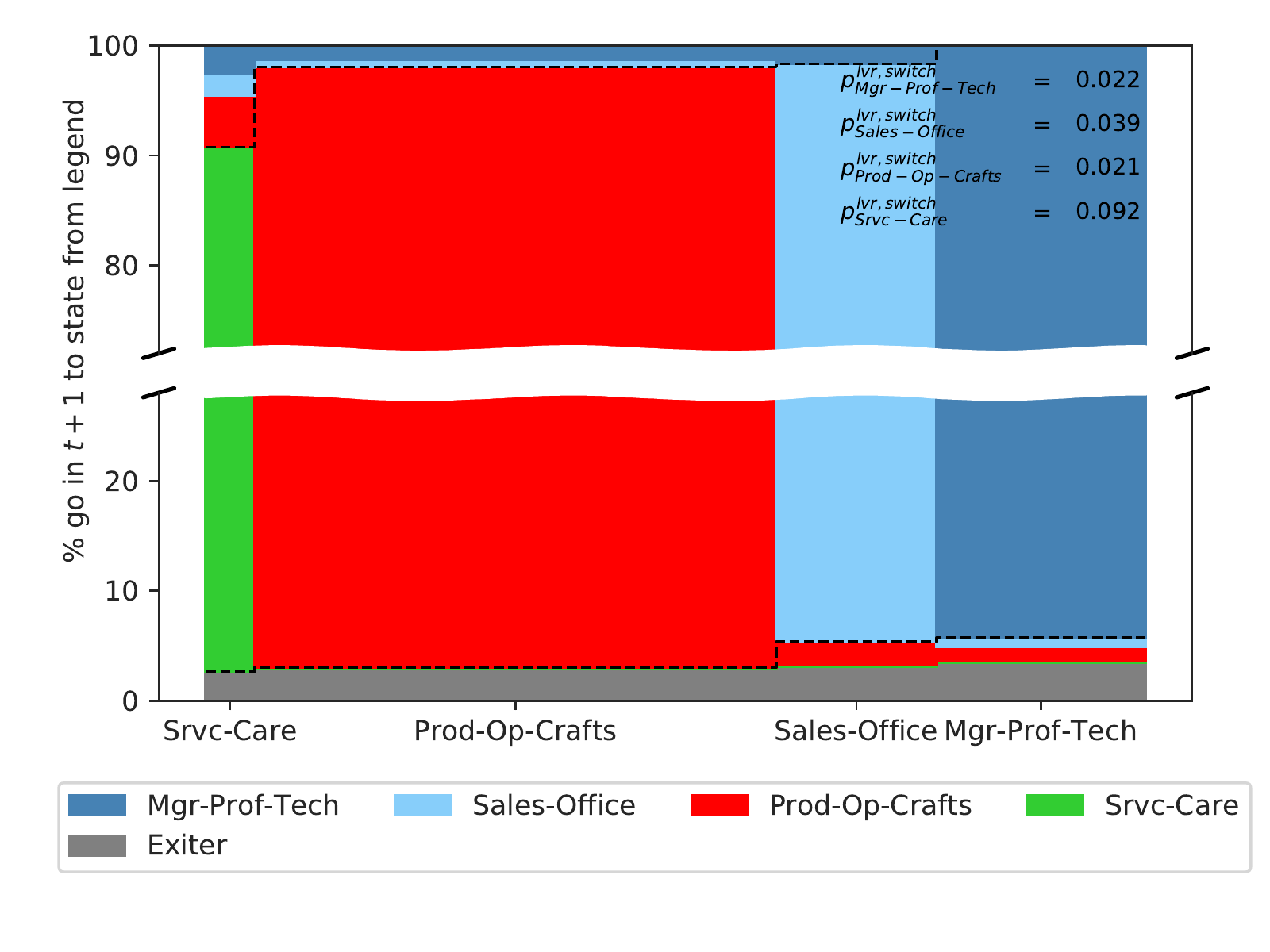}
  \end{minipage}
  \begin{minipage}[t]{\panelwidth}
    \subcaption{Distribution of annual wage growth}
    \label{mc:fig:descriptives-trends-amenities-wage-growth-distribution}
    \includegraphics[width=\textwidth]{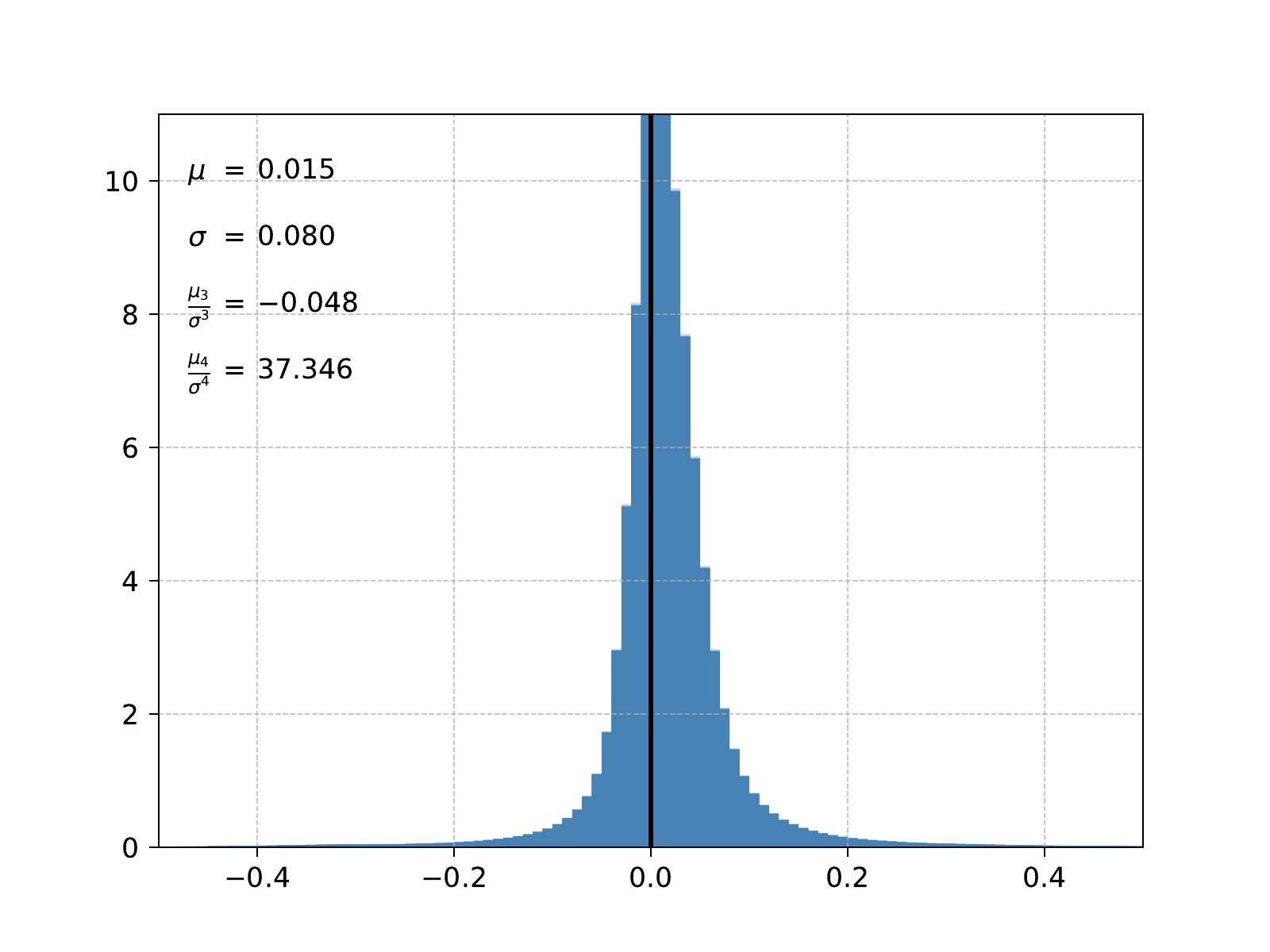}
  \end{minipage}
  \begin{minipage}[t]{\panelwidth}
    \subcaption{Evolution of the wage distribution}
    \label{mc:fig:descriptives-trends-amenities-evolution-wage-inequality}
    \vskip2ex
    \includegraphics[width=\textwidth]{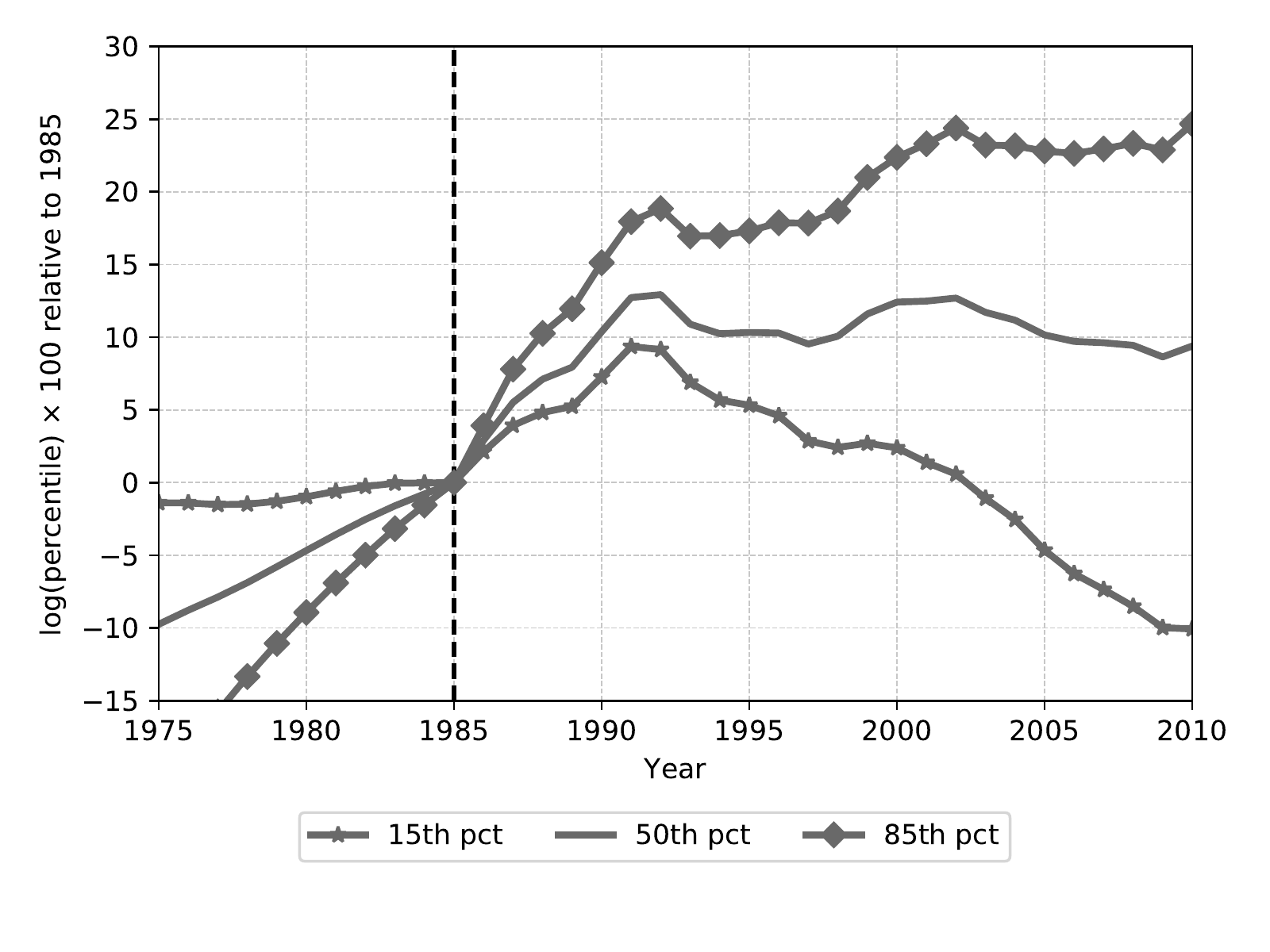}
  \end{minipage}
  \begin{minipage}{\textwidth}
    \scriptsize
    \emph{Notes:} \mcdescriptives
  \end{minipage}
\end{figure}

Here, we introduce trends in non-wage amenities to the worker's decision problem. These trends make some occupations more attractive over time even when prices did not change. We implement them
so that relative to Prod-Op-Crafts, the other occupations became less attractive. This makes workers move into Prod-Op-Crafts despite falling prices.
Figure~\ref{mc:fig:descriptives-trends-amenities} shows the descriptives.

\begin{figure}[ht!]
  \caption{Estimation results, trends in amenities}
  \label{mc:fig:estimation-trends-amenities}
  \centering
  \begin{minipage}[t]{\panelwidth}
    \subcaption{Cumulative prices, saturated OLS}
    \label{mc:fig:estimation-ols-trends-amenities-prices}
    \includegraphics[width=\textwidth]{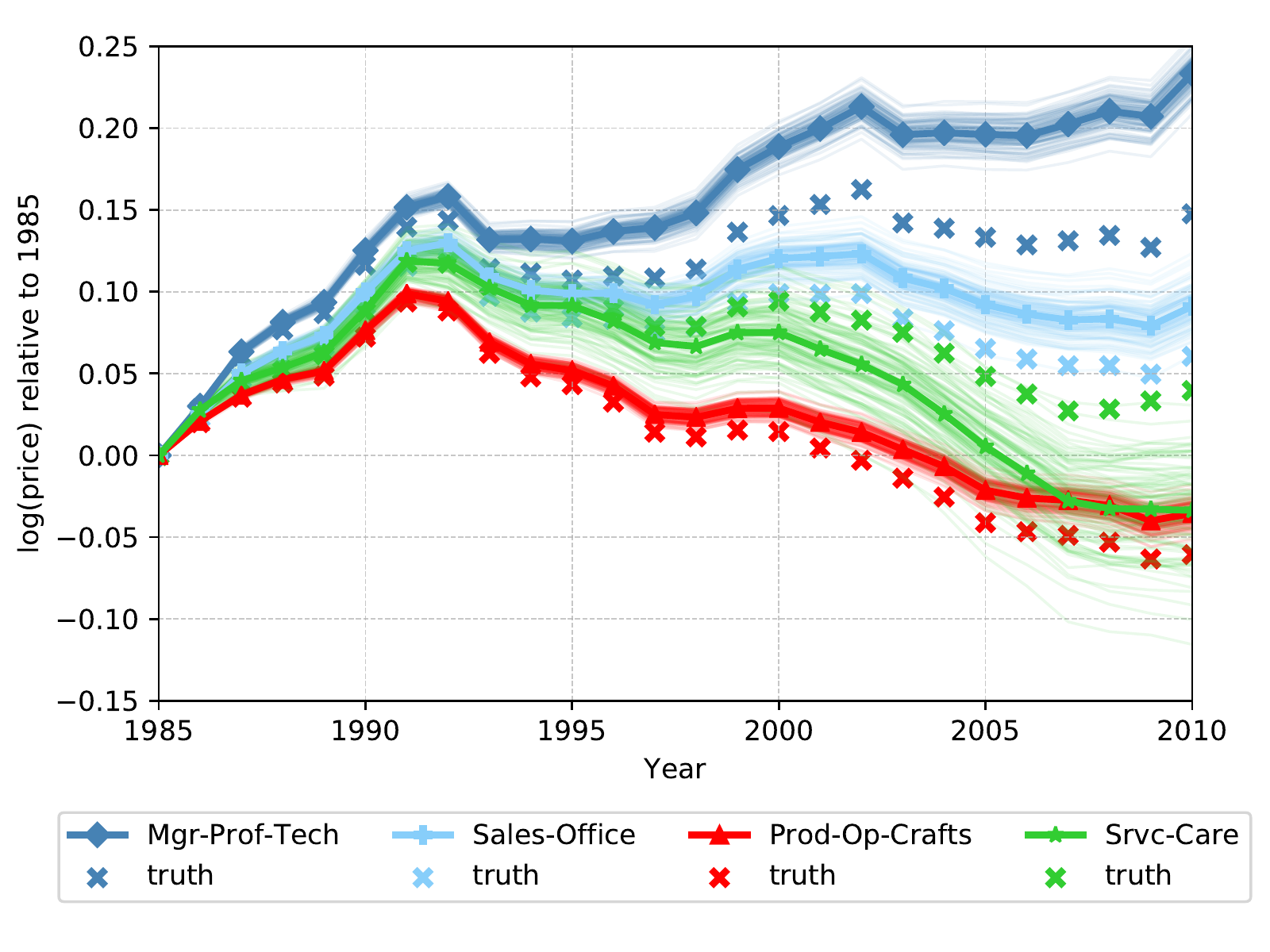}
  \end{minipage}
  \begin{minipage}[t]{\panelwidth}
    \subcaption{Skill accumulation, saturated OLS}
    \label{mc:fig:estimation-ols-trends-amenities-skills}
    \includegraphics[width=\textwidth]{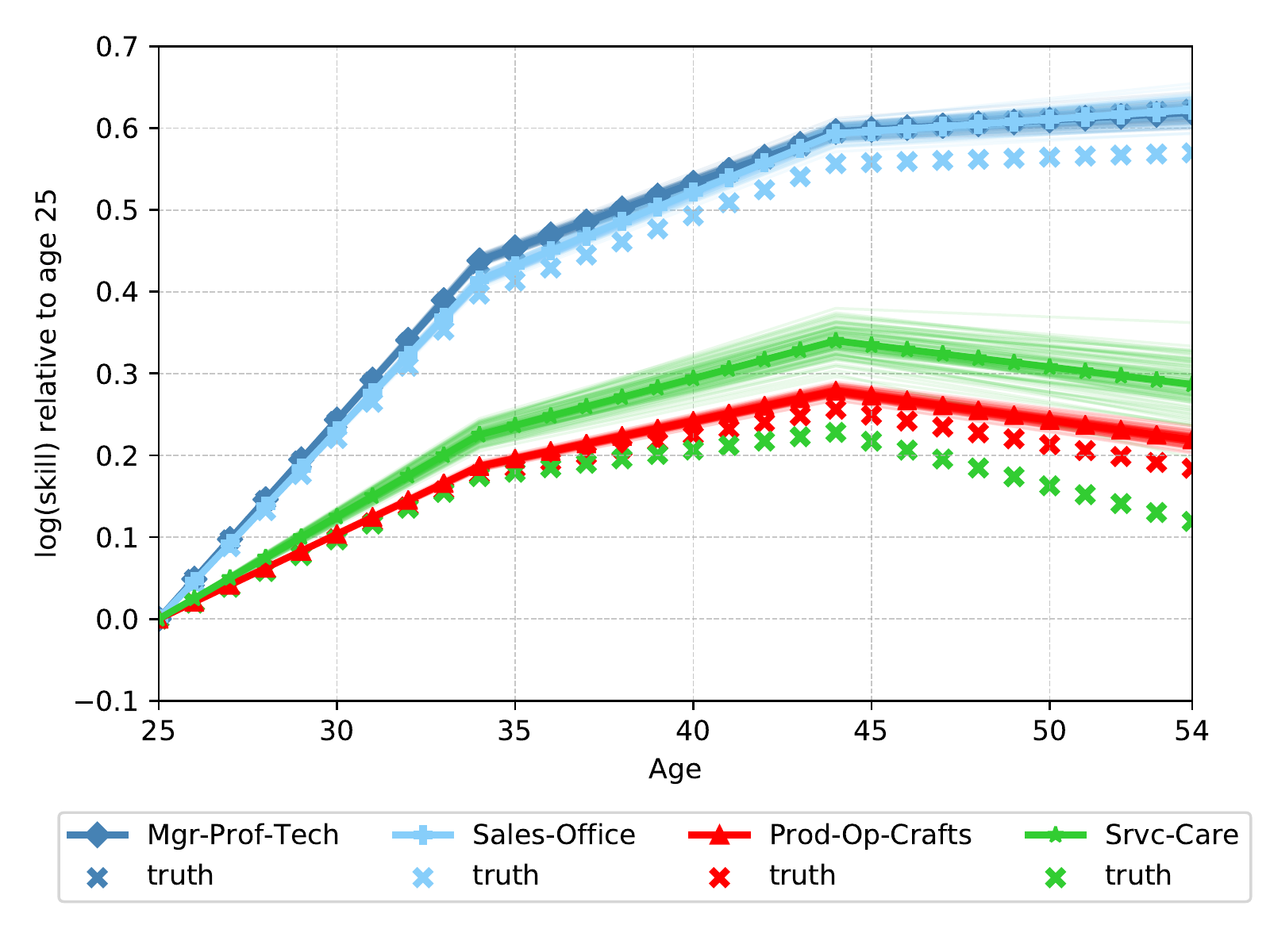}
  \end{minipage}
  \begin{minipage}[t]{\panelwidth}
    \subcaption{Cumulative prices, OLS + Amenities correction}
    \label{mc:fig:estimation-correction-trends-amenities-prices}
    \includegraphics[width=\textwidth]{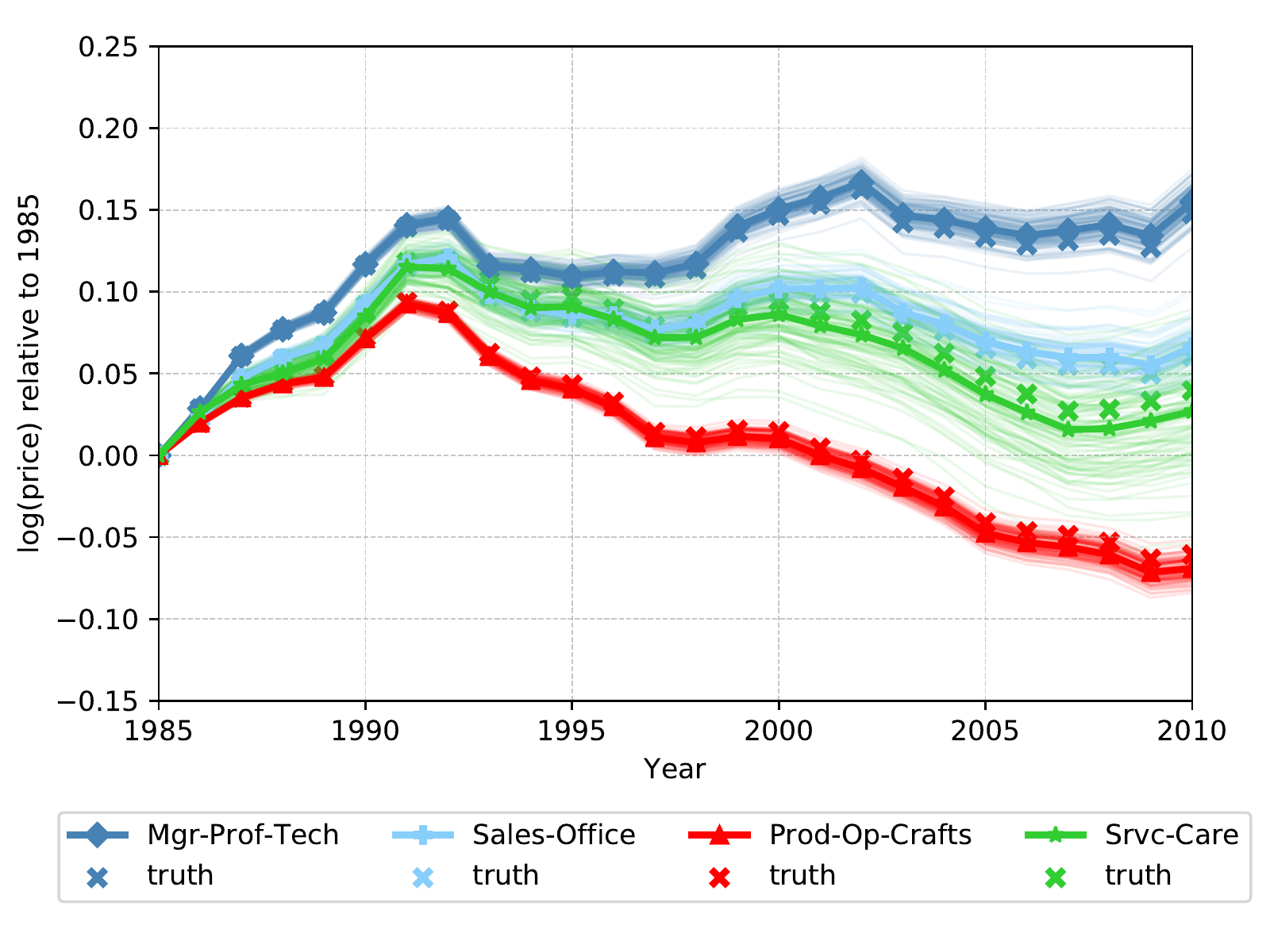}
  \end{minipage}
  \begin{minipage}[t]{\panelwidth}
    \subcaption{Skill accumulation, OLS + Amenities correction}
    \label{mc:fig:estimation-correction-trends-amenities-skills}
    \includegraphics[width=\textwidth]{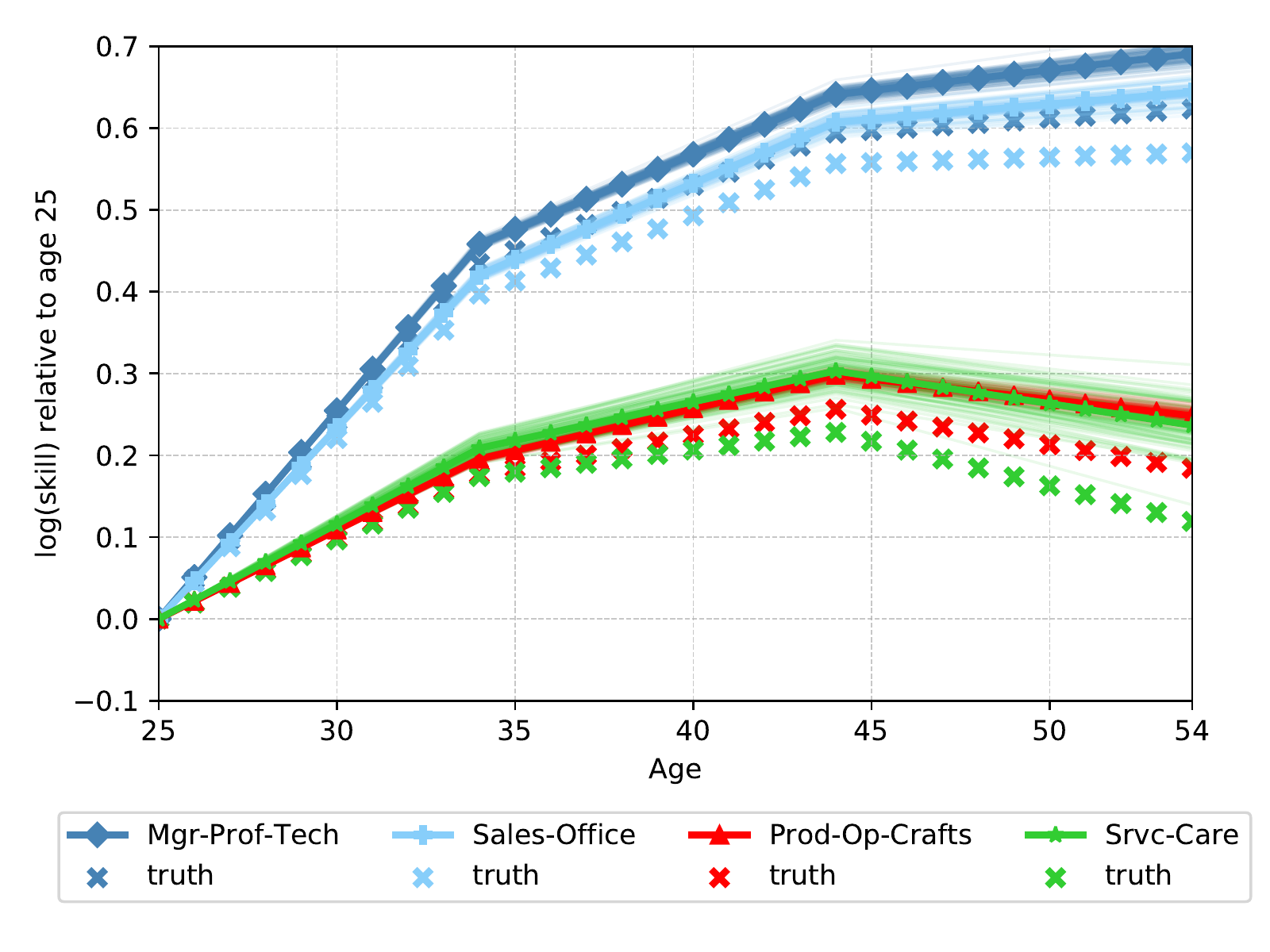}
  \end{minipage}
  \begin{minipage}{\textwidth}
    \scriptsize
    \emph{Notes:} \mcresults {} \olsresultsmc \amenityresults
  \end{minipage}
\end{figure}

Figure~\ref{mc:fig:estimation-trends-amenities} shows the estimation results. The baseline method is biased as skill price and amenity values are confounded by each other, making us overpredict the fall in Srvc-Care prices where amenities fell. The adjustment described in Equation~\eqref{mc:eq:estimation-utility} takes care of this.

\begin{table}[h!]
  \centering
  \caption{Parameters}
  \label{mc:tab:trends-amenities}
  \begin{minipage}[t]{\linewidth}
    \centering
    \begin{tabular}{lp{8cm}} \toprule 
Parameter & Value \tabularnewline
\midrule
$N$ & 50000 \tabularnewline
Repetitions & 100 \tabularnewline
Skill shocks in $k$ & observed wage growth distribution, $(\mu, \sigma_k) = (0, 0.5 \cdot \sigma^{SIAB}_{\Delta \log(w_i)})$ \tabularnewline
Stayers accumulation $\gamma_{{k, k, a}}, k' = k $ & $\hat{{\gamma}}^{{SIAB}}_{{k, k, a}} $  \tabularnewline
Cross accumulation $\gamma_{k', k, a}, k' \ne k $ & $\frac{1}{3} \hat{\gamma}^{SIAB}_{k', k, a}$ \tabularnewline
$\rho$ in $\varepsilon_{i, t} = \rho \varepsilon_{i, t-1} + v_{i, t}$ & 0 \tabularnewline
Switching costs $c$ & $ 0 $ \tabularnewline
Amenity trends, $t = 1985,...,2010$ & $  [\Delta \Psi_{k, t}]_{k = 1,...,4} = [0.02, 0, 0, 0] $ \tabularnewline
\bottomrule
\end{tabular}
  \end{minipage}
\end{table}

\clearpage

\subsection{Occupation-Specific Fixed Effects}
\label{mc:subsub:cortes}

\begin{figure}[ht!]
  \caption{Estimation results, no shocks as in Table~\ref{mc:tab:no-shocks}}
  \label{mc:fig:estimation-cortes-no-shocks}
  \centering
  \begin{minipage}[t]{\panelwidth}
    \subcaption{Cumulative prices, no base period}
    \label{mc:fig:estimation-cortes-no-shocks-prices}
    \includegraphics[width=\textwidth]{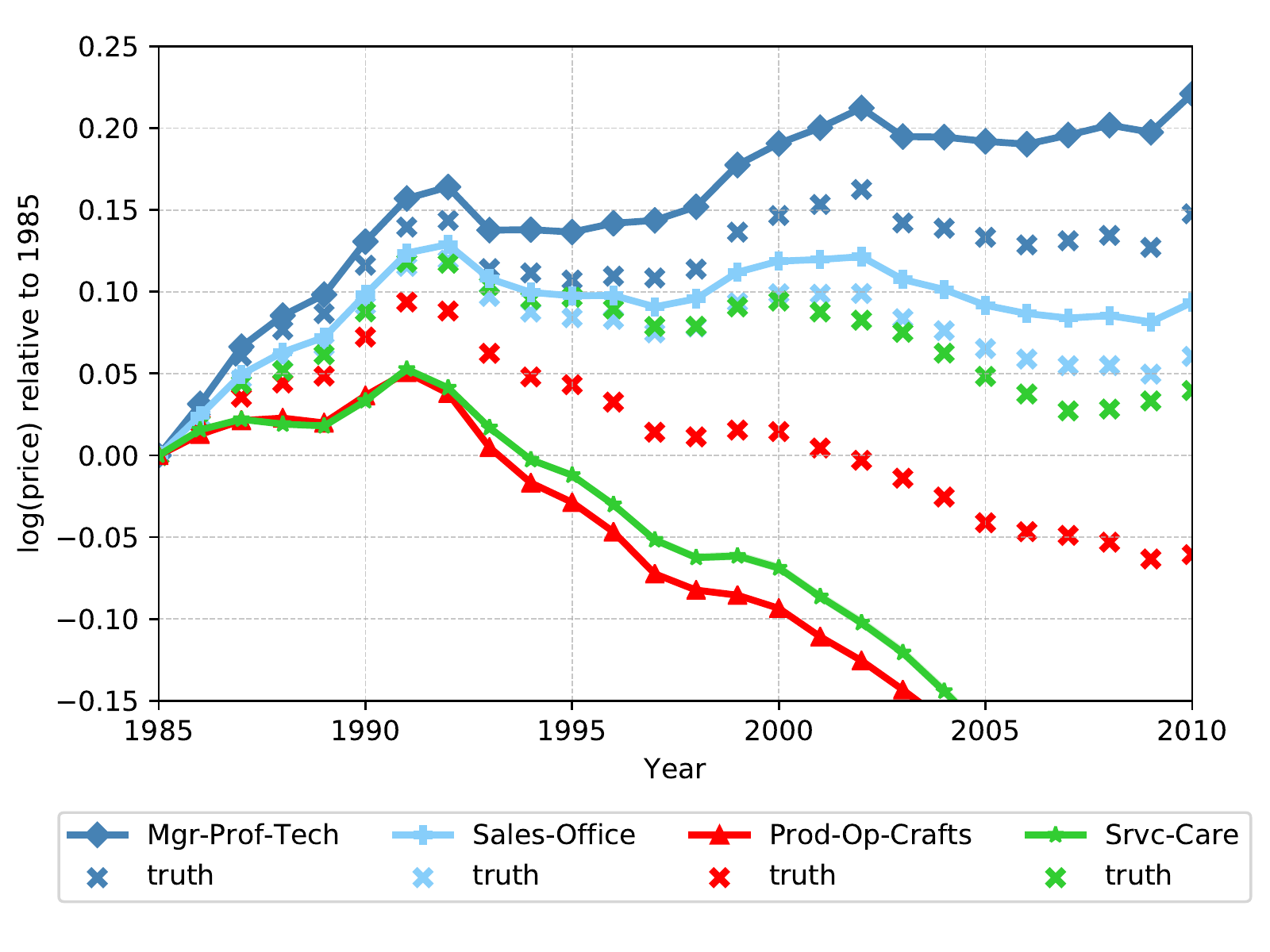}
  \end{minipage}
  \begin{minipage}[t]{\panelwidth}
    \subcaption{Skill accumulation, no base period}
    \label{mc:fig:estimation-cortes-no-shocks-skills}
    \includegraphics[width=\textwidth]{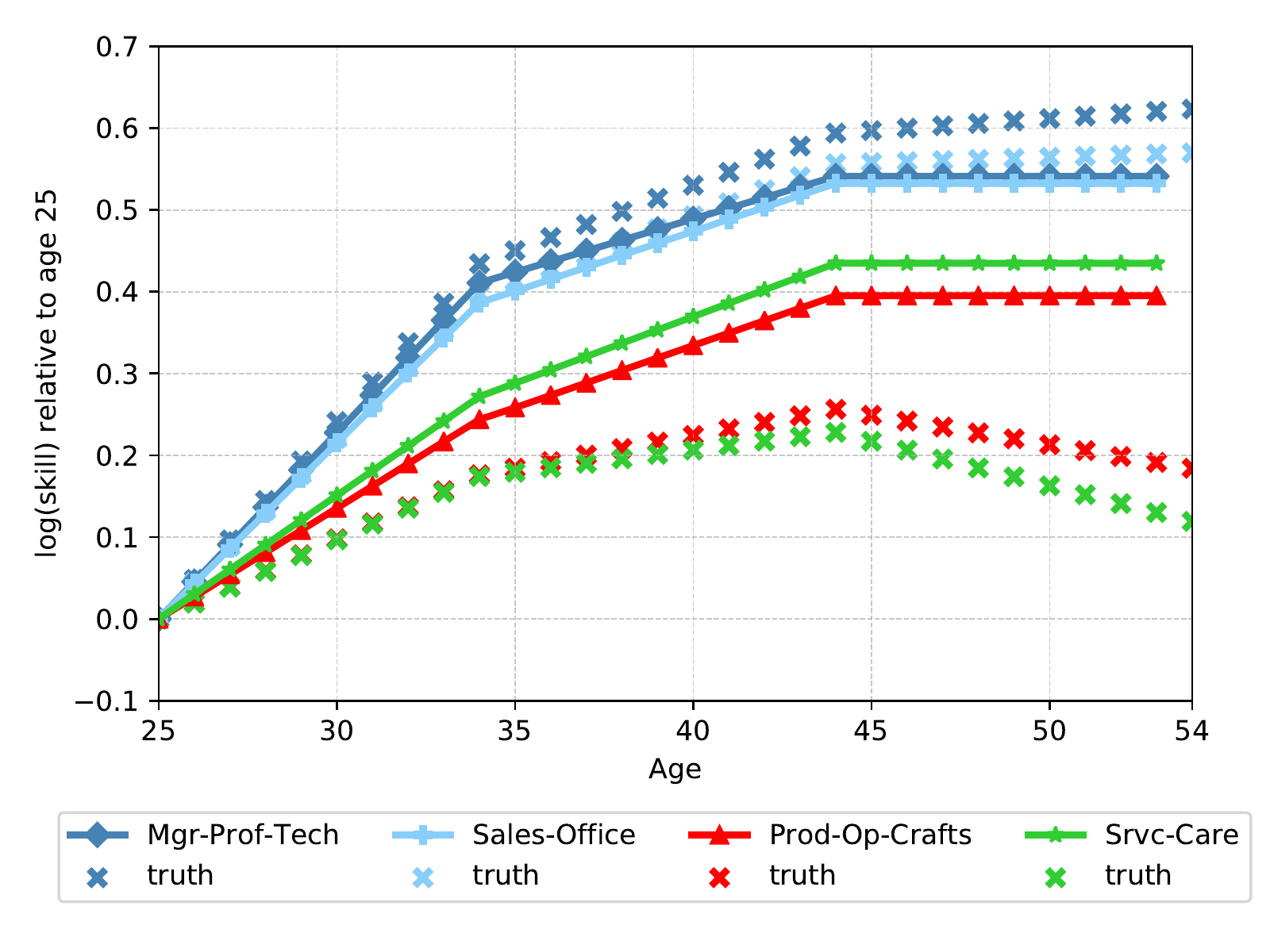}
  \end{minipage}
  \begin{minipage}[t]{\panelwidth}
    \subcaption{Cum.\ prices, occ.-stint fixed effects}
    \label{mc:fig:estimation-cortes-stint-no-shocks-prices}
    \includegraphics[width=\textwidth]{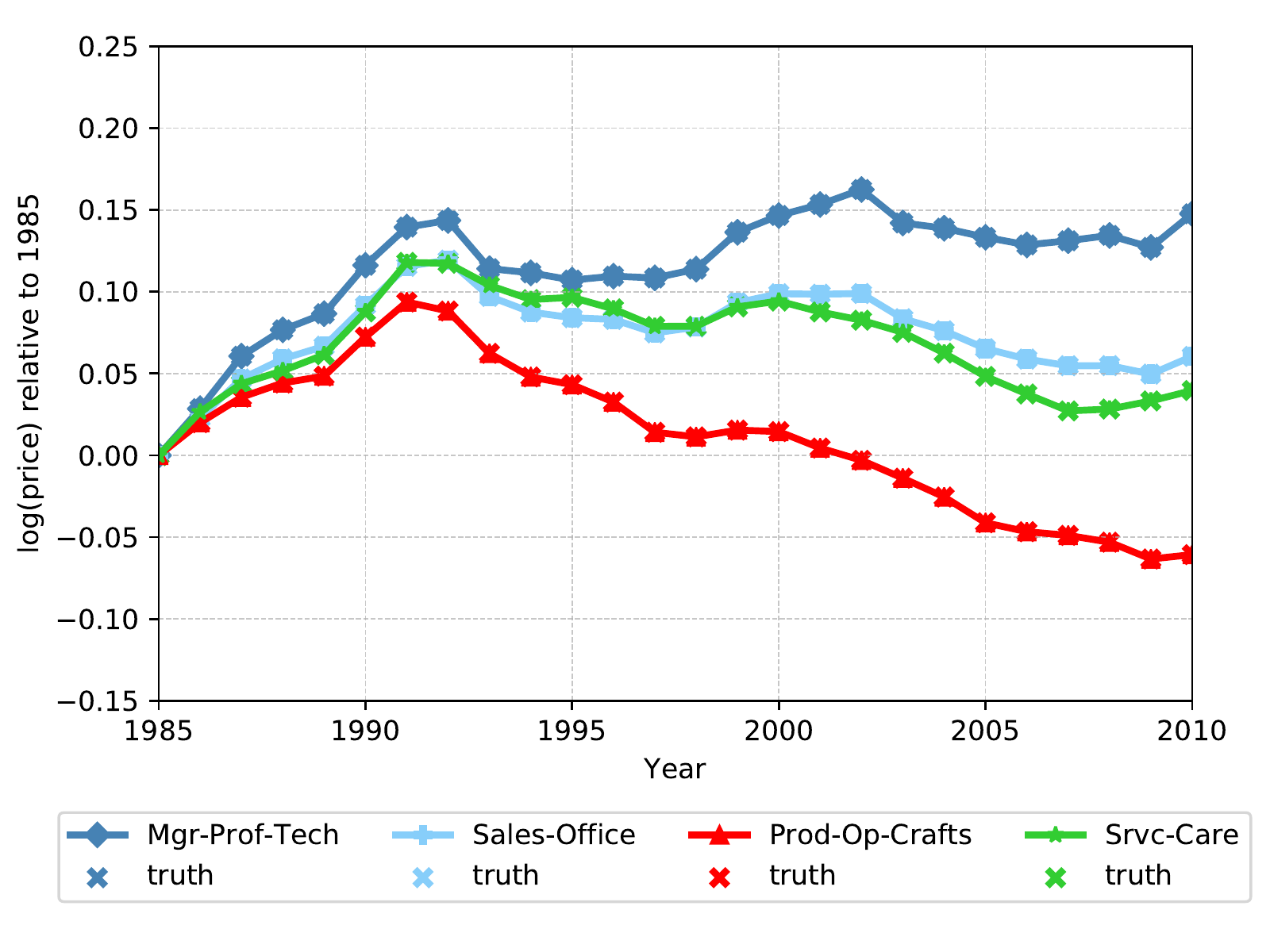}
  \end{minipage}
  \begin{minipage}[t]{\panelwidth}
    \subcaption{Skill acc., occ.-stint fixed effects}
    \label{mc:fig:estimation-cortes-stint-no-shocks-skills}
    \includegraphics[width=\textwidth]{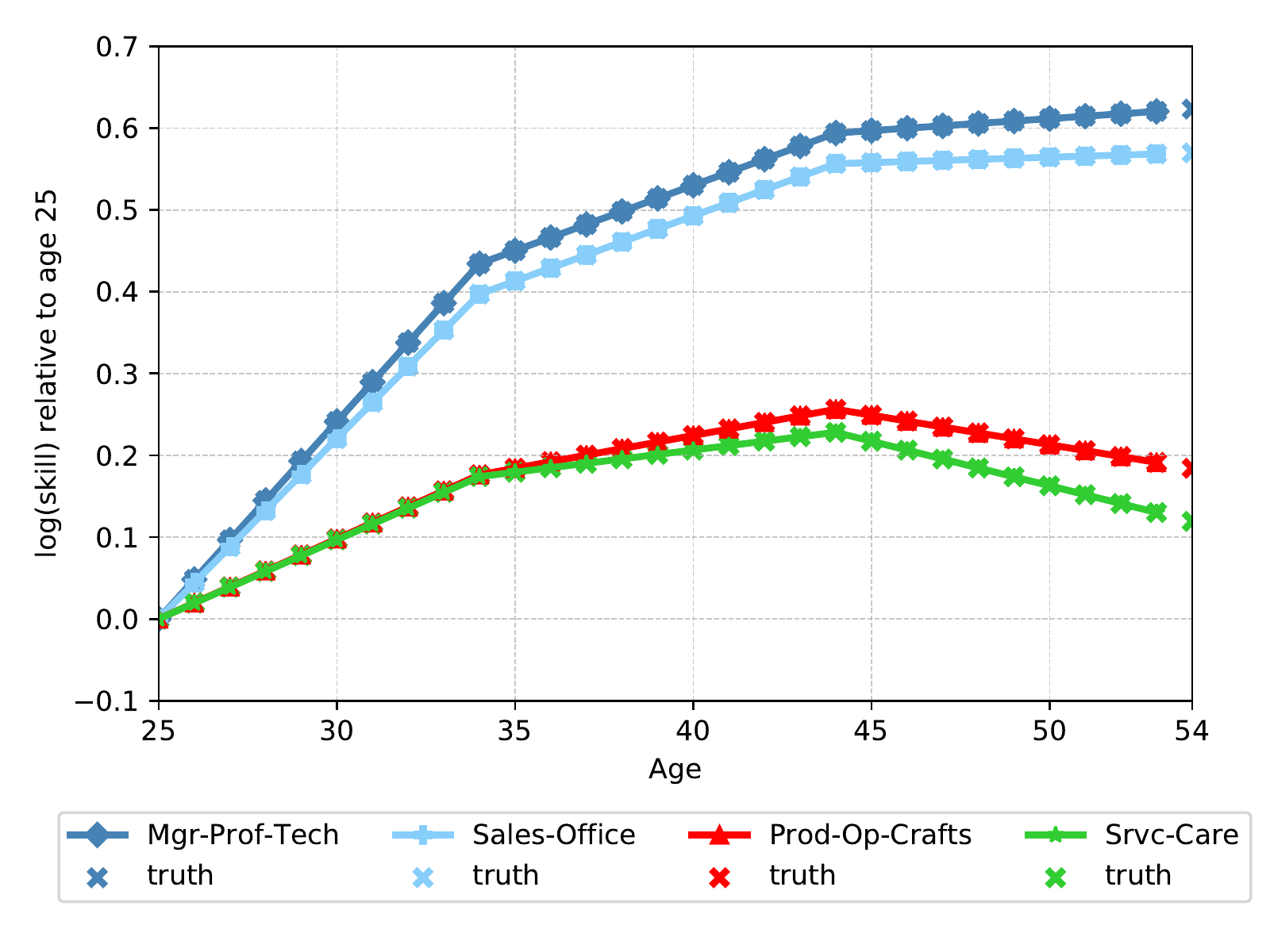}
  \end{minipage}
  \begin{minipage}{\textwidth}
    \scriptsize
    \emph{Notes:} \mcresults \norandomness Panels~\ref{mc:fig:estimation-cortes-no-shocks-prices} and \ref{mc:fig:estimation-cortes-no-shocks-skills} show results from fixed effects estimation when no base period is included as described in Section~\ref{mc:sub:fixed-effects-model-skill-acc}. The remaining two Panels show estimates which are identified from year specific occupation fixed effects while including a separate worker-occupation fixed effects for each time the worker revisits an occupation (after a possible break or after return from another occupation). Additionally, we include controls for age and occupation dependent skill accumulation following Equation~\eqref{mc:eq:wageFEspecific-stint}.
  \end{minipage}
\end{figure}

\begin{figure}[ht!]
  \caption{Estimation results, moderate shocks as in Table~\ref{mc:tab:moderate-shocks}}
  \label{mc:fig:estimation-cortes-moderate-shocks}
  \centering
  \begin{minipage}[t]{\panelwidth}
    \subcaption{Cum.\ prices, occ.-stint fixed effects}
    \label{mc:fig:estimation-cortes-stint-moderate-shocks-prices}
    \includegraphics[width=\textwidth]{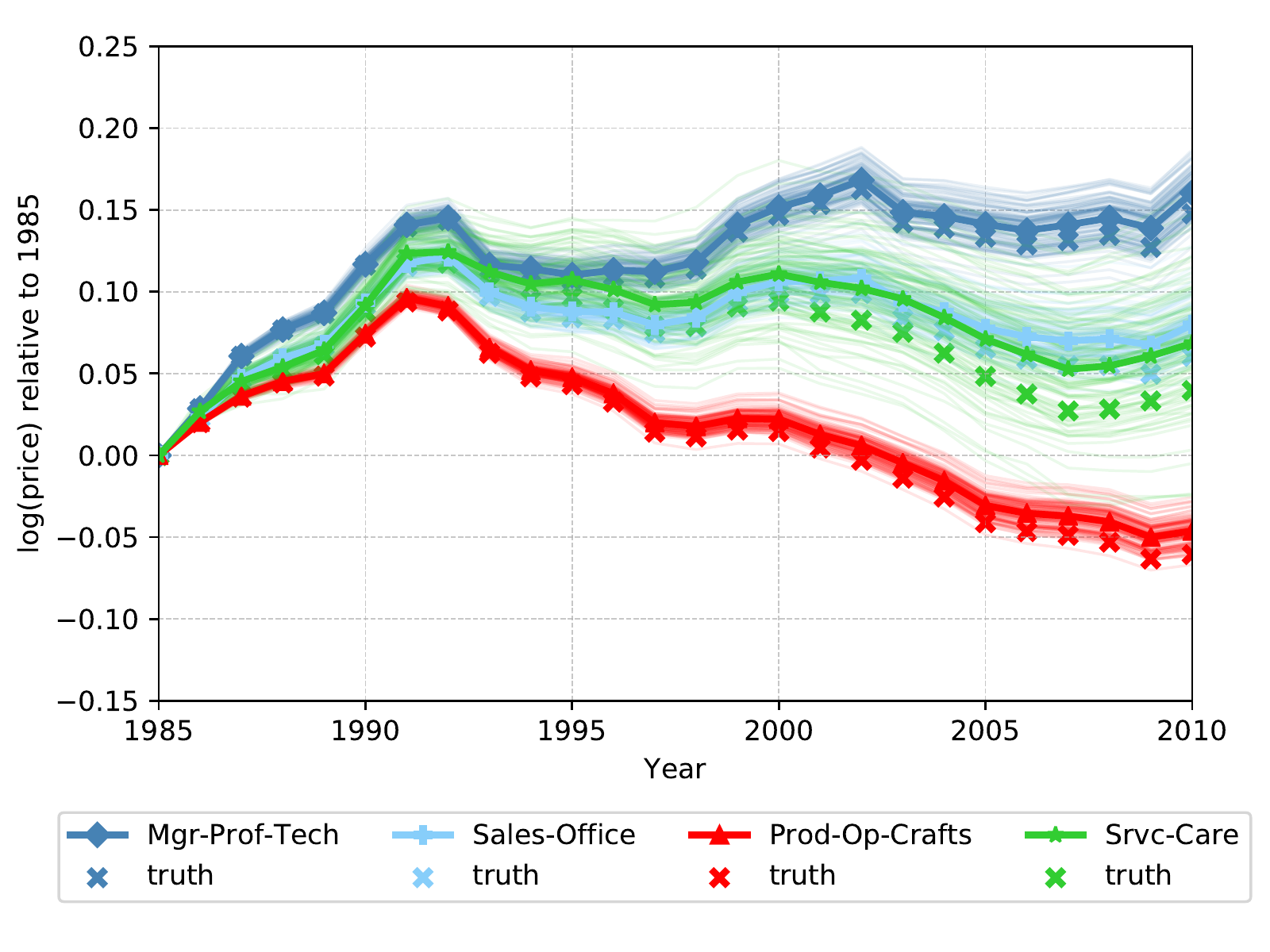}
  \end{minipage}
  \begin{minipage}[t]{\panelwidth}
    \subcaption{Skill acc., occ.-stint fixed effects}
    \label{mc:fig:estimation-cortes-stint-moderate-shocks-skills}
    \includegraphics[width=\textwidth]{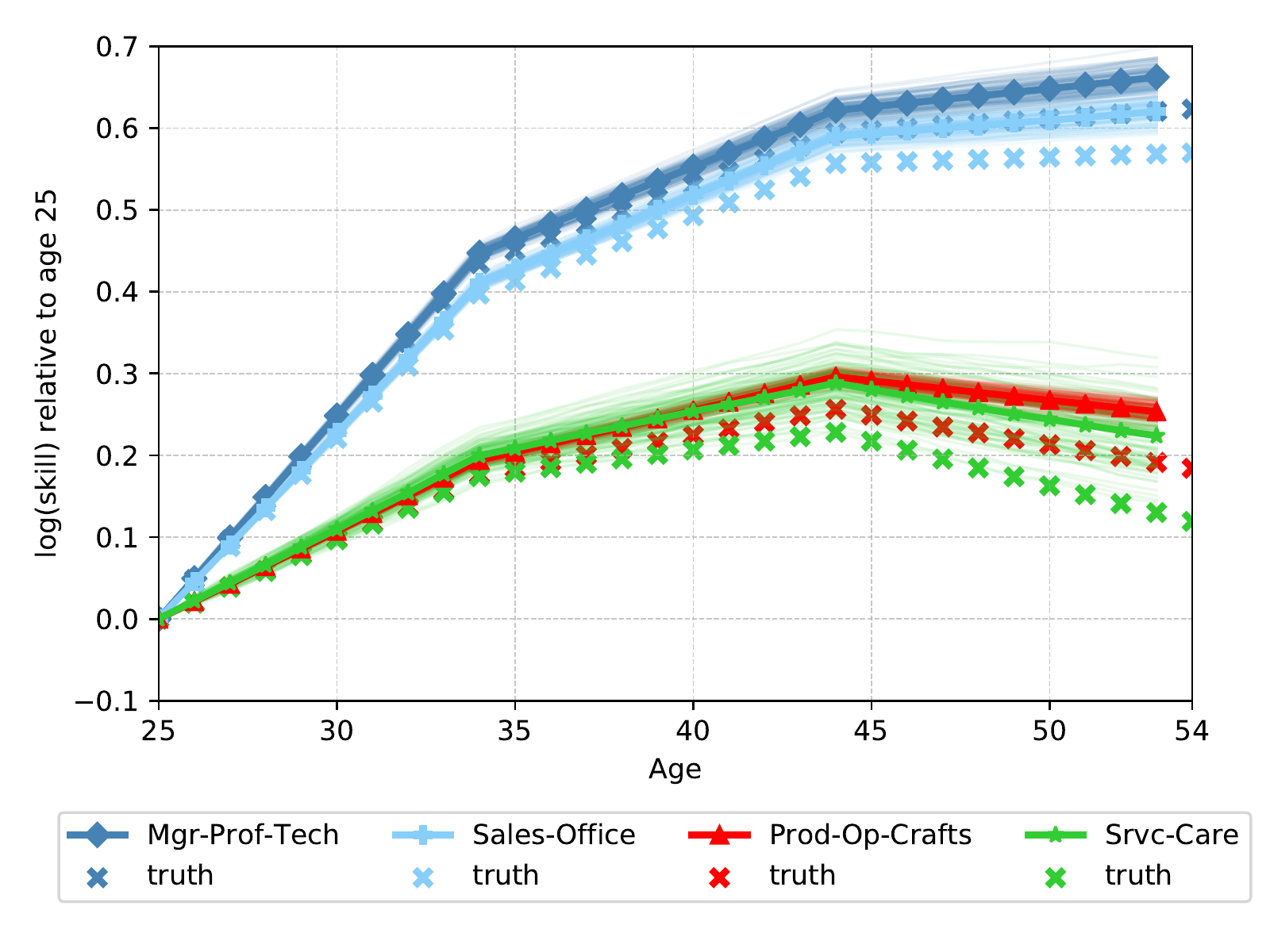}
  \end{minipage}
  \begin{minipage}{\textwidth}
    \scriptsize
    \emph{Notes:} \mcresults \stintfixedeffects
  \end{minipage}
\end{figure}

Finally, we compare our estimation method to an alternative approach proposed by \citet{C2016} who uses occupation-specific fixed effects to estimate changing skill prices. The top row of Figure~\ref{mc:fig:estimation-cortes-no-shocks} implements this approach without a base period. As discussed in Section~\ref{mc:sec:fixed-effects-model}, one age group within ${\Gamma}_{X(i, t-1), k}$ is omitted because of perfect multicollinearity. Therefore, all the estimated parameters have to be interpreted relative to that age group's skill accumulation. This is quite complicated since some skill price changes are loaded on the estimated skill accumulation parameters and vice versa. Indeed, the top row of Figure~\ref{mc:fig:estimation-cortes-moderate-shocks} shows that both $\Delta \hat{\pi}_{t,k}$ and $\hat{\Gamma}_{X(i, t-1), k}$ substantially deviate from the truth and in the respective opposite direction for each occupation.

The bottom row of Figure~\ref{mc:fig:estimation-cortes-no-shocks} implements individuals' occupation-specific fixed effects in the way that we recommend, i.e., with a base period, occupation-\emph{stint} specific fixed effects, and occupation-specific age profiles. This is very similar to the occupation-specific tenure profiles that \citet{C2016} uses in one of his key robustness checks (see again our discussion in Section~\ref{mc:sec:fixed-effects-model}) plus the base period. Supporting this specification, the skill prices and skill accumulation are perfectly identified if there are not any idiosyncratic skill shocks and therefore exogenous mobility holds.

\begin{figure}[ht!]
  \caption{Estimation results, highly dispersed shocks as in Table~\ref{mc:tab:vlarge-shocks}}
  \label{mc:fig:estimation-cortes-vlarge-shocks}
  \centering
  \begin{minipage}[t]{\panelwidth}
    \subcaption{Cum.\ prices, occ.-stint fixed effects}
    \label{mc:fig:estimation-cortes-stint-vlarge-shocks-prices}
    \includegraphics[width=\textwidth]{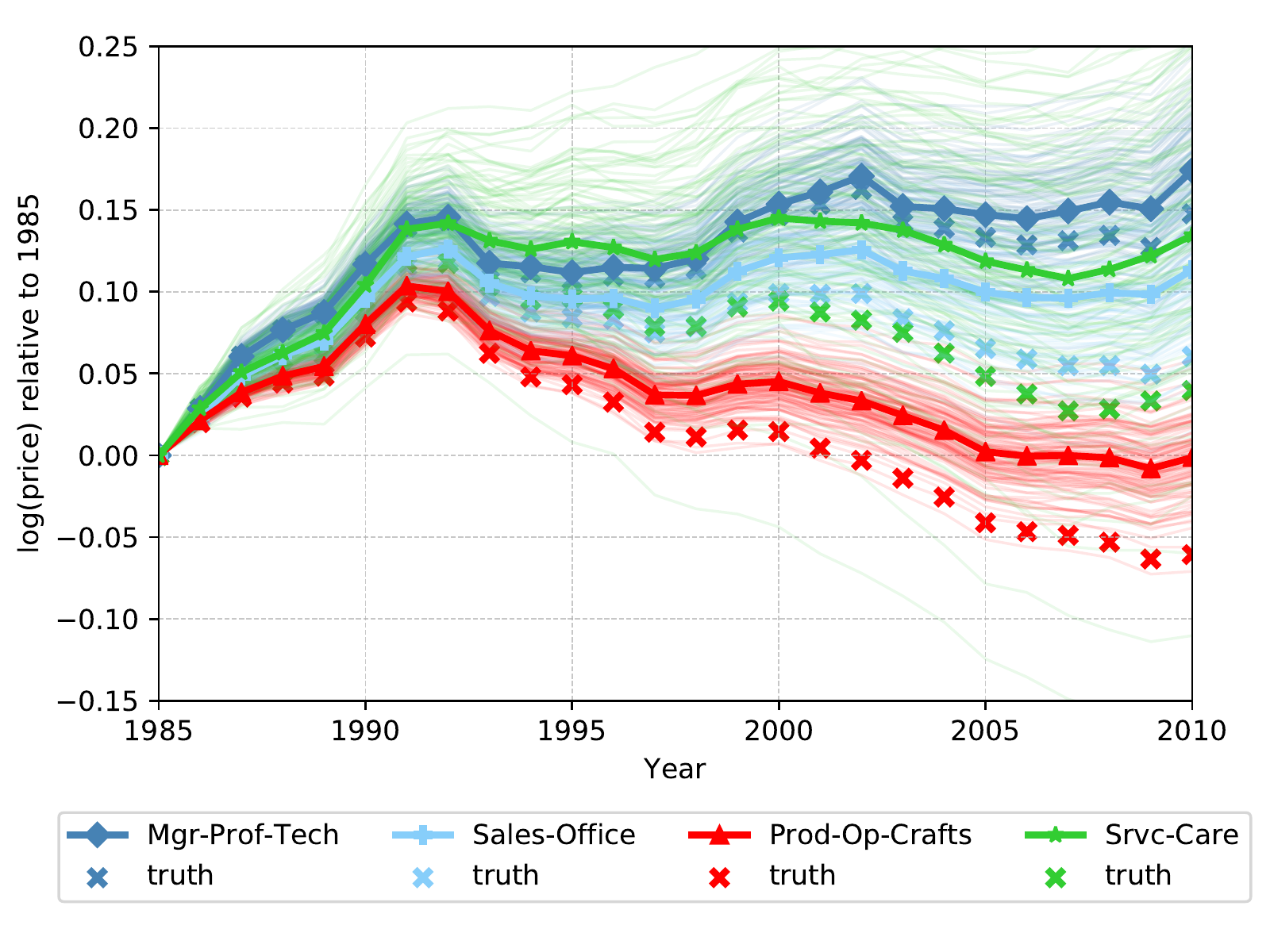}
  \end{minipage}
  \begin{minipage}[t]{\panelwidth}
    \subcaption{Skill acc., occ.-stint fixed effects}
    \label{mc:fig:estimation-cortes-stint-vlarge-shocks-skills}
    \includegraphics[width=\textwidth]{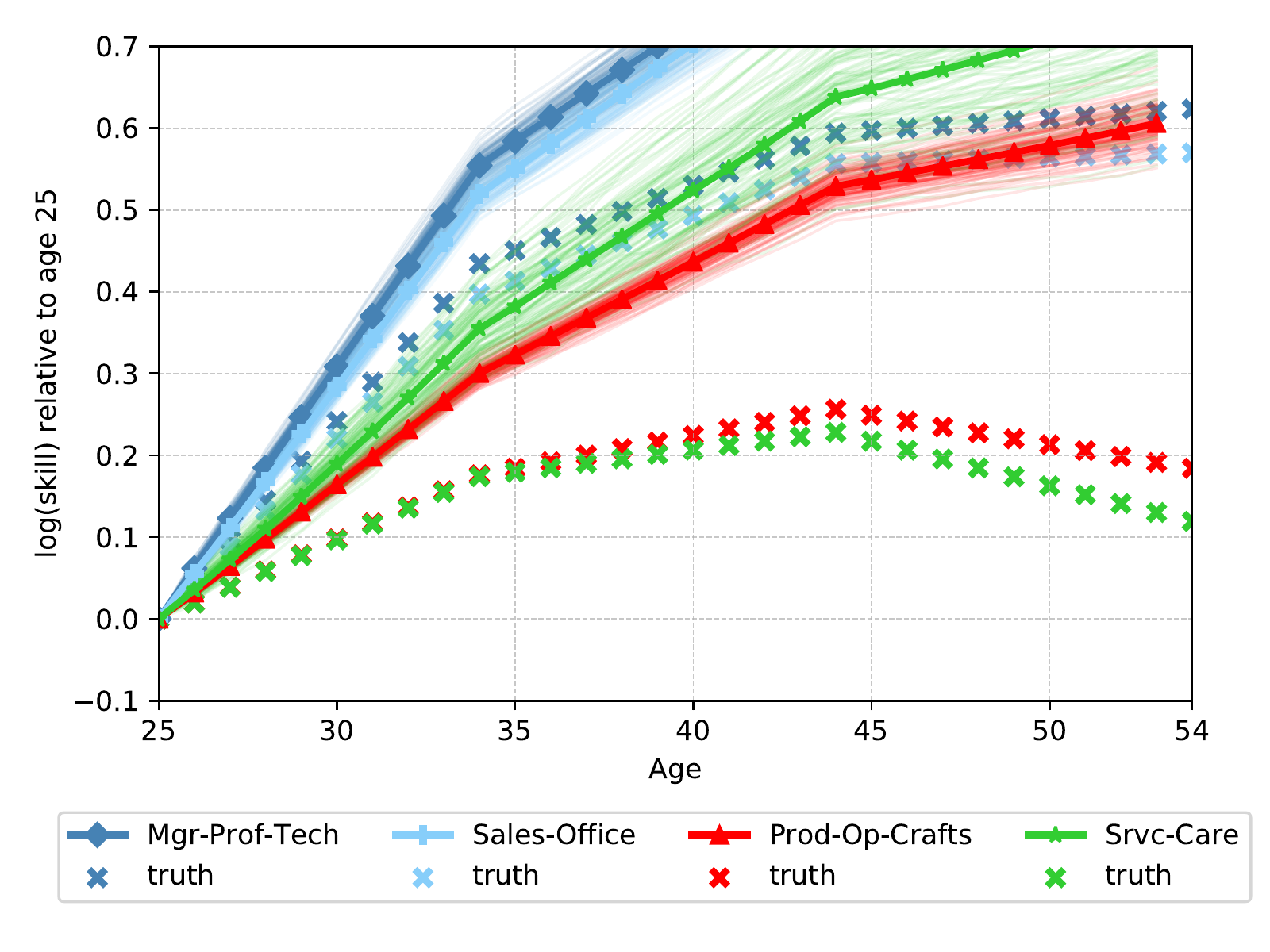}
  \end{minipage}
  \begin{minipage}{\textwidth}
    \scriptsize
    \emph{Notes:} \mcresults \stintfixedeffects
  \end{minipage}
\end{figure}

Next we add moderate skill shocks to the data generating process as detailed in Table~\ref{mc:tab:moderate-shocks} above. We run our recommended regression with occupation-stint specific fixed effects for this sample. Figure~\ref{mc:fig:estimation-cortes-moderate-shocks} depicts the results, showing that fixed effects approach still performs well. Finally, Figure~\ref{mc:fig:estimation-cortes-vlarge-shocks} shows the estimation results for the data generating process with large skill shocks from Table~\ref{mc:tab:vlarge-shocks}. The prices are now substantially off for three out of four occupations, and the skill accumulation estimates are far away from the truth. This was predicted by us in the main text and in Section~\ref{mc:sec:fixed-effects-model}, as now endogenous occupation switching and \emph{staying} becomes quantitatively important. It contrasts especially with the IV implementation of our approach in Figure~\ref{mc:fig:estimation-vlarge-shocks}, which comes very close to the true skill prices and reasonably close to skill accumulation even with large shocks.

\clearpage

\section{Different Empirical Specifications using the Data of \citet{BGS}}
\label{sec:appdx-iv-tables}

\subsection{Instrumental Variable Estimates}

This section repeats \citet{BGS}'s estimation of task prices in the SIAB data using the IV specification developed in Section~\ref{mc:sub:iv-estimation}. We perform this exercise for the four broad occupations only because of the extensive data requirements.

\begin{figure}[ht!]
  \caption{The evolution of skill prices and skill accumulation of stayers}
  \label{fig:appdx:skill-prices-skill-acc-stayers-iv}
  \centering
  \begin{minipage}[t]{\panelwidth}
    \subcaption{Skill prices}
    \label{fig:appdx:skill-prices-skill-acc-stayers-iv-a}
    \includegraphics[width = \textwidth]{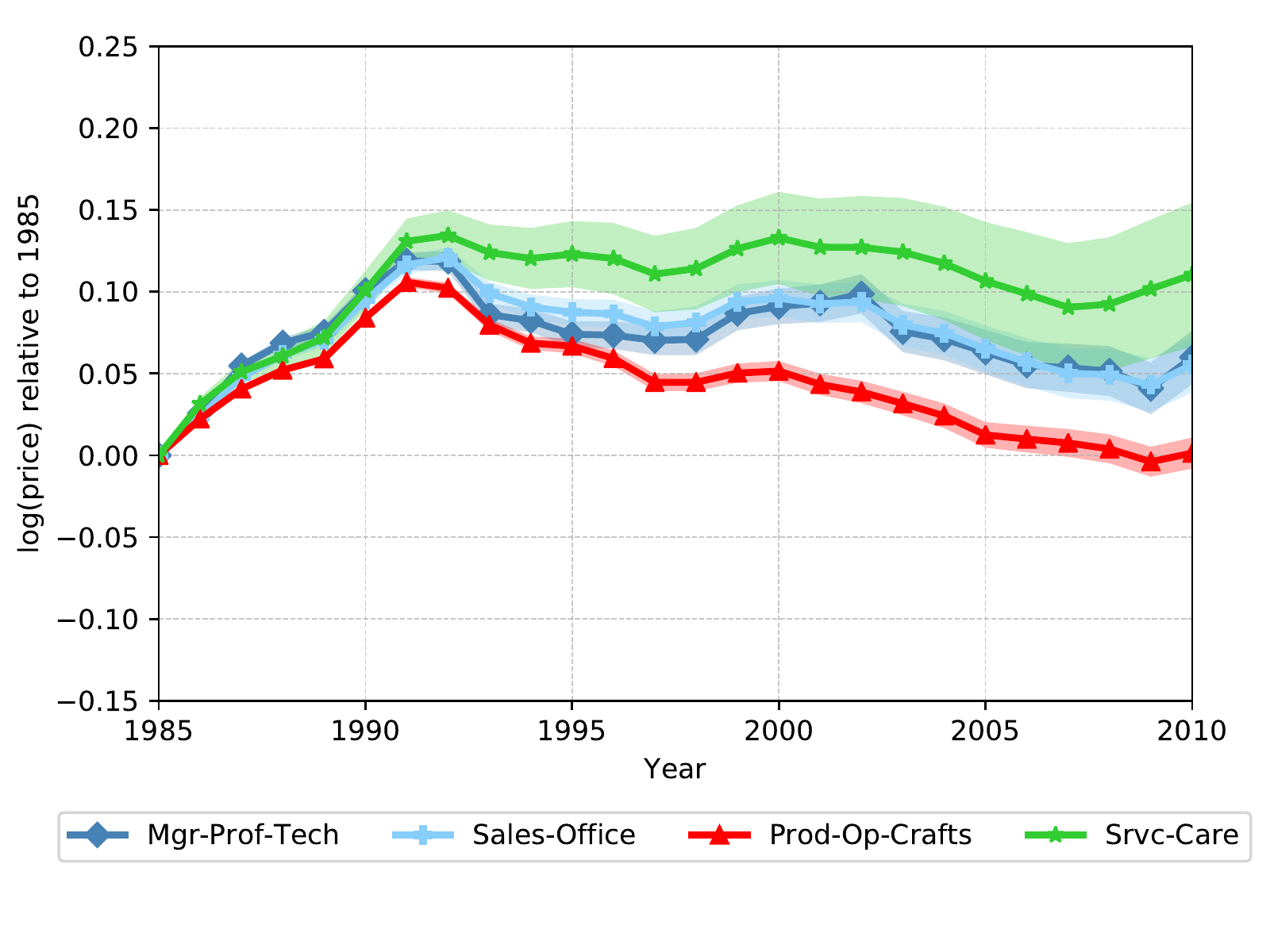}
  \end{minipage}
  \begin{minipage}[t]{\panelwidth}
    \subcaption{Stayers' skill accumulation}
    \label{fig:appdx:skill-prices-skill-acc-stayers-iv-b}
    \includegraphics[width = \textwidth]{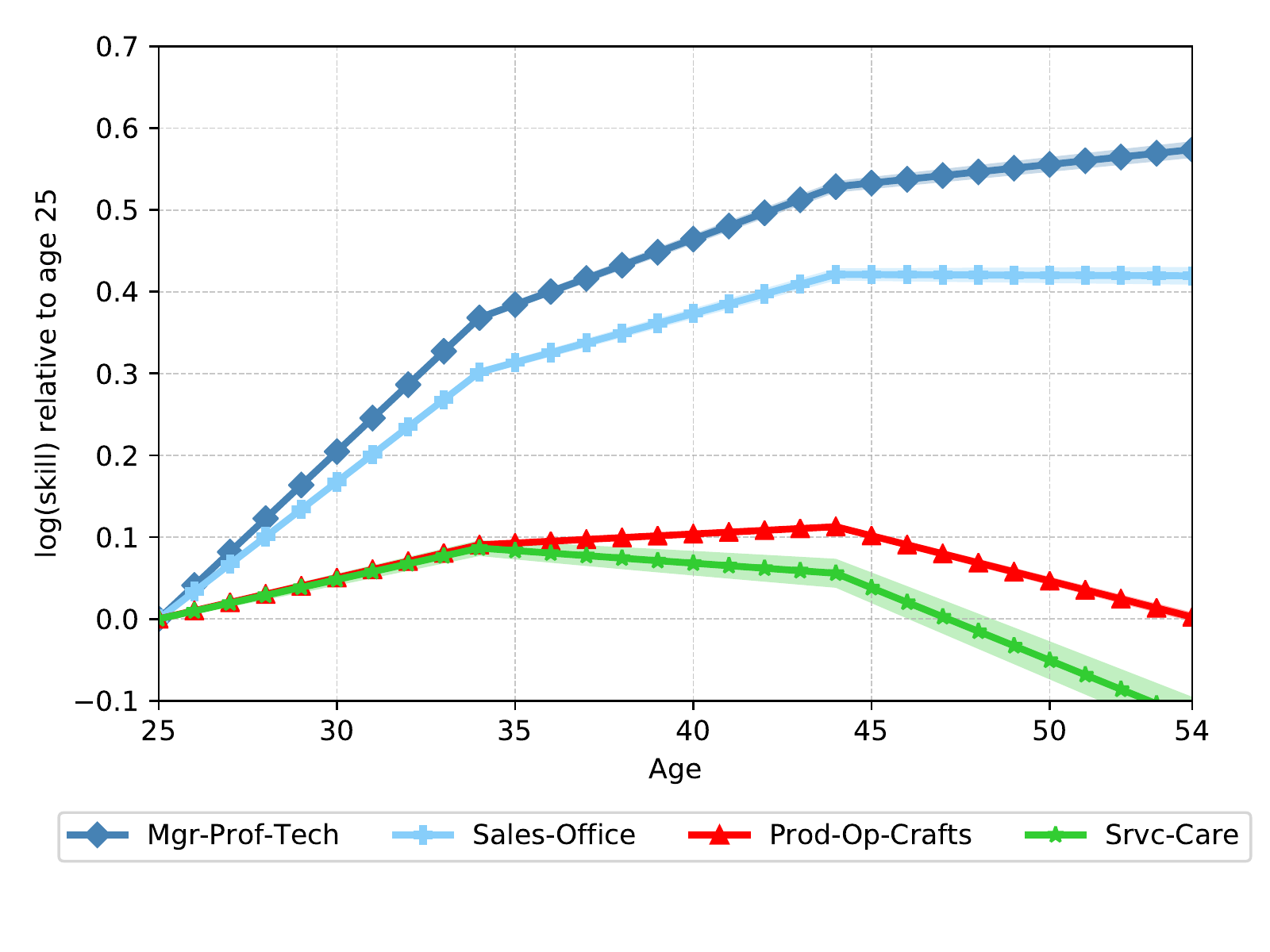}
  \end{minipage}
  \begin{minipage}{\textwidth}
    \scriptsize
    \emph{Notes:} Panel~\ref{fig:appdx:skill-prices-skill-acc-stayers-iv-a} shows changes in skill price IV estimates over time as detailed in Section~\ref{mc:sub:iv-estimation}. Panel~\ref{fig:appdx:skill-prices-skill-acc-stayers-iv-b} shows stayers' skill accumulation profiles estimated with IV. \backgroundnotemc \confidenceintervalls
  \end{minipage}
\end{figure}

In Figure~\ref{fig:appdx:skill-prices-skill-acc-stayers-iv} the broad patterns for skill price changes hold up in the instrumental variable estimates with slightly different numerical values. The skill change estimate, which now should be closer to the structural (ex ante) parameters, for the four occupation groups in Table~\ref{tab:appdx-iv-tables:switchers-gammas-iv} are slightly lower, but none of the broad patterns change.

\clearpage

\begin{table}[h!]
  \centering
  \caption{Estimated skill accumulation coefficients (occupation groups, IV)}
  \label{tab:appdx-iv-tables:switchers-gammas-iv}
  \begin{minipage}[t]{\linewidth}
    \centering
    \begin{tabular}{lllrrr}
\toprule & & \multicolumn{4}{c}{Age group} \\ \cmidrule{4-6}
           Previous sector &           Current sector &                   &  [25, 34] &  [35, 44] &  [45, 54] \\
\midrule
Mgr-Prof-Tech  & Mgr-Prof-Tech & $\gamma$ &     0.041 &     0.016 &     0.005 \\
           &           & $\sigma_{\gamma}$ &     0.001 &     0.000 &     0.001 \\
           & Sales-Office & $\gamma$ &     0.216 &     0.194 &     0.227 \\
           &           & $\sigma_{\gamma}$ &     0.058 &     0.055 &     0.081 \\
           & Prod-Op-Crafts & $\gamma$ &    -0.141 &     0.056 &    -0.116 \\
           &           & $\sigma_{\gamma}$ &     0.045 &     0.045 &     0.065 \\
           & Srvc-Care & $\gamma$ &    -0.400 &    -0.062 &    -0.051 \\
           &           & $\sigma_{\gamma}$ &     0.179 &     0.138 &     0.229 \\ \midrule
Sales-Office  & Mgr-Prof-Tech & $\gamma$ &     0.397 &     0.247 &     0.156 \\
           &           & $\sigma_{\gamma}$ &     0.055 &     0.040 &     0.061 \\
           & Sales-Office & $\gamma$ &     0.034 &     0.012 &    -0.000 \\
           &           & $\sigma_{\gamma}$ &     0.001 &     0.001 &     0.001 \\
           & Prod-Op-Crafts & $\gamma$ &     0.074 &     0.101 &    -0.119 \\
           &           & $\sigma_{\gamma}$ &     0.036 &     0.037 &     0.075 \\
           & Srvc-Care & $\gamma$ &    -0.061 &     0.067 &    -0.284 \\
           &           & $\sigma_{\gamma}$ &     0.168 &     0.185 &     0.291 \\ \midrule
Prod-Op-Crafts  & Mgr-Prof-Tech & $\gamma$ &     0.371 &     0.327 &     0.117 \\
           &           & $\sigma_{\gamma}$ &     0.054 &     0.055 &     0.092 \\
           & Sales-Office & $\gamma$ &     0.343 &     0.337 &     0.361 \\
           &           & $\sigma_{\gamma}$ &     0.071 &     0.070 &     0.109 \\
           & Prod-Op-Crafts & $\gamma$ &     0.010 &     0.002 &    -0.011 \\
           &           & $\sigma_{\gamma}$ &     0.000 &     0.000 &     0.000 \\
           & Srvc-Care & $\gamma$ &     0.372 &     0.193 &    -0.114 \\
           &           & $\sigma_{\gamma}$ &     0.095 &     0.107 &     0.176 \\ \midrule
Srvc-Care  & Mgr-Prof-Tech & $\gamma$ &     0.390 &     0.222 &     0.327 \\
           &           & $\sigma_{\gamma}$ &     0.192 &     0.129 &     0.109 \\
           & Sales-Office & $\gamma$ &     0.498 &     0.619 &     0.419 \\
           &           & $\sigma_{\gamma}$ &     0.185 &     0.240 &     0.243 \\
           & Prod-Op-Crafts & $\gamma$ &     0.198 &     0.346 &     0.216 \\
           &           & $\sigma_{\gamma}$ &     0.050 &     0.051 &     0.077 \\
           & Srvc-Care & $\gamma$ &     0.010 &    -0.003 &    -0.018 \\
           &           & $\sigma_{\gamma}$ &     0.002 &     0.001 &     0.001 \\
\bottomrule
\end{tabular}

  \end{minipage}
  \begin{minipage}{\textwidth}
    \scriptsize
    \emph{Notes:} \crossgammas \broadgroupsmc \ivresults
  \end{minipage}
\end{table}

\clearpage
\subsection{Non-pecuniary benefits}

The next estimation repeats \citet{BGS} but allows for changing relative average occupation-specific amenities or future values by age group over time as described in Section~\ref{mc:sub:nonpecuniary}. In particular, we implement Equation~\eqref{mc:eq:estimation-utility}, adding the change in choices between two periods as a regressor (interacted with age) to the estimation. We again perform this exercise for the four broad occupations only because of the extensive data requirements.

\begin{figure}[ht!]
	\caption{Accounting for non-pecuniary benefits}
	\label{appdx:fig:price-estimates-method-amenities}
	\centering
	\begin{minipage}[t]{\panelwidth}
		\subcaption{Prices}
		\label{appdx:fig:price-estimates-method-amenities-a}
		\includegraphics[width = \linewidth]{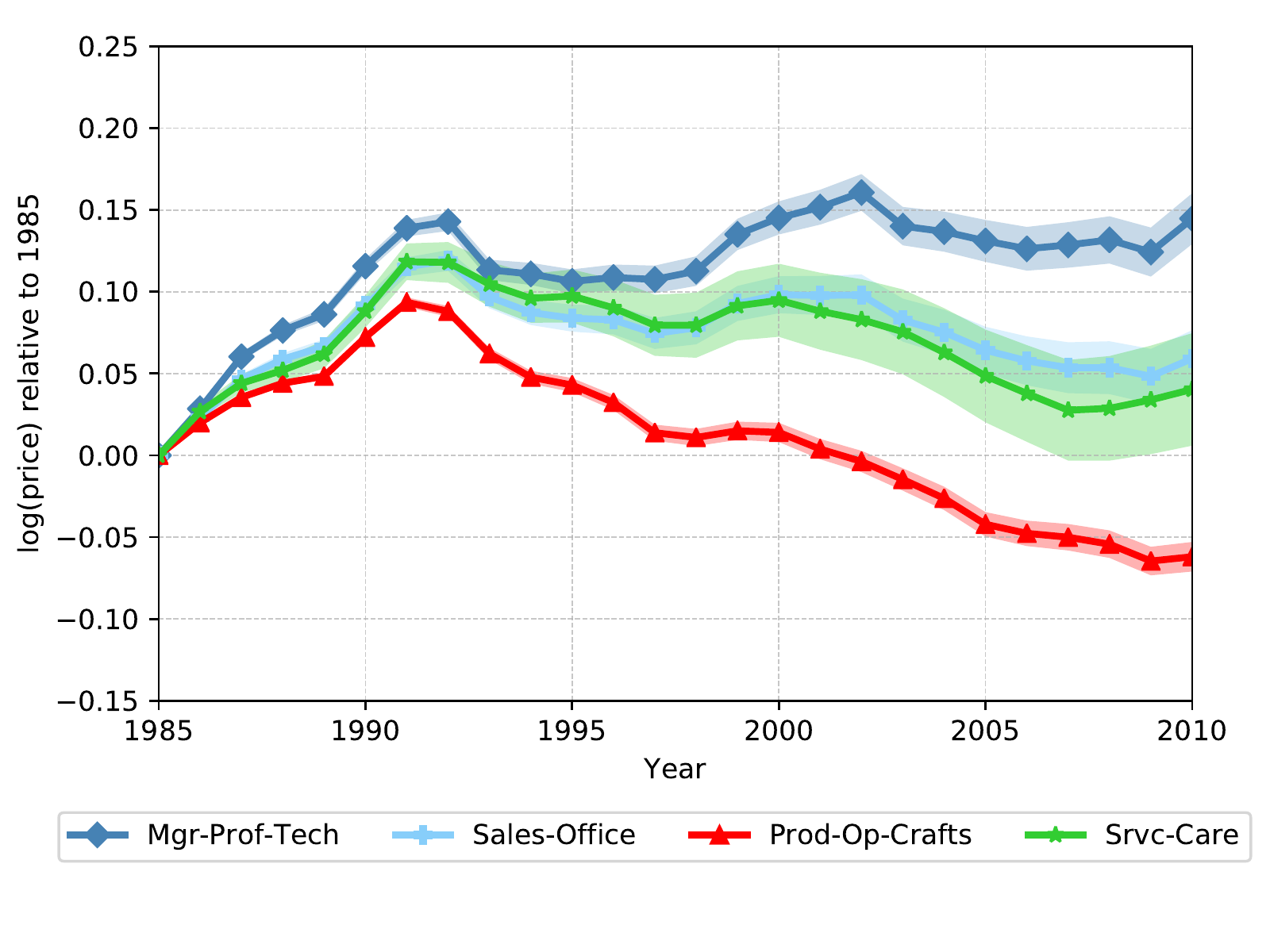}
	\end{minipage}
	\begin{minipage}[t]{\panelwidth}
		\subcaption{Skills}
		\label{appdx:fig:price-estimates-method-amenities-b}
		\includegraphics[width = \linewidth]{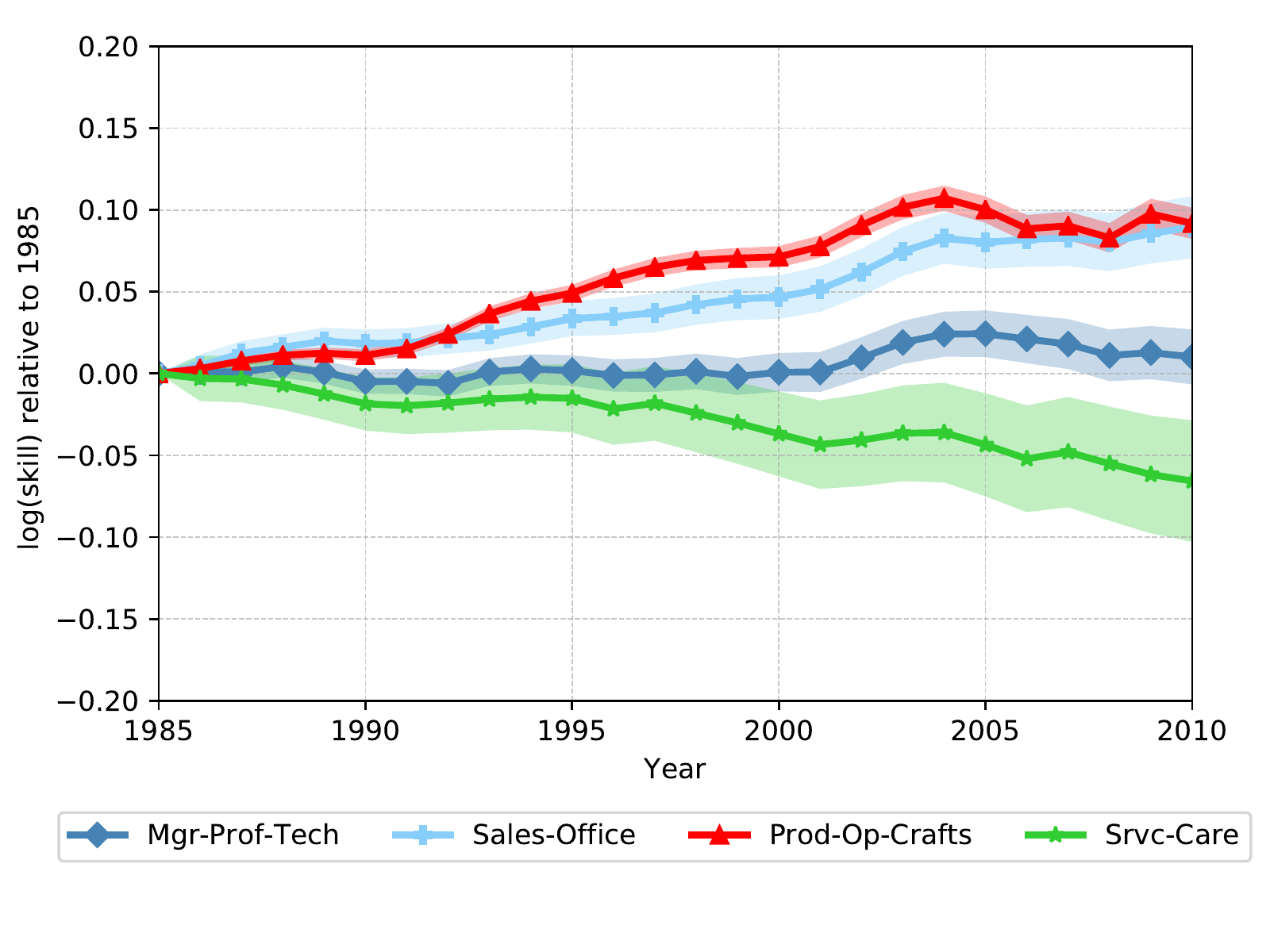}
	\end{minipage}
	\begin{minipage}[t]{\panelwidth}
		\subcaption{Skill accumulation}
		\label{appdx:fig:price-estimates-method-amenities-c}
		\includegraphics[width = \linewidth]{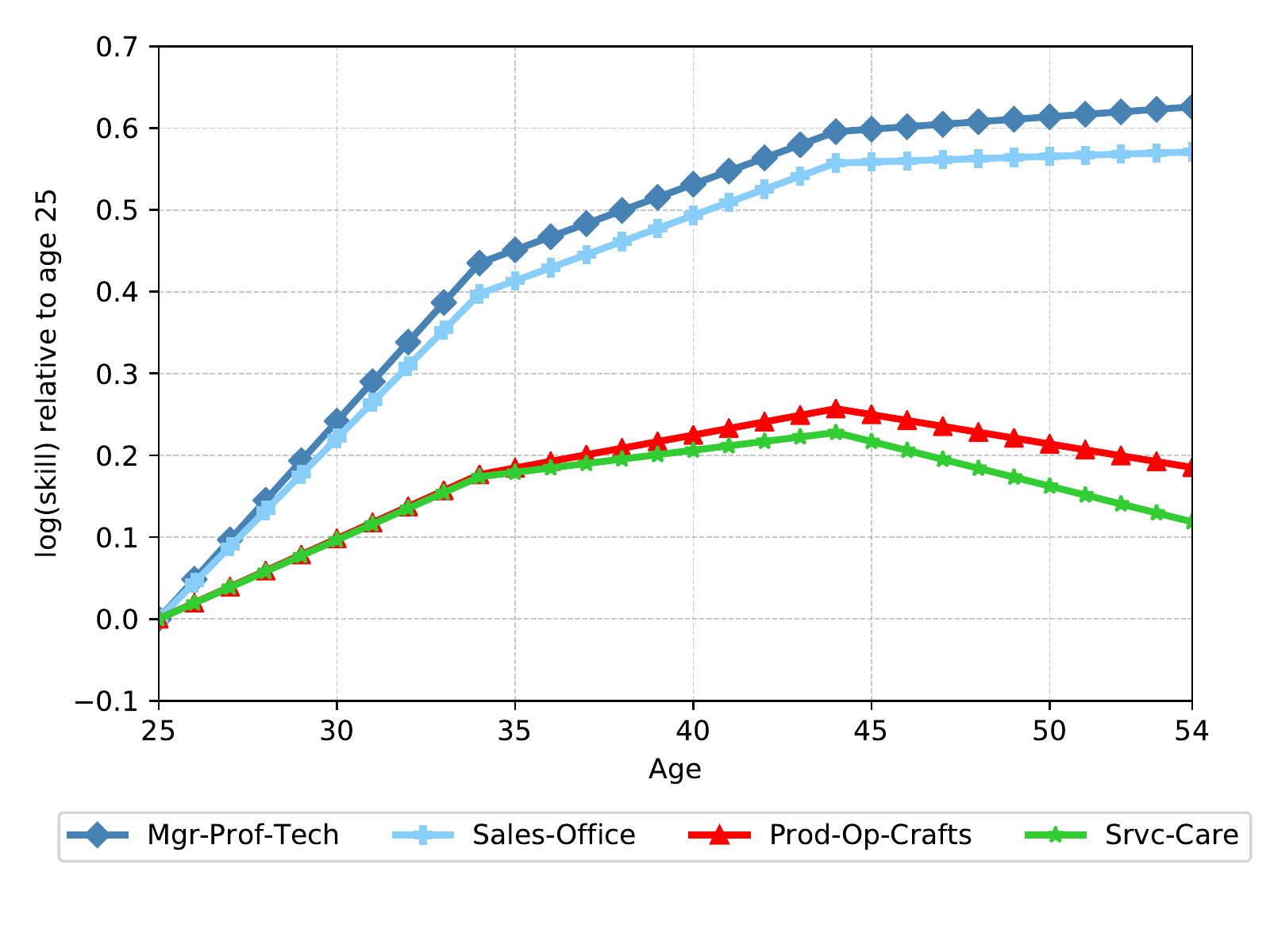}
	\end{minipage}
	\begin{minipage}[t]{\panelwidth}
		\subcaption{Amenity estimates, 25--34 year olds}
		\label{appdx:fig:price-estimates-method-amenities-d}
		\includegraphics[width = \linewidth]{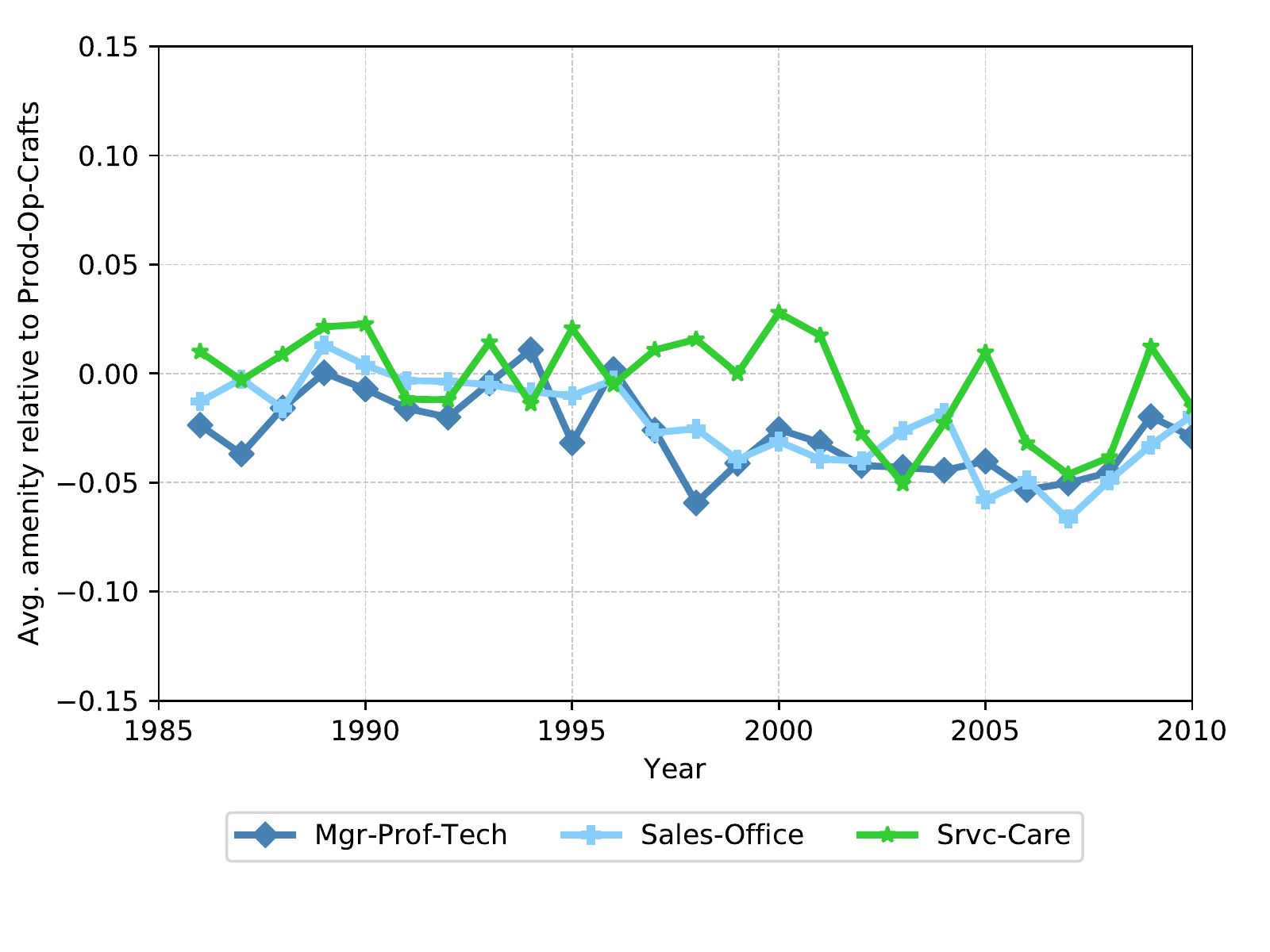}
	\end{minipage}
	\begin{minipage}[t]{\panelwidth}
		\subcaption{Amenity estimates, 35--44 year olds}
		\label{appdx:fig:price-estimates-method-amenities-e}
		\includegraphics[width = \linewidth]{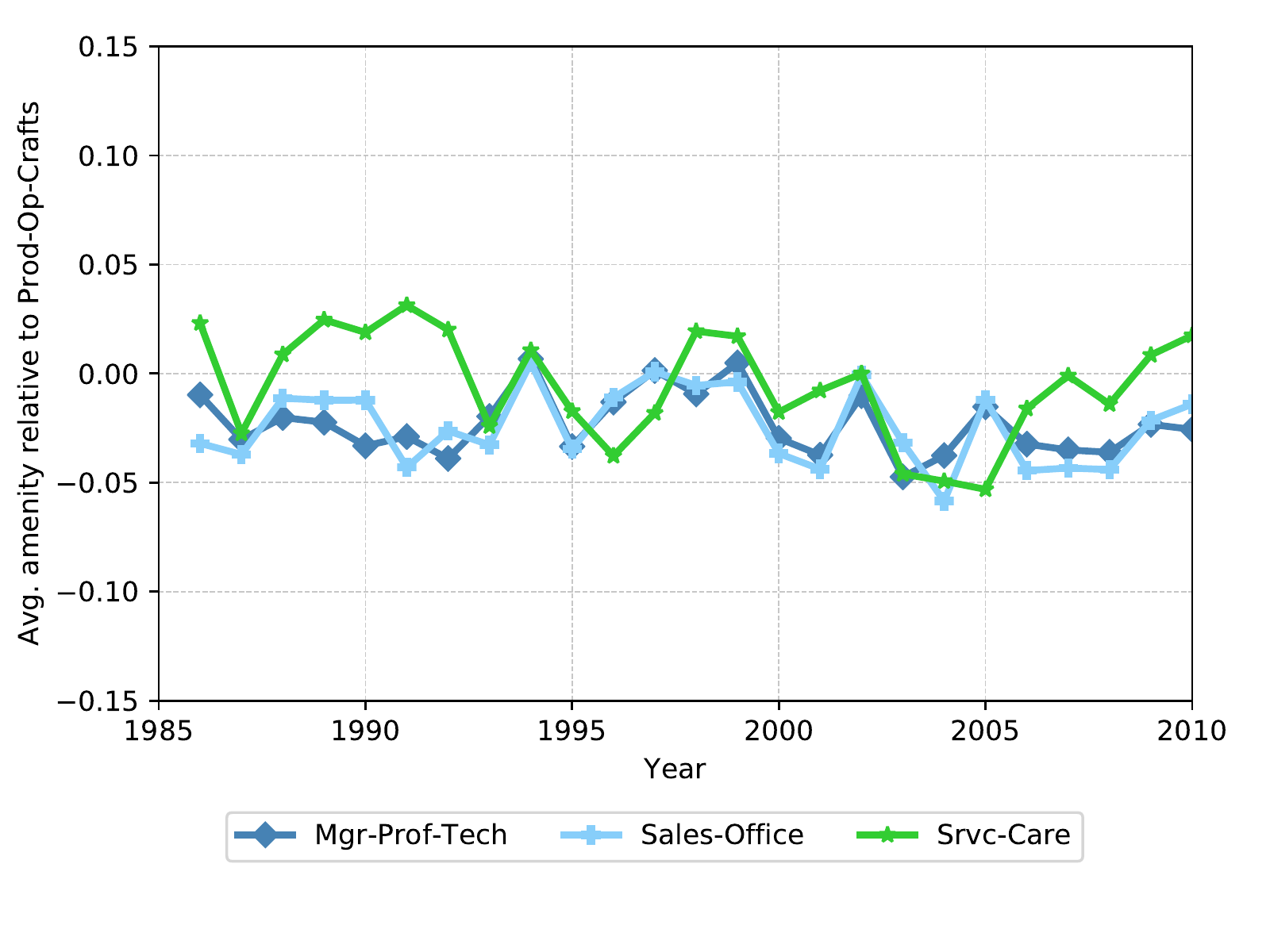}
	\end{minipage}
	\begin{minipage}[t]{\panelwidth}
		\subcaption{Amenity estimates, 45--54 year olds}
		\label{appdx:fig:price-estimates-method-amenities-f}
		\includegraphics[width = \linewidth]{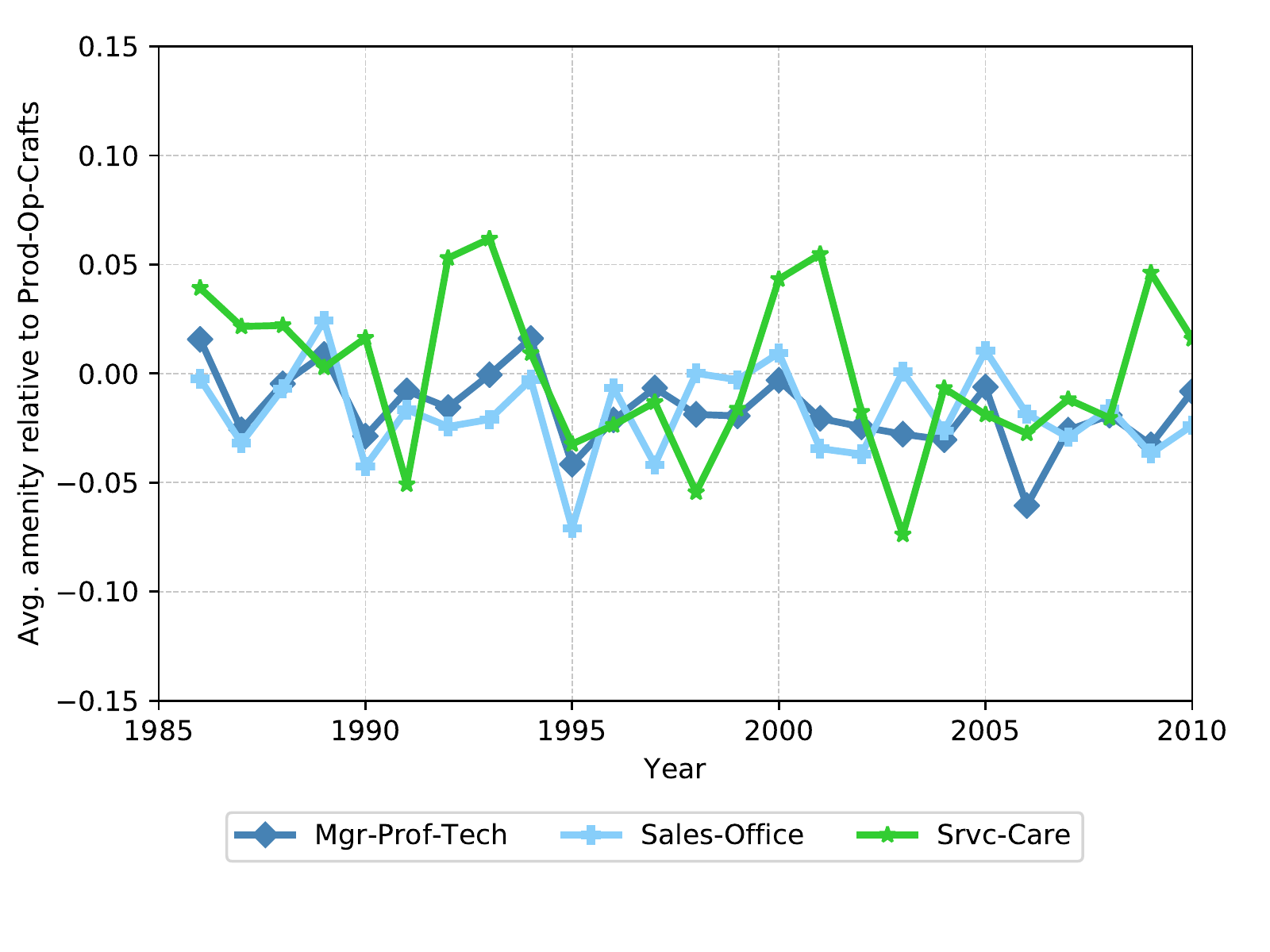}
	\end{minipage}
	\begin{minipage}{\textwidth}
		\scriptsize
		\emph{Notes:} Panels~\ref{appdx:fig:price-estimates-method-amenities-d} to \ref{appdx:fig:price-estimates-method-amenities-f} present the results the three age groups contained in the main sample. \amenityresults
	\end{minipage}
\end{figure}

Figure~\ref{appdx:fig:price-estimates-method-amenities} plots the resulting skill prices and skill accumulation coefficients of the four broad occupation groups, showing that they are hardly affected by this augmented estimation model compared to \citet[][Figure~5]{BGS}. In addition, we can also identify the changes of amenities themselves in this specification. Panels~\ref{appdx:fig:price-estimates-method-amenities-d} to \ref{appdx:fig:price-estimates-method-amenities-f} show the amenities relative to the omitted Prod-Op-Crafts and the base period by age group. We see that for 45--55 year olds these are about zero and pretty stable over time. For young 25--34 olds we do see declining amenities (i.e., rising estimation coefficient on $\Delta{I}_{k(i,t)}$), first for Mgr-Prof-Tech and Sales-Office after the mid-1990s and then for Srvc-Care in the early 2000s. The middle-aged workers are somewhere in between old and young ages with possibly a slight decline of amenities in the other three occupation groups compared to Prod-Op-Crafts toward the end of the sample period.

\clearpage
\subsection{Fixed Effects Estimation}

\begin{figure}[ht!]
	\caption{Stint fixed effects estimation}
	\label{appdx:fig:price-estimates-method-stint-fixed-effects}
	\centering
	\begin{minipage}[t]{\panelwidth}
		\subcaption{Wages}
		\label{appdx:fig:price-estimates-method-stint-fixed-effects-a}
		\includegraphics[width = \linewidth]{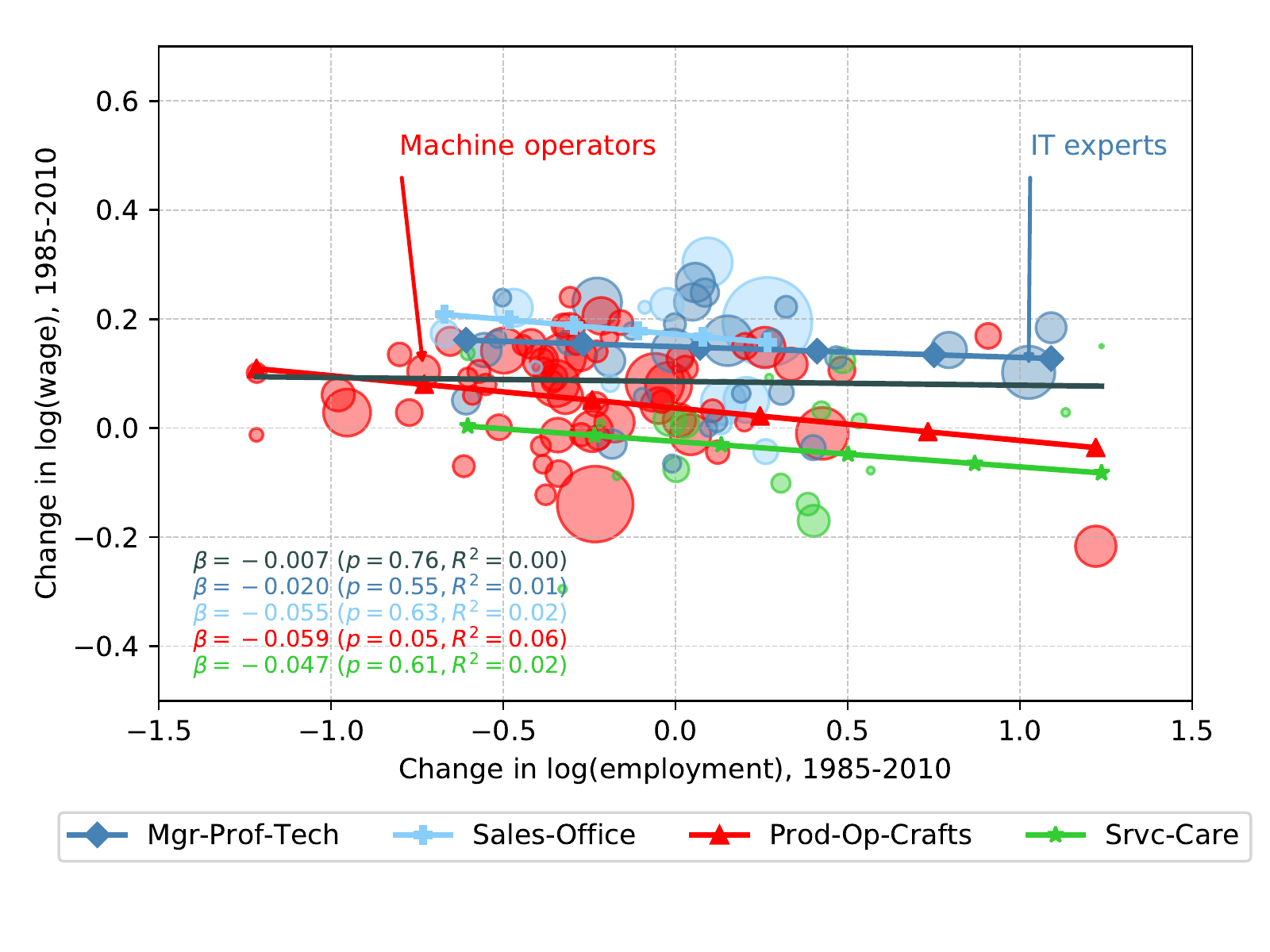}
	\end{minipage}
	\begin{minipage}[t]{\panelwidth}
		\subcaption{Prices}
		\label{appdx:fig:price-estimates-method-stint-fixed-effects-b}
		\includegraphics[width = \linewidth]{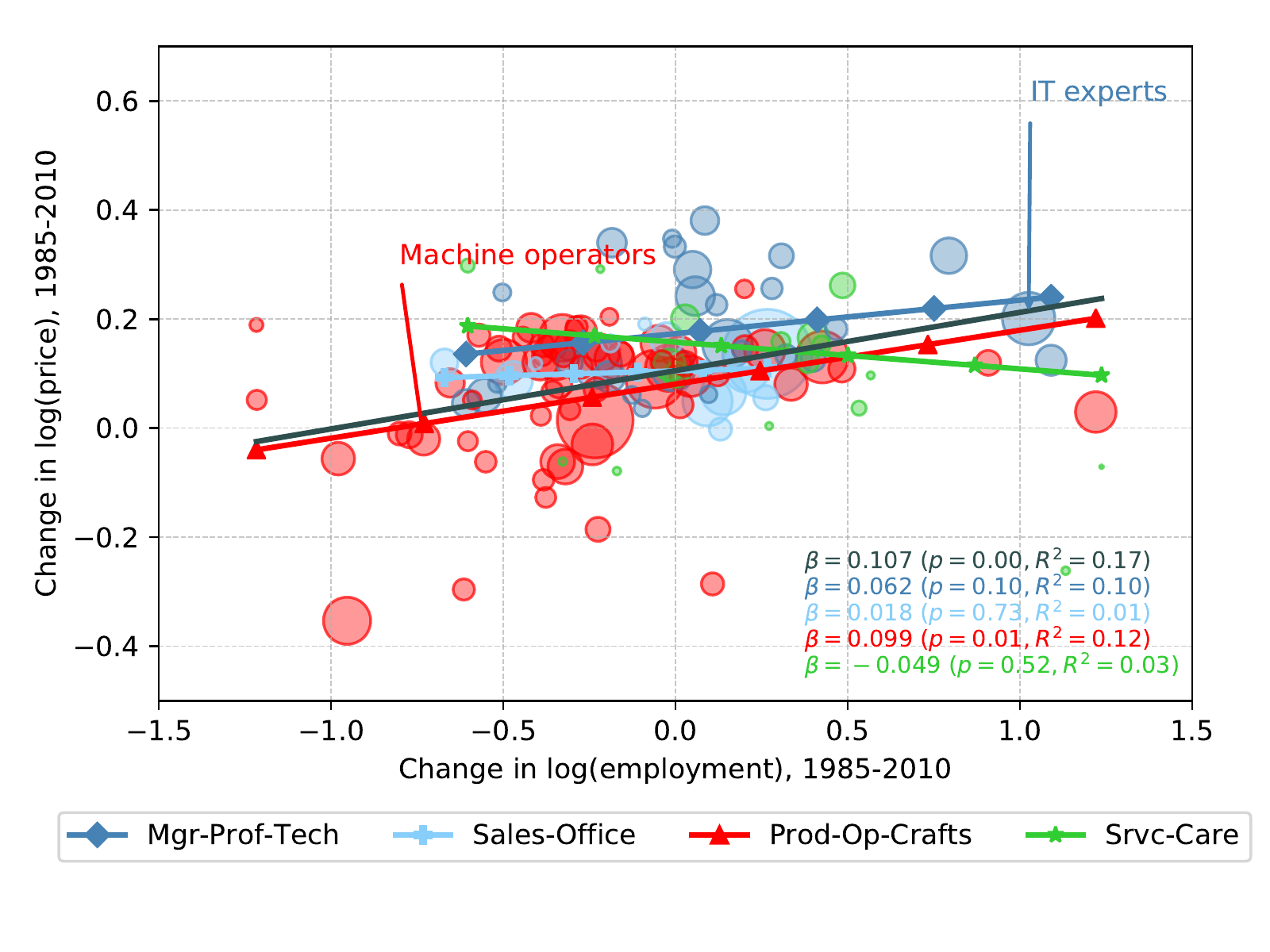}
	\end{minipage}
	\begin{minipage}[t]{\panelwidth}
		\subcaption{Skills}
		\label{appdx:fig:price-estimates-method-stint-fixed-effects-c}
		\includegraphics[width = \linewidth]{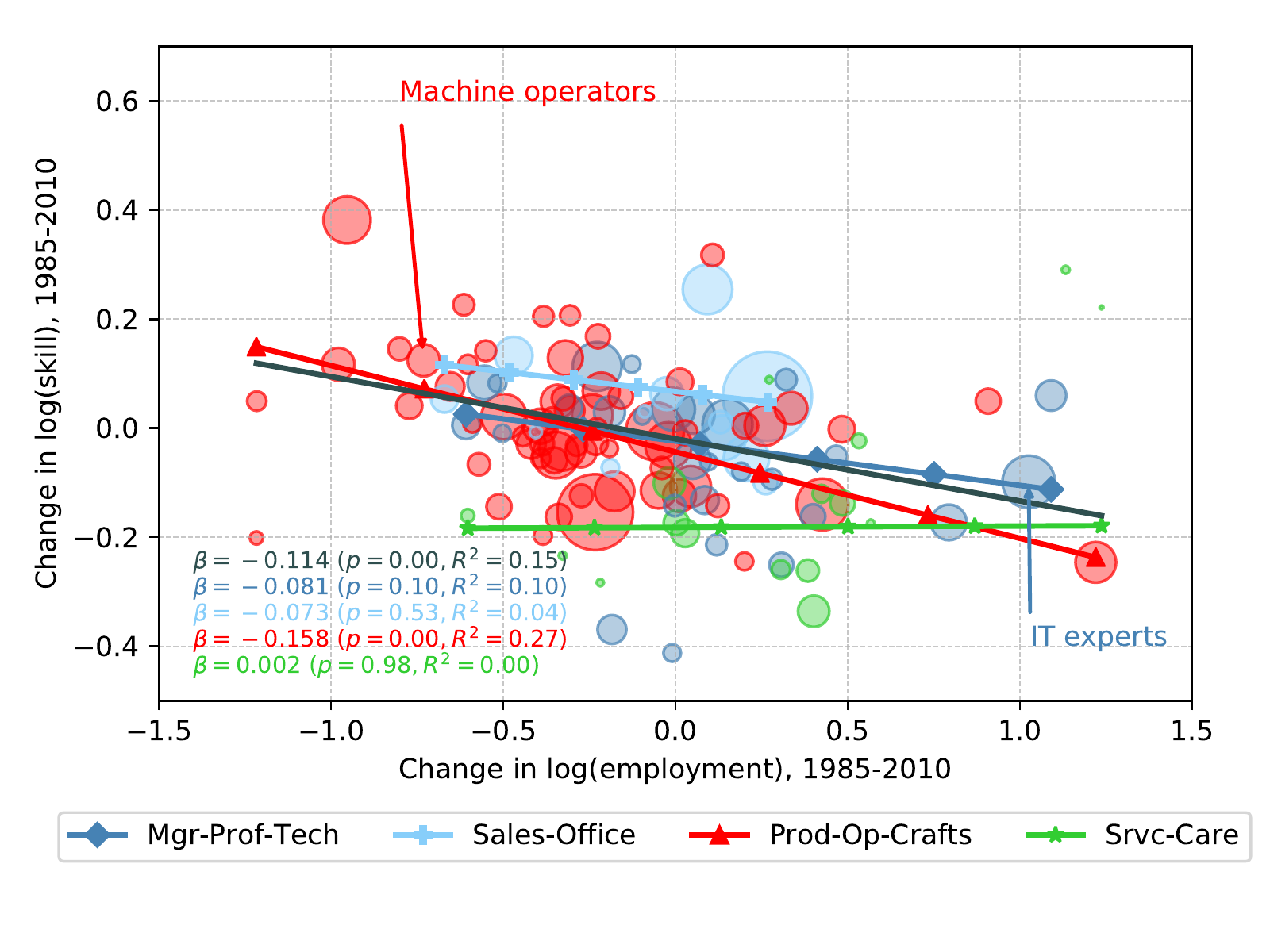}
	\end{minipage}
	\begin{minipage}[t]{\panelwidth}
		\subcaption{Growth-selection}
		\label{appdx:fig:price-estimates-method-stint-fixed-effects-d}
		\includegraphics[width = \linewidth]{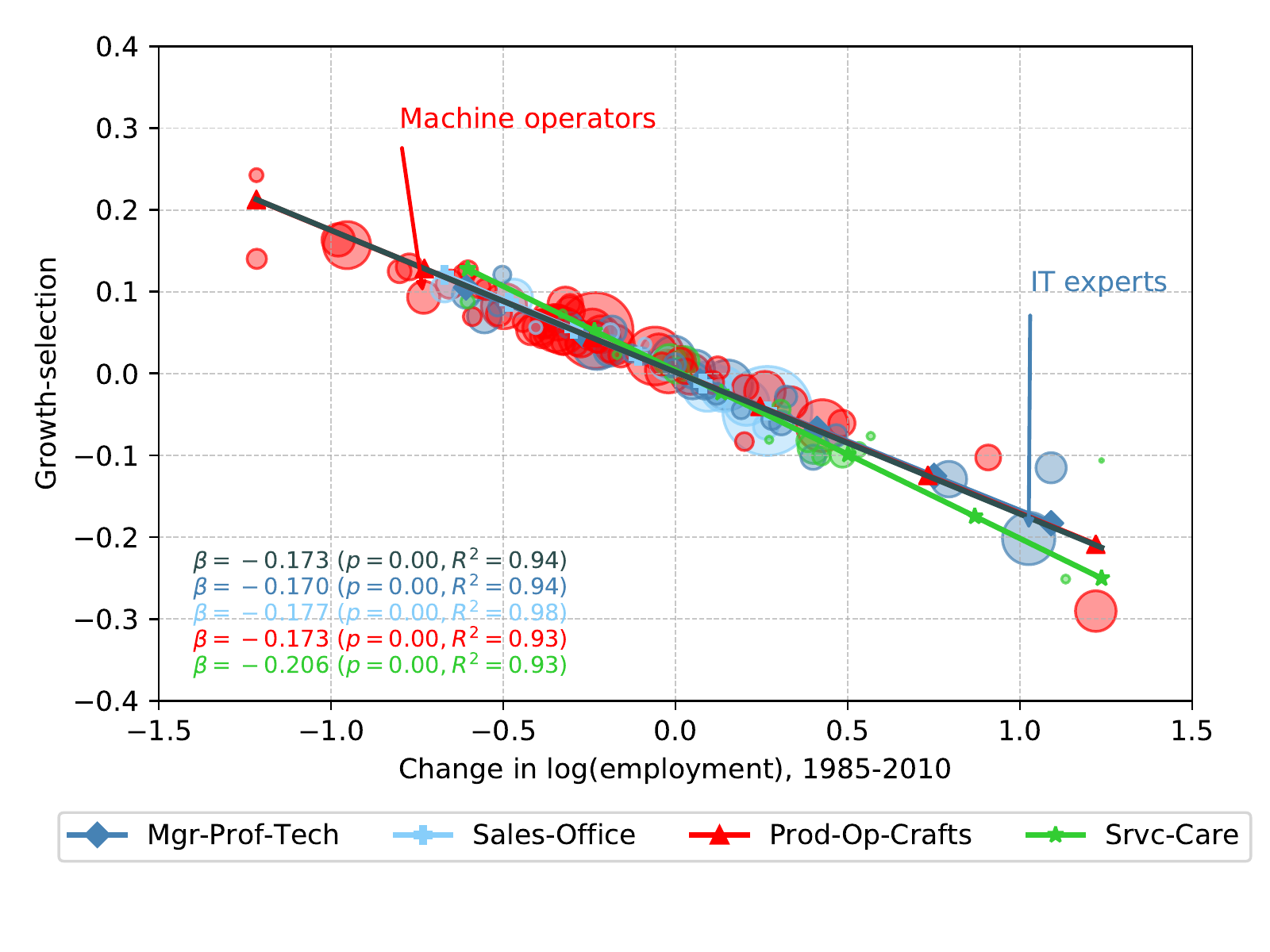}
	\end{minipage}
	\begin{minipage}{\textwidth}
		\scriptsize
		\emph{Notes:} \stintfixedeffects \employmenthorizontalnotemany \bubblenotemc
	\end{minipage}
\end{figure}

Finally, in Figure~\ref{appdx:fig:price-estimates-method-stint-fixed-effects} we also compare results to the alternative estimation method using fixed effects due to \citet{C2016}. As discussed in Section~\ref{mc:sec:fixed-effects-model}, in order not to control for worker's entire labor market history, we implement it with individual fixed effects for each occupation stint \citep[as do][]{CE2020}. That is, we estimate Equation~\eqref{mc:eq:wageFEspecific-stint} in the SIAB data.

With large idiosyncratic skill shocks, there can be some bias in the fixed effects approach, which was illustrated in the Monte Carlo simulations in Section~\ref{mc:subsub:cortes}. Nonetheless, it seems supportive of \citet{BGS}'s empirical results that this alternative estimation method yields qualitatively similar findings in Figure~\ref{appdx:fig:price-estimates-method-stint-fixed-effects}.

\section{Conclusion}
\label{mc:sec:conclusion}
This paper has studied the performance of recent approaches for estimating changes of skill prices in longitudinal data. Overall, the results are very encouraging, as the explicit choice-based approach by \citet{BGS} identifies the paramaters of interest well under various assumptions about the data generating process. Fixed effects-based methods \citep[e.g., employed by][]{McLaughlin2001,C2016,CE2020} also perform well.

\clearpage
\setstretch{1}
\bibliographystyle{ecta}
\bibliography{ms}

\end{document}